\documentclass{elsart}
\usepackage{graphicx}
\usepackage{amssymb}
\usepackage{url,makeidx}
\begin{document}

\newcommand{\beq}{\begin{equation}}
\newcommand{\eeq}{\end{equation}}
\newcommand{\beqa}{\begin{eqnarray}}
\newcommand{\eeqa}{\end{eqnarray}}
\newcommand{\bubble}[1]{\textcircled{\scriptsize {\sf #1}}}
\newcommand{\capsub}[2]{{#1}_{\scriptscriptstyle {\rm #2}}}
\newcommand{\capsup}[2]{{#1}^{\,\scriptscriptstyle {\rm #2}}}
\newcommand{\smallsub}[2]{{#1}_{\mbox{\scriptsize {#2}}}}
\newcommand{\smsubsub}[3]{ {#1}_{\mbox{\scriptsize {#2}},{#3}} }
\newcommand{\subsmsub}[3]{ {#1}_{{#2},\mbox{\scriptsize {#3}}} }
\newcommand{\supsubsmsub}[4]{ {#1}^{#2}_{{#3},\mbox{\scriptsize {#4}}} }
\newcommand{\smallfrac}[2]{\textstyle{\frac {#1}{#2}}}
\newcommand{\bigfrac}[2]{\displaystyle\frac{#1}{#2}}
\newcommand{\bigint}[2]{\displaystyle\int\limits_{#1}^{\;\;\;#2}}
\newcommand{\medfrac}[2]{\textstyle\frac{#1}{#2}}
\newcommand{\darkmath}[1]{\boldmath${#1}$\unboldmath}

\newcommand{\rlum}{\smallsub{\rho}{lum}}
\newcommand{\xo}{\smallsub{x}{0}}

\newcommand{\Qbulb}{\smallsub{Q}{bulb}}
\newcommand{\Qgal}{Q_g}
\newcommand{\Jgal}{J_g}
\newcommand{\ngal}{n_g}
\newcommand{\too}{t_0}
\newcommand{\Ro}{R_{\,0}}
\newcommand{\no}{n_0}
\newcommand{\Rtil}{\tilde{R}}
\newcommand{\Ho}{H_0}
\newcommand{\ho}{h_0}
\newcommand{\Lo}{L_0}
\newcommand{\Htil}{\tilde{H}}
\newcommand{\curlyL}{{\mathcal L}}
\newcommand{\curlyLo}{\curlyL_0}
\newcommand{\Qstar}{Q_{\ast}}
\newcommand{\Qstat}{\smallsub{Q}{stat}}
\newcommand{\Otot}{\smallsub{\Omega}{tot}}
\newcommand{\rtot}{\smallsub{\rho}{tot}}
\newcommand{\Ototo}{\smsubsub{\Omega}{tot}{0}}
\newcommand{\rtoto}{\smsubsub{\rho}{tot}{0}}
\newcommand{\Omat}{\Omega_m}
\newcommand{\Omato}{\Omega_{m,0}}
\newcommand{\Obar}{\smallsub{\Omega}{bar}}
\newcommand{\Olum}{\smallsub{\Omega}{lum}}
\newcommand{\Obdm}{\smallsub{\Omega}{bdm}}
\newcommand{\Ocdm}{\smallsub{\Omega}{cdm}}
\newcommand{\Onu}{\Omega_{\nu}}
\newcommand{\Ogam}{\Omega_{\gamma}}
\newcommand{\Orad}{\Omega_r}
\newcommand{\Orado}{\Omega_{r,0}}
\newcommand{\Olam}{\Omega_{\Lambda}}
\newcommand{\Olamo}{\Omega_{\Lambda,0}}
\newcommand{\Oein}{\Omega_{\Lambda,{\scriptstyle {\rm E}}}}
\newcommand{\plam}{\capsub{p}{\Lambda}}
\newcommand{\rlam}{\capsub{\rho}{\Lambda}}
\newcommand{\pvac}{p_v}
\newcommand{\rvac}{\rho_v}
\newcommand{\gvac}{\gamma_v}
\newcommand{\rvaco}{\rho_{v,0}}
\newcommand{\psti}{p_s}
\newcommand{\rsti}{\rho_s}
\newcommand{\ro}{\rho_0}
\newcommand{\pmat}{p_m}
\newcommand{\rmat}{\rho_m}
\newcommand{\gmat}{\gamma_m}
\newcommand{\rmato}{\rho_{m,0}}
\newcommand{\prad}{p_r}
\newcommand{\rrad}{\rho_r}
\newcommand{\grad}{\gamma_r}
\newcommand{\rrado}{\rho_{r,0}}
\newcommand{\rcrit}{\smallsub{\rho}{crit}}
\newcommand{\rcrito}{\smsubsub{\rho}{crit}{0}}
\newcommand{\zmax}{z_{\mbox{\scriptsize max}}}

\newcommand{\dQlame}{\smallsub{dQ_{\lambda}}{,em}}
\newcommand{\dQlamo}{\smallsub{dQ_{\lambda}}{,obs}}
\newcommand{\LCDM}{$\Lambda$CDM}
\newcommand{\LBDM}{$\Lambda$BDM}
\newcommand{\lamo}{\lambda_0}
\newcommand{\Ilam}{I_\lambda}
\newcommand{\IlamS}{\subsmsub{I}{\lambda}{stat}}
\newcommand{\lamp}{\lambda_p}
\newcommand{\Tsun}{T_{\odot}}
\newcommand{\Ltil}{\tilde{\mathcal L}}
\newcommand{\Id}{I_{\delta}}
\newcommand{\IUs}{\mbox{erg s}^{-1}\mbox{ cm}^{-2}\mbox{ \AA}^{-1}\mbox{ ster}^{-1}}
\newcommand{\siglam}{\sigma_\lambda}
\newcommand{\Ig}{I_g}
\newcommand{\sigSB}{\capsub{\sigma}{SB}}
\newcommand{\To}{T_0}
\newcommand{\Ib}{I_b}
\newcommand{\Lsun}{L_{\odot}}
\newcommand{\Fn}{F_n}
\newcommand{\Fs}{F_s}
\newcommand{\Ln}{L_n}
\newcommand{\Ls}{L_s}
\newcommand{\nN}{n_n}
\newcommand{\nS}{n_s}
\newcommand{\ello}{\ell_0}
\newcommand{\fo}{f_0}
\newcommand{\zp}{z_p}
\newcommand{\In}{I^n_\lambda}
\newcommand{\Is}{I^s_\lambda}
\newcommand{\Ins}{I_{ns}}
\newcommand{\ntil}{\tilde{n}}
\newcommand{\IlamSn}{\supsubsmsub{I}{n}{\lambda}{stat}}
\newcommand{\IlamSs}{\supsubsmsub{I}{s}{\lambda}{stat}}

\newcommand{\Msun}{M_{\odot}}
\newcommand{\MLunits}{M_{\odot}/L_{\odot}}
\newcommand{\MLcrito}{(M/L)_{\mbox{\scriptsize crit},{\scriptscriptstyle {\rm 0}}}}
\newcommand{\Mmw}{\smallsub{M}{mw}}
\newcommand{\Lmw}{\smallsub{L}{mw}}
\newcommand{\MLmw}{\smallsub{(M/L)}{mw}}
\newcommand{\Mbar}{\smallsub{M}{bar}}
\newcommand{\barf}{\smallsub{\Mbar/M}{tot}}
\newcommand{\tnr}{\smallsub{t}{nr}}
\newcommand{\tnuc}{\smallsub{t}{nuc}}
\newcommand{\tdec}{\smallsub{t}{dec}}

\newcommand{\Tmat}{{\mathcal T}_{\mu\nu}}
\newcommand{\Teff}{{\mathcal T}^{\mbox{\scriptsize eff}}_{\mu\nu}}
\newcommand{\Tphi}{{\mathcal T}^{\varphi}_{\mu\nu}}
\newcommand{\Lamo}{\Lambda_0}
\newcommand{\lPl}{\capsub{\ell}{Pl}}
\newcommand{\LamPl}{\capsub{\Lambda}{Pl}}
\newcommand{\reff}{\smallsub{\rho}{eff}}
\newcommand{\peff}{\smallsub{p}{eff}}
\newcommand{\rcon}{\rho_c}
\newcommand{\xr}{x_r}
\newcommand{\xm}{x_m}
\newcommand{\av}{\alpha_v}
\newcommand{\am}{\alpha_m}
\newcommand{\ar}{\alpha_r}
\newcommand{\Sm}{{\mathcal S}_m}
\newcommand{\tao}{\tau_0}
\newcommand{\teq}{\smallsub{t}{eq}}
\newcommand{\cho}{\chi_0}
\newcommand{\tf}{t_f}
\newcommand{\Vo}{V_0}
\newcommand{\roo}{r_0}
\newcommand{\LT}{\smallsub{L}{th}}
\newcommand{\LR}{\smallsub{L}{rel}}
\newcommand{\Tnu}{T_{\nu}}
\newcommand{\rgam}{\rho_{\gamma}}
\newcommand{\rnu}{\rho_{\nu}}
\newcommand{\curlyLvo}{{\mathcal L}_{v,0}}
\newcommand{\Qv}{Q_v}
\newcommand{\lamv}{\lambda_v}
\newcommand{\lamcmb}{\smallsub{\lambda}{cmb}}
\newcommand{\nuInu}{\nu I_{\nu}}
\newcommand{\Ovaco}{\Omega_{v,0}}

\newcommand{\ma}{m_a c^2}
\newcommand{\macrit}{\smallsub{m_a}{,crit} c^2}
\newcommand{\fpq}{\capsub{f}{PQ}}
\newcommand{\gagg}{g_{a\gamma\gamma}}
\newcommand{\taua}{\tau_a}
\newcommand{\mone}{m_1}
\newcommand{\gstar}{g_{\ast F}}
\newcommand{\Oma}{\Omega_a}
\newcommand{\Mtot}{\smallsub{M}{tot}}
\newcommand{\Oabound}{\smallsub{\Omega}{a,bound}}
\newcommand{\lama}{\lambda_a}
\newcommand{\sigfifty}{\sigma_{50}}
\newcommand{\Iobs}{\smallsub{I}{obs}}
\newcommand{\Ith}{\smallsub{I}{th}}

\newcommand{\taunu}{\tau_{\nu}}
\newcommand{\mnu}{m_{\nu}}
\newcommand{\mtau}{m_{\nu_{\tau}}}
\newcommand{\mmu}{m_{\nu_{\mu}}}
\newcommand{\Etau}{m_{\nu_{\tau}} c^2}
\newcommand{\Egam}{E_{\gamma}}
\newcommand{\lamnu}{\lambda_{\nu}}
\newcommand{\sigthirty}{\sigma_{30}}
\newcommand{\nsun}{n_{\odot}}
\newcommand{\rsun}{r_{\odot}}
\newcommand{\xmax}{\smallsub{x}{max}}
\newcommand{\ynu}{y_{\nu}}
\newcommand{\znu}{z_{\nu}}
\newcommand{\ftau}{f_{\tau}}
\newcommand{\Qbound}{\smallsub{Q}{bound}}
\newcommand{\Onubound}{\subsmsub{\Omega}{\nu}{bound}}
\newcommand{\Onufree}{\subsmsub{\Omega}{\nu}{free}}
\newcommand{\curlyLf}{\curlyL_f}
\newcommand{\Qfree}{\smallsub{Q}{free}}
\newcommand{\Inu}{I_{\nu}}
\newcommand{\taugas}{\smallsub{\tau}{gas}}
\newcommand{\taudust}{\smallsub{\tau}{dust}}
\newcommand{\tauZP}{\capsub{\tau}{ZP}}
\newcommand{\lamL}{\capsub{\lambda}{L}}
\newcommand{\tauFP}{\capsub{\tau}{FP}}
\newcommand{\taumax}{\smallsub{\tau}{max}}

\newcommand{\mchi}{m_{\tilde{\chi}} c^2}
\newcommand{\tilchi}{\tilde{\chi}}
\newcommand{\lamann}{\smallsub{\lambda}{ann}}
\newcommand{\mten}{m_{10}}
\newcommand{\signine}{\sigma_9}
\newcommand{\tilmten}{\tilde{m}_{10}}
\newcommand{\Lann}{\subsmsub{L}{h}{ann}}
\newcommand{\Qann}{\subsmsub{Q}{\tilde{\chi}}{ann}}
\newcommand{\Iann}{\subsmsub{I}{\tilde{\chi}}{ann}}
\newcommand{\vR}{\capsub{v}{R}}
\newcommand{\lamloop}{\smallsub{\lambda}{loop}}
\newcommand{\tauchi}{\tau_{\tilde{\chi}}}
\newcommand{\Lloop}{\subsmsub{L}{h}{loop}}
\newcommand{\Iloop}{\subsmsub{I}{\tilde{\chi}}{loop}}
\newcommand{\fR}{\capsub{f}{R}}
\newcommand{\Nics}{\smallsub{N}{ics}}
\newcommand{\Emax}{\smallsub{E}{max}}
\newcommand{\Ecmb}{\smallsub{E}{cmb}}
\newcommand{\Tcmb}{\smallsub{T}{cmb}}
\newcommand{\ncmb}{\smallsub{n}{cmb}}
\newcommand{\Fics}{\smallsub{F}{ics}}
\newcommand{\Ltree}{\subsmsub{L}{h}{tree}}
\newcommand{\Ncasc}{\smallsub{N}{casc}}
\newcommand{\Fcasc}{\smallsub{F}{casc}}
\newcommand{\Itree}{\subsmsub{I}{\tilde{\chi}}{tree}}
\newcommand{\TR}{\capsub{T}{R}}
\newcommand{\taug}{\tau_{\tilde{g}}}
\newcommand{\gamcmb}{\smallsub{\gamma}{cmb}}
\newcommand{\mg}{m_{\tilde{g}} c^2}
\newcommand{\Lgrav}{\subsmsub{L}{h}{grav}}
\newcommand{\Igrav}{I_{\tilde{g}}}
\newcommand{\IE}{\capsub{I}{E}}
\newcommand{\EIE}{E\capsub{I}{E}(>E_0)}
\newcommand{\Istar}{I_{\ast}}
\newcommand{\Estar}{E_{\ast}}

\newcommand{\tpbh}{\smallsub{t}{pbh}}
\newcommand{\Mstar}{M_{\ast}}
\newcommand{\rpbh}{\smallsub{\rho}{pbh}}
\newcommand{\kbeta}{k_{\beta}}
\newcommand{\Opbh}{\smallsub{\Omega}{pbh}}
\newcommand{\lampbh}{\smallsub{\lambda}{pbh}}
\newcommand{\Lpbh}{\smallsub{L}{pbh}}
\newcommand{\tauf}{\tau_f}
\newcommand{\Mcut}{{\mathcal M}_c}
\newcommand{\Mmin}{\smallsub{M}{min}}
\newcommand{\Qpbh}{\smallsub{Q}{pbh}}
\newcommand{\keps}{k_{\varepsilon}}
\newcommand{\tilto}{\tilde{t}_0}
\newcommand{\tiltm}{\tilde{t}_m}
\newcommand{\Ipbh}{\smallsub{I}{pbh}}
\newcommand{\Runits}{\mbox{ pc}^{-3}\mbox{ yr}^{-1}}
\newcommand{\Mgal}{\smallsub{M}{gal}}
\newcommand{\rcmb}{\smallsub{\rho}{cmb}}

\begin{frontmatter}
\title{Dark Matter and Background Light}
\author{J.M. Overduin}
\address{Gravity Probe B, Hansen Experimental Physics Laboratory, Stanford University, Stanford, California, U.S.A. 94305-4085}
\ead{overduin@relgyro.stanford.edu}
\and
\author{P.S. Wesson}
\address{Department of Physics, University of Waterloo, Ontario, Canada N2L 3G1}
\ead{wesson@astro.uwaterloo.ca}
\begin{abstract}
Progress in observational cosmology over the past five years has
established that the Universe is dominated dynamically by dark matter
and dark energy.  Both these new and apparently independent forms of
matter-energy have properties that are inconsistent with anything in
the existing standard model of particle physics, and it appears that
the latter must be extended.  We review what is known about dark matter
and energy from their impact on the light of the night sky.
Most of the candidates that have been proposed so far are not perfectly
black, but decay into or otherwise interact with photons in characteristic
ways that can be accurately modelled and compared with observational data.
We show how experimental limits on the intensity of cosmic background
radiation in the microwave, infrared, optical, ultraviolet, x-ray and
$\gamma$-ray bands put strong limits on decaying vacuum energy,
light axions, neutrinos, unstable weakly-interacting massive particles
(WIMPs) and objects like black holes.  Our conclusion is that the
dark matter is most likely to be WIMPs if conventional cosmology holds;
or higher-dimensional sources if spacetime needs to be extended.
\end{abstract}
\end{frontmatter}

\section{Introduction: the light of the night sky} \label{ch1}

Olbers was one of a long line of thinkers who pondered the paradox:
how can an infinite Universe full of stars not be ablaze with light
in every direction?  Although cosmologists now speak of galaxies (and
other sources of radiation) rather than stars, the question retains its
relevance.  In fact, the explanation of the intensity of the background
radiation at all wavelengths has become recognized as one of the fundamental
keys to cosmology.  We will begin in this review with what is known about
this radiation itself, and then move on to what it tells us about the
dark energy and dark matter.

The optical and near-optical (ultraviolet and infrared) portions of the
background comprise what is known as the extragalactic background light
(EBL), the domain of the ``classical'' Olbers problem.  The observed
intensity of background light in these bands guides our understanding
of the way in which the luminous components of the Universe (i.e. the
galaxies) formed and evolved with time.
We now know what Olbers did not: that the main reason why the sky
is dark at night is that the Universe had a {\em beginning in time\/}.
This can be appreciated qualitatively (and quantitatively to within
a factor of a few) with no relativity at all beyond the fact of
a finite speed of light.  Imagine yourself at the center of a ball
of glowing gas with radius $R$ and uniform luminosity density
$\curlyL(r)=\curlyLo$.  The intensity $Q$ of background radiation
between you and the edge of the ball is just
\beq
Q = \int_0^{R}\curlyL(r)dr = c\too\curlyLo \; ,
\label{Naive}
\eeq
where we have used $R=c\too$ as a naive approximation to the size
of the Universe.  Thus knowledge of the luminosity density
$\curlyLo$ and measurement of the background intensity $Q$ tells us
immediately that the galaxies have been shining only for a time $\too$.

More refined calculations introduce only minor changes to this result.
Expansion stretches the path length $R$, but this is more than offset
by the dilution of the luminosity density $\curlyL(r)$, which drops
by roughly the same factor cubed.  There is a further reduction in
$\curlyL(r)$ due to the redshifting of light from distant sources.
So Eq.~(\ref{Naive}) represents a theoretical upper limit on the
background intensity.  In a fully general relativistic treatment, one
obtains the following expression for $Q$ in standard cosmological models
whose scale factor varies as a power-law function of time
($R \propto t^{\ell}$):
\beq
Q = \frac{c\too\curlyLo}{1 + \ell} \; ,
\label{NotSoNaive}
\eeq
as may be checked using Eq.~(\ref{QtDefn}) in Sec.~\ref{ch2}.
Thus Eq.~(\ref{Naive}) overestimates $Q$ as a function of $\too$ by a
factor of $5/3$ in a universe filled with dust-like matter ($\ell=2/3$).

Insofar as $Q$ and $\curlyLo$ are both known quantities, one can in
principle use them to infer a value for $\too$.
Intensity $Q$, for instance, is obtained by measuring spectral
intensity $\Ilam(\lambda)$ over the wavelengths where starlight is
brightest and integrating: $Q=\int\Ilam(\lambda)d\lambda$.
This typically leads to values of around
$Q \approx 1.4\times10^{-4}$~erg~s$^{-1}$~cm$^{-2}$ \cite{OW03}.
Luminosity density $\curlyLo$ can be determined by counting the number of
faint galaxies in the sky down to some limiting magnitude, and extrapolating
to fainter magnitudes based on assumptions about the true distribution
of galaxy luminosities.  One finds in this way that
$\curlyLo\approx 1.9\times10^{-32}$~erg~s$^{-1}$~cm$^{-3}$ \cite{Fuk98a}.
Alternatively, one can extrapolate from the properties of the Sun,
which emits its energy at a rate per unit mass of
$\epsilon_{\odot}=\Lsun/\Msun=1.9$~erg~s$^{-1}$~g$^{-1}$.
A colour-magnitude diagram for nearby stars shows us that the Sun is
modestly brighter than average, with a more typical rate of stellar
energy emission given by about $1/4$ the Solar value, or
$\epsilon\sim 0.5$~erg~s$^{-1}$~g$^{-1}$.  Multipying this number by
the density of luminous matter in the Universe
($\rlum=4\times 10^{-32}$~g~cm$^{-3}$) gives a figure for mean luminosity
density which is the same as that derived above from galaxy counts:
$\curlyLo=\epsilon\rlum\sim 2\times10^{-32}$~erg~s$^{-1}$~cm$^{-3}$.
Either way, plugging $Q$ and $\curlyLo$ into Eq.~(\ref{Naive}) with
$\ell=2/3$ implies a cosmic age of $\too=13$~Gyr, which differs from
the currently accepted figure by only 5\%.  (The remaining difference
can be accounted for if cosmic expansion is not a simple power-law
function of time; more on this later.)  {\em Thus the brightness of the
night sky tells us not only that there was a big bang, but also roughly
when it occurred.\/}  Conversely, the intensity of background radiation
is largely determined by the age of the Universe.  Expansion merely deepens
the shade of a night sky that is already black.

\begin{figure}[t]
\begin{center}
\includegraphics[width=\textwidth]{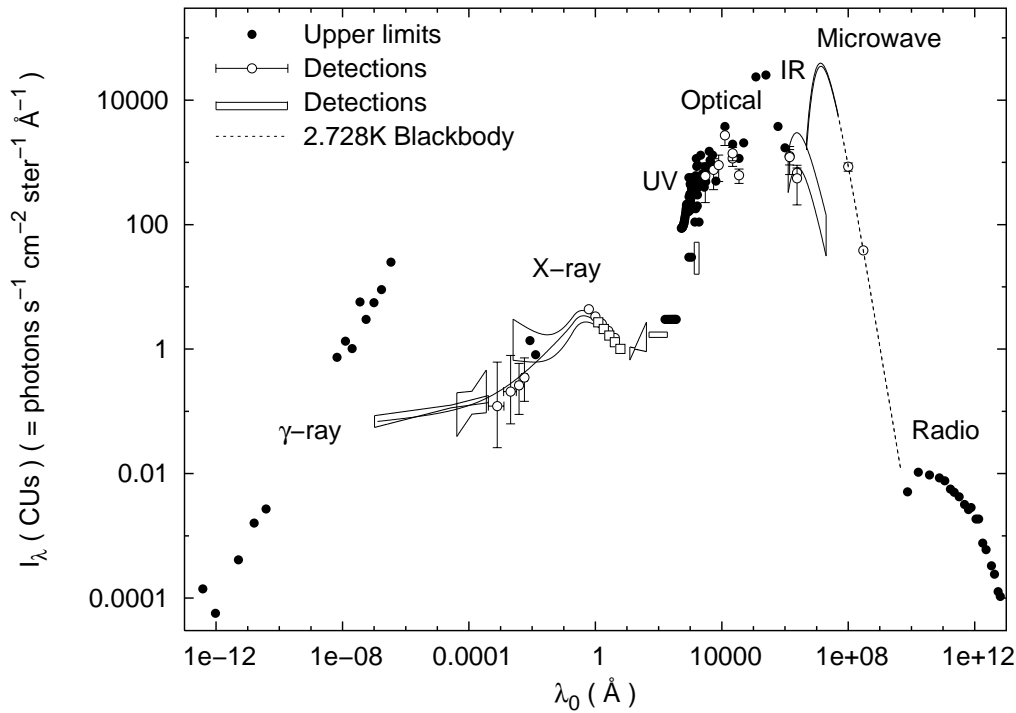}
\caption{A compilation of experimental measurements of the intensity of
   cosmic background radiation at all wavelengths.  This figure and the
   data shown in it will be discussed in more detail in later sections.}
\end{center}
\label{fig1.2}
\end{figure}

We have so far discussed only the bolometric, or integrated intensity
of the background light over all wavelengths, whose significance will
be explored in more detail in Sec.~2.  The {\em spectral\/} background ---
from radio to microwave, infrared, optical, ultraviolet, x-ray and
$\gamma$-ray bands --- represents an even richer store of information
about the Universe and its contents (Fig.~\ref{fig1.2}).
The optical waveband (where galaxies emit most of their light) has been
of particular importance historically, and the infrared band (where the
redshifted light of distant galaxies actually reaches us) has come into
new prominence more recently.  By combining the observational data in
both of these bands, we can piece together much of the evolutionary history
of the galaxy population, make inferences about the nature of the
intervening intergalactic medium, and draw conclusions about the dynamical
history of the Universe itself.  Interest in this subject has exploded
over the past few years as improvements in telescope and detector
technology have brought us to the threshold of the first EBL detection
in the optical and infrared bands.  These developments and their
implications are discussed in Sec.~\ref{ch3}.

In the remainder of the review, we move on to what the background
radiation tells us about the dark matter and energy, whose current
status is reviewed in Sec.~\ref{ch4}.  The leading candidates are
taken up individually in Secs.~\ref{ch5}--\ref{ch9}.  {\em None of them
are perfectly black\/}.  All of them are capable in principle of decaying
into or interacting with ordinary photons, thereby leaving telltale
signatures in the spectrum of background radiation.  We begin
with dark energy, for which there is particularly good reason to suspect
a decay with time.  The most likely place to look for evidence of such
a process is in the cosmic microwave background, and we review the stringent
constraints that can be placed on any such scenario in Sec.~\ref{ch5}.
Axions, neutrinos and weakly interacting massive particles are treated
next: these particles decay into photons in ways that depend on
parameters such as coupling strength, decay lifetime, and rest mass.
As we show in Secs.~\ref{ch6}, \ref{ch7} and \ref{ch8},
data in the infrared, optical, ultraviolet, x-ray and $\gamma$-ray bands
allow us to be specific about the kinds of properties that these
particles must have if they are to make up the dark matter in the Universe.
In Sec.~\ref{ch9}, finally, we turn to black holes.  The observed intensity
of background radiation, especially in the $\gamma$-ray band, is sufficient
to rule out a significant role for standard four-dimensional black holes,
but it may be possible for their higher-dimensional analogs (known as
solitons) to make up all or part of the dark matter.  We wrap up our 
review in Sec.~\ref{ch10} with some final comments and a view toward
future developments.

\section{The intensity of cosmic background radiation} \label{ch2}

\subsection{Bolometric intensity} \label{sec:Q}

\begin{figure}[b!]
\begin{center}
\includegraphics[width=65mm]{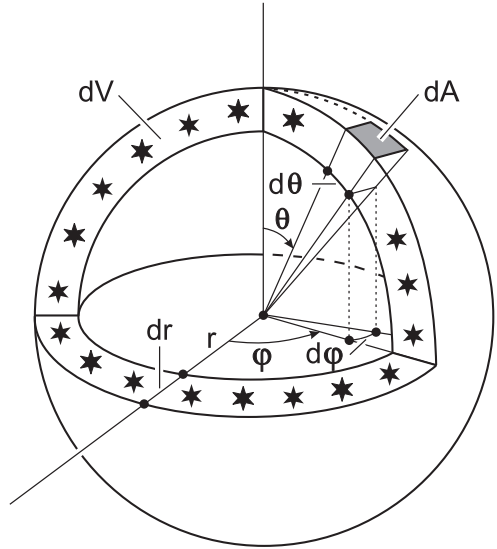}
\end{center}
\caption{Surface area element $dA$ and volume $dV$ of a thin spherical shell,
   containing all the light sources at coordinate distance $r$.}
\label{fig2.1}
\end{figure}

Let us begin with the general problem of adding up the contributions
from many sources of radiation in the Universe in such a way as to arrive
at their combined intensity as received by us in the Milky Way
(Fig.~\ref{fig2.1}).
To begin with, we take the sources to be ordinary galaxies, but the formalism
is general.  Consider a single galaxy at coordinate distance $r$ whose
luminosity, or rate of energy emission per unit time, is given by $L(t)$.
In a standard Friedmann-Lema\^{\i}tre-Roberston-Walker (FLRW) universe,
its energy has been spread over an area
\beq
A = \int dA = \int_{\theta=0}^{\pi} \int_{\phi=0}^{2\pi}
   [\Ro r d\theta][\Ro r \sin\theta d\phi] = 4\pi \Ro^{\,2} r^2 \;
\eeq
by the time it reaches us at $t=\too$.  Here we follow standard practice
and use the subscript ``0'' to denote any quantity taken at the present
time, so $\Ro\equiv R(\too)$ is the present value of the cosmological
scale factor $R$.

The intensity, or flux of energy per unit area reaching us from this galaxy
is given by
\beq
d\Qgal = \left[ \frac{R(t)}{\Ro} \right]^2 \!\! \frac{L(t)}{A} =
   \frac{R^{\,2}(t) L(t)}{4\pi \Ro^{\,4} r^2} \; .
\label{dQgDefn}
\eeq
Here the subscript ``$g$'' denotes a single galaxy, and the two factors
of $R(t)/\Ro$ reflect the fact that expansion increases the wavelength of
the light as it travels toward us (reducing its energy), and also spaces
the photons more widely apart in time (Hubble's energy and number effects).

To describe the whole population of galaxies, distributed through space
with physical number density $\ngal(t)$, it is convenient to use the
four-dimensional galaxy current $\Jgal^{\mu} \equiv \ngal U^{\mu}$
where $U^{\mu}\equiv(1,0,0,0)$ is the galaxy four-velocity \cite{Wei72}.
If galaxies are conserved (i.e. their rates of formation and
destruction by merging or other processes are slow in comparison to the
expansion rate of the Universe), then $\Jgal^{\mu}$ obeys a conservation
equation
\beq
\nabla_{\mu}\Jgal^{\mu} = 0 \; ,
\eeq
where $\nabla_{\mu}$ denotes the covariant derivative.  Using the
Robertson-Walker metric this reduces to:
\beq
\frac{1}{R^3} \frac{d}{dt} \left( R^3 \ngal \right) = 0 \; .
\label{GalCon}
\eeq
In what follows, we will replace $\ngal$ by the {\em comoving
number density\/} ($n$), defined in terms of $\ngal$ by
\beq
n \equiv \ngal \left( \frac{R}{\Ro} \right)^3 \; .
\label{ngDefn}
\eeq
Under the assumption of galaxy conservation, which we shall make for
the most part, this quantity is always equal to its value at present
($n=\no=$~const).  When mergers or other galaxy non-conserving processes
are important, $n$ is no longer constant.  We will allow for this
situation in Sec.~\ref{sec:CompObs}.

We now shift our origin so that we are located at the center of the
spherical shell in Fig.~\ref{fig2.1}, and consider those galaxies located
inside the shell which extends from radial coordinate distance $r$
to $r+dr$.  The volume of this shell is obtained from the
Robertson-Walker metric as
\beq
dV = \int_{\theta=0}^{\pi} \int_{\phi=0}^{2\pi}
   \left[ \frac{R dr}{\sqrt{1-k r^2}} \right]
   \left[ R r d\theta \right] \left[ R r \sin\theta d\phi \right] 
   = \frac{4\pi R^3 r^2 dr}{\sqrt{1-k r^2}} \; .
\eeq
The only trajectories of interest are those of light rays striking our
detectors at origin.  By definition, these are radial ($d\theta = d\phi = 0$)
null geodesics ($ds^2=0$), for which the metric relates time $t$
and coordinate distance $r$ via
\beq
c dt = \frac{R \, dr}{\sqrt{1-k r^2}} \; .
\label{dtdr}
\eeq
Thus the volume of the shell can be re-expressed as
\beq
dV = 4\pi R^{\,2} r^2 c dt \; ,
\label{dVdt}
\eeq
and the latter may now be thought of as extending between
look-back times $\too-t$ and $\too-(t+dt)$,
rather than distances $r$ and $r+dr$.

The total energy received at origin from the galaxies in the shell is then
just the product of their individual intensities (\ref{dQgDefn}), their number
per unit volume (\ref{ngDefn}), and the volume of the shell (\ref{dVdt}):
\beq
dQ = d\Qgal \, \ngal \, dV = c n(t) \Rtil(t) L(t) dt \; .
\label{dQdt}
\eeq
Here we have defined the relative scale factor by $\Rtil\equiv R/\Ro$.
We henceforth use tildes throughout our review to denote dimensionless
quantities {\em taken relative to their present value\/} at $\too$.

Integrating (\ref{dQdt}) over all shells between $\too$ and $\too-t_f$,
where $t_f$ is the source formation time, we obtain
\beq
Q = \int dQ = c \! \int_{t_f}^{\too} n(t) \, L(t) \, \Rtil(t) \, dt \; .
\label{QtDefn}
\eeq
Eq.~(\ref{QtDefn}) defines the {\em bolometric intensity of the
extragalactic background light\/} (EBL).  This is the energy received
by us (over all wavelengths of the electromagnetic spectrum) per unit time,
per unit area, from all the galaxies which have been shining since time
$t_f$.  In principle, if we let $t_f\rightarrow0$, we will encompass
the entire history of the Universe since the big bang.  Although this
sometimes provides a useful mathematical shortcut, we will see in later
sections that it is physically more realistic to cut the integral off
at a finite formation time.  The quantity $Q$ is a measure of the amount
of light in the Universe, and Olbers' ``paradox'' is merely another way
of asking why it is low.

\subsection{Characteristic values} \label{sec:Qstar}

While the cosmic time $t$ is a useful independent variable for theoretical
purposes, it is not directly observable.  In studies aimed at making contact
with eventual observation it is better to work in terms of redshift $z$,
which is the relative shift in wavelength $\lambda$ of a light signal
between the time it is emitted and observed:
\beq
z \equiv \frac{\Delta\lambda}{\lambda} = \frac{\Ro-R(t)}{R(t)} =
   \Rtil^{-1} - 1 \; .
\label{zDefn}
\eeq
Differentiating, and defining Hubble's parameter $H\equiv\dot{R}/R$,
we find that
\beq
dt = -\frac{dz}{(1+z) H(z)} \; ,
\label{dtdz}
\eeq
Hence Eq.~(\ref{QtDefn}) is converted into an integral over $z$ as
\beq
Q = c \int_{0}^{z_f} \frac{n(z) \, L(z) \, dz}{(1+z)^2 \, H(z)} \; ,
\label{QzDefn}
\eeq
where $z_f$ is the redshift of galaxy formation.

For some problems, and for this section in particular, the physics of the
sources themselves are of secondary importance, and it is reasonable to
take $L(z)=\Lo$ and $n(z)=\no$ as constants over the range of redshifts
of interest.  Then Eq.~(\ref{QzDefn}) can be written in the form
\beq
Q = \Qstar \int_{0}^{z_f} \frac{dz}{(1+z)^2 \, \Htil(z)} \; .
\label{QzDefn2}
\eeq
Here $\Htil\equiv H/\Ho$ is the Hubble expansion rate relative to its
present value, and $\Qstar$ is a constant containing all the dimensional
information:
\beq
\Qstar \equiv \frac{c\curlyLo}{\Ho} \; .
\eeq
The quantities $\Ho$, $\curlyLo$ and $\Qstar$ are fundamental to much of
what follows, and we pause briefly to discuss them here.  The value of
$\Ho$ (referred to as Hubble's constant) is still debated, and it is
commonly expressed in the form
\beq
\Ho = 100 \, \ho \mbox{ km s}^{-1}\mbox{Mpc}^{-1}
      = 0.102 \, \ho \mbox{ Gyr}^{-1} \; .
   \label{HoValue}
\eeq
Here 1~Gyr $\equiv 10^9$~yr and the uncertainties have been absorbed
into a dimensionless parameter $\ho$ whose value is now conservatively
estimated to lie in the range $0.6\leqslant\ho\leqslant0.9$
(Sec.~\ref{sec:baryons}).

The quantity $\curlyLo$ is the {\em comoving luminosity density\/} of
the Universe at $z=0$:
\beq
\curlyLo \equiv \no \Lo \; .
\eeq
This can be measured experimentally by counting galaxies down to some
faint limiting apparent magnitude, and extrapolating to fainter ones based
on assumptions about the true distribution of absolute magnitudes.
A recent compilation of seven such studies over the past decade
gives \cite{Fuk98a}:
\beqa
\curlyLo & = & (2.0 \pm 0.2) \times 10^8 \, \ho \, \Lsun \mbox{ Mpc}^{-3} \nonumber \\
         & = & (2.6 \pm 0.3) \times 10^{-32} \, \ho \mbox{ erg s}^{-1} \mbox{ cm}^{-3} \; ,
\label{curlyLoValue}
\eeqa
near 4400\AA\ in the B-band, where galaxies emit most of their light.
We will use this number throughout our review.  Recent measurements by
the Two Degree Field (2dF) team suggest a slightly lower value,
$\curlyLo=(1.82 \pm 0.17) \times 10^8 \, \ho \, \Lsun$~Mpc$^{-3}$
\cite{Nor02}, and this agrees with newly revised figures from the
Sloan Digital Sky Survey ({\sc Sdss}):
$\curlyLo=(1.84 \pm 0.04) \times 10^8 \, \ho \, \Lsun$~Mpc$^{-3}$
\cite{Bla03}.  If the final result inferred from large-scale galaxy 
surveys of this kind proves to be significantly different from that in
(\ref{curlyLoValue}), then our EBL intensities (which are proportional to
$\curlyLo$) would go up or down accordingly.

Using (\ref{HoValue}) and (\ref{curlyLoValue}), we find that the
characteristic intensity associated with the integral~(\ref{QzDefn2})
takes the value
\beq
\Qstar \equiv \frac{c \curlyLo}{\Ho} = (2.5 \pm 0.2) \times 10^{-4} 
   \mbox{ erg s}^{-1} \mbox{ cm}^{-2} \; .
\label{QstarValue}
\eeq
There are two important things to note about this quantity.  First,
because the factors of $\ho$ attached to both $\curlyLo$ and $\Ho$
cancel each other out, $\Qstar$ is {\em independent\/} of the uncertainty
in Hubble's constant.  This is not always appreciated but was first
emphasized by Felten \cite{Fel66}.  Second, the value of $\Qstar$
is {\em very small\/} by everyday standards: more than a million times
fainter than the bolometric intensity that would be produced by a 100~W
bulb in the middle of an ordinary-size living room whose walls,
floor and ceiling have a summed surface area of 100~m$^2$
($\Qbulb=10^3$~erg~s$^{-1}$~cm$^{-2}$). The smallness of $\Qstar$ is
intimately related to the resolution of Olbers' paradox.

\subsection{Matter, energy and expansion} \label{sec:FL}

The remaining unknown in Eq.~(\ref{QzDefn2}) is the relative expansion rate
$\Htil(z)$, which is obtained by solving the field equations of general
relativity.  For standard FLRW models one obtains the following
differential equation:
\beq
\Htil^2 + \frac{k c^2}{\Ho^2 R^{\,2}} = \frac{8\pi G}{3\Ho^2} \, \rtot \; .
\label{FL1}
\eeq
Here $\rtot$ is the total density of all forms of matter-energy, 
including the vacuum energy density associated with the cosmological
constant $\Lambda$ via
\beq
\rlam c^2 \equiv \frac{\Lambda c^4}{8\pi G} = \mbox{ const} \; .
   \label{rlamDefn}
\eeq
It is convenient to define the present {\em critical density\/} by
\beq
\rcrito \equiv \frac{3 \Ho^2}{8 \pi G}
   = (1.88 \times 10^{-29}) \, \ho^2 \mbox{ g cm}^{-3} \; .
\label{rcritoDefn}
\eeq
We use this quantity to re-express all our densities $\rho$ in dimensionless
form, $\Omega\equiv\rho/\rcrito$, and eliminate the unknown $k$ by evaluating
Eq.~(\ref{FL1}) at the present time so that $k c^2/(\Ho\Ro)^2=\Ototo-1$.
Substituting this result back into (\ref{FL1}) puts the latter into the form
\beq
\Htil^2 = \Otot - \left( \Ototo - 1 \right) \Rtil^{-2} \; .
\label{FLtemp}
\eeq
To complete the problem we need only the form of $\Otot(\Rtil)$, which
comes from energy conservation.  Under the usual assumptions of isotropy
and homogeneity, the matter-energy content of the Universe can be modelled
by an energy-momentum tensor of the perfect fluid form
\beq
{\mathcal T}_{\mu\nu} = (\rho + p/c^2) U_{\mu} U_{\nu} + p \, g_{\mu\nu} \; .
\label{PFdefn}
\eeq
Here density $\rho$ and pressure $p$ are related by an equation of state,
which is commonly written as
\beq
p = (\gamma - 1) \rho c^2 \; .
\label{EOS}
\eeq
Three equations of state are of particular relevance to cosmology,
and will make regular appearances in the sections that follow:
\beq
\gamma = \left\{ \begin{array}{llll}
   4/3 \hspace{5mm} & \Rightarrow & \hspace{5mm} \prad = \rrad c^2/3 &
      \hspace{1cm} \mbox{ (radiation)}\\
   1 \hspace{5mm} & \Rightarrow & \hspace{5mm} \pmat = 0 &
      \hspace{1cm} \mbox{ (dust-like matter)}\\
   0 \hspace{5mm} & \Rightarrow & \hspace{5mm} \pvac = -\rvac \, c^2 &
      \hspace{1cm} \mbox{ (vacuum energy)}
   \end{array} \right. \; .
\label{gammaValues}
\eeq
The first of these is a good approximation to the {\em early\/} Universe,
when conditions were so hot and dense that matter and radiation existed
in nearly perfect thermodynamic equilibrium (the radiation era).
The second has often been taken to describe the {\em present\/} 
Universe, since we know that the energy density of electromagnetic 
radiation now is far below that of dust-like matter.
The third may be a good description of the {\em future\/} state of the
Universe, if recent measurements of the magnitudes of high-redshift
Type~Ia supernovae are borne out (Sec.~4).  These indicate that vacuum-like
dark energy is already more important than all other contributions to the
density of the Universe combined, including those from pressureless
dark matter.

Assuming that energy and momentum are neither created nor destroyed,
one can proceed exactly as with the galaxy current $\Jgal^{\mu}$. 
The conservation equation in this case reads
\beq
\nabla^{\mu} {\mathcal T}_{\mu\nu} = 0 \; .
\label{PFcon}
\eeq
With the definition (\ref{PFdefn}) this reduces to
\beq
\frac{1}{R^3} \frac{d}{dt} \left[ R^3 \left( \rho c^2 + p \right)
   \right] = \frac{dp}{dt} \; ,
\label{PFcon2}
\eeq
which may be compared with (\ref{GalCon}) for galaxies.  Eq.~(\ref{PFcon2}) is
solved with the help of the equation of state (\ref{EOS}) to yield
\beq
\rho = \ro \Rtil^{-3\gamma} \; .
\eeq
In particular, for the single-component fluids in (\ref{gammaValues}):
\beq
\begin{array}{llll}
\rrad & = & \rrado \Rtil^{-4} \hspace{1cm} & \mbox{ (radiation)} \\
\rmat & = & \rmato \Rtil^{-3} \hspace{1cm} & \mbox{ (dust-like matter)} \\
\rvac & = & \rvaco = \mbox{ const } \hspace{1cm} & \mbox{ (vacuum energy)}
\end{array} \; .
\label{rhoR}
\eeq
These expressions will frequently prove useful in later sections.  They 
are also applicable to cases in which several components are present,
as long as these components exchange energy sufficiently slowly
(relative to the expansion rate) that each is in effect conserved
separately.

If the Universe contains radiation, matter and vacuum energy, so that
$\Otot=(\rrad+\rmat+\rvac)/\rcrit$, then the expansion rate~(\ref{FLtemp})
can be expressed in terms of redshift $z$ with the help of Eqs.~(\ref{zDefn})
and (\ref{rhoR}) as follows:
\beq
\Htil(z) = \left[ \Orado (1+z)^4 \! + \! \Omato (1+z)^3 \! + \! \Olamo \! -
           (\Ototo-1) (1+z)^2 \right]^{1/2} \!\!\!\!\! .
\label{FL2}
\eeq
Here $\Olamo\equiv\rlam/\rcrito=\Lambda c^2/3 \Ho^2$ from (\ref{rlamDefn})
and (\ref{rcritoDefn}).  Eq.~(\ref{FL2}) is sometimes referred to as the
{\em Friedmann-Lema\^{\i}tre equation\/}.  The radiation ($\Orado$) term
can be negelected in all but the earliest stages of cosmic history
($z\gtrsim100$), since $\Orado$ is some four orders of magnitude
smaller than $\Omato$.

The vacuum ($\Olamo$) term in Eq.~(\ref{FL2}) is independent of redshift,
which means that its influence is not diluted with time.  Any universe
with $\Lambda>0$ will therefore eventually be dominated by vacuum energy.
In the limit $t \rightarrow\infty$, in fact, the Friedmann-Lema\^{\i}tre
equation reduces to $\Olamo=(H_{\infty}/\Ho)^2$, where $H_{\infty}$
is the limiting value of $H(t)$ as $t\rightarrow\infty$ (assuming that
this latter quantity exists; i.e. that the Universe does not recollapse).
It follows that
\beq
\Lambda c^2 = 3 H_{\infty}^2 \; .
\label{WolfsEqun}
\eeq
If $\Lambda >0$, then we will {\em necessarily\/} measure $\Olamo\sim1$
at late times, regardless of the microphysical origin of the vacuum energy.

It was common during the 1980s to work with a simplified version of
Eq.~(\ref{FL2}), in which not only the radiation term was neglected,
but the vacuum ($\Olamo$) and curvature ($\Ototo$) terms as well.
There were four principal reasons for the popularity of this
{\em Einstein-de~Sitter\/} (EdS) model.
First, all four terms on the right-hand side of Eq.~(\ref{FL2}) depend
differently on $z$, so it would seem surprising to find ourselves living
in an era when any two of them were of comparable size.
By this argument, which goes back to Dicke \cite{Dic70},
it was felt that one term ought to dominate at any given time.
Second, the vacuum term was regarded with particular suspicion
for reasons to be discussed in Sec.~\ref{sec:vacenergy}.
Third, a period of inflation was asserted to have driven $\Otot(t)$
to unity.  (This is still widely believed, but depends on the initial
conditions preceding inflation, and does not necessarily hold in all
plausible models \cite{Ell88}.)  And finally, the EdS model was favoured
on grounds of simplicity.  These arguments are no longer compelling today,
and the determination of $\Omato$ and $\Olamo$ has shifted largely back
into the empirical domain.  We discuss the observational status of these
constants in more detail in Sec.~\ref{ch4}, merely assuming here that
radiation and matter densities are positive and not too large
($0\leqslant\Orado\leqslant1.5$ and $0\leqslant\Omato\leqslant1.5$),
and that vacuum energy density is neither too large nor too negative
($-0.5\leqslant\Olamo\leqslant1.5$).

\subsection{Olbers' paradox} \label{sec:olbers}

Eq.~(\ref{QzDefn2}) provides us with a simple integral for the bolometric
intensity $Q$ of the extragalactic background light in terms of the constant
$\Qstar$, Eq.~(\ref{QstarValue}), and the expansion rate $\Htil(z)$,
Eq.~(\ref{FL2}).  On dimensional grounds, we would expect $Q$ to be
close to $\Qstar$ as long as the function $\Htil(z)$ is sufficiently
well-behaved over the lifetime of the galaxies, and we will find that
this expectation is borne out in all realistic cosmological models.

The question of why $Q$ is so small (of order $\Qstar$) has historically
been known as Olbers' paradox.  Its significance can be appreciated when
one recalls that projected area on the sky drops like distance squared,
but that volume (and hence the number of galaxies) increases as roughly
the distance {\em cubed\/}.  Thus one expects to see more and more
galaxies as one looks further out.  Ultimately, in an infinite Universe
populated uniformly by galaxies, every line of sight should end up at
a galaxy and the sky should resemble a ``continuous sea of immense stars,
touching on one another,'' as Kepler put it in arguably the first statement
of the problem in 1606 (see \cite{Jak00} for a review).
Olbers himself suggested in 1823 that most of this light might be absorbed
en route to us by an intergalactic medium, but this explanation does not
stand up since the photons so absorbed would eventually be re-radiated
and simply reach us in a different waveband. Other potential solutions,
such as a non-uniform distribution of galaxies (an idea explored by Charlier
and others), are at odds with observations on the largest scales.
In the context of modern big-bang cosmology the darkness of the night
sky can only be due to two things: the {\em finite age\/} of the 
Universe (which limits the total amount of light that has been produced)
or {\em cosmic expansion\/} (which dilutes the intensity of intergalactic
radiation, and also redshifts the light signals from distant sources).

The relative importance of the two factors continues to be a subject of
controversy and confusion (see \cite{WVS87} for a review).
In particular there is a lingering perception that general relativity
``solves'' Olbers' paradox chiefly because the expansion of
the Universe stretches and dims the light it contains.

There is a simple way to test this supposition using the formalism we have 
already laid out, and that is to ``turn off'' expansion by setting the
scale factor of the Universe equal to a constant value, $R(t)=\Ro$. 
Then $\Rtil=1$ and Eq.~(\ref{QtDefn}) gives the bolometric intensity
of the EBL as
\beq
\Qstat = \Qstar \, \Ho \int_{t_f}^{t_0} dt \; .
\label{QStDefn}
\eeq
Here we have taken $n=\no$ and $L=\Lo$ as before, and used
(\ref{QstarValue}) for $\Qstar$.  The subscript ``stat'' denotes the
{\em static analog\/} of $Q$; that is, the intensity that one would
measure in a universe which did not expand.  Eq.~(\ref{QStDefn}) shows
that this is just the length of time for which the galaxies have been
shining, measured in units of Hubble time $(\Ho^{-1})$ and scaled
by $\Qstar$.

We wish to compare (\ref{QzDefn2}) in the expanding Universe with
its static analog~(\ref{QStDefn}), while keeping all other factors
the same.  In particular, if the comparison is to be meaningful, the
{\em lifetime\/} of the galaxies should be identical.
This is just $\int dt$, which may --- in an expanding Universe ---
be converted to an integral over redshift $z$ by means of (\ref{dtdz}):
\beq
\int_{t_f}^{t_0} dt = \frac{1}{\Ho} \int_0^{\, z_f} \frac{dz}{(1+z)
   \Htil(z)} \; .
\label{AgeDefn}
\eeq
In a static Universe, of course, redshift does not carry its usual
physical significance.  But nothing prevents us from retaining $z$
as an integration variable.  Substitution of (\ref{AgeDefn}) into
(\ref{QStDefn}) then yields
\beq
\Qstat = \Qstar \int_{0}^{z_f} \frac{dz}{(1+z) \Htil(z)} \; .
\label{QSzDefn}
\eeq
We emphasize that $z$ and $\Htil$ are to be seen here as algebraic
parameters whose usefulness lies in the fact that they ensure
consistency in {\em age\/} between the static and expanding pictures.

Eq.~(\ref{QzDefn2}) and its static analog (\ref{QSzDefn}) allow us to
isolate the relative importance of expansion versus lifetime in
determining the intensity of the EBL.  The procedure is straightforward:
evaluate the ratio $Q/\Qstat$ for all reasonable values of the cosmological
parameters $\Orado,\Omato$ and $\Olamo$.  If $Q/\Qstat \ll 1$ over much of
this phase space, then expansion must reduce $Q$ significantly from what
it would otherwise be in a static Universe.  Conversely, values of
$Q/\Qstat \approx 1$ would tell us that expansion has little effect,
and that (as in the static case) the brightness of the night sky is
determined primarily by the length of time for which the galaxies
have been shining.

\subsection{Flat single-component models}

We begin by evaluating $Q/\Qstat$ for the simplest cosmological models,
those in which the Universe has one critical-density component or contains
nothing at all (Table~\ref{table2.1}).
\begin{table}[b!]
\caption{Simple flat and empty models}
\begin{tabular}{@{}lcccc@{}}
\hline
Model Name & $\Orado$ & $\Omato$ & $\Olamo$ & $1-\Ototo$ \\
\hline
Radiation & 1 & 0 & 0 & 0 \\
Einstein-de Sitter & 0 & 1 & 0 & 0 \\
de Sitter & 0 & 0 & 1 & 0 \\
Milne & 0 & 0 & 0 & 1 \\
\hline
\end{tabular}
\label{table2.1}
\end{table}                                                                     
Consider first the {\em radiation model\/} with a critical density of
radiation or ultra-relativistic particles ($\Orado=1$) but $\Omato=\Olamo=0$.
Bolometric EBL intensity in the expanding Universe is, from (\ref{QzDefn2})
\beq
\frac{Q}{\Qstar} = \int_1^{1+z_f} \frac{dx}{x^4}
 = \left\{ \begin{array}{cc}
     21/64 & (z_f=3) \\
     1/3 & (z_f=\infty)
     \end{array} \right. \; ,
\eeq
where $x\equiv1+z$.  The corresponding result for a static model is given
by (\ref{QSzDefn}) as
\beq
\frac{\Qstat}{\Qstar} =\int_1^{1+z_f} \frac{dx}{x^3}
 = \left\{ \begin{array}{cc}
     15/32 & (z_f=3) \\
     1/2 & (z_f=\infty)
     \end{array} \right. \; .
\label{StatRad}
\eeq
Here we have chosen illustrative lower and upper limits on the redshift
of galaxy formation ($z_f=3$ and $\infty$ respectively).  The actual value
of this parameter has not yet been determined, although there are now
indications that $z_f$ may be as high as six.  In any case, it may be seen
that overall EBL intensity is rather insensitive to this parameter.
Increasing $z_f$ lengthens the period over which galaxies radiate,
and this increases both $Q$ and $\Qstat$.  The ratio $Q/\Qstat$,
however, is given by
\beq
\frac{Q}{\Qstat} = \left\{ \begin{array}{cc}
   7/10 & (z_f=3) \\
   2/3 & (z_f=\infty)
   \end{array} \right. \; ,
\eeq
and this changes but little.  We will find this to be true in general.

Consider next the Einstein-de~Sitter model,
which has a critical density of dust-like matter ($\Omato=1$)
with $\Orado=\Olamo=0$.  Bolometric EBL intensity in the expanding
Universe is, from (\ref{QzDefn2})
\beq
\frac{Q}{\Qstar} = \int_1^{1+z_f} \frac{dx}{x^{7/2}}
 = \left\{ \begin{array}{cc}
       31/80 & (z_f=3) \\
       2/5 & (z_f=\infty)
       \end{array} \right. \; .
\label{QQstarEdS}
\eeq
The corresponding static result is given by (\ref{QSzDefn}) as
\beq
\frac{\Qstat}{\Qstar} = \int_1^{1+z_f} \frac{dx}{x^{5/2}}
 = \left\{ \begin{array}{cc}
       7/12 & (z_f=3) \\
       2/3 & (z_f=\infty)
       \end{array} \right. \; .
\eeq
The ratio of EBL intensity in an expanding Einstein-de~Sitter model to that
in the equivalent static model is thus
\beq
\frac{Q}{\Qstat} = \left\{ \begin{array}{cc}
   93/140 & (z_f=3) \\
   3/5 & (z_f=\infty)
   \end{array} \right. \; .
\eeq
A third simple case is the {\em de~Sitter model\/}, which consists entirely
of vacuum energy ($\Olamo=1$), with $\Orado=\Omato=0$.  Bolometric EBL
intensity in the expanding case is, from (\ref{QzDefn2})
\beq
\frac{Q}{\Qstar} = \int_1^{1+z_f} \frac{dx}{x^2}
 = \left\{ \begin{array}{cc}
       3/4 & (z_f=3) \\
       1 & (z_f=\infty)
       \end{array} \right. \; .
\label{ExpdeSit}
\eeq
Eq.~(\ref{QSzDefn}) gives for the equivalent static case
\beq
\frac{\Qstat}{\Qstar} = \int_1^{1+z_f} \frac{dx}{x}
 = \left\{ \begin{array}{cc}
       \ln 4 & (z_f=3) \\
       \infty & (z_f=\infty)
       \end{array} \right. \; .
\eeq
The ratio of EBL intensity in an expanding de~Sitter model to that
in the equivalent static model is then
\beq
\frac{Q}{\Qstat} = \left\{ \begin{array}{cc}
   3/(4\ln 4) & (z_f=3) \\
   0 & (z_f=\infty)
   \end{array} \right. \; .
\eeq
The de~Sitter Universe is older than other models, which means it has
more time to fill up with light, so intensities are higher.  In fact,
$\Qstat$ (which is proportional to the lifetime of the galaxies) goes
to infinity as $z_f \rightarrow \infty$, driving $Q/\Qstat$ to zero in
this limit.  (It is thus possible in principle to ``recover Olbers'
paradox'' in the de~Sitter model, as noted by White and Scott \cite{Whi96}.)
Such a limit is however unphysical in the context of the EBL, since the
ages of galaxies and their component stars are bounded from above.
For realistic values of $z_f$ one obtains values of $Q/\Qstat$ which
are only slightly lower than those in the radiation and matter cases.

Finally, we consider the {\em Milne model\/}, which is empty of all forms
of matter and energy ($\Orado=\Omato=\Olamo=0$), making it an idealization
but one which has often proved useful.  Bolometric EBL intensity in the
expanding case is given by (\ref{QzDefn2}) and turns out to be identical
to Eq.~(\ref{StatRad}) for the static radiation model.
The corresponding static result, as given by (\ref{QSzDefn}), 
turns out to be the same as Eq.~(\ref{ExpdeSit}) for the expanding de~Sitter
model.  The ratio of EBL intensity in an expanding Milne model to that
in the equivalent static model is then
\beq
\frac{Q}{\Qstat} = \left\{ \begin{array}{cc}
   5/8 & (z_f=3) \\
   1/2 & (z_f=\infty)
   \end{array} \right. \; .
\eeq
This again lies close to previous results.
In all cases (except the $z_f \rightarrow \infty$ limit of the de~Sitter
model) the ratio of bolometric EBL intensities with and without expansion
lies in the range $0.4 \lesssim Q/\Qstat \lesssim 0.7$.

\subsection{Curved and multi-component models}

\begin{figure}[t!]
\begin{center}
\includegraphics[width=90mm]{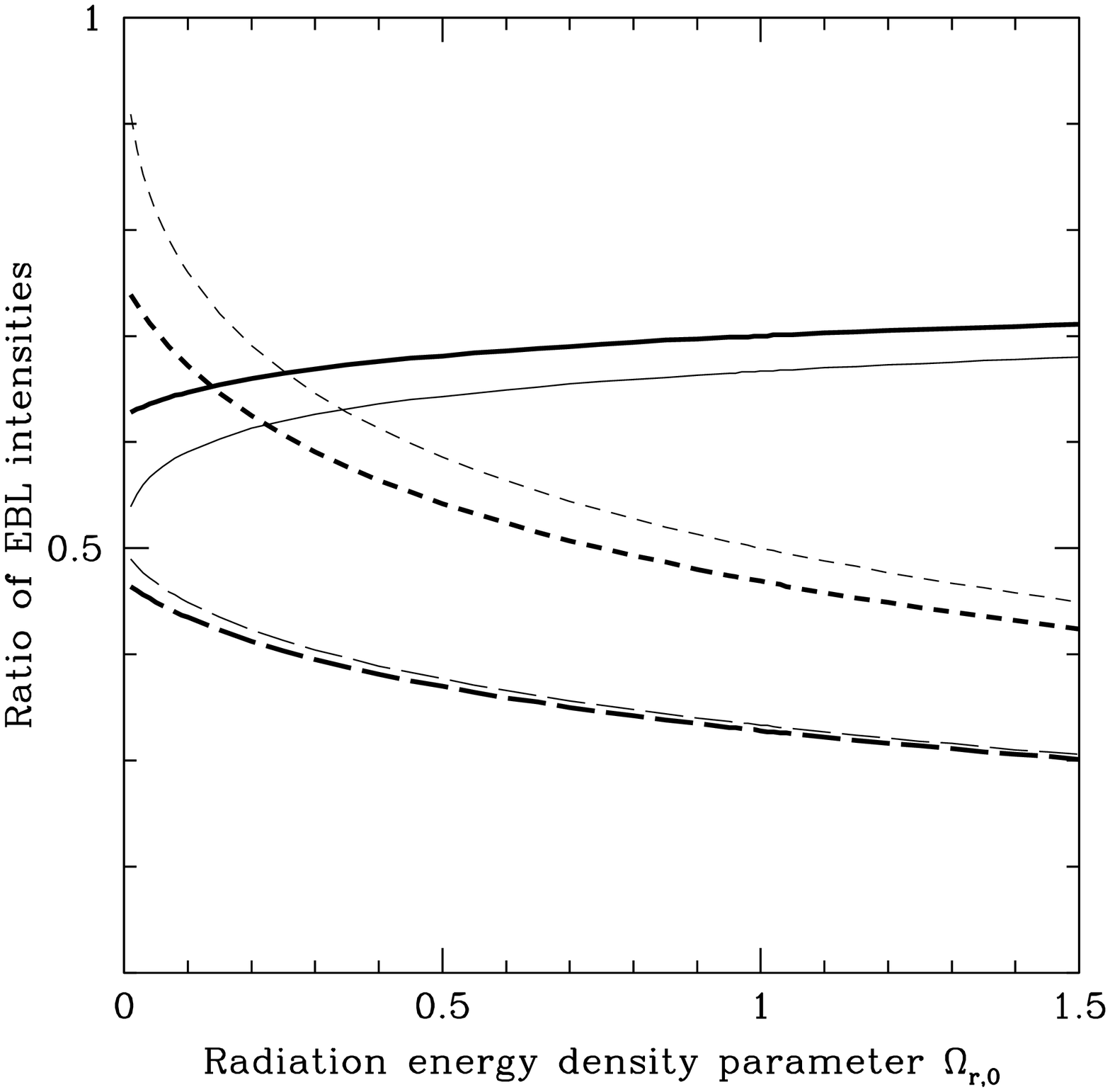}
\end{center}
\caption{Ratios $Q/\Qstar$ (long-dashed lines), $\Qstat/\Qstar$
   (short-dashed lines) and $Q/\Qstat$ (solid lines) as a function of
   radiation density $\Orado$.  Heavier lines are calculated for
   $z_f=3$ while lighter ones have $z_f=\infty$.}
\label{fig2.3}
\end{figure}

\begin{figure}[b!]
\begin{center}
\includegraphics[width=90mm]{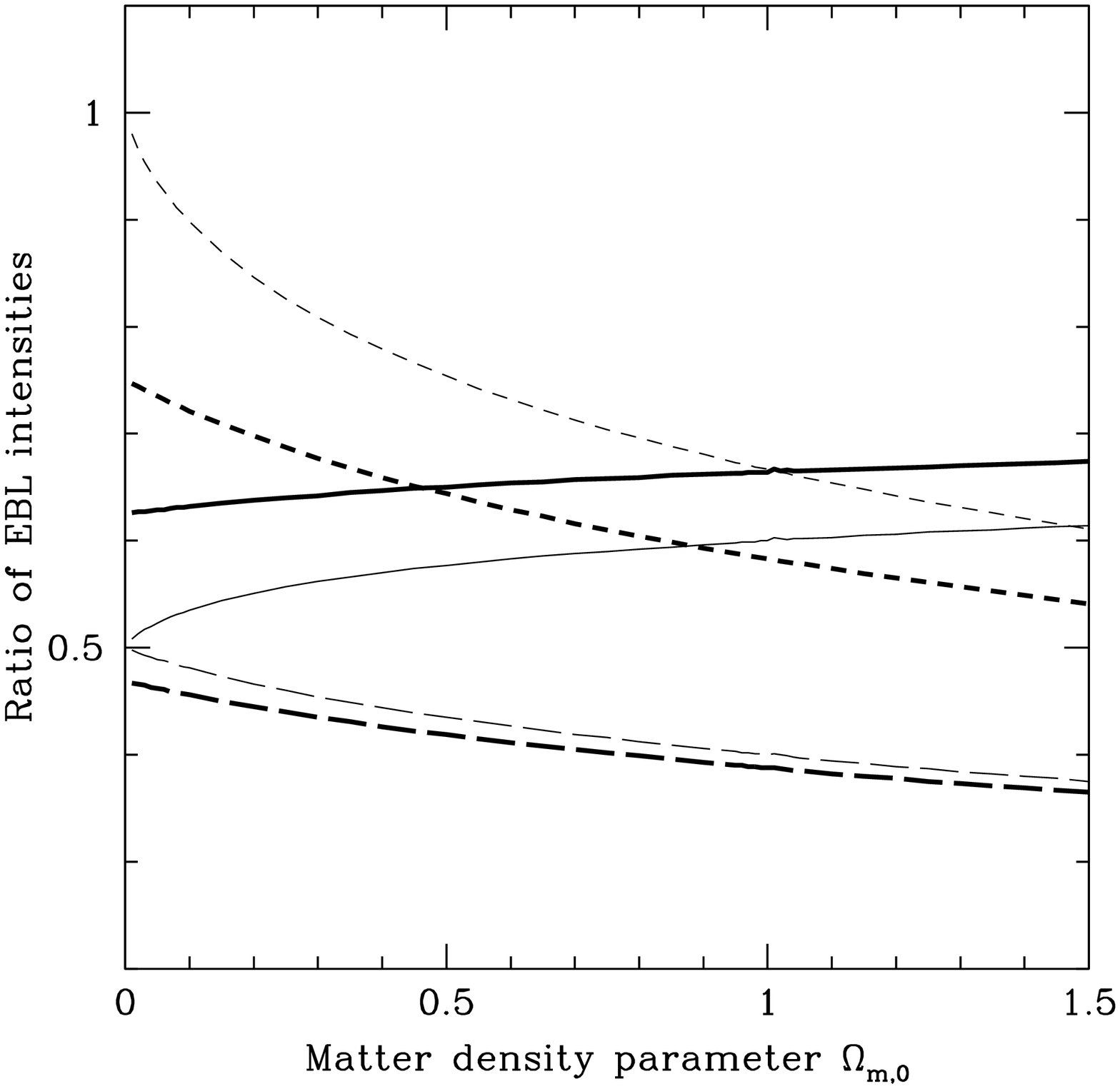}
\end{center}
\caption{Ratios $Q/\Qstar$ (long-dashed lines), $\Qstat/\Qstar$
   (short-dashed lines) and $Q/\Qstat$ (solid lines) as a function of
   matter density $\Omato$.  Heavier lines are calculated for $z_f=3$
   while lighter ones have $z_f=\infty$.}
\label{fig2.4}
\end{figure}

\begin{figure}[t!]
\begin{center}
\includegraphics[width=90mm]{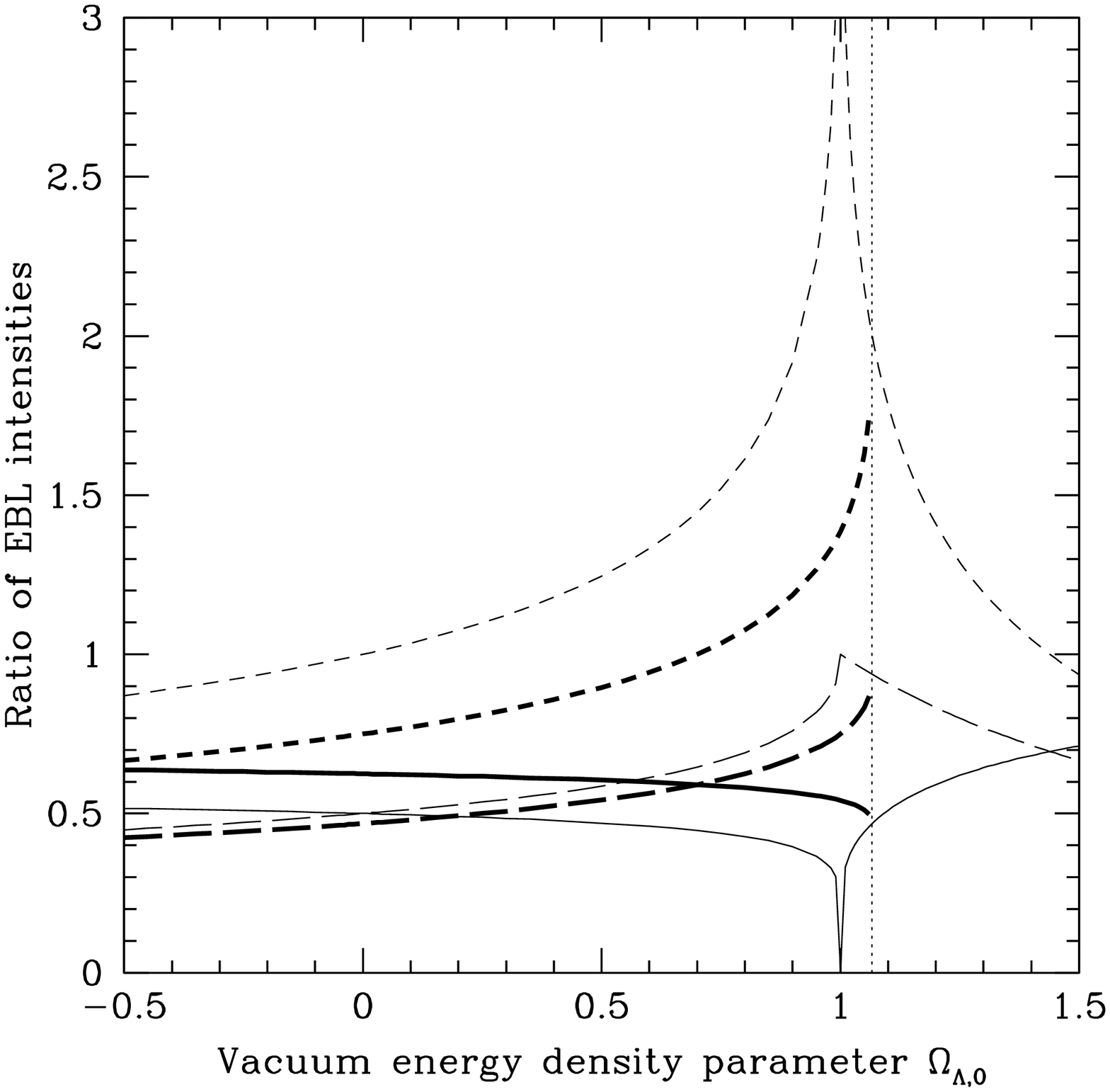}
\end{center}
\caption{Ratios $Q/\Qstar$ (long-dashed lines), $\Qstat/\Qstar$
   (short-dashed lines) and $Q/\Qstat$ (solid lines) as a function of
   vacuum energy density $\Olamo$.  Heavier lines are calculated for
   $z_f=3$ while lighter ones have $z_f=\infty$ for $\Olamo\leqslant1$
   and $z_f=\zmax$ for $\Olamo>1$.  The dotted vertical line marks the
   maximum value of $\Olamo$ for which one can integrate to $z_f=3$.}
\label{fig2.5}
\end{figure}

To see whether the pattern observed in the previous section holds
more generally, we expand our investigation to the wider class of open
and closed models.  Eqs.~(\ref{QzDefn2}) and (\ref{QSzDefn}) may be
solved analytically for these cases, if they are dominated by a single
component \cite{OW03}.  We plot the results in Figs.~\ref{fig2.3},
\ref{fig2.4} and \ref{fig2.5} for radiation-, matter- and
vacuum-dominated models respectively.
In each figure, long-dashed lines correspond to EBL intensity in
expanding models ($Q/\Qstar$) while short-dashed ones show the
equivalent static quantities ($\Qstat/\Qstar$).  The ratio of these
two quantities ($Q/\Qstat$) is indicated by solid lines.  Heavy lines
have $z_f=3$ while light ones are calculated for $z_f=\infty$.


Figs.~\ref{fig2.3} and \ref{fig2.4} show that while the individual
intensities $Q/\Qstar$ and $\Qstat/\Qstar$ do vary significantly with
$\Orado$ and $\Omato$ in radiation- and matter-dominated models,
their ratio $Q/\Qstat$ remains nearly constant across the whole of
the phase space, staying inside the range
$0.5 \lesssim Q/\Qstat \lesssim 0.7$ for both models.

Fig.~\ref{fig2.5} shows that a similar trend occurs in vacuum-dominated
models.  While absolute EBL intensities $Q/\Qstar$ and $\Qstat/\Qstar$
differ from those in the radiation- and matter-dominated models, 
their {\em ratio\/} (solid lines) is again close to flat.
The exception occurs as $\Olamo\rightarrow1$ (de~Sitter model),
where $Q/\Qstat$ dips well below 0.5 for large $z_f$.  For 
$\Olamo > 1$, there is no big bang (in models with $\Orado=\Omato=0$),
and one has instead a ``big bounce'' (i.e. a nonzero scale factor at the
beginning of the expansionary phase).  This implies a maximum possible
redshift $\zmax$ given by
\beq
1 + \zmax = \sqrt{\frac{\Olamo}{\Olamo - 1}} \; .
   \label{zmaxDefn}
\eeq
While such models are rarely considered, it is interesting to note
that the same pattern persists here, albeit with one or two wrinkles.
In light of (\ref{zmaxDefn}), one can no longer integrate out to
arbitarily high formation redshift $z_f$.  If one wants to integrate to
{\em at least\/} $z_f$, then one is limited to vacuum densities less than
$\Olamo<(1+z_f)^2/[(1+z_f)^2-1]$, or $\Olamo<16/15$ for the case
$z_f=3$ (heavy dotted line).
More generally, for $\Olamo>1$ the limiting value of EBL
intensity (shown with light lines) is reached as $z_f \rightarrow \zmax$
rather than $z_f\rightarrow\infty$ for both expanding and static models.
Over the entire parameter space $-0.5\leqslant\Olamo\leqslant1.5$ (except
in the immediate vicinity of $\Olamo=1$), Fig.~\ref{fig2.5} shows that
$0.4 \lesssim Q/\Qstat \lesssim 0.7$ as before.

\begin{figure}[t!]
\begin{center}
\includegraphics[width=90mm]{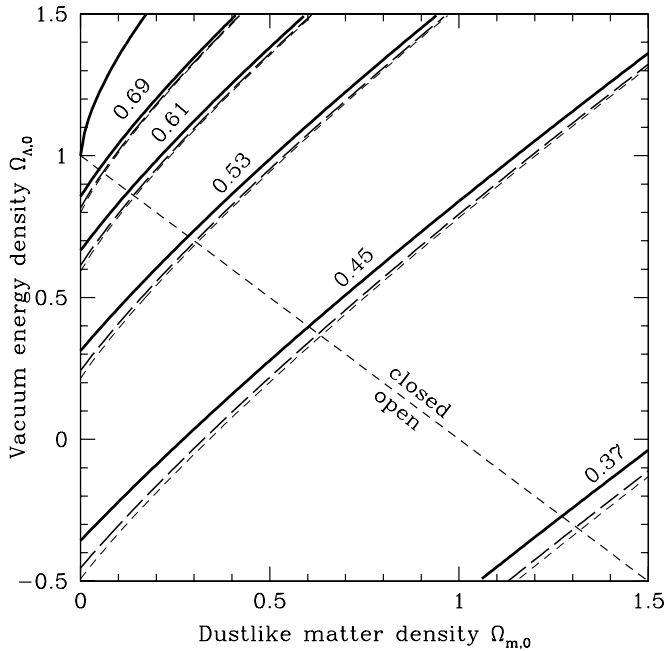}
\end{center}
\caption{The ratio $Q/\Qstar$ in an expanding Universe, plotted as a function
   of matter density parameter $\Omato$ and vacuum density parameter $\Olamo$
   (with the radiation density parameter $\Orado$ set to zero).  Solid lines
   correspond to $z_f=5$, while long-dashed lines assume $z_f=10$ and
   short-dashed ones have $z_f=50$.}
\label{fig2.6}
\end{figure}

When more than one component of matter is present, analytic expressions
can be found in only a few special cases, and the ratios $Q/\Qstar$ and
$\Qstat/\Qstar$ must in general be computed numerically.
We show the results in Fig.~\ref{fig2.6} for the situation which
is of most physical interest: a universe containing both dust-like matter
($\Omato$, horizontal axis) and vacuum energy ($\Olamo$, vertical axis),
with $\Orado=0$.
This is a contour plot, with five bundles of equal-EBL intensity contours
for the expanding Universe (labelled $Q/\Qstar=0.37, 0.45, 0.53, 0.61$ and
$0.69$).  The heaviest (solid) lines are calculated for $z_f=5$, while
medium-weight (long-dashed) lines assume $z_f=10$ and the lightest
(short-dashed) lines have $z_f=50$.  Also shown is the boundary between
big bang and bounce models (heavy solid line in top left corner), and the
boundary between open and closed models (diagonal dashed line).

Fig.~\ref{fig2.6} shows that the bolometric intensity of the EBL is only
modestly sensitive to the cosmological parameters $\Omato$ and $\Olamo$.
Moving from the lower right-hand corner of the phase space ($Q/\Qstar=0.37$)
to the upper left-hand one ($Q/\Qstar=0.69$) changes the value of this
quantity by less than a factor of two.  Increasing the redshift of
galaxy formation from $z_f=5$ to 10 has little effect, and increasing
it again to $z_f=50$ even less.  This means that, regardless of the redshift
at which galaxies actually form, {\em essentially all of the light reaching
us from outside the Milky Way comes from galaxies at\/} $z<5$.

\begin{figure}[t!]
\begin{center}
\includegraphics[width=90mm]{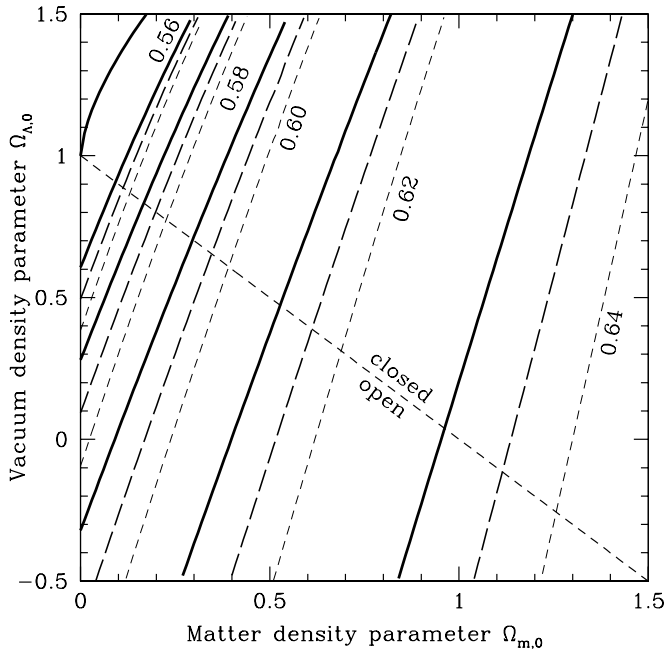}
\end{center}
\caption{The ratio $Q/\Qstat$ of EBL intensity in an expanding Universe
   to that in a static universe with the same values of the matter 
   density parameter $\Omato$ and vacuum density parameter $\Olamo$
   (with the radiation density parameter $\Orado$ set to zero).
   Solid lines correspond to $z_f=4.5$, while long-dashed lines assume
   $z_f=5$ and short-dashed ones have $z_f=5.5$.}
\label{fig2.7}
\end{figure}                                                                    

While Fig.~\ref{fig2.6} confirms that the night sky is dark in any reasonable
cosmological model, Fig.~\ref{fig2.7} shows {\em why\/}.  It is a contour plot
of $Q/\Qstat$, the value of which varies so little across the phase space
that we have restricted the range of $z_f$-values to keep the diagram from
getting too cluttered.  The heavy (solid) lines are calculated for
$z_f=4.5$, the medium-weight (long-dashed) lines for $z_f=5$,
and the lightest (short-dashed) lines for $z_f=5.5$.  The spread in contour
values is extremely narrow, from $Q/\Qstat=0.56$ in the upper left-hand
corner to 0.64 in the lower right-hand one --- a difference of less than 15\%.
Fig.~\ref{fig2.7} confirms our previous analytical results and leaves no
doubt about the resolution of Olbers' paradox:
{\em the brightness of the night sky is determined to order
of magnitude by the lifetime of the galaxies, and reduced by a factor of
only $0.6 \pm 0.1$ due to the expansion of the Universe.}

\subsection{A look ahead}

In this section, we inquire into the evolution of bolometric EBL
intensity $Q(t)$ with time, as specified mathematically by 
Eq.~(\ref{QtDefn}).  Evaluation of this integral requires a knowledge
of $R(t)$, which is not well constrained by observation.  Exact
solutions are however available under the assumption that the Universe
is spatially flat, as currently suggested by data on the power spectrum
of fluctuations in the cosmic microwave background (Sec.~\ref{ch4}).
In this case $k=0$ and (\ref{FL1}) simplifies to
\beq
\left( \! \frac{\dot{R}}{R} \! \right)^2 \! \! \! = \frac{8\pi G}{3} \left(
   \rrad + \rmat + \rlam \right) \; ,
\label{FLflat}
\eeq
where we have taken $\rho = \rrad + \rmat$ in general and used 
(\ref{rlamDefn}) for $\rlam$.  If one of these three components is dominant
at a given time, then we can make use of (\ref{rhoR}) to obtain
\beq
\left( \! \frac{\dot{R}}{R} \! \right)^2 \! \! \! = \frac{8\pi G}{3} \! 
   \times \!  \left\{ \begin{array}{ll}
   \rrado (R/\Ro)^{-4} & \hspace{1cm} \mbox{(radiation)} \\
   \rmato (R/\Ro)^{-3} & \hspace{1cm} \mbox{(matter)} \\
   \rlam & \hspace{1cm} \mbox{(vacuum)}
   \end{array} \right. \; .
   \label{FL1comp}
\eeq
These differential equations are solved to give
\beq
\frac{R(t)}{\Ro} = \left\{ \begin{array}{ll}
   (t/\too)^{1/2} & \hspace{1cm} \mbox{(radiation)} \\
   (t/\too)^{2/3} & \hspace{1cm} \mbox{(matter)} \\
   \exp[\Ho(t-\too)] & \hspace{1cm} \mbox{(vacuum)}
   \end{array} \right. \; .
   \label{Rt1comp}
\eeq
We emphasize that these expressions assume (i)~spatial flatness, and
(ii)~a single-component cosmic fluid which must have the critical density.

Putting (\ref{Rt1comp}) into (\ref{QtDefn}), we can solve for
the bolometric intensity under the assumption that the luminosity of the
galaxies is constant over their lifetimes, $L(t)=\Lo$:
\beq
\frac{Q(t)}{\Qstar} = \left\{ \begin{array}{ll}
   (1/3) (t/\too)^{3/2} & \hspace{1cm} \mbox{(radiation)} \\
   (2/5) (t/\too)^{5/3} & \hspace{1cm} \mbox{(matter)} \\
   \exp\left[\Ho\too ( t/\too-1) \right] & \hspace{1cm} \mbox{(vacuum)}
   \end{array} \right. \; ,
   \label{QtValues}
\eeq
where we have used (\ref{QstarValue}) and assumed that $t_f\ll\too$
and $t_f\ll t$.

The intensity of the light reaching us from intergalactic space climbs
as $t^{\, 3/2}$ in a radiation-filled Universe, $t^{\, 5/3}$ in a
matter-dominated one, and $\exp(\Ho t)$ in one which contains only
vacuum energy.  This happens because the horizon of the Universe
expands to encompass more and more galaxies, and hence more photons.
Clearly it does so at a rate which more than compensates for the dilution
and redshifting of existing photons due to expansion.  Suppose for
argument's sake that this state of affairs could continue indefinitely.
How long would it take for the night sky to become as bright as, say,
the interior of a typical living-room containing a single 100~W bulb
and a summed surface area of 100~m$^2$ ($\Qbulb=1000$~erg~cm$^{-2}$~s$^{-1}$)?

The required increase of $Q(t)$ over $\Qstar$
$(=2.5 \times 10^{-4}$~erg~cm$^{-2}$~s$^{-1}$) is 4~million times.
Eq.~(\ref{QtValues}) then implies that
\beq
t \approx \left\{ \begin{array}{ll}
   730\,000 \mbox{ Gyr} & \hspace{1cm} \mbox{(radiation)} \\
   220\,000 \mbox{ Gyr} & \hspace{1cm} \mbox{(matter)} \\
   230 \mbox{ Gyr} & \hspace{1cm} \mbox{(vacuum)}
   \end{array} \right. \; ,
\label{Ages}
\eeq
where we have taken $\Ho\too\approx 1$ and $\too\approx 14$~Gyr, as
suggested by the observational data (Sec.~\ref{ch4}).
The last of the numbers in (\ref{Ages}) is particularly intriguing.
In a vacuum-dominated universe in which the luminosity of the galaxies
could be kept constant indefinitely, the night sky would fill up with light
over timescales of the same order as the theoretical hydrogen-burning
lifetimes of the longest-lived stars.
Of course, the luminosity of galaxies {\em cannot\/} stay constant
over these timescales, because most of their light comes from much
more massive stars which burn themselves out after tens of Gyr or less.

To check whether the situation just described is perhaps an artefact of
the empty de~Sitter universe, we require an expression for $R(t)$ in models
with both dust-like matter and vacuum energy.  An analytic solution does
exist for such models, if they are flat (i.e. $\Olamo=1-\Omato$).
It reads \cite{OW03}:
\beq
\Rtil(t) = \left[ \sqrt{\frac{\Omato}{1-\Omato}} \sinh
   \left( \smallfrac{3}{2} \sqrt{1-\Omato} \, \Ho t \right)
   \right]^{2/3} \; .
\label{Rt2comp}
\eeq
This formula receives surprisingly little attention, given the importance
of vacuum-dominated models in modern cosmology.  Differentiation with
respect to time gives the Hubble expansion rate:
\beq
\Htil(t) = \sqrt{1-\Omato} \, \coth
   \left( \smallfrac{3}{2} \sqrt{1-\Omato} \, \Ho t \right) \; .
\label{Ht2comp}
\eeq
This goes over to $\Olamo^{1/2}$ as $t\rightarrow\infty$, a result which
(as noted in Sec.~\ref{sec:FL}) holds quite generally for models with
$\Lambda>0$.  Alternatively, setting $\Rtil=(1+z)^{-1}$ in (\ref{Rt2comp})
gives the age of the Universe at a redshift $z$:
\beq
t = \frac{2}{3\Ho\sqrt{1-\Omato}} \, \sinh^{-1}
   \!\!\sqrt{\frac{1-\Omato}{\Omato(1+z)^3}} \; .
\label{t2comp}
\eeq
Setting $z=0$ in this equation gives the present age of the Universe ($\too$).
Thus a model with (say) $\Omato=0.3$ and $\Olamo=0.7$ has an age of
$\too=0.96\Ho^{-1}$ or, using (\ref{HoValue}), $\too=9.5\ho^{-1}$~Gyr.
Alternatively, in the limit $\Omato\rightarrow1$, Eq.~(\ref{t2comp})
gives back the standard result for the age of an EdS Universe,
$\too=2/(3\Ho)=6.5\ho^{-1}$~Gyr.

\begin{figure}[t!]
\begin{center}
\includegraphics[width=90mm]{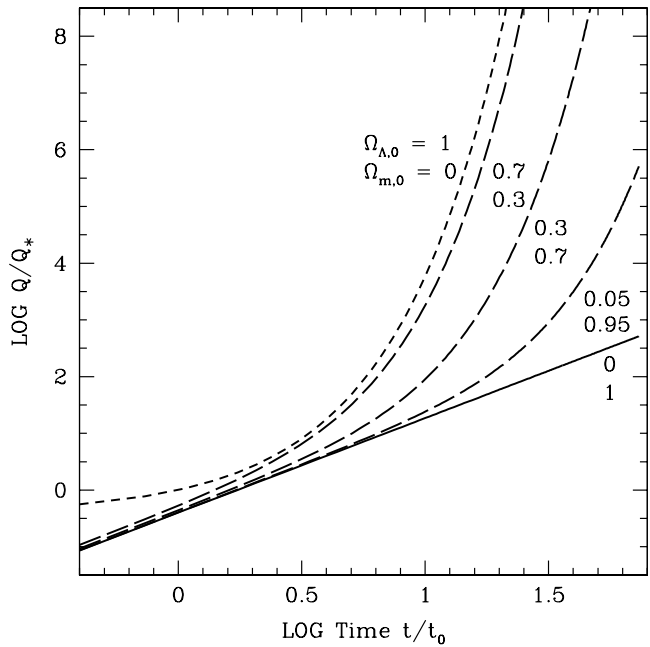}
\end{center}
\caption{Plots of $Q(t)/\Qstar$ in flat models containing both dust-like
   matter ($\Omato$) and vacuum energy ($\Olamo$).  The solid line is
   the Einstein-de~Sitter model, while the short-dashed line is pure
   de~Sitter, and long-dashed lines represent intermediate models.
   The curves do not meet at $t=\too$ because $Q(\too)$ differs from
   model to model, with $Q(\too)=\Qstar$ only for the pure de~Sitter case.}
\label{fig2.8}
\end{figure}                                                                    

Putting (\ref{Rt2comp}) into Eq.~(\ref{QtDefn}) with $L(t)=\Lo=$~constant,
and integrating over time, we obtain the plots of bolometric EBL intensity
shown in Fig.~\ref{fig2.8}.  This diagram shows that the specter of a
rapidly-brightening Universe does not occur only in the pure de~Sitter
model (short-dashed line).  A model with an admixture of dust-like matter
and vacuum energy such that $\Omato=0.3$ and $\Olamo=0.7$, for instance,
takes only slightly longer (280~Gyr) to attain the ``living-room'' intensity
of $\Qbulb=1000$~erg~cm$^{-2}$~s$^{-1}$.

In theory then, it might be thought that our remote descendants could
live under skies perpetually ablaze with the light of distant galaxies,
in which the rising and setting of their home suns would barely be noticed.
This will not happen in practice, of course, because galaxy luminosities
change with time as their brightest stars burn out.
The lifetime of the galaxies is critical, in other words,
{\em not only in the sense of a finite past, but a finite
future\/}.  A proper assessment of this requires that we move from
considerations of background cosmology to the astrophysics of the
sources themselves.

\section{The spectrum of cosmic background radiation} \label{ch3}

\subsection{Spectral intensity}

The spectra of real galaxies depend strongly on wavelength and also
evolve with time.  How might these facts alter the conclusion obtained
in Sec.~\ref{ch2}; namely, that the brightness of the night sky
is overwhelmingly determined by the age of the Universe,
with expansion playing only a minor role?

The significance of this question is best appreciated in the microwave
portion of the electromagnetic spectrum (at wavelengths from about
1~mm to 10~cm) where we know from decades of radio astronomy that the
``night sky'' is brighter than its optical counterpart (Fig.~\ref{fig1.2}).
The majority of this microwave background radiation is thought to come, not
from the redshifted light of distant galaxies, but from the fading glow
of the big bang itself --- the ``ashes and smoke'' of creation in
Lema\^{\i}tre's words.  Since its nature and suspected origin are different
from those of the EBL, this part of the spectrum has its own name,
the {\em cosmic microwave background\/} (CMB).  Here expansion is of
paramount importance, since the source radiation in this case was emitted
at more or less a single instant in cosmological history (so that the
``lifetime of the sources'' is negligible).  Another way to see this is
to take expansion out of the picture, as we did in Sec.~\ref{sec:olbers}:
the CMB intensity we would observe in this ``equivalent static model''
would be that of the primordial fireball and would roast us alive.

While Olbers' paradox involves the EBL, not the CMB, this example is still
instructive because it prompts us to consider whether similar (though less
pronounced) effects could have been operative in the EBL as well.
If, for instance, galaxies emitted most of their light in a relatively
brief burst of star formation at very early times, this would be a galactic
approximation to the picture just described, and could conceivably boost
the importance of expansion relative to lifetime, at least in some
wavebands.  To check on this, we need a way to calculate EBL intensity
as a function of wavelength.  This is motivated by other considerations
as well.  Olbers' paradox has historically been concerned primarily
with the {\em optical\/} waveband (from approximately 4000\AA\ to
8000\AA), and this is still what most people mean when they refer to
the ``brightness of the night sky.''  And from a practical standpoint,
we would like to compare our theoretical predictions with observational
data, and these are necessarily taken using detectors which are optimized
for finite portions of the electromagnetic spectrum.

We therefore adapt the bolometric formalism of Sec.~\ref{ch2}.  Instead of
total luminosity $L$, consider the energy emitted by a source per unit time
between wavelengths $\lambda$ and $\lambda+d\lambda$.  Let us write this
in the form $dL_{\lambda} \equiv F(\lambda,t) \, d\lambda$
where $F(\lambda,t)$ is the {\em spectral energy distribution\/} (SED),
with dimensions of energy per unit time per unit wavelength.
Luminosity is recovered by integrating the SED over all wavelengths:
\beq
L(t) = \int_0^\infty dL_{\lambda} = \int_0^\infty F(\lambda,t) \, d\lambda \; .
\label{Fnorm}
\eeq
We then return to (\ref{dQdt}), the bolometric intensity of the
spherical shell of galaxies depicted in Fig.~\ref{fig2.1}.
Replacing $L(t)$ with $dL_{\lambda}$ in this equation gives the
intensity of light emitted between $\lambda$ and $\lambda+d\lambda$:
\beq
\dQlame = c n(t) \Rtil(t) [ F(\lambda,t) \, d\lambda ] dt \; .
\label{dQe}
\eeq
This light reaches us at the redshifted wavelength $\lamo=\lambda/\Rtil(t)$.
Redshift also stretches the wavelength interval by the same factor,
$d\lamo=d\lambda/\Rtil(t)$.  So the intensity of light {\em observed\/}
by us between $\lamo$ and $\lamo+d\lamo$ is
\beq
\dQlamo = c n(t) \Rtil^2(t) F[\Rtil(t) \lamo,t] \, d\lamo dt \; .
\label{dQo}
\eeq
The intensity of the shell {\em per unit wavelength\/}, as observed at
wavelength $\lamo$, is then given simply by
\beq
4\pi \, d\Ilam(\lamo) \equiv \frac{\dQlamo}{d\lamo} =
   c n(t) \Rtil^2(t) F[\Rtil(t) \lamo,t] \, d\lamo dt \; ,
\eeq
where the factor $4\pi$ converts from an all-sky intensity to one
measured per steradian.  (This is merely a convention, but has become
standard.)  Integrating over all the spherical shells corresponding to
times $\too$ and $\too-t_f$ (as before) we obtain the spectral analog of
our earlier bolometric result, Eq.~(\ref{QtDefn}):
\beq
\Ilam(\lamo) = \frac{c}{4\pi} \int_{t_f}^{\too} n(t) F[\Rtil(t) \lamo,t] 
   \Rtil^2(t) \, dt \; .
\label{ItDefn}
\eeq
This is the integrated light from many galaxies, which has been emitted
at various wavelengths and redshifted by various amounts, but which is all
in the waveband centered on $\lamo$ when it arrives at us.  We refer to this
as the {\em spectral intensity of the EBL\/} at $\lamo$.
Eq.~(\ref{ItDefn}), or ones like it, have been considered from the
theoretical side principally by McVittie and Wyatt \cite{McV59},
Whitrow and Yallop \cite{Whi64,Whi65} and Wesson \etal\ \cite{WVS87,Wes91a}.

Eq.~(\ref{ItDefn}) can be converted from an integral over $t$ to one over
$z$ by means of Eq.~(\ref{dtdz}) as before.  This gives
\beq
\Ilam(\lamo) = \frac{c}{4\pi\Ho} \int_0^{\, z_f} \frac{n(z)
   F[\lamo/(1+z),z] \, dz} {(1+z)^3 \Htil(z)} \; .
\label{IzDefn}
\eeq
Eq.~(\ref{IzDefn}) is the spectral analog of (\ref{QzDefn}).  It may be
checked using (\ref{Fnorm}) that bolometric intensity is just the integral
of spectral intensity over all observed wavelengths,
$Q=\int_0^\infty I(\lamo)d\lamo$.  Eqs.~(\ref{ItDefn}) and (\ref{IzDefn})
provide us with the means to constrain any kind of radiation source by
means of its contributions to the background light, once its number density
$n(z)$ and energy spectrum $F(\lambda,z)$ are known.  In subsequent
sections we will apply them to various species of dark (or not so dark)
energy and matter.

In this section, we return to the question of lifetime and the EBL.
The static analog of Eq.~(\ref{ItDefn}) (i.e. the equivalent spectral EBL
intensity in a universe without expansion, but with the properties of the
galaxies unchanged) is obtained exactly as in the bolometric case by setting
$\Rtil(t)=1$ (Sec.~\ref{sec:olbers}):
\beq
\IlamS(\lamo) = \frac{c}{4\pi} \int_{t_f}^{\too} n(t) F(\lamo,t) \, dt \; .
\label{IStDefn}
\eeq
Just as before, we may convert this to an integral over $z$ if we choose.
The latter parameter no longer represents physical redshift (since this
has been eliminated by hypothesis), but is now merely an algebraic way
of expressing the age of the galaxies.  This is convenient because it
puts (\ref{IStDefn}) into a form which may be directly compared with
its counterpart (\ref{IzDefn}) in the expanding Universe:
\beq
\IlamS(\lamo) = \frac{c}{4\pi\Ho} \int_0^{\, z_f} \frac{n(z) F(\lamo,z) \, dz}
   {(1+z) \Htil(z)} \; .
\label{ISzDefn}
\eeq
If the same values are adopted for $\Ho$ and $z_f$, and the same
functional forms are used for $n(z),F(\lambda,z)$ and $\Htil(z)$, then
Eqs.~(\ref{IzDefn}) and (\ref{ISzDefn}) allow us to compare model
universes which are alike in every way, except that one is expanding
while the other stands still.

Some simplification of these expressions is obtained as before in
situations where the comoving source number density can be taken as
constant, $n(z)=\no$.  However, it is not possible to go farther and pull
all the dimensional content out of these integrals, as was done in the
bolometric case, until a specific form is postulated for the
SED $F(\lambda,z)$.

\subsection{Comoving luminosity density} \label{sec:LumDens}

The simplest possible source spectrum is one in which all the energy is
emitted at a single peak wavelength $\lamp$ at each redshift $z$, thus
\beq
F(\lambda,z) = F_p(z) \, \delta \left(\frac{\lambda}{\lamp}-1\right) \; .
\label{deltaSED}
\eeq
SEDs of this form are well-suited to sources of electromagnetic radiation
such as elementary particle decays, which are characterized by specific
decay energies and may occur in the dark-matter halos surrounding galaxies.
The $\delta$-function SED is not a realistic approximation for the spectra
of galaxies themselves, but we will apply it here in this context to lay
the foundation for later sections.

The function $F_p(z)$ is obtained in terms of the total source
luminosity $L(z)$ by normalizing over all wavelengths
\beq
L(z) \equiv \int_0^\infty F(\lambda,z) \, d\lambda = F_p(z) \lamp \; ,
\eeq
so that $F_p(z)=L(z)/\lamp$.  In the case of galaxies, a logical choice
for the characteristic wavelength $\lamp$ would be the peak wavelength
of a blackbody of ``typical'' stellar temperature.  Taking the Sun as
typical ($T=\Tsun=5770$K), this would be
$\lamp = (2.90$~mm~K)/T = 5020\AA\ from Wiens' law.  Distant galaxies
are seen chiefly during periods of intense starburst activity when many
stars are much hotter than the Sun, suggesting a shift toward shorter
wavelengths.  On the other hand, most of the short-wavelength light
produced in large starbursting galaxies (as much as 99\% in the most
massive cases) is absorbed within these galaxies by dust and re-radiated
in the infrared and microwave regions ($\lambda \gtrsim 10,000$\AA).
It is also important to keep in mind that while distant starburst galaxies
may be hotter and more luminous than local spirals and ellipticals, the
latter contribute most to EBL intensity by virtue of their numbers at
low redshift.  The best that one can do with a single characteristic
wavelength is to locate it somewhere within the B-band ($3600-5500$\AA).
For the purposes of this exercise we associate $\lamp$ with the nominal
center of this band, $\lamp = 4400$\AA, corresponding to a blackbody
temperature of 6590~K.

Substituting the SED (\ref{deltaSED}) into the spectral intensity
integral (\ref{IzDefn}) leads to
\beq
\Ilam(\lamo) = \frac{c}{4\pi\Ho\,\lamp} \int_0^{\, z_f} \frac{{\mathcal L}(z)}
   {(1+z)^3 \Htil(z)} \, \delta \left[ \frac{\lamo}{\lamp (1+z)} - 1
   \right] \, dz \; ,
\label{Idelta1}
\eeq
where we have introduced a new shorthand for the {\em comoving luminosity
density\/} of galaxies:
\beq
{\mathcal L}(z) \equiv n(z) L(z) \; .
\eeq
At redshift $z=0$ this takes the value $\curlyLo$, as given
by (\ref{curlyLoValue}).  Numerous studies have shown that the product of
$n(z)$ and $L(z)$ is approximately conserved with redshift, even when
the two quantities themselves appear to be evolving markedly.
So it would be reasonable to take ${\mathcal L}(z)=\curlyLo=$~const.
However, recent analyses have been able to benefit from
observational work at deeper redshifts, and a consensus is emerging
that ${\mathcal L}(z)$ does rise slowly but steadily with $z$, peaking in
the range $2 \lesssim z \lesssim 3$, and falling away sharply thereafter
\cite{Fuk96}.  This is consistent with a picture in which the first
generation of massive galaxy formation occurred near $z\sim3$, being
followed at lower redshifts by galaxies whose evolution proceeded 
more passively.

\begin{figure}[t!]
\begin{center}
\includegraphics[width=100mm]{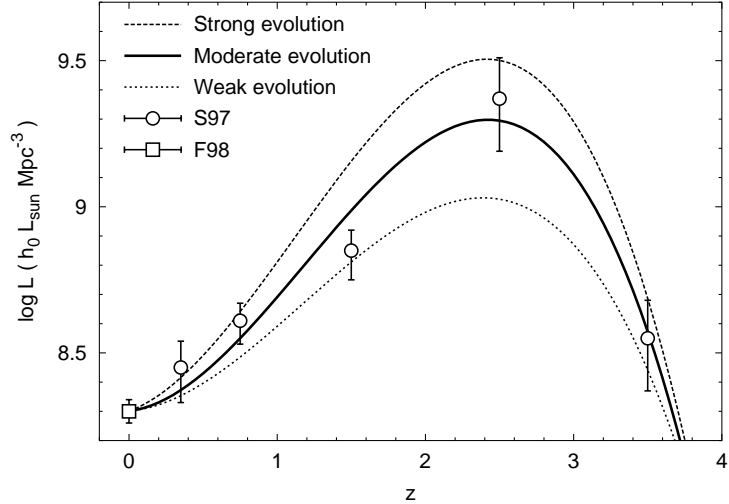}
\end{center}
\caption{The comoving luminosity density of the Universe $\curlyL(z)$
   (in $\ho L_{\odot}$~Mpc$^{-3}$), as observed at $z=0$ (square) and
   extrapolated to higher redshifts based on analysis of the Hubble
   Deep Field (circles).  The solid curve is a least-squares fit to the
   data; dashed lines represent upper and lower limits.}
\label{fig3.1}
\end{figure}

Fig.~\ref{fig3.1} shows the value of $\curlyLo$ from (\ref{curlyLoValue})
at $z=0$ \cite{Fuk98a} together with the extrapolation of ${\mathcal L}(z)$
to five higher redshifts from an analysis of photometric galaxy redshifts
in the Hubble Deep Field ({\sc Hdf}) \cite{Saw97}.  We define a
{\em relative comoving luminosity density\/} $\Ltil(z)$ by
\beq
\Ltil(z) \equiv {\mathcal L}(z)/\curlyLo \; ,
\label{curlyLtilDefn}
\eeq
and fit this to the data with a cubic
[$\log \Ltil(z) = \alpha z + \beta z^2 + \gamma z^3$].  The best
least-squares fit is plotted as a solid line in Fig.~\ref{fig3.1} along
with upper and lower limits (dashed lines).  We refer to these cases in
what follows as the ``moderate,'' ``strong'' and ``weak'' galaxy evolution
scenarios respectively.

\subsection{The delta-function spectrum}

Inserting (\ref{curlyLtilDefn}) into (\ref{Idelta1}) puts the latter into
the form
\beq
\Ilam(\lamo) = \Id \int_0^{\, z_f} \frac{\Ltil(z)}
   {(1+z)^3 \Htil(z)} \, \delta \left[ \frac{\lamo}{\lamp (1+z)} - 1
   \right] \, dz \; .
\label{Idelta2}
\eeq
The dimensional content of this integral has been concentrated into
a prefactor $\Id$, defined by
\beq
\Id = \frac{c \curlyLo}{4\pi \Ho \lamp} =
   4.4 \times 10^{-9} \IUs \left(\frac{\lamp}{4400\mbox{\AA}}\right)^{-1}
   \!\!\!\!\! .
\label{IdeltaValue}
\eeq
This constant shares two important properties of its bolometric counterpart
$\Qstar$ (Sec.~\ref{sec:Qstar}).  First, it is independent of the
uncertainty $\ho$ in Hubble's constant.  Second, it is {\em low\/} by
everyday standards.  It is, for example, far below the intensity of the
zodiacal light, which is caused by the scattering of sunlight by dust in
the plane of the solar system.
This is important, since the value of $\Id$ sets the scale of
the integral (\ref{Idelta2}) itself.  Indeed, existing observational
{\em bounds\/} on $\Ilam(\lamo)$ at $\lamo\approx$~4400\AA\ are of the
same order as $\Id$.  Toller, for example, set an upper limit of
$\Ilam(4400$\AA$)<4.5\times10^{-9}\IUs$ using data from the
Pioneer~10 photopolarimeter \cite{Tol83}.

Dividing $\Id$ of (\ref{IdeltaValue}) by the photon energy
$E_0=hc/\lamo$ (where $hc=1.986 \times 10^{-8}$~erg~\AA) puts 
the EBL intensity integral (\ref{Idelta2}) into new units, sometimes 
referred to as {\em continuum units\/} (CUs):
\beq
\Id = \Id(\lamo) = \frac{\curlyLo}{4\pi h \Ho} \left( 
   \frac{\lamo}{\lamp} \right) = 970 \mbox{ CUs } \!\! \left(
   \frac{\lamo}{\lamp} \right) \; ,
\eeq
where 1~CU~$\equiv 1$~photon s$^{-1}$~cm$^{-2}$~\AA$^{-1}$~ster$^{-1}$.
While both kinds of units (CUs and $\IUs$) are in common use for reporting
spectral intensity at near-optical wavelengths, CUs appear most frequently.
They are also preferable from a theoretical point of view, because they most
faithfully reflect the {\em energy content\/} of a spectrum \cite{Hen99}.
A third type of intensity unit, the $S_{10}$ (loosely, the equivalent of
one tenth-magnitude star per square degree) is also occasionally
encountered but will be avoided in this review as it is wavelength-dependent
and involves other subtleties which differ between workers.

If we let the redshift of formation $z_f\rightarrow\infty$ then
Eq.~(\ref{Idelta2}) reduces to
\beq
\Ilam(\lamo) = \left\{ \begin{array}{lc}
   \Id\!\left(\bigfrac{\lamo}{\lamp}\right)^{-2}\!\!\!
   \bigfrac{\Ltil(\lamo/\lamp-1)}{\Htil(\lamo/\lamp-1)}
      & (\mbox{if } \lamo\geqslant\lamp) \\
   0 
      & (\mbox{if } \lamo<\lamp) \\
   \end{array} \right. \; .
\eeq
The comoving luminosity density $\Ltil(\lamo/\lamp-1)$ which appears here
is fixed by the fit (\ref{curlyLtilDefn}) to the {\sc Hdf} data in
Fig.~\ref{fig3.1}.  The Hubble parameter is given by (\ref{FL2}) as
$\Htil(\lamo/\lamp-1)=
[\Omato(\lamo/\lamp)^3+\Olamo-(\Omato+\Olamo-1)(\lamo/\lamp)^2]^{1/2}$
for a universe containing dust-like matter and vacuum energy with density
parameters $\Omato$ and $\Olamo$ respectively.

Turning off the luminosity density evolution (so that $\Ltil=1=$~const.),
one obtains three trivial special cases:
\beq
\Ilam(\lamo) = \Id \times \left\{ \begin{array}{ll}
   (\lamo/\lamp)^{-7/2}
      & \mbox{ (Einstein-de Sitter) } \\
   (\lamo/\lamp)^{-2}
      & \mbox{ (de Sitter) } \\
   (\lamo/\lamp)^{-3}
      & \mbox{ (Milne) } \\
   \end{array} \right. \; .
\eeq
These are taken at $\lamo\geqslant\lamp$, where $(\Omato,\Olamo)=(1,0), (0,1)$
and $(0,0)$ respectively for the three models cited (Table~\ref{table2.1}).
The first of these is the ``$\smallfrac{7}{2}$-law'' which often appears
in the particle-physics literature as an approximation to the spectrum of EBL
contributions from decaying particles.  But the second (de~Sitter) probably
provides a better approximation, given current thinking regarding the
values of $\Omato$ and $\Olamo$.

To evaluate the spectral EBL intensity (\ref{Idelta2}) and other quantities
in a general situation, it will be helpful to define a suite of
{\em cosmological test models\/} which span the widest range possible in the
parameter space defined by $\Omato$ and $\Olamo$.  We list four such models
in Table~\ref{table3.1} and summarize the main rationale for each here
(see Sec.~\ref{ch4} for more detailed discussion).
The Einstein-de~Sitter (EdS) model has long been favoured on grounds of
simplicity, and still sometimes referred to as the
``standard cold dark matter'' or SCDM model.
It has come under increasing pressure, however, as evidence mounts for
levels of $\Omato\lesssim0.5$, and most recently from observations of
Type~Ia supernovae (SNIa) which indicate that $\Olamo>\Omato$.
The {\em Open Cold Dark Matter\/} (OCDM) model is more consistent with data
on $\Omato$ and holds appeal for those who have been reluctant to
accept the possibility of a nonzero vacuum energy.
It faces the considerable challenge, however, of explaining data on
the spectrum of CMB fluctuations, which imply that $\Omato+\Olamo\approx1$.
The {\em $\Lambda+$Cold Dark Matter\/} (\LCDM) model has rapidly
become the new standard in cosmology because it agrees best with both the
SNIa and CMB observations.  However, this model suffers from a
``coincidence problem,'' in that $\Omat(t)$ and $\Olam(t)$ evolve so
differently with time that the probability of finding ourselves at a
moment in cosmic history when they are even of the same order of magnitude
appears unrealistically small.  This is addressed to some extent in the
last model, where we push $\Omato$ and $\Olamo$ to their lowest and
highest limits, respectively.  In the case of $\Omato$ these limits are set
by big-bang nucleosynthesis, which requires a density of at least
$\Omato\approx0.03$ in baryons (hence the {\em $\Lambda+$Baryonic
Dark Matter\/} or \LBDM\ model).
Upper limits on $\Olamo$ come from various arguments, such as the observed
frequency of gravitational lenses and the requirement that the Universe
began in a big-bang singularity.  Within the context of isotropic and
homogeneous cosmology, these four models cover the full range of what
would be considered plausible by most workers.

\begin{table}[t!]
\caption{Cosmological test models}
\begin{tabular}{@{}lcccc@{}}
\hline
         & EdS/SCDM & OCDM & \LCDM\ & \LBDM\ \\
\hline
$\Omato$ & 1          & 0.3  & 0.3   & 0.03 \\
$\Olamo$ & 0          & 0    & 0.7   & 1 \\
\hline
\end{tabular}
\label{table3.1}
\end{table}

\begin{figure}[t!]
\begin{center}
\includegraphics[width=\textwidth]{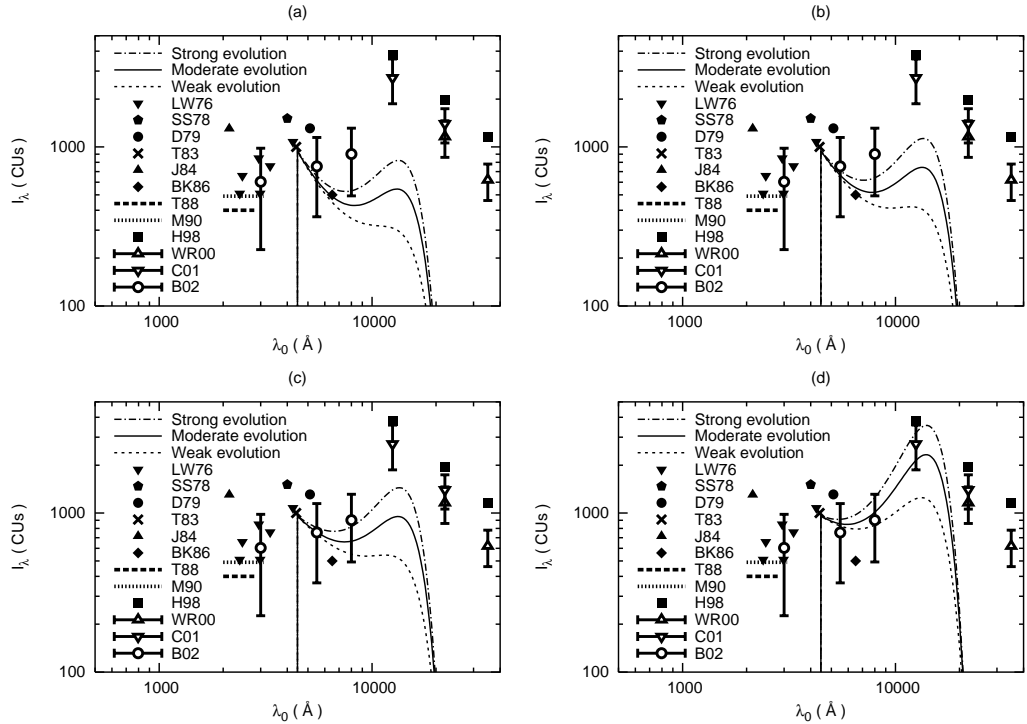}
\end{center}
\caption{The spectral EBL intensity of galaxies whose radiation is modelled
   by $\delta$-functions at a rest frame wavelength of 4400\AA, calculated
   for four different cosmological models: (a)~EdS, (b)~OCDM, (c)~\LCDM\
   and (d)~\LBDM\ (Table~\ref{table3.1}).  Also shown are observational
   upper limits (solid symbols and heavy lines) and reported detections
   (empty symbols) over the waveband 2000-40,000\AA.}
\label{fig3.2}
\end{figure}

Fig.~\ref{fig3.2} shows the solution of the full integral (\ref{Idelta2})
for all four test models, superimposed on a plot of available experimental
data at near-optical wavelengths (i.e. a close-up of Fig.~\ref{fig1.2}).
The short-wavelength cutoff in these plots is an artefact of the
$\delta$-function SED, but the behaviour of $\Ilam(\lamo)$
at wavelengths above $\lamp=4400$~\AA\ is quite revealing, even in a model
as simple as this one.  In the EdS case~(a), the rapid fall-off in intensity
with $\lamo$ indicates that {\em nearby\/} (low-redshift) galaxies dominate. 
There is a secondary hump at $\lamo\approx10,000$~\AA, which is an ``echo''
of the peak in galaxy formation, redshifted into the near infrared.
This hump becomes progressively larger relative to the optical peak at
4400~\AA\ as the ratio of $\Olamo$ to $\Omato$ grows.  Eventually
one has the situation in the de~Sitter-like model~(d), where the
galaxy-formation peak entirely dominates the observed EBL signal, despite
the fact that it comes from distant galaxies at $z\approx3$.  This is 
because a large $\Olamo$-term (especially one which is large relative to
$\Omato$) inflates comoving volume at high redshifts.  Since the comoving
{\em number density\/} of galaxies is fixed by the fit to observational
data on $\Ltil(z)$ (Fig.~\ref{fig3.1}), the number of galaxies at these
redshifts must go up, pushing up the infrared part of the spectrum.
Although the $\delta$-function spectrum is an unrealistic one,
we will see that this trend persists in more sophisticated models,
providing a clear link between observations of the EBL and the
cosmological parameters $\Omato$ and $\Olamo$.

Fig.~\ref{fig3.2} is plotted over a broad range of wavelengths from the near
ultraviolet (NUV; 2000-4000\AA) to the near infrared (NIR; 8000-40,000\AA).
The upper limits in this plot (solid symbols and heavy lines) come from
analyses of {\sc Oao}-2 satellite data (LW76 \cite{Lil76}),
ground-based telescopes (SS78 \cite{Spi78}, D79 \cite{Dub79},
BK86 \cite{Bou86}), Pioneer~10 (T83 \cite{Tol83}), sounding rockets
(J84 \cite{Jak84}, T88 \cite{Ten88}), the shuttle-borne
Hopkins~{\sc Uvx} (M90 \cite{Mur90}) and --- in the near infrared ---
the {\sc Dirbe} instrument aboard the {\sc Cobe} satellite (H98 \cite{Hau98}).
The past few years have also seen the first widely-accepted
{\em detections\/} of the EBL (Fig.~\ref{fig3.2}, open symbols).
In the NIR these have come from continued analysis of {\sc Dirbe}
data in the K-band (22,000\AA) and L-band (35,000\AA; WR00 \cite{Wri00}),
as well as the J-band (12,500\AA; C01 \cite{Cam01}).  Reported detections
in the optical using a combination of Hubble Space Telescope ({\sc Hst})
and Las~Campanas telescope observations (B02 \cite{Ber02a}) are
preliminary \cite{Mat03} but potentially very important.

Fig.~\ref{fig3.2} shows that EBL intensities based on the simple
$\delta$-function spectrum are in rough agreement with these data.
Predicted intensities come in at or just below the optical limits in the
low-$\Olamo$ cases~(a) and (b), and remain consistent with the infrared
limits even in the high-$\Olamo$ cases (c) and (d).  Vacuum-dominated
models with even higher ratios of $\Olamo$ to $\Omato$ would, however,
run afoul of {\sc Dirbe} limits in the J-band.

\subsection{The Gaussian spectrum} \label{sec:gauss}

The Gaussian distribution provides a useful generalization of the
$\delta$-function for modelling sources whose spectra, while essentially
monochromatic, are broadened by some physical process.  For example, 
photons emitted by the decay of elementary particles inside dark-matter
halos would have their energies Doppler-broadened by circular velocities
$v_c\approx220$~km~s$^{-1}$, giving rise to a spread of order
$\siglam(\lambda)=(2v_c/c)\lambda\approx0.0015\lambda$ in the SED.
In the context of galaxies, this extra degree of freedom provides a
simple way to model the width of the bright part of the spectrum.
If we take this to cover the B-band (3600-5500\AA) then
$\siglam\sim1000$\AA.  The Gaussian SED reads
\beq
F(\lambda,z) = \frac{L(z)}{\sqrt{2\pi}\,\siglam} \exp \left[ -\frac{1}{2}
   \left( \frac{\lambda-\lamp}{\siglam} \right)^2 \right] \; ,
\label{gaussSED}
\eeq
where $\lamp$ is the wavelength at which the galaxy emits most of its light.
We take $\lamp=4400$\AA\ as before, and note that integration over $\lamo$
confirms that $L(z)=\int_0^\infty F(\lambda,z)d\lambda$ as required.  Once
again we can make the simplifying assumption that $L(z)=\Lo=$~const., or
we can use the empirical fit $\Ltil(z)\equiv n(z)L(z)/\curlyLo$ to the
{\sc Hdf} data in Fig.~\ref{fig3.1}.  Taking the latter course and
substituting (\ref{gaussSED}) into (\ref{IzDefn}), we obtain
\beq
\Ilam(\lamo) = \Ig \int_0^{\, z_f} \frac{\Ltil(z)}
   {(1+z)^3 \Htil(z)} \, \exp \left\{ -\frac{1}{2} \left[
   \frac{\lamo/(1+z)-\lamp}{\siglam} \right]^2 \right\} dz \; .
\label{Igauss}
\eeq
The dimensional content of this integral has been pulled into a
prefactor $\Ig=\Ig(\lamo)$, defined by
\beq
\Ig = \frac{\curlyLo}{\sqrt{32\pi^3}\, h \Ho} \left( 
   \frac{\lamo}{\siglam} \right) = 390 \mbox{ CUs } \!\!\! \left(
   \frac{\lamo}{\siglam} \right) \; .
\eeq
Here we have divided (\ref{Igauss}) by the photon energy
$E_0=hc/\lamo$ to put the result into CUs, as before.

\begin{figure}[t!]
\begin{center}
\includegraphics[width=\textwidth]{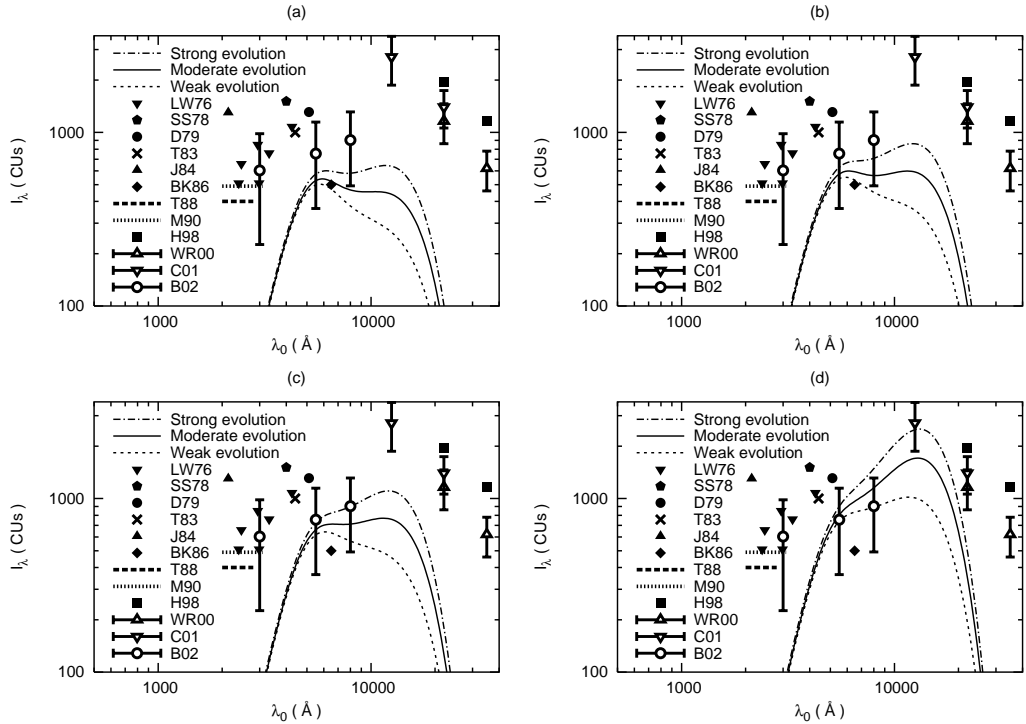}
\end{center}
\caption{The spectral EBL intensity of galaxies whose spectra has been
   represented by Gaussian distributions with rest frame peak wavelength
   4400\AA\ and standard deviation 1000\AA, calculated for the (a)~EdS,
   (b)~OCDM, (c)~\LCDM\ and (d)~\LBDM\ cosmologies and compared with
   observational upper limits (solid symbols and heavy lines) and
   reported detections (empty symbols).}
\label{fig3.3}
\end{figure}

Results are shown in Fig.~\ref{fig3.3}, where we have taken $\lamp=4400$\AA,
$\siglam=1000$\AA\ and $z_f=6$.  Aside from the fact that the
short-wavelength cutoff has disappeared, the situation is qualitatively
similar to that obtained using a $\delta$-function approximation.
(This similarity becomes formally exact as $\siglam$ approaches zero.)
One sees, as before, that the expected EBL signal is brightest at
optical wavelengths in an EdS Universe~(a),
but that the infrared hump due to the redshifted peak of galaxy formation
begins to dominate for higher-$\Olamo$ models~(b) and (c), becoming
overwhelming in the de~Sitter-like model~(d).  Overall, the best
agreement between calculated and observed EBL levels occurs in
the \LCDM\ model (c).  The matter-dominated EdS (a) and OCDM (b) models
contain too little light (requiring one to postulate an
additional source of optical or near-optical background radiation
besides that from galaxies), while the \LBDM\ model (d) comes
uncomfortably close to containing {\em too much\/} light.
This is an interesting situation, and one which motivates us to
reconsider the problem with more realistic models for the galaxy SED.

\subsection{The Planckian spectrum}

The simplest nontrivial approach to a galaxy spectrum is to model it as
a blackbody, and this was done by previous workers such as McVittie and
Wyatt \cite{McV59}, Whitrow and Yallop \cite{Whi64,Whi65} and Wesson
\cite{Wes91a}.  Let us suppose that the galaxy SED is a product of
the Planck function and some wavelength-independent parameter $C(z)$:
\beq
F(\lambda,z) = \frac{2\pi hc^2}{\sigSB} \, \frac{C(z)/\lambda^5}
   {\exp \left[ hc/kT(z)\lambda \right] - 1} \; .
\label{bbodySED}
\eeq
Here $\sigSB\equiv2\pi^5k^4/15c^2h^3 = 5.67 \times
10^{-5}$~erg~cm$^{-2}$~s$^{-1}$~K$^{-1}$ is the Stefan-Boltzmann constant.
The function $F$ is normally regarded as an increasing function of redshift
(at least out to the redshift of galaxy formation).  This can in principle
be accommodated by allowing $C(z)$ or $T(z)$ to increase with $z$ in 
(\ref{bbodySED}).  The former choice would correspond to a situation in which
galaxy luminosity decreases with time while its spectrum remains unchanged,
as might happen if stars were simply to die.  The second choice corresponds
to a situation in which galaxy luminosity decreases with time as its
spectrum becomes redder, as may happen when its stellar population ages.
The latter scenario is more realistic, and will be adopted here.  The
luminosity $L(z)$ is found by integrating $F(\lambda,z)$ over all wavelengths:
\beq
L(z) = \frac{2\pi hc^2}{\sigSB} \, C(z) \int_0^\infty
   \frac{\lambda^{-5} d\lambda}{\exp \left[ hc/kT(z)\lambda \right] - 1} =
   C(z) \left[ T(z) \right]^4 \; ,
\eeq
so that the unknown function $C(z)$ must satisfy $C(z)=L(z)/[T(z)]^{\,4}$.
If we require that Stefan's law ($L\propto T^4$)
hold at each $z$, then
\beq
C(z) = \mbox{ const. } = \Lo/\To^4 \; ,
\label{CzDefn}
\eeq
where $\To$ is the present ``galaxy temperature'' (i.e. the blackbody
temperature corresponding to a peak wavelength in the B-band).  Thus the
evolution of galaxy luminosity in this model is just that which is
required by Stefan's law for blackbodies whose {\em temperatures\/}
evolve as $T(z)$.  This is reasonable, since galaxies are made up of stellar
populations which cool and redden with time as hot massive stars die out.

Let us supplement this with the assumption of constant comoving number
density, $n(z)=\no=$~const.  This is sometimes referred to as the pure
luminosity evolution or PLE scenario, and while there is some controversy
on this point, PLE has been found by many workers to be roughly consistent
with observed numbers of galaxies at faint magnitudes, especially if there
is a significant vacuum energy density $\Olamo>0$.  Proceeding on this
assumption, the comoving galaxy luminosity density can be written
\beq
\Ltil(z) \equiv \frac{n(z)L(z)}{\curlyLo} = \frac{L(z)}{\Lo} =
   \left[ \frac{T(z)}{\To} \right]^4 \; .
\eeq
This expression can then be inverted for blackbody temperature $T(z)$
as a function of redshift, since the form of $\Ltil(z)$ is fixed by
Fig.~\ref{fig3.1}:
\beq
T(z) = \To [\Ltil(z)]^{1/4} \; .
\label{TzDefn}
\eeq
We can check this by choosing $\To=6600$K (i.e. a present peak wavelength
of 4400\AA) and reading off values of $\Ltil(z)=\curlyL(z)/\curlyLo$ at
the peaks of the curves marked ``weak,'' ``moderate'' and ``strong''
evolution in Fig.~\ref{fig3.1}.  Putting these numbers into (\ref{TzDefn})
yields blackbody temperatures (and corresponding peak wavelengths) of 10,000K
(2900\AA), 11,900K (2440\AA) and 13,100K (2210\AA) respectively at the
galaxy-formation peak.  These numbers are consistent with the idea that
galaxies would have been dominated by hot UV-emitting stars at this early time.

Inserting the expressions (\ref{CzDefn}) for $C(z)$ and (\ref{TzDefn}) for
$T(z)$ into the SED (\ref{bbodySED}), and substituting the latter into the
EBL integral (\ref{IzDefn}), we obtain
\beq
\Ilam(\lamo) = \Ib \int_0^{\, z_f} \frac{(1+z)^2 dz} {\left\{ 
   \exp \left[ hc\,(1+z)/kT(z)\lamo \right] - 1 \right\} \Htil(z)} \; .
\label{Ibbody}
\eeq
The dimensional prefactor $\Ib=\Ib(\To,\lamo)$ reads in this case
\beq
\Ib = \frac{c^2\curlyLo}{2\Ho\,\sigSB\To^4\lamo^4} =
   90,100 \mbox{ CUs } \!\! \left( \frac{\To}{6600\mbox{K}} \right)^{-4}
   \!\! \left( \frac{\lamo}{4400\mbox{\AA}} \right)^{-4} \!\!\!\! .
\eeq
\begin{figure}[t!]
\begin{center}
\includegraphics[width=\textwidth]{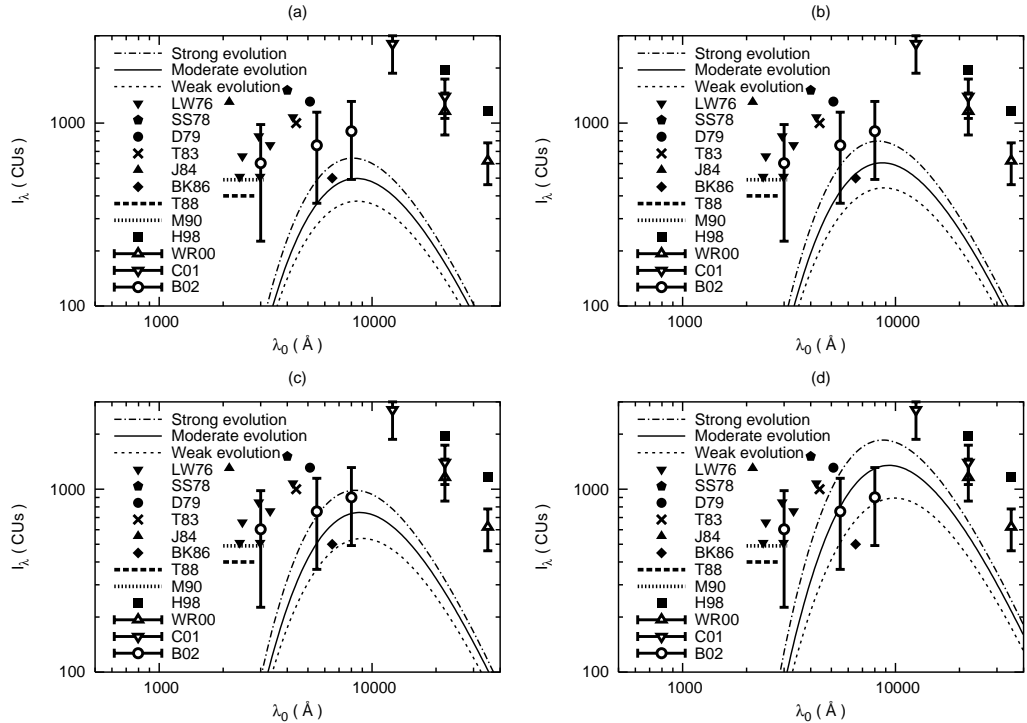}
\end{center}
\caption{The spectral EBL intensity of galaxies, modelled as blackbodies
   whose characteristic temperatures are such that their luminosities
   $L\propto T^4$ combine to produce the observed comoving luminosity
   density $\curlyL(z)$ of the Universe.  Results are shown for the (a)~EdS,
   (b)~OCDM, (c)~\LCDM\ and (d)~\LBDM\ cosmologies.  Also shown are
   observational upper limits (solid symbols and heavy lines) and
   reported detections (open symbols).}
\label{fig3.4}
\end{figure}
This integral is evaluated and plotted in Fig.~\ref{fig3.4}, where we have
set $z_f=6$ following recent observational hints of an epoch of
``first light'' at this redshift \cite{Ell01a}.  Overall EBL intensity
is insensitive to this choice, provided that $z_f\gtrsim3$.
Between $z_f=3$ and $z_f=6$, $\Ilam(\lamo)$ rises by less than 1\%
below $\lamo=$~10,000\AA\ and less than $\sim5$\% at $\lamo=$~20,000\AA\
(where most of the signal originates at high redshifts).  There is no
further increase beyond $z_f>6$ at the three-figure level of precision.

Fig.~\ref{fig3.4} shows some qualitative differences from our earlier results
obtained using $\delta$-function and Gaussian SEDs.  Most noticeably, the
prominent ``double-hump'' structure is no longer apparent.  The key
evolutionary parameter is now blackbody temperature $T(z)$ and this goes
as $[\curlyL(z)]\,,^{\!\!\!1/4}$ so that individual features in the
comoving luminosity density profile are suppressed.  (A similar effect
can be achieved with the Gaussian SED by choosing larger values of $\siglam$.)
As before, however, the long-wavelength part of the spectrum climbs steadily
up the right-hand side of the figure as one moves from the $\Olamo=0$
models~(a) and (b) to the $\Olamo$-dominated models~(c) and (d), whose
light comes increasingly from more distant, redshifted galaxies.

Absolute EBL intensities in each of these four models are consistent with
what we have seen already.  This is not surprising, because changing the
shape of the SED merely shifts light from one part of the spectrum to
another.  It cannot alter the {\em total amount\/} of light in the EBL,
which is set by the comoving luminosity density $\Ltil(z)$ of sources
once the background cosmology (and hence the source lifetime) has been
chosen.  As before, the best match between calculated EBL intensities
and the observational detections is found for the $\Olamo$-dominated
models~(c) and (d).  The fact that the EBL is now spread across a broader
spectrum has pulled down its peak intensity slightly, so that the \LBDM\
model~(d) no longer threatens to violate observational limits and in fact
fits them rather nicely.  The zero-$\Olamo$ models~(a) and (b) again
appear to require some additional source of background radiation (beyond
that produced by galaxies) if they are to contain enough light to make up
the levels of EBL intensity that have been reported.

\subsection{Normal and starburst galaxies} \label{sec:galtypes}

The previous sections have shown that simple models of galaxy spectra,
combined with data on the evolution of comoving luminosity density in
the Universe, can produce levels of spectral EBL intensity in rough
agreement with observational limits and reported detections, and even
discriminate to a degree between different cosmological models.
However, the results obtained up to this point are somewhat unsatisfactory
in that they are sensitive to theoretical input parameters, such as
$\lamp$ and $\To$, which are hard to connect with the properties of
the actual galaxy population.

A more comprehensive approach would use observational data in conjunction
with theoretical models of galaxy evolution to build up an ensemble of
evolving galaxy SEDs $F(\lambda,z)$ and comoving number densities $n(z)$
which would depend not only on redshift but on {\em galaxy type\/} as well.
Increasingly sophisticated work has been carried out along these lines over
the years by Partridge and Peebles \cite{Par67}, Tinsley \cite{Tin73},
Bruzual \cite{Bru81}, Code and Welch \cite{Cod82}, Yoshii and Takahara
\cite{Yos88} and others.  The last-named authors, for instance, divided
galaxies into five morphological types (E/SO, Sab, Sbc, Scd and Sdm),
with a different evolving SED for each type, and found that their
collective EBL intensity at NIR wavelengths was about an order
of magnitude below the levels suggested by observation.

Models of this kind, however, are complicated while at the same time 
containing uncertainties.  This makes their use somewhat incompatible with
our purpose here, which is primarily to obtain a first-order estimate
of EBL intensity so that the importance of {\em expansion\/} can be properly
ascertained.  Also, observations have begun to show that the above
morphological classifications are of limited value at redshifts
$z \gtrsim 1$, where spirals and ellipticals are still in the process
of forming \cite{Abr01}.  As we have already seen, this is precisely
where much of the EBL may originate, especially if luminosity density
evolution is strong, or if there is a significant $\Olamo$-term.

What is needed, then, is a simple model which does not distinguish too
finely between the spectra of galaxy types as they have traditionally
been classified, but which can capture the essence of broad trends in
luminosity density evolution over the full range of redshifts
$0\leqslant z\leqslant z_f$.  For this purpose we will group together
the traditional classes (spiral, elliptical, etc.) under the single
heading of quiescent or {\em normal galaxies\/}.  At higher redshifts
($z\gtrsim1$), we will allow a second class of objects to play a role:
the active or {\em starburst galaxies\/}.  Whereas normal galaxies
tend to be comprised of older, redder stellar populations,
starburst galaxies are dominated by newly-forming stars whose energy
output peaks in the ultraviolet (although much of this is absorbed by
dust grains and subsequently reradiated in the infrared).  One signature
of the starburst type is thus a decrease in $F(\lambda)$ as a function of
$\lambda$ over NUV and optical wavelengths, while normal types show
an increase \cite{Kin96}.  Starburst galaxies also tend to be brighter,
reaching bolometric luminosities as high as $10^{12}-10^{13}\Lsun$,
versus $10^{10}-10^{11}\Lsun$ for normal types.

\begin{figure}[t!]
\begin{center}
\includegraphics[width=\textwidth]{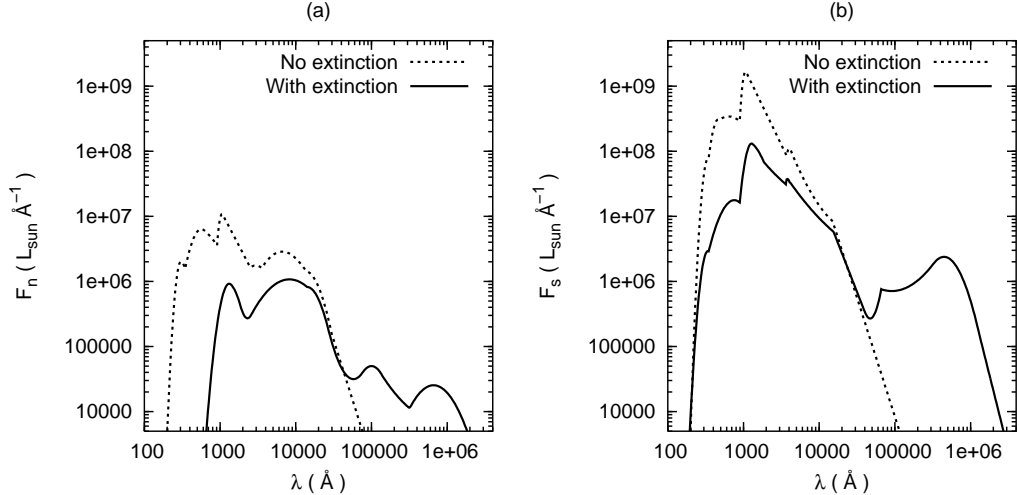}
\end{center}
\caption{Typical galaxy SEDs for (a)~normal and (b)~starburst type
   galaxies with and without extinction by dust.
   These figures are adapted from Figs.~9 and 10 of Devriendt \etal
   \cite{Dev99}.  For definiteness we have normalized
   (over $100-3\times10^{\,4}$\AA) such that
   $\Ln=1\times10^{10}\ho^{-2}\Lsun$ and
   $\Ls=2\times10^{11}\ho^{-2}\Lsun$ with $\ho=0.75$.  
   (These values are consistent with what we will later call ``model~0''
   for a comoving galaxy number density of $\no=0.010 \ho^3$~Mpc$^{-3}$.)}
\label{fig3.5}
\end{figure}

There are two ways to obtain SEDs for these objects:
by reconstruction from observational data, or as output from
theoretical models of galaxy evolution.  The former approach has had some
success, but becomes increasingly difficult at short wavelengths, so that
results have typically been restricted to $\lambda\gtrsim1000$\AA\
\cite{Kin96}.  This represents a serious limitation if we want to
integrate out to redshifts $z_f\sim6$ (say), since it means that our
results are only strictly reliable down to $\lamo=\lambda(1+z_f)\sim7000$\AA.
In order to integrate out to $z_f\sim6$ and still go down as far as the
NUV ($\lamo\sim2000$\AA), we require SEDs which are good to
$\lambda\sim300$\AA\ in the galaxy rest-frame.  For this purpose we will
make use of theoretical galaxy-evolution models, which have advanced
to the point where they cover the entire spectrum from the far ultraviolet
to radio wavelengths.  This broad range of wavelengths involves diverse
physical processes such as star formation, chemical evolution, and
(of special importance here) dust absorption of ultraviolet light
and re-emission in the infrared.  Typical normal and starburst
galaxy SEDs based on such models are now available down to $\sim100$\AA\
\cite{Dev99}.  These functions, displayed in Fig.~\ref{fig3.5},
will constitute our normal and starburst galaxy SEDs,
$\Fn(\lambda)$ and $\Fs(\lambda)$.

Fig.~\ref{fig3.5} shows the expected increase in $\Fn(\lambda)$ with $\lambda$
at NUV wavelengths ($\sim2000$\AA) for normal galaxies, as well as the
corresponding decrease for starbursts.  What is most striking about both
templates, however, is their overall multi-peaked structure.  These objects
are far from pure blackbodies, and the primary reason for this is {\em dust}.
This effectively removes light from the shortest-wavelength peaks
(which are due mostly to star formation), and transfers it to the
longer-wavelength ones.  The dashed lines in Fig.~\ref{fig3.5} show what
the SEDs would look like if this dust reprocessing were ignored.
The main difference between normal and starburst types lies in the
relative importance of this process.  Normal galaxies emit as little
as 30\% of their bolometric intensity in the infrared, while the
equivalent fraction for the largest starburst galaxies can reach 99\%.
Such variations can be incorporated by modifying input parameters
such as star formation timescale and gas density, leading to spectra which
are broadly similar in shape to those in Fig.~\ref{fig3.5} but differ in
normalization and ``tilt'' toward longer wavelengths.  The results have
been successfully matched to a wide range of real galaxy spectra \cite{Dev99}.

\subsection{Comparison with observation} \label{sec:CompObs}

We proceed to calculate the spectral EBL intensity using $\Fn(\lambda)$ and
$\Fs(\lambda)$, with the characteristic luminosities of these two types
found as usual by normalization, $\int\Fn(\lambda)d\lambda=\Ln$ and
$\int\Fs(\lambda)d\lambda=\Ls$.
Let us assume that the comoving luminosity density of the Universe
at any redshift $z$ is a combination of normal and starburst components
\beq
\curlyL(z) = \nN(z) \Ln + \nS(z) \Ls \; ,
\label{curlyL2comp}
\eeq
where comoving number densities are
\beq
\nN(z) \equiv [1-f(z)] \, n(z) \qquad \nS(z) \equiv f(z) \, n(z) \; .
\label{n2compDefns}
\eeq
In other words, we will account for evolution in $\curlyL(z)$ solely
in terms of a changing {\em starburst fraction\/} $f(z)$,
and a single comoving number density $n(z)$ as before.
$\Ln$ and $\Ls$ are awkward to work with for dimensional reasons,
and we will find it more convenient to specify the SED instead by
two dimensionless parameters, the local starburst fraction $\fo$
and luminosity ratio $\ello$:
\beq
\fo \equiv f(0) \qquad \ello \equiv \Ls/\Ln \; .
\eeq
Observations indicate that $\fo\approx0.05$ in the local population
\cite{Kin96}, and the SEDs shown in Fig.~\ref{fig3.5} have been fitted to
a range of normal and starburst galaxies with
$40 \lesssim \ello \lesssim 890$ \cite{Dev99}.  We will allow these two
parameters to vary in the ranges
$0.01 \leqslant \fo \leqslant 0.1$ and $10 \leqslant \ello \leqslant 1000$.
This, in combination with our ``strong'' and ``weak'' limits on
luminosity-density evolution, gives us the flexibility to obtain
upper and lower bounds on EBL intensity.

The functions $n(z)$ and $f(z)$ can now be fixed by equating $\curlyL(z)$
as defined by (\ref{curlyL2comp}) to the comoving luminosity-density curves
inferred from {\sc Hdf} data (Fig.~\ref{fig3.1}), and requiring that
$f\rightarrow1$ at peak luminosity (i.e. assuming that the galaxy
population is entirely starburst-dominated at the redshift $\zp$ of peak
luminosity).  These conditions are not difficult to set up.  One finds
that modest number-density evolution is required in general,
if $f(z)$ is not to over- or under-shoot unity at $\zp$.
We follow \cite{Tot00} and parametrize this with the function
$n(z)=\no(1+z)^\eta$ for $z\leqslant\zp$.  Here $\eta$ is often termed
the {\em merger parameter\/} since a value of $\eta>0$ would imply that
the comoving number density of galaxies decreases with time.

\begin{figure}[t!]
\begin{center}
\includegraphics[width=\textwidth]{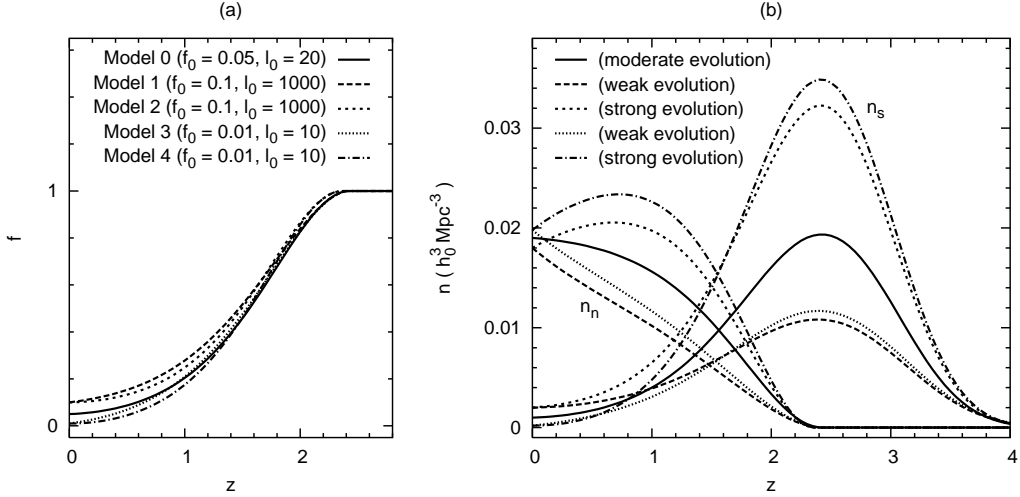}
\end{center}
\caption{Evolution of (a)~starburst fraction $f(z)$ and (b)~comoving
   normal and starburst galaxy number densities $\nN(z)$ and $\nS(z)$,
   where total comoving luminosity density $\curlyL(z)=\nN(z)\Ln+\nS(z)\Ls$
   is matched to the ``moderate,'' ``weak'' and ``strong'' evolution curves
   in Fig.~\ref{fig3.1}.  Each model has different values of the two
   adjustable parameters $\fo\equiv f(0)$ and $\ello\equiv\Ls/\Ln$.}
\label{fig3.6}
\end{figure}

Pulling these requirements together, one obtains a model with
\beqa
f(z) & = & \left\{ \begin{array}{ll}
   \left( \bigfrac{1}{\ello-1} \right) \left[
   \ello (1+z)^{-\eta} {\mathcal N}(z) -1 \right]
      & \qquad (z \leqslant \zp) \\
   \qquad 1 
      & \qquad (z > \zp) \\
   \end{array} \right. \nonumber \\
n(z) & = & \no \times \left\{ \begin{array}{cc}
   (1+z)^\eta
      & \qquad (z \leqslant \zp) \\
   {\mathcal N}(z) 
      & \qquad (z > \zp) \\
   \end{array} \right. \; .
\eeqa
Here ${\mathcal N}(z)\equiv[1/\ello+(1-1/\ello)\fo]\,\Ltil(z)$ and
$\eta=\ln[{\mathcal N}(\zp)]/\ln(1+\zp)$.
The evolution of $f(z)$, $\nN(z)$ and $\nS(z)$ is plotted in
Fig.~\ref{fig3.6} for five models: a best-fit Model~0,
corresponding to the moderate evolution curve in Fig.~\ref{fig3.1}
with $\fo=0.05$ and $\ello=20$, and four other models chosen to produce
the widest possible spread in EBL intensities across the optical band.
Models~1 and 2 are the {\em most\/} starburst-dominated,
with initial starburst fraction and luminosity ratio at their upper
limits ($\fo=0.1$ and $\ello=1000$).  Models~3 and 4 are the
{\em least\/} starburst-dominated, with the same quantities at their
lower limits ($\fo=0.01$ and $\ello=10$).  Luminosity density evolution 
is set to ``weak'' in the odd-numbered Models 1 and 3, and ``strong''
in the even-numbered Models 2 and 4.  (In principle one could identify
four other ``extreme'' combinations, such as maximum $\fo$ with
minimum $\ello$, but these will be intermediate to Models~1-4.)
We find merger parameters $\eta$ between $+0.4,0.5$ in the
strong-evolution Models~2 and 4, and $-0.5,-0.4$ in the weak-evolution
Models~1 and 3, while $\eta=0$ for Model~0.  These are well within
the normal range \cite{Tot98}.

The information contained in Fig.~\ref{fig3.6} can be summarized in words as
follows: starburst galaxies formed near $z_f\sim4$ and increased in
comoving number density until $\zp\sim2.5$ (the redshift of peak comoving
luminosity density in Fig.~\ref{fig3.1}).  They then gave way to a steadily
growing population of fainter normal galaxies which began to dominate
between $1\lesssim z\lesssim 2$ (depending on the model) and now make
up 90-99\% of the total galaxy population at $z=0$.  This scenario is
in good agreement with others that have been constructed to explain the
observed faint blue excess in galaxy number counts \cite{Pea96}.

We are now in a position to compute the total spectral EBL intensity by
substituting the SEDs ($\Fn,\Fs$) and comoving number
densities~(\ref{n2compDefns}) into Eq.~(\ref{IzDefn}).
Results can be written in the form $\Ilam(\lamo) = \In(\lamo)+\Is(\lamo)$
where:
\beqa
\In(\lamo) & = & \Ins \int_0^{\, z_f} \ntil(z) [1-f(z)] \, \Fn 
   \left( \frac{\lamo}{1+z} \right) \frac{dz}{(1+z)^3 \Htil(z)} \; , 
   \nonumber \\
\Is(\lamo) & = & \Ins \int_0^{\, z_f} \ntil(z) f(z) \, \Fs 
   \left( \frac{\lamo}{1+z} \right) \frac{dz}{(1+z)^3 \Htil(z)} \; .
\label{I2comp}
\eeqa
Here $\In$ and $\Is$ represent contributions from normal and starburst
galaxies respectively and $\ntil(z)\equiv n(z)/\no$ is the
{\em relative comoving number density\/}.  The dimensional content
of both integrals has been pulled into a prefactor
\beq
\Ins = \Ins(\lamo) = \frac{\curlyLo}{4\pi h \Ho} \! \left( 
   \frac{\lamo}{\mbox{\AA}} \right) = 970 \mbox{ CUs } \!\!\!
   \left( \frac{\lamo}{\mbox{\AA}} \right) \; .
\eeq
This is independent of $\ho$, as before, because the factor of $\ho$ in
$\curlyLo$ cancels out the one in $\Ho$.  The quantity $\curlyLo$ appears
here when we normalize the galaxy SEDs $\Fn(\lambda)$ and $\Fs(\lambda)$
to the observed comoving luminosity density of the Universe.  To see this,
note that Eq.~(\ref{curlyL2comp}) reads $\curlyLo=\no\Ln[1+(\ello-1)\fo]$
at $z=0$.  Since $\curlyLo\equiv\no\Lo$, it follows that
$\Ln=\Lo/[1+(\ello-1)\fo]$ and $\Ls=\Lo\ello/[1+(\ello-1)\fo]$.
Thus a factor of $\Lo$ can be divided out of the functions $\Fn$ and $\Fs$
and put directly into Eq.~(\ref{I2comp}) as required.

\begin{figure}[t!]
\begin{center}
\includegraphics[width=\textwidth]{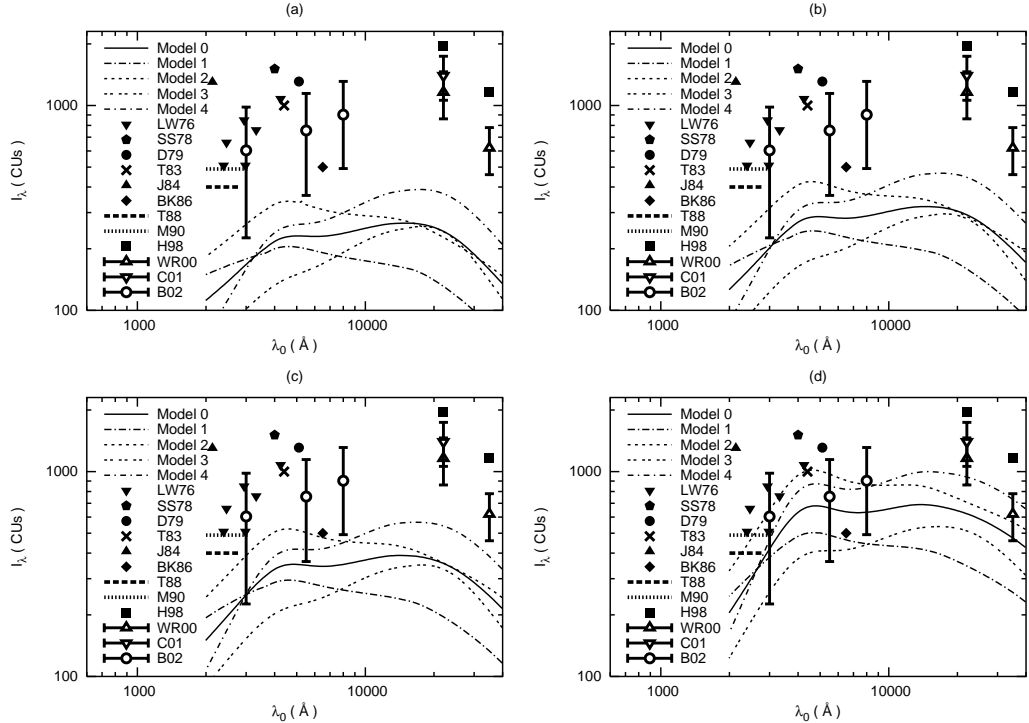}
\end{center}
\caption{The spectral EBL intensity of a combined population of normal and
   starburst galaxies, with SEDs as shown in Fig.~\ref{fig3.5}.  The evolving
   number densities are such as to reproduce the total comoving luminosity
   density seen in the {\sc Hdf} (Fig.~\ref{fig3.1}).  Results are shown for
   the (a)~EdS, (b)~OCDM, (c)~\LCDM\  and (d)~\LBDM\ cosmologies.  Also shown
   are observational upper limits (solid symbols and heavy lines) and reported
   detections (open symbols).}
\label{fig3.7}
\end{figure}

The spectral intensity (\ref{I2comp}) is plotted in Fig.~\ref{fig3.7},
where we have set $z_f=6$ as usual.  (Results are insensitive to this choice,
increasing by less than 5\% as one moves from $z_f=3$ to $z_f=6$, with
no further increase for $z_f\geqslant6$ at three-figure precision.)
These plots show that the most starburst-dominated models (1 and 2)
produce the bluest EBL spectra, as might be expected.  For these two models,
EBL contributions from normal galaxies remain well below those from
starbursts at all wavelengths, so that the bump in the observed spectrum
at $\lamo\sim4000$\AA\ is essentially an echo of the peak at $\sim1100$\AA\
in the starburst SED (Fig.~\ref{fig3.5}), redshifted by a factor $(1+\zp)$
from the epoch $\zp\approx2.5$ of maximum comoving luminosity density.
By contrast, in the least starburst-dominated models (3 and 4),
EBL contributions from normal galaxies catch up to and exceed those from
starbursts at $\lamo\gtrsim10,000$\AA, giving rise to the bump seen at
$\lamo\sim20,000$\AA\ in these models.  Absolute EBL intensities are
highest in the strong-evolution models (2 and 4) and lowest in the
weak-evolution models (1 and 3).  We emphasize that the {\em total\/}
amount of light in the EBL is determined by the choice of luminosity
density profile (for a given cosmological model).  The choice of SED
merely shifts this light from one part of the spectrum to another.
Within the context of the simple two-component model described above,
and the constraints imposed on luminosity density by the {\sc Hdf} data
(Sec.~\ref{sec:LumDens}), the curves in Fig.~\ref{fig3.7} represent
upper and lower limits on the spectral intensity of the EBL
at near-optical wavelengths.

These curves are spread over a broader range of wavelengths than those
obtained earlier using single-component Gaussian and blackbody spectra.
This leads to a drop in overall intensity, as we can appreciate by noting
that there now appears to be a significant gap between theory and observation
in all but the most vacuum-dominated cosmology, \LBDM\ (d).  This is so even
for the models with the strongest luminosity density evolution (models 2 and
4).  In the case of the EdS cosmology~(a), this gap is nearly an order of
magnitude, as reported by Yoshii and Takahara \cite{Yos88}.  Similar
conclusions have been reached more recently from an analysis of
Subaru Deep Field data by Totani \etal\ \cite{Tot01}, who suggest
that the shortfall could be made up by a very diffuse, previously
undetected component of background radiation not associated with galaxies.
Other workers have argued that existing galaxy populations are
enough to explain the data if different assumptions are made about
their SEDs \cite{Jim99}, or if allowance is made for faint low surface
brightness galaxies below the detection limit of existing surveys
\cite{Ber02b}.

\subsection{Spectral resolution of Olbers' paradox}

Having obtained quantitative estimates of the spectral EBL intensity
which are in reasonable agreement with observation,
we return to the question posed in Sec.~\ref{sec:olbers}: why precisely is
the sky dark at night?  By ``dark'' we now mean specifically dark at
{\em near-optical wavelengths\/}.  We can provide a quantitative answer
to this question by using a spectral version of our previous bolometric
argument.  That is, we compute the EBL intensity $\IlamS$ in model
universes which are equivalent to expanding ones in every way
{\em except\/} expansion, and then take the ratio $\Ilam/\IlamS$.  
If this is of order unity, then expansion plays a minor role and the
darkness of the optical sky (like the bolometric one) must be
attributed mainly to the fact that the Universe is too young to
have filled up with light.  If $\Ilam/\IlamS\ll1$, on the other hand,
then we would have a situation qualitatively different from the
bolometric one, and expansion would play a crucial role in the resolution
to Olbers' paradox.

The spectral EBL intensity for the equivalent static model is obtained by
putting the functions $\ntil(z),f(z),\Fn(\lambda),\Fs(\lambda)$ and
$\Htil(z)$ into (\ref{ISzDefn}) rather than (\ref{IzDefn}).  This
results in $\IlamS(\lamo) = \IlamSn(\lamo)+\IlamSs(\lamo)$ where
normal and starburst contributions are given by
\beqa
\IlamSn(\lamo) & = & \Ins \Fn(\lamo) \int_0^{\, z_f} \frac{ \ntil(z) [1-f(z)]
   \, dz} {(1+z) \Htil(z)} \nonumber \\
\IlamSs(\lamo) & = & \Ins \Fs(\lamo) \int_0^{\, z_f} \frac{ \ntil(z) f(z)
   \, dz} {(1+z) \Htil(z)} \; .
\eeqa
Despite a superficial resemblance to their counterparts (\ref{I2comp})
in the expanding Universe, these are vastly different expressions.  Most
importantly, the SEDs $\Fn(\lamo)$ and $\Fs(\lamo)$ no longer depend on $z$
and have been pulled out of the integrals.  The quantity $\IlamS(\lamo)$
is effectively a {\em weighted mean\/} of the SEDs $\Fn(\lamo)$ and
$\Fs(\lamo)$.  The weighting factors (i.e. the integrals over $z$)
are related to the age of the galaxies, $\int_0^{\, z_f}dz/(1+z)\Htil(z)$,
but modified by factors of $\nN(z)$ and $\nS(z)$ under the integral.
This latter modification is important because it prevents the integrals
from increasing without limit as $z_f$ becomes arbitrarily large,
a problem that would otherwise introduce considerable uncertainty into
any attempt to put bounds on the ratio $\IlamS/\Ilam$ \cite{Wes91a}.
A numerical check confirms that $\IlamS$ is nearly as insensitive to
the value of $z_f$ as $\Ilam$, increasing by up to 8\% as one
moves from $z_f=3$ to $z_f=6$, but with no further increase
for $z_f\geqslant6$ at the three-figure level.

\begin{figure}[t!]
\begin{center}
\includegraphics[width=\textwidth]{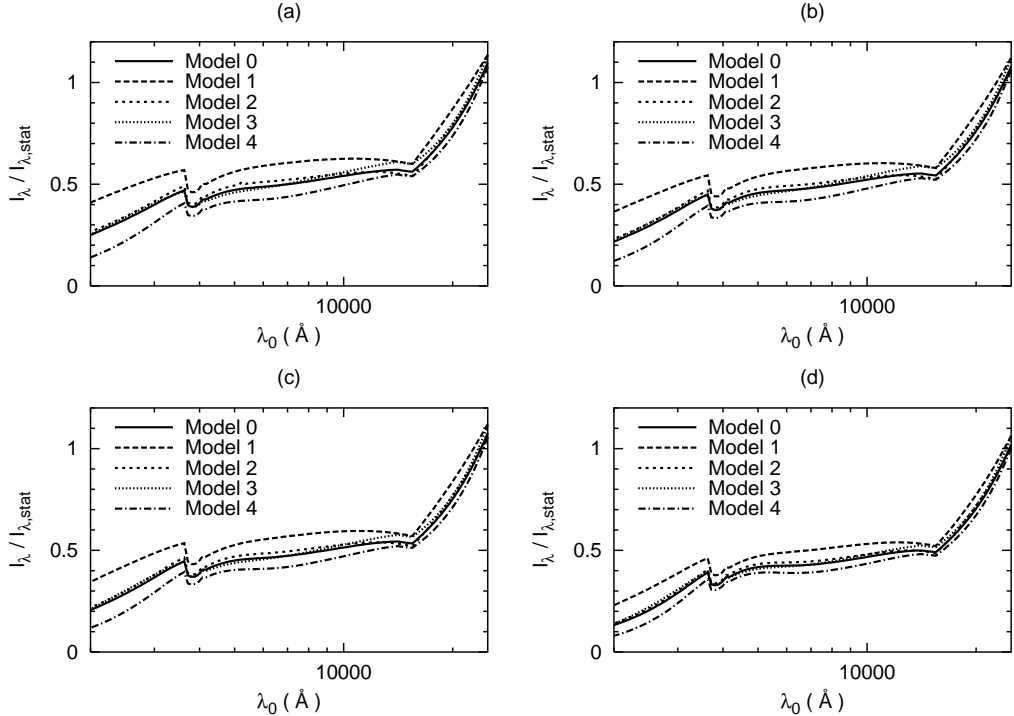}
\end{center}
\caption{The ratio $\Ilam/\IlamS$ of spectral EBL intensity in expanding models
   to that in equivalent static models, for the (a)~EdS, (b)~OCDM, (c)~\LCDM\
   and (d)~\LBDM\ models.  The fact that this ratio lies between 0.3 and 0.6
   in the B-band (4000-5000\AA) tells us that expansion reduces the
   intensity of the night sky at optical wavelengths by a factor of
   between two and three.}
\label{fig3.8}
\end{figure}

The ratio of $\Ilam/\IlamS$ is plotted over the waveband 2000-25,000\AA\ in
Fig.~\ref{fig3.8}, where we have set $z_f=6$.  (Results are insensitive to
this choice, as we have mentioned above, and it may be noted that they are
also independent of uncertainty in constants such as $\curlyLo$ since these
are common to both $\Ilam$ and $\IlamS$.)  Several features in this figure
deserve notice.  First, the average value of $\Ilam/\IlamS$ across
the spectrum is about 0.6, consistent with bolometric expectations
(Sec.~\ref{ch2}).  Second, the diagonal, bottom-left to top-right
orientation arises largely because $\Ilam(\lamo)$ drops off at
short wavelengths, while $\IlamS(\lamo)$ does so at long ones.
The reason why $\Ilam(\lamo)$ drops off at short wavelengths is that 
ultraviolet light reaches us only from the nearest galaxies; anything
from more distant ones is redshifted into the optical.  The reason why
$\IlamS(\lamo)$ drops off at long wavelengths is because it is a weighted
mixture of the galaxy SEDs, and drops off at exactly the same place that
they do: $\lamo\sim3\times10^4$\AA.  In fact, the weighting is heavily
tilted toward the dominant starburst component, so that the two sharp
bends apparent in Fig.~\ref{fig3.8} are essentially (inverted) reflections
of features in $\Fs(\lamo)$; namely, the small bump at $\lamo\sim4000$\AA\
and the shoulder at $\lamo\sim11,000$\AA\ (Fig.~\ref{fig3.5}).

Finally, the numbers: Fig.~\ref{fig3.8} shows that the ratio of $\Ilam/\IlamS$
is remarkably consistent across the B-band (4000-5000\AA) in all four
cosmological models, varying from a high of $0.46\pm0.10$ in the EdS model
to a low of $0.39\pm0.08$ in the \LBDM\ model.  These numbers should be
compared with the bolometric result of $Q/\Qstat\approx0.6\pm0.1$ from
Sec.~\ref{ch2}.  They tell us that expansion {\em does\/} play a greater
role in determining B-band EBL intensity than it does across the spectrum
as a whole --- but not by much.  If its effects were removed, the night
sky at optical wavelengths would be anywhere from two times brighter (in
the EdS model) to three times brighter (in the \LBDM\ model).  These results
depend modestly on the makeup of the evolving galaxy population, and
Fig.~\ref{fig3.8} shows that $\Ilam/\IlamS$ in every case is highest for the
weak-evolution model~1, and lowest for the strong-evolution model~4.
This is as we would expect, based on our discussion at the beginning
of this section: models with the strongest evolution effectively
``concentrate'' their light production over the shortest possible
interval in time, so that the importance of the lifetime factor drops
relative to that of expansion.  Our numerical results, however,
prove that this effect cannot qualitatively alter the resolution of
Olbers' paradox.  Whether expansion reduces the background intensity
by a factor of two or three, its {\em order of magnitude\/} is still
set by the lifetime of the Universe.

There is one factor which we have not considered in this section, and that
is the extinction of photons by intergalactic dust and neutral hydrogen,
both of which are strongly absorbing at ultraviolet wavelengths.
The effect of this would primarily be to remove ultraviolet light
from high-redshift galaxies and transfer it into the infrared ---
light that would otherwise be redshifted into the optical and contribute
to the EBL.  The latter's intensity $\Ilam(\lamo)$ would therefore drop,
and one could expect reductions over the B-band in particular.  The importance
of this effect is difficult to assess because we have limited data on the
character and distribution of dust beyond our own galaxy.  We will find
indications in Sec.~\ref{ch7}, however, that the reduction could be
significant at the shortest wavelengths considered here 
($\lamo\approx$~2000\AA) for the most extreme dust models.
This would further widen the gap between observed and predicted
EBL intensities noted at the end of Sec.~\ref{sec:galtypes}.

Absorption plays far less of a role in the equivalent static models,
where there is no redshift.  (Ultraviolet light is still absorbed, but
the effect does not carry over into the optical).  Therefore, the ratio
$\Ilam/\IlamS$ would be expected to drop in nearly direct proportion to the
drop in $\Ilam$.  In this sense Olbers had part of the solution after all ---
not (as he thought) because intervening matter ``blocks'' the light from
distant sources, but because it transfers it out of the optical. 
The importance of this effect, which would be somewhere below that of
expansion, is a separate issue from the one we have concerned ourselves
with in this section.  We have shown that expansion reduces EBL intensity
by a factor of between two and three, depending on background cosmology
and the evolutionary properties of the galaxy population.
Thus the optical sky, like the bolometric one, is dark at night
{\em primarily because it has not had enough time to fill up with light
from distant galaxies}.

\section{Dark matter and dark energy} \label{ch4}

\subsection{The four elements of modern cosmology} \label{sec:DMintro}

Observations of the intensity of extragalactic background light
effectively allow us to take a census of one component of the Universe:
its luminous matter.  In astronomy, where nearly everything we know
comes to us in the form of light signals, one might
be forgiven for thinking that luminous matter was the only kind that
counted.  This supposition, however, turns out to be spectacularly wrong.
The density $\Olum$ of luminous matter is now thought to comprise less
than one percent of the total density $\Otot$ of all forms of matter
and energy put together.  [Here as in Secs.~\ref{ch2} and \ref{ch3},
we express densities in units of the critical density,
Eq.~(\ref{rcritoDefn}), and denote them with the symbol $\Omega$.]
The remaining 99\% or more consists of dark matter and energy which,
while not seen directly, are inferred to exist from their gravitational
influence on luminous matter as well as the geometry of the Universe.

The identity of this unseen substance, whose existence was first
suspected by astronomers such as Kapteyn \cite{Kap22}, Oort \cite{Oor32}
and Zwicky \cite{Zwi37}, has become the central mystery of modern cosmology.
Indirect evidence over the past few years has increasingly suggested that
there are in fact {\em four distinct categories\/} of dark matter and energy,
three of which imply new physics beyond the existing standard model of
particle interactions.  This is an extraordinary claim, and one whose
supporting evidence deserves to be carefully scrutinized.  We devote
Sec.~\ref{ch4} to a critical review of this evidence, beginning here with a
brief overview of the current situation and followed by a closer look at
the arguments for all four parts of nature's ``dark side.''

At least some of the dark matter, such as that contained in planets and
``failed stars'' too dim to see, must be composed of ordinary atoms and
molecules.  The same applies to dark gas and dust (although these can
sometimes be seen in absorption, if not emission).  Such contributions
comprise {\em baryonic dark matter\/} (BDM), which combined together
with luminous matter gives a total baryonic matter density of
$\Obar\equiv\Olum+\Obdm$.  If our understanding of big-bang theory and
the formation of the light elements is correct, then we will see that
$\Obar$ cannot represent more than 5\% of the critical density.

Besides the dark baryons,
it now appears that three other varieties of dark matter play a role.
The first of these is {\em cold dark matter\/} (CDM), the existence of
which has been inferred from the behaviour of visible matter on scales
larger than the solar system (e.g., galaxies and clusters of galaxies).
CDM is thought to consist of particles (sometimes referred to as
``exotic'' dark-matter particles) whose interactions with ordinary matter
are so weak that they are seen primarily via their gravitational influence.
While they have not been detected (and are indeed hard to detect
by definition), such particles are predicted in plausible extensions of
the standard model.  The overall CDM density $\Ocdm$ is believed by
many cosmologists to exceed that of the baryons ($\Obar$) by
at least an order of magnitude.

Another piece of the puzzle is provided by {\em neutrinos\/}, particles
whose existence is unquestioned but whose collective density ($\Onu$)
depends on their rest mass, which is not yet known.  If neutrinos are
massless, or nearly so, then they remain relativistic throughout the
history of the Universe and behave for dynamical purposes like photons.
In this case neutrino contributions combine with those of photons ($\Ogam$)
to give the present radiation density as $\Orado=\Onu+\Ogam$.
This is known to be very small.  If on the other hand neutrinos are
sufficiently massive, then they are no longer relativistic on average,
and belong together with baryonic and cold dark matter under the category of
pressureless matter, with present density $\Omato=\Obar+\Ocdm+\Onu$.
These neutrinos could play a significant dynamical role, especially in
the formation of large-scale structures in the early Universe, where they
are sometimes known as hot dark matter (HDM).  Recent experimental
evidence suggests that neutrinos do contribute to $\Omato$ but at levels
below those of the baryons.

Influential only over the largest scales --- those of the cosmological
horizon itself --- is the final component of the unseen Universe:
{\em dark energy\/}.  Its many alternative names (the zero-point field,
vacuum energy, quintessence and the cosmological constant $\Lambda$) testify
to the fact that there is currently no consensus as to where dark energy
originates, or how to calculate its energy density ($\Olam$) from first
principles.  Existing theoretical estimates of this latter quantity range
over some 120 orders of magnitude, prompting many cosmologists until very
recently to disregard it altogether.  Observations of distant supernovae
and CMB fluctuations, however, increasingly imply that dark energy is
not only real but that its present energy density ($\Olamo$) exceeds that
of all other forms of matter ($\Omato$) and radiation ($\Orado$) put together.

\begin{figure}[t!]
\begin{center}
\includegraphics[height=90mm]{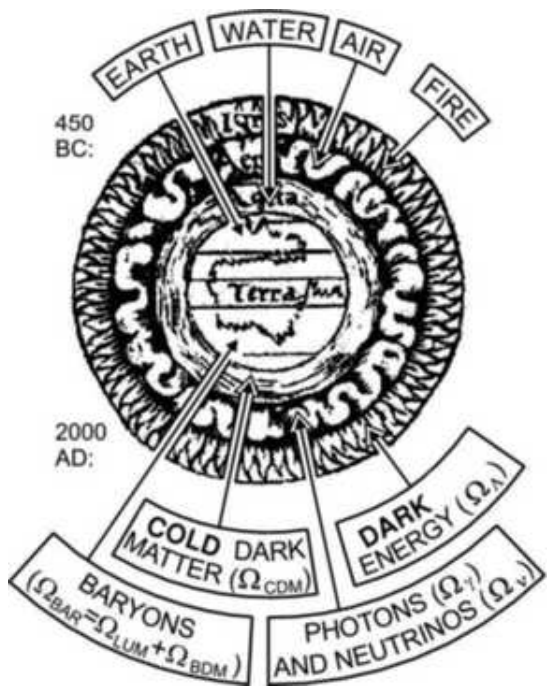}
\end{center}
\caption{Top: earth, water, air and fire, the four elements of ancient
   cosmology (attributed to the Greek philosopher Empedocles).  Bottom:
   their modern counterparts (figure taken from the review in \cite{OP01}).}
\label{fig4.1}
\end{figure}

The Universe described above hardly resembles the one we see.
It is composed to a first approximation of invisible dark energy whose
physical origin remains obscure.  Most of what remains is in the form of
CDM particles, whose ``exotic'' nature is also not yet understood.
Close inspection is needed to make out the further contribution of
neutrinos, although this too is nonzero.  And baryons, the stuff of
which we are made, are little more than a cosmic afterthought.
This picture, if confirmed, constitutes a revolution of Copernican
proportions, for it is not only our location in space which turns
out to be undistinguished, but our very makeup.  The ``four elements''
of modern cosmology are shown schematically in Fig.~\ref{fig4.1}.

\subsection{Baryonic dark matter} \label{sec:baryons}

Let us now go over the evidence for these four species of dark matter
more carefully, beginning with the baryons.  The total present density of
{\em luminous\/} baryonic matter can be inferred from the observed
luminosity density of the Universe, if various reasonable assumptions
are made about the fraction of galaxies of different morphological type,
their ratios of disk-type to bulge-type stars, and so on.  A recent and
thorough such estimate is \cite{Fuk98a}:
\beq
\Olum = (0.0027 \pm 0.0014) \ho^{-1} \; .
\label{OlumValue}
\eeq
Here $\ho$ is as usual the value of Hubble's constant expressed
in units of 100~km~s$^{-1}$~Mpc$^{-1}$.  While this parameter (and hence
the experimental uncertainty in $\Ho$) factored out of the EBL intensities
in Secs.~\ref{ch2} and \ref{ch3}, it must be squarely faced where densities
are concerned.  We therefore digress briefly to discuss the observational
status of $\ho$.

Using various relative-distance methods, all calibrated against the
distance to Cepheid variables in the Large Magellanic Cloud (LMC), the
Hubble Key Project ({\sc Hkp}) team has determined that
$\ho = 0.72 \pm 0.08$ \cite{Fre01}.  Independent ``absolute'' methods
(e.g., time delays in gravitational lenses, the Sunyaev-Zeldovich effect
and the Baade-Wesselink method applied to supernovae)
have higher uncertainties but are roughly consistent with this,
giving $\ho\approx 0.55 - 0.74 $ \cite{Pea99}.  This level of agreement is
a great improvement over the factor-two discrepancies of previous decades.

There are signs, however, that we are still some way from ``precision''
values with uncertainties of less than ten percent.  A recalibrated LMC
Cepheid period-luminosity relation based on a much larger sample (from the
{\sc Ogle} microlensing survey) leads to considerably higher values,
namely $\ho = 0.85 \pm 0.05$ \cite{Wil01}.  A purely geometric
technique, based on the use of long-baseline radio interferometry
to measure the transverse velocity of water masers \cite{Her99}, also
implies that the traditional calibration is off, raising all Cepheid-based
estimates by $12 \pm 9$\% \cite{Mao99}.  This would boost the {\sc Hkp}
value to $\ho = 0.81 \pm 0.09$.  There is some independent support for
such a recalibration in observations of ``red clump stars'' \cite{Sta98}
and eclipsing binaries \cite{Fit03} in the LMC.  New observations at
multiple wavelengths, however, suggest that early conclusions based on
these arguments may be premature \cite{Alv02}.

\begin{figure}[t!]
\begin{center}
\includegraphics[width=100mm]{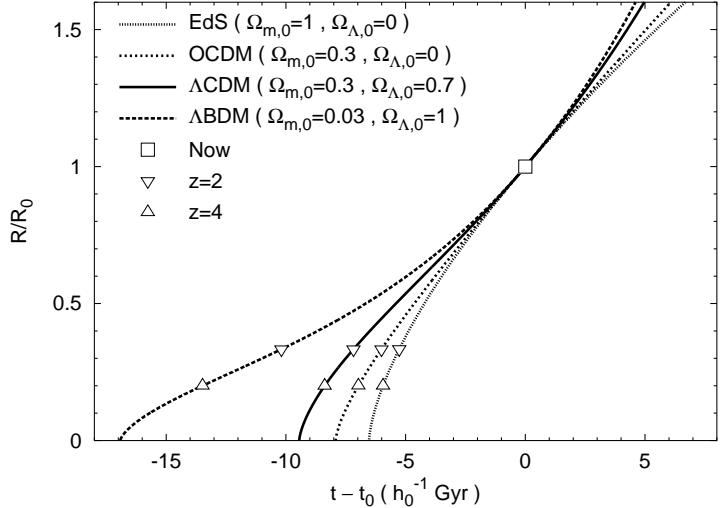}
\end{center}
\caption{Evolution of the cosmological scale factor $\Rtil(t)\equiv R(t)/\Ro$
   as a function of time (in Hubble times) for the cosmological test models
   introduced in Table~\ref{table3.1} (Sec.~\ref{ch3}).  Triangles indicate
   the range $2\leqslant z_f \leqslant4$ where the bulk of galaxy formation
   may have taken place.}
\label{fig4.2}
\end{figure}

On this subject, history encourages caution.  Where it is necessary to
specify the value of $\ho$ in this review, we will adopt:
\beq
\ho = 0.75 \pm 0.15 \; .
\label{hoValue}
\eeq
Values at the edges of this range can discriminate powerfully between
different cosmological models.  This is largely a function of their
{\em ages\/}, which can be computed by integrating (\ref{AgeDefn}) or
(for flat models) directly from Eq.~(\ref{t2comp}).
Alternatively, one can integrate the
Friedmann-Lema\^{\i}tre equation~(\ref{FL2}) numerically backward
in time.  Since this equation defines the expansion rate $H\equiv\dot{R}/R$,
its integral gives the scale factor $R(t)$.  We plot the results in
Fig.~\ref{fig4.2} for the four cosmological ``test models'' in
Table~\ref{table3.1} (EdS, OCDM, \LCDM, and \LBDM).
These are seen to have ages of $7\ho^{-1}$, $8\ho^{-1}$, $10\ho^{-1}$ and
$17\ho^{-1}$~Gyr respectively.  A firm lower limit of 11~Gyr
can be set on the age of the Universe by means of certain metal-poor halo
stars whose ratios of radioactive $^{232}$Th and $^{238}$U imply that they
formed between $14.1 \pm 2.5$~Gyr \cite{Wan02} and $15.5 \pm 3.2$~Gyr ago
\cite{Sch02}.  If $\ho$ lies at the upper end of the above range ($\ho=0.9$),
then the EdS and OCDM models would be ruled out on the basis that they are
not old enough to contain these stars (this is known as the {\em age crisis\/}
in low-$\Olamo$ models).  With $\ho$ at the bottom of the range ($\ho=0.6$),
however, only EdS comes close to being excluded.  The EdS model thus
defines one edge of the spectrum of observationally viable models.

The \LBDM\ model faces the opposite problem:
Fig.~\ref{fig4.2} shows that its age is $17\ho^{-1}$~Gyr, or as high as
28 Gyr (if $\ho=0.6$).  The latter number in particular is well beyond the
age of anything seen in our Galaxy.  Of course, upper limits on the age
of the Universe are not as secure as lower ones.  But following Copernican
reasoning, we do not expect to live in a galaxy which is unusually young.
To estimate the age of a ``typical'' galaxy, we
recall from Sec.~\ref{sec:LumDens} that most galaxies appear to have
formed at redshifts $2\lesssim z_f\lesssim 4$.
The corresponding range of scale factors, from Eq.~(\ref{zDefn}), 
is $0.33\gtrsim R/\Ro\gtrsim 0.2$.  In the \LBDM\ model, Fig.~\ref{fig4.2}
shows that $R(t)/\Ro$ does not reach these values until
$(5\pm2)\,\ho^{-1}$~Gyr after the big bang.  Thus galaxies would have
an age of about $(12\pm2)\,\ho^{-1}$~Gyr in the \LBDM\ model,
and not more than $21$~Gyr in any case.  This is close to upper limits
which have been set on the age of the Universe in models of this type,
$\too<24\pm2$~Gyr \cite{Gri01}.  Thus the \LBDM\ model, or something
close to it, probably defines a position opposite that of EdS
on the spectrum of observationally viable models.

For the other three models, Fig.~\ref{fig4.2} shows that galaxy formation
must be accomplished within less than 2~Gyr after the big bang.  The reason
this is able to occur so quickly is that these models all contain
significant amounts of CDM, which decouples from the primordial plasma
before the baryons and prepares potential wells for the baryons to
fall into.  This, as we will see, is one of the main motivations for CDM.

Returning now to the density of luminous matter, we find with our
values~(\ref{hoValue}) for $\ho$ that Eq.~(\ref{OlumValue}) gives
$\Olum = 0.0036 \pm 0.0020$.  This is the basis for our statement
(Sec.~\ref{sec:DMintro}) that the visible components of the Universe
account for less than 1\% of its density.

It may however be that most of the baryons are not visible.
How significant could such dark baryons be?  The theory of primordial
big-bang nucleosynthesis provides us with an independent method for
determining the density of {\em total\/} baryonic matter in the Universe,
based on the assumption that the light elements we see today were
forged in the furnace of the hot big bang.  Results using different
light elements are roughly consistent, which is impressive
in itself.  The primordial abundances of $^4$He (by mass) and $^7$Li
(relative to H) imply a baryon density of $\Obar = (0.010 \pm 0.004)
\ho^{-2}$ \cite{Suz00}.  By contrast, measurements based exclusively
on the primordial D/H abundance give a higher value with lower uncertainty:
$\Obar = (0.019 \pm 0.002) \ho^{-2}$ \cite{Tyt00}.  Since it appears
premature at present to exclude either of these results, we choose an
intermediate value of $\Obar = (0.016 \pm 0.005) \ho^{-2}$.  Combining
this with our range of values~(\ref{hoValue}) for $\ho$, we conclude that
\beq
\Obar = 0.028 \pm 0.012 \; .
\label{ObarValue}
\eeq
This agrees very well with independent estimates obtained by adding up
individual mass contributions from all known repositories of baryonic
matter via their estimated mass-to-light ratios \cite{Fuk98a}.
It also provides the rationale for our choice of $\Omato=0.03$ in the
\LBDM\ model.  Eq.~(\ref{ObarValue}) implies that all the atoms and
molecules in existence make up less than 5\% of the critical density.

The vast majority of these baryons, moreover, are invisible.
Using Eqs.~(\ref{OlumValue}) and (\ref{hoValue}) together with the
above-mentioned value of $\Obar\ho^2$, we infer a baryonic dark matter
fraction $\Obdm/\Obar=1-\Olum/\Obar=(87\pm8)$\%.
Where do these dark baryons reside?  One possibility is that they are
smoothly distributed in a gaseous intergalactic medium, which would have
to be strongly ionized in order to explain why it has not left a more
obvious absorption signature in the light from distant quasars.
Observations using {\sc Ovi} absorption lines as a tracer of
ionization suggest that the contribution of such material to $\Obar$
is at least $0.003 \ho^{-1}$ \cite{Tri00}, comparable to $\Olum$.
Simulations are able to reproduce many observed features of the
``forest'' of Lyman-$\alpha$ (Ly$\alpha$) absorbers with
as much as $80-90$\% of the baryons in this form \cite{Mir96}.

Dark baryonic matter could also be bound up in clumps of matter such as
substellar objects (jupiters, brown dwarfs) or stellar remnants
(white, red and black dwarfs, neutron stars, black holes).
Substellar objects are not likely to make a large contribution,
given their small masses.  Black holes are limited in the opposite sense:
they cannot be more massive than about $10^5 \Msun$ since this would
lead to dramatic tidal disruptions and lensing effects which are
not seen \cite{Car98a}.  The baryonic dark-matter clumps of most interest
are therefore ones whose mass is within a few orders of $\Msun$.
Gravitational microlensing constraints based on quasar variability
do not seriously limit such objects at present, setting an upper bound of 0.1
(well above $\Obar$) on their combined contributions to $\Omato$
in an EdS Universe \cite{Sch93}.

The existence of at least one class of compact dark objects, the
{\em massive compact halo objects\/} (MACHOs), has been confirmed
within our own galactic halo by the {\sc Macho} microlensing survey of
LMC stars \cite{Alc00}.  The inferred lensing masses lie in the range
$(0.15-0.9)\Msun$ and would account for between 8\% and 50\% of the
high rotation velocities seen in the outer parts of the Milky Way,
depending on a choice of halo model.  If the halo is spherical,
isothermal and isotropic, then at most 25\% of its mass can be
ascribed to MACHOs, according to a complementary survey ({\sc Eros})
of microlensing in the direction of the SMC \cite{Afo03}.
The identity of the lensing bodies discovered in these surveys has
been hotly debated.  White dwarfs are unlikely candidates, since we see
no telltale metal-rich ejecta from their massive progenitors \cite{Fie00}.
Low-mass red dwarfs would have to be older and/or fainter than usually
assumed, based on the numbers seen so far \cite{Fly96}.  Degenerate
``beige dwarfs'' that could form above the theoretical hydrogen-burning
mass limit of $0.08\Msun$ without fusing have been proposed as an
alternative \cite{Han99}, but it appears that such stars would form
far too slowly to be important \cite{Lyn01}.

\subsection{Cold dark matter} \label{sec:cdm}

The introduction of a second species of unseen dark matter into the
Universe has been justified on three main grounds:
(1)~a range of observational arguments imply that the total density parameter
of gravitating matter exceeds that provided by baryons and bound neutrinos;
(2)~our current understanding of the growth of large-scale structure (LSS)
requires the process to be helped along by large quantities of
non-relativistic, weakly interacting matter in the early Universe,
creating the potential wells for infalling baryons; and
(3)~theoretical physics supplies several plausible (albeit still undetected)
candidate CDM particles with the right properties.

Since our ideas on structure formation may change, and the candidate
particles may not materialize, the case for cold dark matter turns at present
on the observational arguments.  At one time, these were compatible with
$\Ocdm\approx1$, raising hopes that CDM would resolve two of the biggest
challenges in cosmology at a single stroke:  accounting for LSS formation
{\em and\/} providing all the dark matter necessary to make
$\Omato=1$, vindicating the EdS model (and with it, the simplest models
of inflation).  Observations, however, no longer support values of
$\Omato$ this high, and independent evidence now points to the existence
of at least two other forms of matter-energy beyond the standard model
(neutrinos and dark energy).  The CDM hypothesis is therefore no longer
as compelling as it once was.  With this in mind we will pay special
attention to the observational arguments in this section.  The
{\em lower limit\/} on $\Omato$ is crucial:
only if $\Omato > \Obar+\Onu$ do we require $\Ocdm > 0$.

The arguments can be broken into two classes: those which are purely 
empirical, and those which assume in addition the validity of the 
{\em gravitational instability\/} (GI) theory of structure formation.
Let us begin with the empirical arguments.
The first has been mentioned already in Sec.~\ref{sec:baryons}:
the spiral galaxy rotation curve.  If the {\sc Macho} and {\sc Eros}
results are taken at face value, and if the Milky Way is typical, then
compact objects make up less than 50\% of the mass of the halos of spiral
galaxies.  If, as has been argued \cite{Heg86}, the remaining halo mass
cannot be attributed to baryonic matter in known forms such as dust,
rocks, planets, gas, or hydrogen snowballs, then a more exotic form
of dark matter is required.

The total mass of dark matter in galaxies, however, is limited.
The easiest way to see this is to compare the {\em mass-to-light ratio\/}
($M/L$) of our own Galaxy to that of the Universe as a whole.  If the
latter has the critical density, then its $M/L$-ratio is just the ratio of
the critical density to its luminosity density: $\MLcrito=
\rcrito/\curlyLo=(1040 \pm 230) \MLunits$, where we have used 
(\ref{curlyLoValue}) for $\curlyLo$, (\ref{rcritoDefn}) for $\rcrito$
and (\ref{hoValue}) for $\ho$.  The corresponding value for the Milky Way
is $\MLmw=(21\pm7)\MLunits$, since the latter's luminosity is
$\Lmw=(2.3 \pm 0.6) \times 10^{10} \Lsun$ (in the B-band) and its
total dynamical mass (including that of any unseen halo component) is
$\Mmw=(4.9 \pm 1.1) \times 10^{11} \Msun$ inside 50~kpc from the
motions of Galactic satellites \cite{Koc96a}.  The ratio of $\MLmw$
to $\MLcrito$ is thus less than 3\%, and even if we multiply this
by a factor of a few to account for possible halo mass outside 50~kpc,
it is clear that galaxies like our own cannot make up more than
10\% of the critical density.

Most of the mass of the Universe, in other words, is spread over scales 
larger than galaxies, and it is here that the arguments for CDM take on
the most force.  The most straightforward of these involve further
applications of the mass-to-light ratio: one measures $M/L$ for a chosen
region, corrects for the corresponding value in the ``field,'' and divides
by $\MLcrito$ to obtain $\Omato$.  Much, however, depends on the choice of
region.  A widely respected application of this approach is that of the
{\sc Cnoc} team \cite{Car97}, which uses rich clusters of galaxies.  These
systems sample large volumes of the early Universe, have dynamical masses
which can be measured by three independent methods (the virial theorem,
x-ray gas temperatures and gravitational lensing), and are subject to
reasonably well-understood evolutionary effects.  They are found to have
$M/L\sim200 \MLunits$ on average, giving $\Omato = 0.19 \pm 0.06$ when
$\Olamo = 0$ \cite{Car97}.  This result scales as $(1-0.4\Olamo)$
\cite{Car99a}, so that $\Omato$ drops to $0.11 \pm 0.04$ in a model
with $\Olamo = 1$.

The weak link in this chain of inference is that rich clusters may not
be characteristic of the Universe as a whole.  Only about 10\% of galaxies
are found in such systems.  If {\em individual\/} galaxies (like the
Milky Way, with $M/L\approx21\MLunits$) are substituted for clusters,
then the inferred value of $\Omato$ drops by a factor of ten, approaching
$\Obar$ and removing the need for CDM.  An effort to address the impact
of scale on $M/L$ arguments has led to the conclusion that
$\Omato = 0.16 \pm 0.05$ (for flat models) when regions of all scales
are considered from individual galaxies to superclusters \cite{Bah00}.

Another line of argument is based on the {\em cluster baryon fraction\/},
or ratio of baryonic-to-total mass ($\barf$) in galaxy clusters.
Baryonic matter is defined as the sum of visible galaxies and hot gas
(the mass of which can be inferred from x-ray temperature data).
Total cluster mass is measured by the above-mentioned methods
(virial theorem, x-ray temperature, or gravitational lensing).
At sufficiently large radii, the cluster may be taken as representative
of the Universe as a whole, so that $\Omato = \Obar/(\barf)$, where
$\Obar$ is fixed by big-bang nucleosynthesis (Sec.~\ref{sec:baryons}).
Applied to various clusters, this procedure leads to $\Omato = 0.3 \pm 0.1$
\cite{Bah99}.  This result is probably an upper limit, partly because
baryon enrichment is more likely to take place inside the cluster than out,
and partly because {\em dark\/} baryonic matter (such as MACHOs) is
not taken into account; this would raise $\Mbar$ and lower $\Omato$.

Other direct methods of constraining the value of $\Omato$ are rapidly
becoming available, including those based on the evolution of galaxy
cluster x-ray temperatures \cite{Hen00}, radio galaxy lobes as
``standard rulers'' in the classical angular~size-distance relation
\cite{Gue00} and distortions in the images of distant galaxies due
to weak gravitational lensing by intervening large-scale structures
\cite{Hoe02}.  In combination with other evidence, especially that
based on the SNIa magnitude-redshift relation and the CMB power
spectrum (to be discussed shortly), these techniques show considerable
promise for reducing the uncertainty in the matter density.

We move next to measurements of $\Omato$ based on the assumption that
the growth of LSS proceeded via gravitational instability from a Gaussian
spectrum of primordial density fluctuations (GI theory for short).
These argmuents are circular in the sense that such a process could
not have taken place as it did {\em unless\/} $\Omato$ is considerably
larger than $\Obar$.  But inasmuch as GI theory is the only
structure-formation theory we have which is both fully worked out and
in good agreement with observation (with some difficulties on small
scales \cite{Ost03}), this way of determining $\Omato$ should be
taken seriously.

According to GI theory, the formation of large-scale structures is more
or less complete by $z \approx \Omato^{-1} - 1$ \cite{Pad93}.  Therefore,
one way to constrain $\Omato$ is to look for {\em number density evolution\/}
in large-scale structures such as galaxy clusters.
In a low-matter-density Universe, this would be
relatively constant out to at least $z\sim1$, whereas in a
high-matter-density Universe one would expect the abundance
of clusters to drop rapidly with $z$ because they are still in the process
of forming.  The fact that massive clusters are seen at redshifts as high
as $z=0.83$ has been used to infer that $\Omato = 0.17^{+0.14}_{-0.09}$
for $\Olamo=0$ models, and $\Omato = 0.22^{+0.13}_{-0.07}$
for flat ones \cite{Bah98}.

Studies of the {\em power spectrum\/} $P(k)$ of the distribution of 
galaxies or other structures can be used in a similar way.
In GI theory, structures of a given mass form by the collapse of
large volumes in a low-matter-density Universe,
or smaller volumes in a high-matter-density Universe.  Thus $\Omato$
can be constrained by changes in $P(k)$ between one redshift and another.
Comparison of the mass power spectrum of Ly$\alpha$ absorbers
at $z\approx2.5$ with that of local galaxy clusters
at $z=0$ has led to an estimate of $\Omato = 0.46^{+0.12}_{-0.10}$ for
$\Olamo=0$ models \cite{Wei99}.  This result goes as approximately
$(1-0.4 \Olamo)$, so that the central value of $\Omato$ drops to 0.34
in a flat model, and 0.28 if $\Olamo=1$.  One can also constrain $\Omato$
from the local galaxy power spectrum alone, although this involves some
assumptions about the extent to which ``light traces mass'' (i.e. to
which visible galaxies trace the underlying density field).  Results
from the 2dF survey give $\Omato=0.29^{+0.12}_{-0.11}$ assuming
$\ho=0.7\pm0.1$ \cite{Pea01} (here and elsewhere we quote 95\% or
$2\sigma$ confidence levels where these can be read from the data,
rather than the more commonly reported 68\% or $1\sigma$ limits).
A preliminary best fit from the Sloan Digital Sky Survey ({\sc Sdss})
is $\Omato=0.19^{+0.19}_{-0.11}$ \cite{Bah03}.  As we discuss below,
there are good prospects for reducing this uncertainty by combining data
from galaxy surveys of this kind with CMB data \cite{Teg04}, though such
results must be interpreted with care at present \cite{Bri03}.

A third group of measurements, and one which has traditionally yielded 
the highest estimates of $\Omato$, comes from the analysis of
{\em galaxy peculiar velocities\/}.  These are generated by the
gravitational potential of locally over or under-dense regions
relative to the mean matter density.
The power spectra of the velocity and density distributions can be related
to each other in the context of GI theory in a way which depends explicitly
on $\Omato$.  Tests of this kind probe relatively small volumes and are
hence insensitive to $\Olamo$, but they can depend significantly on $\ho$
as well as the spectral index $n$ of the density distribution.  In
\cite{Zeh99}, where the latter is normalized to CMB fluctuations, results
take the form $\Omato\ho^{1.3}n^2 \approx 0.33\pm0.07$ or (taking $n=1$
and using our values of $\ho$) $\Omato\approx0.48 \pm 0.15$.

In summarizing these results, one is struck by the fact that
arguments based on gravitational instability (GI) theory favour values of
$\Omato\gtrsim0.2$ and {\em higher\/}, whereas purely empirical arguments
require $\Omato\lesssim0.4$ and {\em lower\/}.  The latter are in
fact compatible in some cases with values of $\Omato$ as low as $\Obar$,
raising the possibility that CDM might not in fact be necessary.  The
results from GI-based arguments, however, cannot be stretched this far.
What is sometimes done is to ``go down the middle'' and blend the results
of both kinds of arguments into a single bound of the form
$\Omato\approx0.3\pm0.1$.  Any such bound with $\Omato>0.05$ constitutes
a proof of the existence of CDM, since $\Obar\leqslant0.04$ from
(\ref{ObarValue}).  (Neutrinos only strengthen this argument, as we
note in Sec.~\ref{sec:neutrinos}.)  A more conservative
interpretation of the data, bearing in mind the full range of
$\Omato$ values implied above ($\Obar\lesssim\Omato\lesssim0.6$), is
\beq
\Ocdm = 0.3 \pm 0.3 \; .
\label{OcdmValue}
\eeq
But it should be stressed that values of $\Ocdm$ at the bottom of
this range carry with them the (uncomfortable) implication that the
conventional picture of structure formation via gravitational instability
is incomplete.  Conversely, {\em if our current understanding of structure
formation is correct, then CDM must exist} and $\Ocdm>0$.

The question, of course, becomes moot if CDM is discovered in the
laboratory.  From a large field of theoretical particle candidates,
two have emerged as frontrunners:  axions and supersymmetric
weakly-interacting massive particles (WIMPs).
The plausibility of both candidates rests on three properties: they are
(1)~{\em weakly interacting\/} (i.e. ``noticed'' by ordinary matter
primarily via their gravitational influence); (2)~{\em cold\/} (i.e.
non-relativistic in the early Universe, when structures began to form);
and (3)~{\em expected\/} on theoretical grounds to have a collective
density within a few orders of magnitude of the critical one.  We will
return to these particles in Secs.~\ref{ch6} and \ref{ch8} respectively.

\subsection{Massive neutrinos} \label{sec:neutrinos}

Since neutrinos indisputably exist in great numbers, they have been
leading dark-matter candidates for longer than axions or WIMPs.
They gained prominence in 1980 when teams in the U.S.A. and the Soviet Union
both reported evidence of nonzero neutrino rest masses.
While these claims did not stand up, a new round of experiments once again
indicates that $m_{\nu}$ (and hence $\Onu)>0$. 
 
The neutrino number density $n_{\nu}$ per species is $3/11$ that of the
CMB photons.  Since the latter are in thermal equilibrium, their number
density is $\ncmb=2\zeta(3)(k\Tcmb/\hbar c)^3/\pi^2$ \cite{Pad00} where
$\zeta(3)=1.202$.  Multiplying by $3/11$ and dividing through by the
critical density~(\ref{rcritoDefn}), we obtain
\beq
\Onu = \frac{{\textstyle \sum} m_{\nu} c^2}{(93.8 \mbox{ eV}) \ho^2} \; ,
\label{OnuDefn}
\eeq
where the sum is over three neutrino species.  We follow
convention and specify particle masses in units of eV/$c^2$, where 
1~eV/$c^2=1.602\times10^{-12}$~erg$/c^2=1.783\times10^{-33}$~g.
The calculations in this section are strictly valid only for
$m_{\nu} c^2 \lesssim 1$~MeV.  More massive neutrinos with
$m_{\nu} c^2\sim 1$~GeV were once considered as CDM candidates but
are no longer viable since experiments at the LEP collider
rule out additional neutrino species with masses up to at least half of
that of the $Z_0$ ($\capsub{m}{Z_0} c^2=91$~GeV).

Current laboratory upper bounds on neutrino rest masses are 
$m_{\nu_e}c^2<3$~eV, $m_{\nu_{\mu}}c^2<0.19$~MeV and 
$m_{\nu_{\tau}}c^2<18$~MeV, so it would appear feasible in principle 
for these particles to close the Universe.  In fact $m_{\nu_{\mu}}$ and 
$m_{\nu_{\tau}}$ are limited far more stringently by (\ref{OnuDefn}) 
than by laboratory bounds.  Perhaps the best-known theory along these
lines is that of Sciama \cite{Sci93a}, who postulated a population of
$\tau$-neutrinos with $m_{\nu_{\tau}}c^2\approx29$~eV.
Eq.~(\ref{OnuDefn}) shows that such neutrinos would account for much
of the dark matter, contributing a {\em minimum\/} collective density
of $\Onu\geqslant0.38$ (assuming as usual that $\ho\leqslant0.9$).
We will consider decaying neutrinos further in Sec.~\ref{ch7}.

Strong upper limits can be set on $\Onu$ within the context of the
gravitational instability picture.  Neutrinos constitute
{\em hot dark matter\/} (i.e. they are relativistic when they
decouple from the primordial fireball) and are therefore able
to stream freely out of density perturbations in the early Universe,
erasing them before they have a chance to grow and suppressing the
power spectrum $P(k)$ of density fluctuations on small scales $k$.
Agreement between LSS theory and observation can be achieved in models
with $\Onu$ as high as 0.2, but only in combination with values of the
other cosmological parameters that are no longer considered realistic
(e.g. $\Obar+\Ocdm=0.8$ and $\ho=0.5$) \cite{Gaw98}.
A recent 95\% confidence-level upper limit on the neutrino density
based on data from the 2dF galaxy survey reads \cite{Elg02}
\beq
\Onu < 0.24 (\Obar+\Ocdm) \; ,
\label{OnuMax}
\eeq
if no prior assumptions are made about the values of $\Omato$, $\Obar$,
$\ho$ and $n$.  When combined with Eqs.~(\ref{ObarValue}) and
(\ref{OcdmValue}) for $\Obar$ and $\Ocdm$, this result implies that
$\Onu < 0.15$.  Thus if structure grows by gravitational instability 
as generally assumed, then neutrinos may still play a significant,
but not dominant role in cosmological dynamics.  Note that neutrinos
lose energy after decoupling and become {\em non\/}relativistic on
timescales $\tnr \approx 190,000$~yr~$(m_{\nu}c^2/$eV)$^{-2}$ \cite{Kol90},
so that Eq.~(\ref{OnuMax}) is quite consistent with our neglect of
relativistic particles in the Friedmann-Lema\^{\i}tre
equation (Sec.~\ref{sec:FL}).

{\em Lower\/} limits on $\Onu$ follow from a spectacular string of neutrino
experiments employing particle accelerators ({\sc Lsnd} \cite{Ath98}),
cosmic rays in the upper atmosphere (Super-Kamiokande \cite{Fuk98b}),
the flux of neutrinos from the Sun ({\sc Sage} \cite{Abd99}, Homestake
\cite{Cle98}, {\sc Gallex} \cite{Ham99}, {\sc Sno} \cite{Ahm02}),
nuclear reactors (Kam{\sc Land} \cite{Egu03}) and, most recently,
directed neutrino beams (K2K \cite{Ahn03}).  The evidence in each case
points to interconversions between species known as neutrino oscillations,
which can only take place if all species involved have nonzero rest masses.
It now appears that oscillations occur between at least two neutrino mass
eigenstates whose masses squared differ by
$\Delta_{21}^2\equiv|m_2^2-m_1^2| = 6.9^{+1.5}_{-0.8} \times 10^{-5}
\mbox{ eV}^2/c^4$ and $\Delta_{31}^2\equiv|m_3^2-m_1^2| = 2.3^{+0.7}_{-0.9}
\times 10^{-3} \mbox{ eV}^2/c^4$ \cite{Mal04}.
Scenarios involving a fourth ``sterile'' neutrino were once thought to be
needed but are disfavoured by global fits to all the data.  Oscillation
experiments are sensitive to mass differences, and cannot fix the mass
of any one neutrino flavour unless they are combined with another
experiment such as neutrinoless double-beta decay \cite{Kla01}.
Nevertheless, if neutrino masses are hierarchical, like those of
other fermions, then one can take $m_3 \gg m_2 \gg m_1$ so that the
above measurements impose a lower limit on total neutrino mass:
$\Sigma m_{\nu} c^2 > 0.045$~eV.  Putting this number into (\ref{OnuDefn}),
we find that
\beq
\Onu > 0.0005 \; ,
\label{OnuMin}
\eeq
with $\ho\leqslant0.9$ as usual.  If, instead, neutrino masses are nearly
degenerate, then $\Onu$ could in principle be larger than this, but will
in any case still lie below the upper bound~(\ref{OnuMax}) imposed by
structure formation.  The neutrino contribution to $\Ototo$ is thus anywhere
from about one-tenth that of luminous matter (Sec.~\ref{sec:baryons})
to about one-quarter of that attributed to CDM (Sec.~\ref{sec:cdm}).
We emphasize that, if $\Ocdm$ is small, then $\Onu$ must be small also.
In theories (like that to be discussed in Sec.~\ref{ch7}) where the density
of neutrinos exceeds that of other forms of matter, one would need to
modify the standard gravitational instability picture by encouraging the
growth of structure in some other way, as for instance by ``seeding''
them with loops of cosmic string.

\subsection{Dark energy} \label{sec:vacenergy}

There are at least four reasons to include a cosmological constant
($\Lambda$) in Einstein's field equations.  The first
is mathematical: $\Lambda$ plays a role in these equations
similar to that of the additive constant in an indefinite integral
\cite{Rin77}.  The second is dimensional: $\Lambda$ specifies
the radius of curvature $R_\Lambda \equiv \Lambda^{-1/2}$ in closed
models at the moment when the matter density parameter $\Omat$ goes
through its maximum, providing a fundamental length scale for
cosmology \cite{Pri95}.  The third is dynamical: $\Lambda$
determines the asymptotic expansion rate of the Universe according to
Eq.~(\ref{WolfsEqun}).  And the fourth is material:
$\Lambda$ is related to the energy density of the
vacuum via Eq.~(\ref{rlamDefn}).

With so many reasons to take this term seriously, why was it ignored
for so long?  Einstein himself set $\Lambda=0$ in 1931
``for reasons of logical economy,'' because he saw no hope of measuring
this quantity experimentally at the time.  He is often quoted as adding that
its introduction in 1915 was the ``biggest blunder'' of his life.
This comment (which was attributed to him by Gamow \cite{Gam70} but does
not appear anywhere in his writings), is sometimes interpreted as a
rejection of the very idea of a cosmological constant.  It more likely
represents Einstein's rueful recognition that, by invoking the
$\Lambda$-term solely to obtain a static solution of the field equations,
he had narrowly missed what would surely have been one of the greatest
{\em triumphs\/} of his life: the prediction of cosmic expansion.

The relation between $\Lambda$ and the energy density of the vacuum
has led to a quandary in more recent times: modern quantum field theories
such as quantum chromodynamics (QCD), electroweak (EW) and grand unified
theories (GUTs) imply values for $\rlam$ and $\Olamo$ that are impossibly
large (Table~\ref{table4.1}).  This ``cosmological-constant problem''
has been reviewed by many people, but there is no consensus on how to
solve it \cite{Car01}.  It is undoubtedly another reason why some
cosmologists have preferred to set $\Lambda = 0$, rather than deal with
a parameter whose physical origins are still unclear.

Setting $\Lambda$ to zero, however, is no longer an appropriate response
because observations (reviewed below) now indicate that $\Olamo$ is in fact
of order {\em unity\/}.  The cosmological constant problem has therefore
become more baffling than before, in that an explanation of this parameter
must apparently contain a cancellation mechanism which is almost --- but not
quite --- exact, failing at precisely the 123rd decimal place.

\begin{table}[t!]
\caption{Theoretical Estimates of $\Olamo$}
\begin{tabular}{@{}ccc@{}}
\hline
Theory & Predicted value of $\rlam$ & $\Olamo$ \\
\hline
QCD & (0.3 GeV)$^4 \hbar^{-3} c^{-5} = 10^{16}$~g~cm$^{-3}$ & $10^{44}\ho^{-2}$ \\
EW & (200 GeV)$^4 \hbar^{-3} c^{-5} = 10^{26}$~g~cm$^{-3}$ & $10^{55}\ho^{-2}$ \\
GUTs & ($10^{19}$ GeV)$^4 \hbar^{-3} c^{-5} = 10^{93}$~g~cm$^{-3}$ & $10^{122}\ho^{-2}$ \\
\hline
\end{tabular}
\label{table4.1}
\end{table}

One suggestion for understanding the possible nature of such a cancellation
has been to treat the vacuum energy field literally as an Olbers-type
summation of contributions from different places in the Universe \cite{Put89}.
It can then be handled with the same formalism that we have developed in
Secs.~\ref{ch2} and \ref{ch3} for background radiation.  This has the
virtue of framing the problem in concrete terms, and raises some interesting
possibilities, but does not in itself explain why the energy density
inherent in such a field does not gravitate in the conventional way
\cite{Wes91b}.

Another idea is that theoretical expectations for the
value of $\Lambda$ might refer only to the latter's ``bare'' value,
which could have been progressively ``screened'' over time.
The cosmological constant then becomes a {\em variable\/} cosmological
term \cite{Ove98}.  In such a scenario the ``low'' value of $\Olamo$
merely reflects the fact that the Universe is old.
In general, however, this means modifying Einstein's field equations
and/or introducing new forms of matter such as scalar fields.
We look at this suggestion in more detail in Sec.~\ref{ch5}.

A third possibility occurs in higher-dimensional gravity,
where the cosmological constant can arise as an artefact of dimensional
reduction (i.e. in extracting the appropriate four-dimensional limit from
the theory).  In such theories the ``effective'' $\Lambda_{\,4}$ may be
small while its $N$-dimensional analog $\Lambda_N$ is large \cite{Wes01}.
We consider some aspects of higher-dimensional gravity in Sec.~\ref{ch9}.

As a last recourse, some workers have argued that $\Lambda$ might be small
``by definition,'' in the sense that a universe in which $\Lambda$ was large
would be incapable of giving rise to intelligent observers like ourselves
\cite{Wei01}.  This is an application of the anthropic principle whose
status, however, remains unclear.

\subsection{Cosmological concordance}

Let us pass to what is known about the value of $\Olamo$ from cosmology.
It is widely believed that the Universe originated in a
big bang singularity rather than passing through a ``big bounce'' at the
beginning of the current expansionary phase.  By differentiating the
Friedmann-Lema\^{\i}tre equation~(\ref{FL2}) and setting the
expansion rate and its time derivative to zero, one obtains
an upper limit (sometimes called the Einstein limit $\Oein$)
on $\Olamo$ as a function of $\Omato$.  For $\Omato=0.3$ the
requirement that $\Olamo<\Oein$ implies $\Olamo<1.71$, a limit that
tightens to $\Olamo<1.16$ for $\Omato=0.03$ \cite{OW03}.

A slightly stronger constraint can be formulated (for closed models) in
terms of the antipodal redshift.  The antipodes are the set of points
located at $\chi=\pi$, where $\chi$ (radial coordinate distance)
is related to $r$ by $d\chi=(1-k r^2)^{-1/2}\,dr$.  Using (\ref{dtdr})
and (\ref{dtdz}) this can be rewritten in the form
$d\chi=-(c/\Ho\Ro)\,dz/\Htil(z)$
and integrated with the help of (\ref{FL2}).
Gravitational lensing of sources beyond the antipodes
cannot give rise to normal (multiple) images \cite{Got89}, so the
redshift $z_a$ of the antipodes must exceed that of the most distant
normally-lensed object, currently a protogalaxy at $z=10.0$ \cite{Pel04}.
Requiring that $z_a > 10.0$ leads to the upper bound $\Olamo<1.51$ if
$\Omato=0.3$.  This tightens to $\Olamo<1.13$ for \LBDM-type models
with $\Omato=0.03$.  

The statistics of gravitational lenses lead to a different and stronger
upper limit which applies regardless of geometry.
The increase in path length to a given redshift in vacuum-dominated
models (relative to, say, EdS) means that there are more sources to
be lensed, and presumably more lensed objects to be seen.
The observed frequency of lensed quasars, however, is rather modest,
leading to an early bound of $\Olamo<0.66$ for flat models \cite{Koc96b}.
Dust could hide distant sources \cite{Mal97}.  However, radio lenses
should be far less affected, and these give only slightly weaker constraints:
$\Olamo < 0.73$ (for flat models) or $\Olamo \lesssim 0.4 + 1.5\Omato$
(for curved ones) \cite{Fal98}.  Recent indications are that this method
loses much of its sensitivity to $\Olamo$ when assumptions about the
lensing population are properly normalized to galaxies at high redshift
\cite{Kee02}.  A new limit from radio lenses in the Cosmic Lens All-Sky
Survey ({\sc Class}) is $\Olamo<0.89$ for flat models \cite{Mit04}.

Tentative {\em lower\/} limits have been set on $\Olamo$ using faint galaxy
number counts.  This premise is similar to that behind lensing statistics:
the enhanced comoving volume at large redshifts in vacuum-dominated
models should lead to greater (projected) galaxy number densities at
faint magnitudes.  In practice, it has proven difficult to disentangle
this effect from galaxy luminosity evolution.  Early claims of a best fit
at $\Olamo\approx 0.9$ \cite{Fuk90} have been disputed on the basis that
the steep increase seen in numbers of blue galaxies is not matched in the
K-band, where luminosity evolution should be less important
\cite{Gar93}.  Attempts to account more fully for evolution have
subsequently led to a lower limit of $\Olamo>0.53$ \cite{Tot97},
and most recently a reasonable fit (for flat models) with a vacuum
density parameter of $\Olamo=0.8$ \cite{Tot00}.

\begin{figure}[t!]
\begin{center}
\includegraphics[width=100mm]{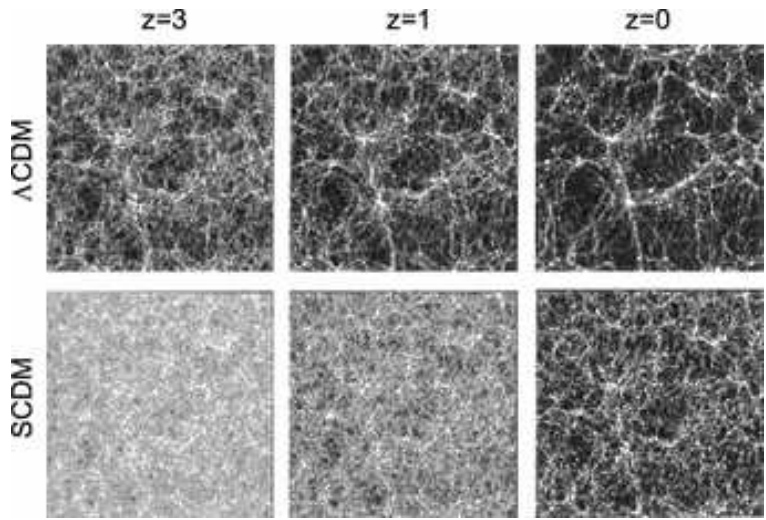}
\end{center}
\caption{Numerical simulations of structure formation.
   In the top row is the \LCDM\ model with $\Omato=0.3, \Olamo=0.7$
   and $\ho=0.7$.  The bottom row shows the EdS (``SCDM'') model with
   $\Omato=1, \Olamo=0$ and $\ho=0.5$.  The panel size is comoving with
   the Hubble expansion, and time runs from left ($z=3$) to right ($z=0$).
   (Images courtesy of J.~Colberg and the {\sc Virgo} Consortium).}
\label{fig4.3}
\end{figure}

Other evidence for a significant $\Olamo$-term has come from numerical
simulations of large-scale structure formation.
Fig.~\ref{fig4.3} shows the evolution of massive structures between $z=3$
and $z=0$ in simulations by the {\sc Virgo}~Consortium \cite{Jen98}. 
The \LCDM\ model (top row) provides a qualitatively better match to the
observed distribution of galaxies than EdS (``SCDM,'' bottom row).
The improvement is especially marked at higher redshifts (left-hand panels).
Power spectrum analysis, however, reveals that the match is not particularly
good in either case \cite{Jen98}.  This could reflect {\em bias} (i.e.
systematic discrepancy between the distributions of mass and light).
Different combinations of $\Omato$ and $\Olamo$ might also provide better
fits.  Simulations of closed \LBDM-type models would be of particular
interest \cite{Whi96,OP01,Fel93}.

The first measurements to put both lower {\em and\/} upper bounds on
$\Olamo$ have come from Type Ia supernovae (SNIa).  These objects are
very bright, with luminosities that are consistent (when calibrated against
rise time), and they are not thought to evolve significantly with redshift.
All of these properties make them ideal standard candles for use in the
magnitude-redshift relation.  In 1998 and 1999 two independent groups
({\sc Hzt} \cite{Rie98} and {\sc Scp} \cite{Per99}) reported a systematic
dimming of SNIa at $z\approx0.5$ by about 0.25~magnitudes relative to
that expected in an EdS model, suggesting that space at these redshifts
is ``stretched'' by dark energy.  These programs have now expanded to
encompass more than 200 supernovae, with results that can be summarized
in the form of a 95\% confidence-level relation between $\Omato$ and
$\Olamo$ \cite{Ton03}:
\beq
\Olamo = 1.4 \, \Omato + 0.35 \pm 0.28 \; .
\label{OlamSNdefn}
\eeq
Such a relationship is inconsistent with the EdS and OCDM models, which have
large values of $\Omato$ with $\Olamo=0$.  To extract quantitative limits on
$\Olamo$ alone, we recall that
$\Omato\geqslant\Obar\geqslant0.02$ (Sec.~\ref{sec:baryons}) and
$\Omato\leqslant0.6$ (Sec.~\ref{sec:cdm}).  Interpreting these constraints
as conservatively as possible (i.e. $\Omato=0.31\pm0.29$), we infer from
(\ref{OlamSNdefn}) that
\beq
\Olamo = 0.78 \pm 0.49 \; .
\label{OlamSNvalue}
\eeq
This is not a high-precision measurement, but it is enough to establish
that $\Olamo\geqslant0.29$ and hence that {\em the dark energy is real\/}.
New supernovae observations continue to reinforce this conclusion
\cite{Kno03}.  Not all cosmologists are yet convinced, however, and 
a healthy degree of caution is still in order regarding a quantity whose
physical origin is so poorly understood.  Alternative explanations can 
be constructed that fit the observations by means of ``grey dust''
\cite{Agu99} or luminosity evolution \cite{Dre00}, though these must
be increasingly fine-tuned to match new SNIa data at higher redshifts
\cite{Rie04}.  Much also remains to be learned about the physics of
supernova explosions.  The shape of the magnitude-redshift relation
suggests that observations may have to reach $z\sim2$ routinely in order
to be able to discriminate statistically between models (like \LCDM\ and
\LBDM) with different ratios of $\Omato$ to $\Olamo$.

Further support for the existence of dark energy has arisen from a
completely independent source: the angular power spectrum of CMB fluctuations.
These are produced by density waves in the primordial plasma, 
``the oldest music in the Universe'' \cite{Lin01}.  The first peak
in their power spectrum picks out the angular size of the largest
fluctuations in this plasma at the moment when the Universe became
transparent to light.  Because it is seen through the ``lens'' of a
curved Universe, the location of this peak is sensitive to the latter's
total density $\Ototo=\Olamo+\Omato$.  Beginning in 2000, a series of
increasingly precise measurements of $\Ototo$ have been reported from 
experiments including {\sc Boomerang} \cite{deB00}, {\sc Maxima}
\cite{Han00}, {\sc Dasi} \cite{Pry01} and most recently the {\sc Wmap}
satellite.  The latter's results can be summarized as follows
\cite{Spe03} (at the 95\% confidence level, assuming $\ho>0.5$):
\beq
\Omato + \Olamo = 1.03 \pm 0.05 \; .
\label{OlamCMBdefn}
\eeq
{\em The Universe is therefore spatially flat, or very close to it.}
To extract a value for $\Olamo$ alone, we can do as in the SNIa case and
substitute our matter density bounds ($\Omato=0.31\pm0.29$) into
(\ref{OlamCMBdefn}) to obtain
\beq
\Olamo = 0.72 \pm 0.29 \; .
\label{OlamCMBvalue}
\eeq
This is consistent with (\ref{OlamSNvalue}), but has error bars which have
been reduced by almost half, and are now due entirely to the uncertainty in
$\Omato$.  This measurement is impervious to most of the uncertainties of
the earlier ones, because it leapfrogs ``local'' systems whose interpretation
is complex (supernovae, galaxies, and quasars), going directly back to the
radiation-dominated era when physics was simpler.  Eq.~(\ref{OlamCMBvalue})
is sufficient to establish that $\Olamo\geqslant0.43$, and hence that
{\em dark energy not only exists, but probably dominates the energy density
of the Universe.}

The CMB power spectrum favours vacuum-dominated models, but is not yet
resolved with sufficient precision to discriminate (on its own) between
models which have exactly the critical density (like \LCDM) and those 
which are close to the critical density (like \LBDM).  As it stands,
the location of the first peak in these data actually hints at
``marginally closed'' models, although the implied departure from
flatness is not statistically significant and could also be explained
in other ways \cite{Whi00}.

Much attention is focused on the higher-order peaks of the spectrum,
which contain valuable clues about the matter component.
Odd-numbered peaks are produced by regions of the primordial plasma
which have been maximally compressed by infalling material, and even ones
correspond to maximally rarefied regions which have rebounded due to
photon pressure.  A high baryon-to-photon ratio enhances the compressions
and retards the rarefractions, thus suppressing the size of the second
peak relative to the first.  The strength of this effect depends on the
fraction of baryons (relative to the more weakly-bound neutrinos
and CDM particles) in the overdense regions.  The {\sc Boomerang} and
{\sc Maxima} data show an unexpectedly weak second peak.  While there are
a number of ways to account for this in \LCDM\ models (e.g., by ``tilting''
the primordial spectrum), the data are fit most naturally by a ``no-CDM''
\LBDM\ model with $\Ocdm=0$, $\Omato=\Obar$ and $\Olamo\approx1$ \cite{McG00}.
Models of this kind have been discussed for some time in connection with
analyses of the Lyman-$\alpha$ forest of quasar absorption lines
\cite{OP01,Lie92}.  The {\sc Wmap} data show a stronger second peak and
are fit by both \LCDM\ and \LBDM-type models \cite{McG04}.  Data from the
upcoming {\sc Planck} satellite should settle this issue.

\begin{figure}[t!]
\begin{center}
\includegraphics[angle=90,width=100mm]{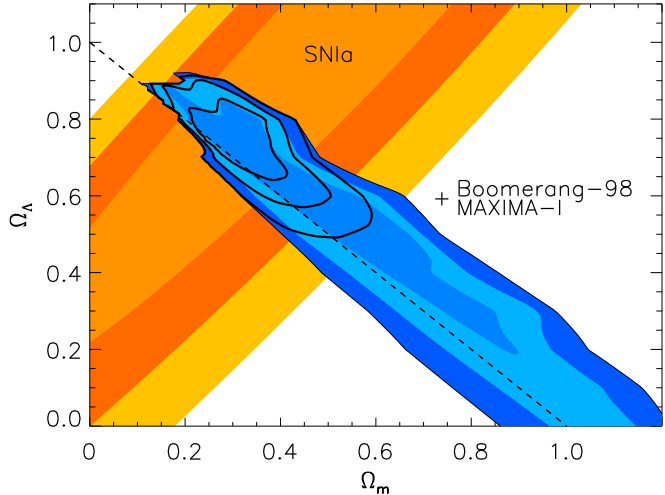}
\end{center}
\caption{Observational constraints on the values of $\Omato$ and $\Olamo$
   from both SNIa and CMB observations ({\sc Boomerang, Maxima}).  Shown
   are 68\%, 95\% and 99.7\% confidence intervals inferred both separately
   and jointly from the data.  (Reprinted from \cite{Jaf00} by permission
   of A.~Jaffe and P.~L.~Richards.)}
\label{fig4.4}
\end{figure}

The best constraints on $\Olamo$ come from taking {\em both\/} the
supernovae and microwave background results at face value, and
substituting one into the other.  This provides a valuable cross-check
on the matter density, because the SNIa and CMB constraints are very
nearly orthogonal in the $\Omato$-$\Olamo$ plane (Fig.~\ref{fig4.4}).
Thus, forgetting all about our conservative bounds on $\Omato$ and merely
substituting (\ref{OlamCMBdefn}) into (\ref{OlamSNdefn}), we find
\beq
\Olamo = 0.75 \pm 0.12 \; .
\label{OlamBoth}
\eeq
Alternatively, extracting the matter density parameter, we obtain
\beq
\Omato = 0.28 \pm 0.12 \; .
\label{OmatBoth}
\eeq
These results further tighten the case for a universe dominated by dark
energy.  Eq.~(\ref{OlamBoth}) also implies that $\Olamo\leqslant 0.87$,
which begins to put pressure on models of the \LBDM\ type.  Perhaps most
importantly, Eq.~(\ref{OmatBoth}) establishes that $\Omato\geqslant 0.16$,
which is inconsistent with \LBDM\ and {\em requires the existence of CDM}.
Moreover, the fact that the range of values picked out by (\ref{OmatBoth})
agrees so well with that derived in Sec.~\ref{sec:cdm} constitutes solid
evidence for the \LCDM\ model in particular, and for the gravitational
instability picture of large-scale structure formation in general.

The depth of the change in thinking that has been triggered by these
developments on the observational side can hardly be exaggerated.  Only a
few years ago, it was still routine to set $\Lambda=0$ and cosmologists had
two main choices: the ``one true faith'' (flat, with $\Omato\equiv1$),
or the ``reformed'' (open,
with individual believers being free to choose their own values near
$\Omato\approx0.3$).  All this has been irrevocably altered by the CMB
experiments.  If there is a guiding principle now, it is no longer
$\Omato\approx0.3$, and certainly not $\Olamo=0$; it is $\Ototo\approx1$
from the power spectrum of the CMB.  Cosmologists have been obliged to
accept a $\Lambda$-term, and it is not so much a question of whether or
not it dominates the energy budget of the Universe, but by {\em how much\/}.

\subsection{The coincidental Universe} \label{sec:preposterous}

The observational evidence reviewed in the foregoing sections has led
us into the corner of parameter space occupied by vacuum-dominated models
with close to (or exactly) the critical density.
The resulting picture is self-consistent, and agrees with nearly all
the data.  Major questions, however, remain on the theoretical side.
Prime among these is the problem of the cosmological constant,
which (as described above) is particularly acute in models with nonzero
values of $\Lambda$, because one can no longer hope that a simple symmetry
of nature will eventually be found which requires $\Lambda=0$.

\begin{figure}[t!]
\begin{center}
\includegraphics[width=\textwidth]{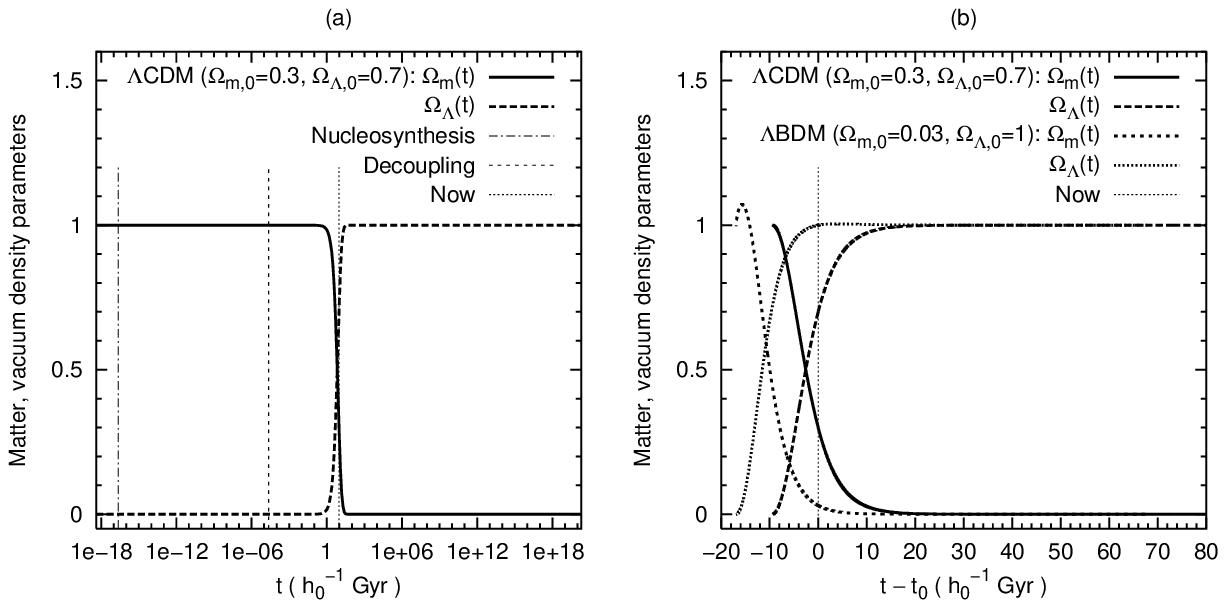}
\end{center}
\caption{The evolution of $\Omat(t)$ and $\Olam(t)$ in vacuum-dominated
   models.  Panel~(a) shows a single model (\LCDM) over twenty powers
   of time in either direction.  Plotted this way, we are seen to live
   at a very special time (marked ``Now'').  Standard nucleosynthesis
   ($\tnuc\sim1$~s) and matter-radiation decoupling times
   ($\tdec\sim10^{11}$~s) are included for comparison.  Panel~(b)
   shows both the \LCDM\ and \LBDM\ models on a linear rather than
   logarithmic scale, for the first 100$\ho^{-1}$~Gyr after the big
   bang (i.e. the lifetime of the stars and galaxies).}
\label{fig4.5}
\end{figure}

A related concern has to do with the {\em evolution\/} of the matter
and dark-energy density parameters $\Omat$ and $\Olam$ over time.
Eqs.~(\ref{rcritoDefn}) and (\ref{rhoR}) can be combined to give
\beq
\Omat(t) \equiv \frac{\rmat(t)}{\rcrit(t)} =
   \frac{\Omato}{\Rtil^3(t) \, \Htil^2(t)} \qquad
\Olam(t) \equiv \frac{\rlam(t)}{\rcrit(t)} =
   \frac{\Olamo}{\Htil^2(t)} \; .
\label{OmatOlam}
\eeq
Here $\Htil[z(t)]$ is given by (\ref{FL2}) as usual and
$z(t)=1/\Rtil(t)-1$ from (\ref{zDefn}).  Eqs.~(\ref{OmatOlam})
can be solved exactly for flat models using (\ref{Rt2comp}) and
(\ref{Ht2comp}) for $\Rtil(t)$ and $\Htil(t)$.  Results for the
\LCDM\ model are illustrated in Fig.~\ref{fig4.5}(a).
At early times, dark energy is insignificant relative to matter
($\Olam\sim0$ and $\Omat\sim1$), but the situation is reversed at late
times when $\Olam\sim1$ and $\Omat\sim0$.

What is remarkable in this figure is the location of the present
(marked ``Now'') in relation to the values of $\Omat$ and $\Olam$.
{\em We have apparently arrived on the scene at the precise moment when
these two parameters are in the midst of switching places.\/}  (We have not
considered radiation density $\Orad$ here, but similar considerations apply
to it.) This has come to be known as the coincidence problem, and Carroll
\cite{Car01} has aptly described such a universe as ``preposterous,''
writing: ``This scenario staggers under the burden of its unnaturalness,
but nevertheless crosses the finish line well ahead of any of its competitors
by agreeing so well with the data.''  Cosmology may be moving toward
a position like that of particle physics, where a standard model accounts
for all observed phenomena to high precision, but appears to be founded
on a series of finely-tuned parameter values which leave one with the
distinct impression that the underlying reality has not yet been grasped.

Fig.~\ref{fig4.5}(b) is a close-up view of Fig.~\ref{fig4.5}(a),
with one difference: it is plotted on a {\em linear\/} scale in time
for the first $100\ho^{-1}$~Gyr after the big bang,
rather than a logarithmic scale over $10^{\pm20}\ho^{-1}$~Gyr.
The rationale for this is simple: 100~Gyr is approximately the lifespan
of the galaxies (as determined by their main-sequence stellar populations).
One would not, after all, expect observers to appear on the scene long
after all the galaxies had disappeared, or in the early stages of the
expanding fireball.
Seen from the perspective of Fig.~\ref{fig4.5}(b), the coincidence,
while still striking, is perhaps no longer so preposterous.
The \LCDM\ model still appears fine-tuned, in that ``Now'' follows
rather quickly on the heels of the epoch of matter-vacuum equality.
In the \LBDM\ model, $\Omato$ and $\Olamo$ are closer to the cosmological
time-averages of $\Omat(t)$ and $\Olam(t)$ (namely zero and one
respectively).  In such a picture it might be easier to believe that we
have not arrived on the scene at a special time, but merely a
{\em late\/} one.  Whether or not this is a helpful way to approach
the coincidence problem is, to a certain extent, a matter of taste.

To summarize the contents of this section: what
can be seen with our telescopes constitutes no more than one percent of
the density of the Universe.  The rest is dark.  A small portion (no more
than five percent) of this dark matter is made up of ordinary baryons.
Many observational arguments hint at the existence of a second, more
exotic species known as cold dark matter (though they do not quite
establish its existence unless they are combined with ideas about the
formation of large-scale structure).  Experiments also imply the
existence of a third dark-matter species, the massive neutrino,
although its role appears to be more limited.  Finally, all these 
components are dwarfed in importance by a newcomer whose physical
origin remains shrouded in obscurity: the dark energy of the vacuum.

In the remainder of this review, we explore the leading candidates
for dark energy and matter in more detail: the cosmological vacuum,
elementary particles such as axions, neutrinos and weakly-interacting
massive particles (WIMPs), and black holes.  Our approach is to
constrain each proposal by calculating its contributions to the 
background radiation, and comparing these with observational data
at wavelengths ranging from radio waves to $\gamma$-rays.
In the spirit of Olbers' paradox and what we have done so far, our main
question for each candidate will thus be: Just {\em how\/} dark is it?

\section{Dark energy} \label{ch5}

\subsection{The variable cosmological ``constant''} \label{sec:VarLam}

The cosmological-constant problem is essentially the problem of reconciling
the very high vacuum-energy densities expected on the basis of quantum
field theory with the small (but nonzero) dark-energy density now inferred
from cosmological observation (Sec.~\ref{sec:vacenergy}).
Many authors have sought to bridge the gap by looking for a mechanism
that would allow the energy density $\rvac$ of the vacuum to {\em decay\/}
with time.  Since $\Lambda c^2 = 8\pi G\rvac$ from (\ref{rlamDefn}), this
means replacing Einstein's cosmological constant by a variable
``cosmological term.''  With such a mechanism in hand, the problem 
would be reduced to explaining why the Universe is of intermediate age:
old enough that $\Lambda$ has relaxed from primordial values like those
suggested by quantum field theory to the values which we measure now,
but young enough that $\Olam\equiv\rvac/\rcrit$ has not yet reached its
asymptotic value of unity.

Energy conservation requires that any decrease in the energy density
of the vacuum be made up by a corresponding increase somewhere else.
In some scenarios, dark energy goes into the kinetic energy of new forms
of matter such as scalar fields, which have yet to be observed in nature.
In others it is channelled instead into baryons, photons or neutrinos.
Baryonic decays would produce equal amounts of matter and antimatter,
whose subsequent annihilation would flood the Universe with $\gamma$-rays.
Radiative decays would similarly pump photons into intergalactic space,
but are harder to constrain because they could in principle involve
any part of the electromagnetic spectrum.  As we will see, however,
robust limits can be set on any such process under conservative assumptions.  

But how can $\Lambda$, originally introduced by Einstein in 1917 as a
constant of nature akin to $c$ and $G$, be allowed to vary?  To answer this,
we go back to the field equations of general relativity:
\beq
{\mathcal R}_{\mu\nu} - \frac{1}{2} \, {\mathcal R} \, g_{\mu\nu} - \Lambda \,
g_{\mu\nu} = -(8\pi G/c^{\,4})\, {\mathcal T}_{\mu\nu} \; .
\label{EFEs}
\eeq
The covariant derivative of these equations can be written in the following
form with the help of the Bianchi identities, which read
$\nabla^{\nu}({\mathcal R}_{\mu\nu} - \frac{1}{2} \, {\mathcal R} \,
g_{\mu\nu})=0$:
\beq
\partial_{\mu} \Lambda = \frac{8\pi G}{c^{\,4}} \, \nabla^{\nu} \Tmat \; .
\label{diffEFEs}
\eeq
Within Einstein's theory, it follows that $\Lambda=$~constant as long
as matter and energy (as contained in $\Tmat$) are conserved.

In variable-$\Lambda$ theories, one must therefore do one of three things:
abandon matter-energy conservation, modify general relativity, or stretch
the definition of what is conserved.  The first of these routes was explored
as early as 1933 by Bronstein \cite{Bro33}, who sought to connect energy
non-conservation with the cosmological arrow of time.  Bronstein was executed
in Stalin's Soviet Union a few years later, and his work is not widely
known \cite{Kra96}.

Today, few physicists would be willing to sacrifice energy conservation
outright.  Some, however, would be willing to modify general relativity,
or to consider new forms of matter and energy.  Historically, these two
approaches have sometimes been seen as distinct, with one being a change
to the ``geometry of nature'' while the other is concerned with the
material content of the Universe.  The modern tendency, however, is to
regard them as equivalent.  This viewpoint is best personified by Einstein,
who in 1936 compared the left-hand (geometrical) and right-hand (matter)
sides of his field equations to ``fine marble'' and ``low-grade wooden''
wings of the same house \cite{Ein36}.  In a more complete theory, he
argued, matter fields of all kinds would be seen to be just as
geometrical as the gravitational one.

\subsection{Models based on scalar fields}

Let us see how this works in one of the oldest and simplest
variable-$\Lambda$ theories:  a modification of general relativity in 
which the metric tensor $g_{\mu\nu}$ is supplemented by a scalar field
$\varphi$ whose coupling to matter is determined by a parameter $\omega$.
Ideas of this kind go back to Jordan in 1949 \cite{Jor49},
Fierz in 1956 \cite{Fie56} and Brans and Dicke in 1961 \cite{Bra61}.
In those days, of course, new scalar fields were not bandied about
as freely as they are today, and all these authors sought to associate
$\varphi$ with a known quantity.
Various lines of argument (notably Mach's principle) pointed to an
identification with Newton's gravitational ``constant'' such that
$G\sim 1/\varphi$.  By 1968 it was appreciated that $\Lambda$ and
$\omega$ too would depend on $\varphi$ in general \cite{Ber68}.
The original Brans-Dicke theory (with $\Lambda=0$) has subsequently
been extended to {\em generalized scalar-tensor theories\/}
in which $\Lambda=\Lambda(\varphi)$ \cite{End77},
$\Lambda=\Lambda(\varphi)$, $\omega=\omega(\varphi)$ \cite{Bar90} and
$\Lambda=\Lambda(\varphi,\psi)$, $\omega=\omega(\varphi)$ where
$\psi\equiv\partial^{\mu}\varphi\,\partial_{\mu}\varphi$ \cite{Fuk02}.
In the last and most general of these cases, the field equations read
\beqa
{\mathcal R}_{\mu\nu} & - & \frac{1}{2} \, {\mathcal R} \, g_{\mu\nu} +
   \frac{1}{\varphi} \left[ \nabla_{\mu} ( \partial_{\nu}\varphi ) -
   \Box\,\varphi \, g_{\mu\nu} \right] + \frac{\omega(\varphi)}{\varphi^2} 
   \left( \partial_{\mu}\varphi \, \partial_{\nu}\varphi - \frac{1}{2} \,
   \psi \, g_{\mu\nu} \right) \nonumber \\
& - & \Lambda(\varphi,\psi) \, g_{\mu\nu} + 2 \,
   \frac{\partial\Lambda(\varphi,\psi)}{\partial\psi} \, 
   \partial_{\mu}\varphi \,\partial_{\nu}\varphi = -
   \frac{8\pi}{\varphi\,c^{\,4}}\,{\mathcal T}_{\mu\nu} \; ,
\label{GEFEs}
\eeqa
where $\Box\,\varphi\equiv \nabla^{\mu}(\partial_{\mu}\varphi)$ is the
D'Alembertian.  These reduce to Einstein's equations~(\ref{EFEs}) when
$\varphi=$~const~$=1/G$.

If we now repeat the exercise on the previous page and take the 
covariant derivative of the field equations~(\ref{GEFEs}) with the
Bianchi identities, we obtain a generalized version of the
equation~(\ref{diffEFEs}) faced by Bronstein:
\beqa
& & \partial_{\mu}\varphi \left\{\!\frac{\mathcal R}{2} +
   \frac{\omega(\varphi)}{2\varphi^2} \psi -
   \frac{\omega(\varphi)}{\varphi} \Box\,\varphi + \Lambda(\varphi,\psi) +
   \varphi \, \frac{\partial\Lambda(\varphi,\psi)}{\partial\varphi} -
   \frac{\psi}{2\varphi} \frac{d\omega(\varphi)}{d\varphi} \right. \nonumber \\
& & \left. \hspace{7mm} - 2\varphi\,\Box\,\varphi
   \frac{\partial\Lambda(\varphi,\psi)}{\partial\psi} -
   2\partial^{\kappa}\varphi\,\partial_{\kappa}\!\!\left[ \varphi
   \frac{\partial\Lambda(\varphi,\psi)}{\partial\psi}\right]\!\right\} =
   \frac{8\pi}{c^4} \nabla^{\nu}\,{\mathcal T}_{\mu\nu} \; .
\label{diffGEFEs}
\eeqa
Now energy conservation ($\nabla^{\nu}\Tmat=0$) no longer requires
$\Lambda=$~const.  In fact, it is generally {\em incompatible\/} with
constant $\Lambda$, unless an extra condition is imposed on the terms inside
the curly brackets in (\ref{diffGEFEs}).  (This cannot come from the wave
equation for $\varphi$, which merely confirms that the terms inside the
curly brackets sum to zero, in agreement with energy conservation.)
Similar conclusions hold for other scalar-tensor theories in
which $\varphi$ is no longer associated with $G$.  Examples include
models with non-minimal couplings between $\varphi$ and the curvature scalar
${\mathcal R}$ \cite{Mad88}, conformal rescalings of the metric tensor
by functions of $\varphi$ \cite{Mae89} and nonzero potentials
$V(\varphi)$ \cite{Bar87,Pee88,Wet88}.  (Theories of this last kind
are now known as {\em quintessence\/} models \cite{Cal98}).
In each of these scenarios, the cosmological ``constant''
becomes a dynamical variable.

In the modern approach to variable-$\Lambda$ cosmology, which goes back
to Zeldovich in 1968 \cite{Zel68}, all extra terms of
the kind just described --- including $\Lambda$ --- are moved to the
right-hand side of the field equations~(\ref{GEFEs}), leaving only the
Einstein tensor
(${\mathcal R}_{\mu\nu} - \smallfrac{1}{2} \, {\mathcal R} \, g_{\mu\nu}$)
to make up the ``geometrical'' left-hand side.  The cosmological term,
along with scalar (or other) additional fields, are thus effectively
reinterpreted as {\em new kinds of matter\/}.  Eqs.~(\ref{GEFEs}) then read
\beq
{\mathcal R}_{\mu\nu} - \frac{1}{2} \, {\mathcal R} \, g_{\mu\nu} =
   -\frac{8\pi}{\varphi\,c^{\,4}}\, \Teff +  \Lambda(\varphi)
   \, g_{\mu\nu} \; .
\label{EffEFEs}
\eeq
Here $\Teff$ is an effective energy-momentum tensor describing
the sum of ordinary matter plus whatever scalar (or other) fields have
been added to the theory.  For generalized scalar-tensor theories as
described above, this could be written as $\Teff\equiv\Tmat+\Tphi$ where
$\Tmat$ refers to ordinary matter and $\Tphi$ to the scalar field.
For the case with $\Lambda=\Lambda(\varphi)$ and $\omega=\omega(\varphi)$,
for instance, the latter would be defined by (\ref{GEFEs}) as
\beq
\Tphi \equiv \frac{1}{\varphi} \left[ \nabla_{\mu} ( \partial_{\nu}\varphi ) -
   \Box\,\varphi \, g_{\mu\nu} \right] + \frac{\omega(\varphi)}{\varphi^2} \!
   \left( \partial_{\mu}\varphi \, \partial_{\nu}\varphi - \frac{1}{2} \,
   \psi \, g_{\mu\nu} \right) \; .
\eeq
The covariant derivative of the field equations~(\ref{EffEFEs}) now reads
\beq
0 = \nabla^{\nu} \!\! \left[ \frac{8\pi}{\varphi\,c^{\,4}}\, \Teff -
   \Lambda(\varphi) \, g_{\mu\nu} \right] \; .
\label{modifEFEs}
\eeq
Eq.~(\ref{modifEFEs}) carries the same physical content as (\ref{diffGEFEs}),
but is more general in form and can readily be extended to other theories.
Physically, it says that energy {\em is\/}
conserved in variable-$\Lambda$ cosmology ---
where ``energy'' is now understood to refer to the energy of ordinary
matter along with that in any additional fields which may be present,
{\em and along with that in the vacuum\/}, as represented by $\Lambda$.
In general, the latter parameter can vary as it likes, so long as the
conservation equation~(\ref{modifEFEs}) is satisfied.

It was noted at least as early as 1977 by End\={o} and Fukui \cite{End77}
that the evolution of $\Lambda$ in theories of this kind can help with
the cosmological ``constant'' problem, in the sense of dropping from large
primordial values to ones like those seen today.  These authors found
solutions for $\varphi(t)$ such that $\Lambda\propto t^{-2}$ when
$\Lambda=\Lambda(\varphi)$ and $\omega=$~constant.
In precursors to the modern quintessence scenarios, Barr \cite{Bar87}
found models in which $\Lambda\propto t^{-\ell}$ at late times,
while Peebles and Ratra \cite{Pee88} discussed a theory in which
$\Lambda\propto R^{-m}$ at early ones (here $\ell$ and $m$ are powers).
There is now a rich literature on $\Lambda$-decay laws of this kind
\cite{Ove98}.  Their appeal is easy to understand, and can be
illustrated with a simple dimensional argument for the case with
$\Lambda\propto R^{-2}$ \cite{Che90}.  Since $\Lambda$ already has
dimensions of $L^{-2}$, the proportionality factor in this case is a
pure number ($\alpha$, say) which is presumably of order unity.
Taking $\alpha\sim1$ and identifying $R$ with a suitable length scale
in cosmology (namely the Hubble distance $c/\Ho$), one finds that
$\Lamo\sim\Ho^2/c^2$.  The present vacuum density parameter is then
$\Olamo\equiv\Lamo c^2/3\Ho^2\sim1/3$, close to the values implied
by by the supernovae data (Sec.~\ref{sec:vacenergy}).  The most
natural choice $R\sim\lPl$ gives a primordial $\Lambda$-term
of $\LamPl\sim\alpha\lPl^{-2}$.  It then follows that
$\LamPl/\Lamo\sim(c/\Ho\lPl)^2\sim10^{122}$, in good agreement
with the values suggested by Table~\ref{table4.1}.

\subsection{Theoretical and observational challenges}

While this would seem to be a promising approach, two cautions must be
kept in mind.  The first is theoretical.  Insofar as the mechanisms 
discussed so far are entirely classical, they do not address the
underlying problem.  For this, one would also need to explain why net
contributions to $\Lambda$ from the {\em quantum vacuum\/} do not remain
at the primordial level, or how they are suppressed with time. 
Polyakov \cite{Pol82} and S.L.~Adler \cite{Adl82} in 1982 were the first
to speculate explicitly that such a suppression might come about if the
``bare'' cosmological term implied by quantum field theory were
progressively screened by an ``induced'' counterterm of opposite sign,
driving the effective value of $\Lambda(t)$ toward zero at late times.
Many theoretical adjustment mechanisms have now been identified
as potential sources of such a screening effect, beginning with a 1983
suggestion by Dolgov \cite{Dol83} based on non-minimally coupled scalar
fields.  Subsequent proposals have involved scalar fields
\cite{Ban85,Pec87,Fri95}, fields of higher spin
\cite{Haw84,Bro87,Dol97}, quantum effects during inflation
\cite{Mot85,Tsa93,Muk97} and other phenomena \cite{Ban88,Col88,Gue97}.
In most of these cases, no analytic expression is found for $\Lambda$ in
terms of time or other cosmological parameters; the intent is merely
to demonstrate that decay (and preferably near-cancellation) of the 
cosmological term is possible in principle.  None of these mechanisms
has been widely accepted as successful to date.  In fact, there
is a general argument due to Weinberg to the effect that a successful
mechanism based on scalar fields would necessarily be so finely-tuned 
as to be just as mysterious as the original problem \cite{Wei89}.
Similar concerns have been raised in the case of vector and tensor-based
proposals \cite{Dol98}.  Nevertheless, the idea of the adjustment
mechanism remains feasible in principle, and continues to attract
more attention than any other approach to the cosmological-constant
problem.

The second caution is empirical.  Observational data place increasingly
strong restrictions on the way in which $\Lambda$ can vary with time.
Among the most important are early-time bounds on the dark-energy
density $\rlam c^2=\Lambda c^4/8\pi G$.  The success of standard primordial
nucleosynthesis theory implies that $\rlam$ was smaller than $\rrad$
and $\rmat$ during the radiation-dominated era, and large-scale structure
formation could not have proceeded in the conventional way unless
$\rlam<\rmat$ during the early matter-dominated era.
Since $\rrad\propto R^{-4}$ and $\rmat\propto R^{-3}$ from (\ref{rhoR}),
these requirements mean in practice that the dark-energy density must
climb {\em less steeply than\/} $R^{-3}$ in the past direction, if it is
comparable to that of matter or radiation at present \cite{Fre87,Bir97}.
The variable-$\Lambda$ term must also satisfy late-time bounds like those
which have been placed on the cosmological constant
(Sec.~\ref{sec:vacenergy}).  Tests of this kind have been carried out
using data on the age of the Universe \cite{Ols87,Mat95},
structure formation \cite{Rat88,Sil94,Via98},
galaxy number counts \cite{Yos92},
the CMB power spectrum \cite{Sug92,Sil97},
gravitational lensing statistics \cite{Sil97,Rat92,Blo96}
and Type~Ia supernovae \cite{Sil97,Pod00}.
Some of these tests are less restrictive in the case of a variable
$\Lambda$-term than they are for $\Lambda=$~const, and this can open up
new regions of parameter space.  Observation may even be compatible with
some nonsingular models whose expansion originates in a hot, dense
``big bounce'' rather than a big bang \cite{Ove99a}, a possibility which
can be ruled out on general grounds if $\Lambda=$~constant.

A third group of limits comes from asking what the vacuum decays
{\em into\/}.  In quintessence theories, dark energy is transferred
to the kinetic energy of a scalar field as it ``rolls'' down a gradient
toward the minimum of its potential.  This may have observable consequences 
if the scalar field is coupled strongly to ordinary matter, but is hard 
to constrain in general.  A simpler situation is that in which the 
vacuum decays into known particles such as baryons, photons or neutrinos.
The baryonic decay channel would produce excessive levels of
$\gamma$-ray background radiation due to matter-antimatter annihilation
unless the energy density of the vacuum component is less than
$3 \times 10^{-5}$ times that of matter \cite{Fre87}.
This limit can be weakened if the decay process violates baryon number,
or if it takes place in such a way that matter and antimatter are
segregated on large scales, but such conditions are hard to arrange
in a natural way.  The radiative decay channel is more promising,
but also faces a number of tests.  The decay process should meet
certain criteria of thermodynamic stability \cite{Pav91} and adiabaticity
\cite{Lim96}.  The {\em shape\/} of the spectrum of decay photons must
not differ too much from that of pre-existing background radiation, or
distortions will arise.  Freese \etal\ have argued on this basis that
the energy density of a vacuum decaying primarily into low-energy photons 
could not exceed $4\times10^{-4}$~times that of radiation \cite{Fre87}.

It may be, however, that vacuum-decay photons blend into the spectrum
of background radiation without distorting it.  Fig.~\ref{fig1.2}
shows that the best place to ``hide'' the evidence of such a process
would be the microwave region, where the energy density of background
radiation is highest.  Could all or part of the CMB be due to dark-energy
decay?  We know from the {\sc Cobe} satellite that its spectrum is very
nearly that of a perfect blackbody \cite{Fix96}.
Freese \etal\ pointed out that vacuum-decay photons
would be thermalized efficiently by brehmsstrahlung and double-Compton
scattering in the early Universe, and might continue to assume a
blackbody spectrum at later times if pre-existing CMB photons played
a role in ``inducing'' the vacuum to decay \cite{Fre87}.  Subsequent
work has shown that this would require a special combination of
thermodynamical parameters \cite{Hu93}.  This possibility is important
in practice, however, because it leads to the most conservative limits
on the theory.  Even if the radiation produced by decaying dark energy
does not distort the background, it will contribute to the latter's
{\em absolute intensity\/}.  We can calculate the size of these
contributions to the background radiation using the methods that have
been laid out in Secs.~\ref{ch2} and \ref{ch3}.

\subsection{A phenomenological model}

The first step in this problem is to solve the field equations and
conservation equations for the energy density of the decaying vacuum.
We will do this in the context of a general phenomenological model.
This means that we retain the field equations~(\ref{EffEFEs}) and the
conservation law~(\ref{modifEFEs}) without specifying the form of the
effective energy-momentum tensor in terms of scalar (or other) fields.
These equations may be written
\beqa
{\mathcal R}_{\mu\nu} - \frac{1}{2} \, {\mathcal R} \, g_{\mu\nu} & = & -
   \frac{8\pi G}{c^{\,4}} \! \left( \Teff - \rlam c^{\,2} \, g_{\mu\nu} \right)
\label{PhenEFEs} \\
0 & = & \nabla^{\nu} \! \left( \Teff - \rlam c^{\,2} \, g_{\mu\nu} \right) \; .
\label{PhenCon}
\eeqa
Here $\rlam c^{\,2} \equiv \Lambda c^{\,4}/8\pi G$ from (\ref{rlamDefn})
and we have replaced $1/\varphi$ with $G$ (assumed to be constant in what
follows).
Eqs.~(\ref{PhenEFEs}) and (\ref{PhenCon}) have the same form as their
counterparts~(\ref{EFEs}) and (\ref{PFcon}) in standard cosmology, the key
difference being that the cosmological term has migrated to the right-hand
side and is {\em no longer necessarily constant\/}.  Its evolution
is now governed by the conservation equations~(\ref{PhenCon}), which require
only that any change in $\rlam c^2 g_{\mu\nu}$ be balanced by an equal and
opposite change in the energy-momentum tensor $\Teff$.

While the latter is model-dependent in general, it is reasonable to
assume in the context of isotropic and homogeneous cosmology that its form
is that of a perfect fluid,
as given by (\ref{PFdefn}):
\beq
\Teff = (\reff + \peff/c^2) U_{\mu} U_{\nu} + \peff \, g_{\mu\nu} \; .
\label{EffTmunu}
\eeq
Comparison of Eqs.~(\ref{PhenCon}) and (\ref{EffTmunu}) shows that
the conserved quantity in (\ref{PhenCon}) must then {\em also\/} have the
form of a perfect-fluid energy-momentum tensor, with density and
pressure given by
\beq
\rho = \reff + \rlam \qquad p = \peff - \rlam c^{\,2} \; .
\eeq
The conservation law~(\ref{PhenCon}) may then be simplified at once
by analogy with Eq.~(\ref{PFcon}):
\beq
\frac{1}{R^3} \frac{d}{dt} \left[ R^3 \left( \reff c^2 + \peff \right)
   \right] = \frac{d}{dt} \left( \peff - \rlam c^{\,2} \right) \; .
\label{PhenEMCon}
\eeq
This reduces to the standard result~(\ref{PFcon2}) for the case of a constant
cosmological term, $\rlam=$~const.  Throughout Sec.~\ref{ch5}, we allow
the cosmological term to contain both a constant part {\em and\/}
a time-varying part so that
\beq
\rlam = \rcon + \rvac(t) \qquad \rcon = \mbox{ const} \; .
\label{TwoVacs}
\eeq
Let us assume in addition that the perfect fluid described by $\Teff$
consists of a mixture of dust-like matter
($\pmat=0$) and radiation
($\prad=\smallfrac{1}{3}\rrad c^{\,2}$):
\beq
\reff = \rmat + \rrad \qquad \peff = \smallfrac{1}{3} \rrad c^{\,2} \; .
\label{reffDefn}
\eeq
The conservation equation~(\ref{PhenEMCon}) then reduces to
\beq
\frac{1}{R^{\,4}} \, \frac{d}{dt} \left( R^{\,4} \rrad \right) + 
   \frac{1}{R^{\,3}} \, \frac{d}{dt} \left( R^{\,3} \rmat \right) + 
   \frac{d\rvac}{dt} = 0 \; .
\label{MainVacDE}
\eeq
From this equation it is clear that one (or both) of the radiation and
matter densities can no longer obey the usual relations
$\rrad\propto R^{-4}$ and $\rmat\propto R^{-3}$ in a theory with
$\Lambda\neq$~const.  Any change in $\Lambda$ (or $\rlam$)
must be accompanied by a change in radiation and/or matter densities.

To go further, some simplifying assumptions must be made.  Let us take
to begin with:
\beq
\frac{d}{dt} \left( R^{\,3} \rmat \right) = 0 \; .
\label{PartNumCon}
\eeq
This is just conservation of particle number, as may be seen by
replacing ``galaxies'' with ``particles'' in Eq.~(\ref{GalCon}).  Such an
assumption is well justified during the matter-dominated era by the
stringent constraints on matter creation discussed in Sec.~\ref{sec:VarLam}.
It is equally well justified during the radiation-dominated era,
when the matter density is small so that the $\rmat$ term is of
secondary importance compared to the other terms in (\ref{MainVacDE})
in any case.

In light of Eqs.~(\ref{MainVacDE}) and (\ref{PartNumCon}), the vacuum
can exchange energy only with radiation.  As a model for this process,
let us follow Pollock in 1980 \cite{Pol80} and assume that it takes
place in such a way that the energy density of the decaying vacuum
component remains proportional to that of radiation, $\rvac\propto\rrad$.
We adopt the notation of Freese \etal\ in 1987 and write the
proportionality factor as $x/(1-x)$ with $x$ the coupling parameter
of the theory \cite{Fre87}.  If this is allowed to take
(possibly different) constant values during the radiation
and matter-dominated eras, then
\beq
x \equiv \frac{\rvac}{\rrad + \rvac} = \left\{ \begin{array}{ll}
   \xr & \qquad (t<\teq) \\
   \xm & \qquad (t\geqslant\teq) \; .
\end{array} \right.
\label{xDefn}
\eeq
Here $\teq$ refers to the epoch of matter-radiation equality
when $\rrad=\rmat$.  Standard cosmology is recovered in the limit
$x\rightarrow0$.  The most natural situation is that in which the value
of $x$ stays constant, so that $\xr = \xm$.  However, since observational
constraints on $x$ are in general different for the radiation and
matter-dominated eras, the most conservative limits on the theory are
obtained by letting $\xr$ and $\xm$ take different values.  Physically,
this would correspond to a phase transition or sudden change in the
expansion rate $\dot{R}/R$ of the Universe at $t=\teq$.

With Eqs.~(\ref{PartNumCon}) and (\ref{xDefn}), the conservation
equation~(\ref{MainVacDE}) reduces to
\beq
\frac{\dot{\rvac}}{\rvac} + 4 (1-x) \frac{\dot{R}}{R} = 0 \; ,
\label{EntCon}
\eeq
where overdots denote derivatives with respect to time.  Integration gives
\beq
\rvac(R) = \av R^{-4(1-x)} \; ,
\label{rvacR}
\eeq
where $\av$ is a constant.  The cosmological term $\Lambda$ is thus an
inverse power-law function of the scale factor $R$, a scenario that has
received wide attention also in models where vacuum energy is not
proportional to that of radiation \cite{Ove98}.  Eq.~(\ref{rvacR})
shows that the conserved quantity in this theory has a form
intermediate between that of ordinary radiation entropy ($R^{\,4}\rrad$)
and particle number ($R^{\,3}\rmat$) when $0<x<\frac{1}{4}$.

The fact that $\rrad\propto\rvac\propto R^{-4(1-x)}$
places an immediate upper limit of $\smallfrac{1}{4}$ on $x$ (in both eras),
since higher values would erase the dynamical distinction between radiation
and matter.  With $x\leqslant\smallfrac{1}{4}$ it then follows from
(\ref{xDefn}) that $\rvac\leqslant\smallfrac{1}{3}\rrad$.  This is
consistent with Sec.~\ref{sec:VarLam}, where we noted that a vacuum
component whose density climbs more steeply than $R^{-3}$ in the past
direction cannot have an energy density greater than that of radiation
at present.  Freese \etal\ \cite{Fre87} set a stronger bound by
showing that $x\leqslant0.07$ if the baryon-to-photon ratio $\eta$
is to be consistent with both primordial nucleosynthesis and present-day
CMB observations.  (This argument assumes that $x=\xr=\xm$.)  
As a guideline in what follows, then, we will allow $\xr$ and $\xm$
to take values between zero and 0.07, and consider in addition the
theoretical possibility that $\xm$ could increase to $0.25$ in the
matter-dominated era.

\subsection{Energy density}

With $\rmat(R)$ specified by (\ref{PartNumCon}), $\rrad$ related to $\rvac$
by (\ref{xDefn}) and $\rvac(R)$ given by (\ref{rvacR}), we can solve
for all three components as functions of time if the scale factor $R(t)$
is known.  This comes as usual from the field equations~(\ref{PhenEFEs}).
Since these are the same as Eqs.~(\ref{EFEs}) for standard cosmology,
they lead to the same result, Eq.~(\ref{FL1}):
\beq
\left( \frac{\dot{R}}{R} \right)^2 =
   \frac{8\pi G}{3} \, (\rmat + \rrad + \rvac + \rcon) \; .
\label{FLvac}
\eeq
Here we have used Eqs.~(\ref{TwoVacs}) to replace $\rlam$ with
$\rvac+\rcon$ and (\ref{reffDefn}) to replace $\reff$ with $\rmat+\rrad$.
We have also set $k=0$ since observations indicate that these components
together make up very nearly the critical density (Sec.~\ref{ch4}).

Eq.~(\ref{FLvac}) can be solved analytically in the three cases which are of
greatest physical interest: (1)~the {\em radiation-dominated regime\/},
for which $t<\teq$ and $\rrad+\rvac\gg\rmat+\rcon$; (2)~the
{\em matter-dominated regime\/}, which has $t\geqslant\teq$ and
$\rrad+\rvac\ll\rmat$ (if $\rcon=0$); and (3)~the
{\em vacuum-dominated regime\/}, for which $t\geqslant\teq$
and $\rrad+\rvac\ll\rmat+\rcon$.  The distinction between 
regimes~2 and 3 allows us to model both matter-only universes like EdS
and vacuum-dominated cosmologies like \LCDM\ or \LBDM\
(Table~\ref{table3.1}).  The definitions of these terms should
be amended slightly for this section, since we now consider flat models
containing not only matter and a cosmological constant, but radiation and
a decaying-vacuum component as well.  The densities of the latter two 
components are, however, at least four orders of magnitude below that
of matter at present.  Thus models with $\rcon=0$, for example, have
$\Omato=1$ to four-figure precision or better and are dynamically
indistinguishable from EdS during all but the first fraction (of order
$10^{-4}$ or less) of their lifetimes.  For definiteness, we will use
the terms ``EdS,'' ``\LCDM'' and ``\LBDM'' in this section to refer to
flat models in which $\Omato=1$, 0.3 and 0.03 respectively.
In all cases, the present dark-energy density (if any) comes almost
entirely from its constant-density component.

Eqs.~(\ref{PartNumCon}), (\ref{xDefn}), (\ref{rvacR}) and (\ref{FLvac})
can be solved analytically for $R, \rmat,\rrad$ and $\rvac$ in terms of
$\Ro$, $\rmato$, $\rrado$, $\xr$ and $\xm$ (see \cite{OW03} for details).
The normalized scale factor is found to read
\beq
\Rtil(t) = \left\{ \begin{array}{ll}
   \left( \bigfrac{t}{\too} \right)^{1/2(1-\xr)}
      & \qquad (t<\teq) \nonumber \\
   \left[ \bigfrac{\Sm(t)}{\Sm(\too)} \right]^{2/3}
      & \qquad (t\geqslant\teq) \; .
\end{array} \right.
\label{Rtvac}
\eeq
The dark-energy density is given by
\beq
\rvac(t) = \left\{ \begin{array}{ll}
   \bigfrac{\alpha\xr}{(1-\xr)^{\,2}} \, t^{-2}
      & \qquad (t<\teq) \nonumber \\
   \left( \bigfrac{\xm}{1-\xm} \right) \rrad(t)
      & \qquad (t\geqslant\teq) \; ,
\end{array} \right.
\label{rvacDefn}
\eeq
where $\alpha=3/(32\pi G)=4.47\times10^5$~g~cm$^{-2}$~s$^2$.
The densities of radiation and matter are
\beqa
\rrad(t) & = & \left\{ \begin{array}{ll}
   \left( \bigfrac{1-\xr}{\xr} \right) \rvac(t)
      & \qquad (t<\teq) \\
   \rrado \left[ \bigfrac{\Sm(t)}{\Sm(\too)} \right]^{-8(1-\xm)/3}
      & \qquad (t\geqslant\teq)
   \end{array} \right. \; .
   \label{rradDefn} \\
\frac{\rmat(t)}{\rmato} & = & \left\{ \begin{array}{ll}
   \left[ \bigfrac{\Sm(\teq)}{\Sm(\too)} \right]^{-2} \!\!
      \left( \bigfrac{t}{\teq} \right)^{-3/2(1-\xr)}
      & (t<\teq) \\
   \left[ \bigfrac{\Sm(t)}{\Sm(\too)} \right]^{-2}
      & (t\geqslant\teq)
\end{array} \right. \nonumber
\eeqa
Here we have applied $\rmato=\Omato\,\rcrito$ and $\rrado=\Orado\,\rcrito$
as boundary conditions.  The function $\Sm(t)$ is defined as
\beq
\Sm(t) \equiv \left\{ \begin{array}{ll} t
      & \qquad (\Omato=1) \\
   \sinh ( t/\tao )
      & \qquad (0<\Omato<1) \; ,
\end{array} \right.
\eeq
where $\tao\equiv2/(3\Ho\sqrt{1-\Omato})$.  The age of of the Universe is
\beq
\too = \left\{ \begin{array}{ll} 2/(3\Ho)
      & \qquad (\Omato=1) \\
   \tao\,\sinh^{-1}\!\cho
      & \qquad (0<\Omato<1) \; ,
\end{array} \right.
\eeq
where $\cho\equiv\sqrt{(1-\Omato)/\Omato}$ and we have used
Eq.~(\ref{t2comp}).  Corrections from the radiation-dominated era can be
ignored since $\too\gg\teq$ in all cases.

The parameter $\teq$ is obtained as in standard cosmology by setting
$\rrad(\teq)=\rmat(\teq)$ in Eqs.~(\ref{rradDefn}).  This leads to
\beq
\teq = \left\{ \begin{array}{ll}
   \too \, \Orado^{3/2(1-4\xm)}
      & (\Omato=1) \\
   \tao \sinh^{-1} \left[ \cho \left(
      \bigfrac{\Orado}{\Omato} \right)^{3/2(1-4\xm)} \right]
      & (0<\Omato<1) \; .
\end{array} \right.
\label{teqDefn}
\eeq
The epoch of matter-radiation equality plays a crucial role because it is
at about this time that the Universe became transparent to radiation (the
two events are not simultaneous but the difference between them is minor
for our purposes).  Decay photons created before $\teq$ would simply
have been thermalized by the primordial plasma and eventually re-emitted
as part of the CMB.  It is the decay photons emitted {\em after\/} this
time which can contribute to the extragalactic background radiation,
and whose contributions we wish to calculate.
The quantity $\teq$ is thus analogous to the galaxy formation time
$t_f$ in previous sections.

\begin{figure}[t!]
\begin{center}
\includegraphics[width=\textwidth]{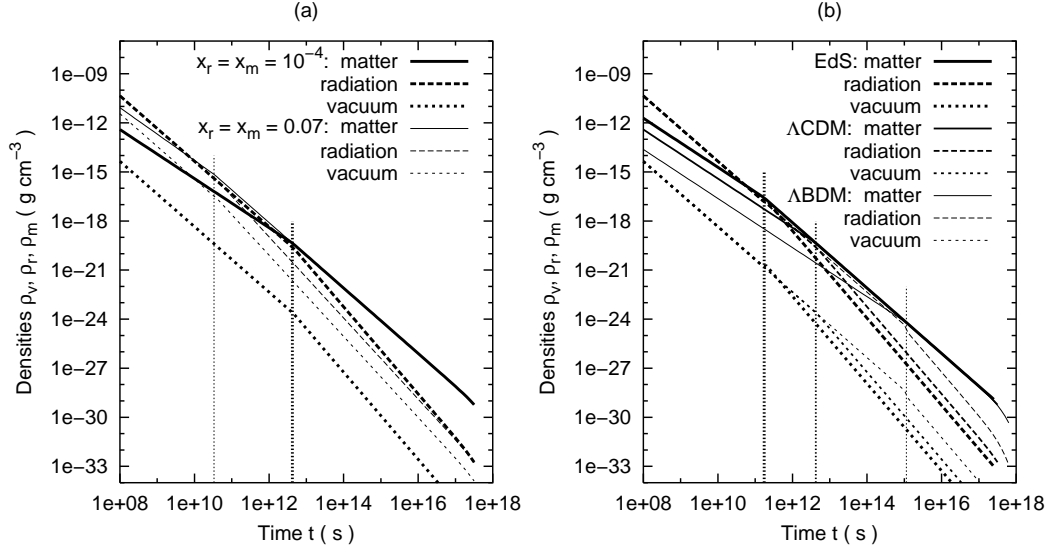}
\end{center}
\caption{The densities of decaying dark energy ($\rvac$), radiation
   ($\rrad$) and matter ($\rmat$) as functions of time.  Panel~(a)
   shows the effects of changing the values of $\xr$ and $\xm$,
   assuming a model with $\Omato=0.3$ (similar to \LCDM).  Panel~(b)
   shows the effects of changing the cosmological model, assuming
   $\xr=\xm=10^{-4}$.  The vertical lines indicate the epochs when the
   densities of matter and radition are equal ($\teq$).  All curves
   assume $\ho=0.75$.}
\label{fig5.1}
\end{figure}

The densities $\rmat(t)$, $\rrad(t)$ and $\rvac(t)$ are plotted as functions
of time in Fig.~\ref{fig5.1}.
The left-hand panel~(a) shows the effects of varying the parameters
$\xr$ and $\xm$ within a given cosmological model (here, \LCDM).
Raising the value of $\xm$ leads to a proportionate increase in $\rvac$
and a modest drop in $\rrad$.  It also flattens the slope of both components.
The change in slope (relative to that of the matter component)
pushes the epoch of equality back toward the big bang (vertical lines).
Such an effect could in principle allow more time for structure to form
during the early matter-dominated era \cite{Fre87}, although the
``compression'' of the radiation-dominated era rapidly becomes unrealistic
for values of $\xm$ close to $\smallfrac{1}{4}$.  Thus Fig.~\ref{fig5.1}(a)
shows that the value of $\teq$ is reduced by a factor of over 100 in going
from a model with $\xm=10^{-4}$ to one with $\xm=0.07$.  In the limit
$\xm\rightarrow\smallfrac{1}{4}$, the radiation-dominated era disappears
altogether, as remarked above and as shown explicitly by Eqs.~(\ref{teqDefn}).

Fig.~\ref{fig5.1}(b) shows the effects of changes in cosmological model
for fixed values of $\xr$ and $\xm$ (here both set to $10^{-4}$).
Moving from the matter-filled EdS model toward vacuum-dominated ones
such as \LCDM\ and \LBDM\ does three things.  The first is to increase
the age ($\too$) of the Universe.  This increases the density of
radiation at any given time, since the latter is fixed at present and
climbs at the same rate in the past direction.  Based on our experience
with the galactic EBL in previous sections, we may expect that this should
lead to significantly higher levels of background radiation when
integrated over time.  However, there is a second effect in the present
theory which acts in the opposite direction: smaller values of $\Omato$
boost the value of $\teq$ as well as $\too$, thus delaying the onset of
the matter-dominated era (vertical lines).  As we will see, these two
changes all but cancel each other out as far as dark-energy contributions
to the background are concerned.  The third consequence of
vacuum-dominated cosmologies is ``late-time inflation,'' the sharp
increase in the expansion rate at recent times (Fig.~\ref{fig4.2}).
This translates in Fig.~\ref{fig5.1}(b) into the drop-off in the densities
of all three components at the right-hand edge of the figure for the
\LCDM\ and \LBDM\ models.

\subsection{Source luminosity} \label{sec:vacLum}

In order to make use of the formalism we have developed in Secs.~\ref{ch2}
and \ref{ch3}, we need to define discrete ``sources'' of radiation from
dark-energy decay, analogous to the galaxies of previous sections.
For this purpose we carve up the Universe into hypothetical regions of
arbitrary comoving volume $\Vo$.  The comoving number density of these
source regions is just 
\beq
n(t) = \no = \Vo^{-1} = \mbox{const} \; .
\label{nvacDefn}
\eeq
These regions are introduced for convenience, and are not physically
significant since dark energy decays uniformly throughout space.  We
therefore expect that the parameter $\Vo$ will not appear in our final
results.

The next step is to identify the ``source luminosity.''  There are at
least two ways to approach this question \cite{Ove93a}.  One could simply
regard the source region as a ball of physical volume
$V(t)=\Rtil^{\,3}(t)\Vo$ filled with fluctuating dark energy.  As the
density of this energy drops by $-d\rvac$ during time $dt$, the ball loses
energy at a rate $-d\rvac/dt$.  If some fraction $\beta$ of this energy
flux goes into photons, then the luminosity of the ball is
\beq
L_v(t) = - \beta \, c^{\,2} \dot{\rvac}(t) \, V(t) \; .
\label{LTdefn}
\eeq
This is the definition of vacuum luminosity which has been assumed
implicitly by workers such as Pav\'{o}n \cite{Pav91},
who investigated the thermodynamical stability of the vacuum decay process
by requiring that fluctuations in $\dot{\rvac}$ not grow larger than the
mean value of $\dot{\rvac}$ with time.  For convenience we will refer
to (\ref{LTdefn}) as the {\em thermodynamical definition\/} of vacuum
luminosity ($\LT$).

A second approach is to treat this as a problem involving spherical symmetry
within general relativity.  The assumption of spherical symmetry
allows the total mass-energy ($M c^{\,2}$) of a localized region of perfect
fluid to be identified unambiguously.  Luminosity can then be related to
the {\em time rate of change\/} of this mass-energy.  Assuming once
again that the two are related by a factor $\beta$, one has
\beq
L_v(t) = \beta \, \dot{M}(t) \, c^2 \; .
\label{LRdefn1}
\eeq
Application of Einstein's field equations leads to the following expression
\cite{Mis73} for the rate of change of mass-energy in terms of the
pressure $\pvac$ at the region's surface:
\beq
\dot{M}(t) \, c^2 = -4\pi \, \pvac(t) \, [r(t)]^2 \dot{r}(t) \; ,
\label{Mdot}
\eeq
where $r(t)=\Rtil(t)\roo$ is the region's physical radius.  Taking
$V=\smallfrac{4}{3}\pi r^3$, applying the vacuum equation of state
$\pvac=-\rvac c^{\,2}$ and substituting (\ref{Mdot}) into (\ref{LRdefn1}),
we find that the latter can be written in the form
\beq
L_v(t) = \beta \, c^{\,2} \rvac(t) \, \dot{V}(t) \; .
\label{LRdefn}
\eeq
This is just as appealing dimensionally as Eq.~(\ref{LTdefn}),
and shifts the emphasis physically from fluctuations in the material
content of the source region toward changes in its geometry.
We will refer to (\ref{LRdefn}) for convenience as the
{\em relativistic definition\/} of vacuum luminosity ($\LR$).

It is not obvious which of the two definitions~(\ref{LTdefn}) and
(\ref{LRdefn}) more correctly describes the luminosity of decaying dark
energy; this is a conceptual issue.  Before choosing between them,
let us inquire whether the two expressions might not be equivalent.
We can do this by taking the ratio
\beq
\frac{\LT}{\LR} = -\frac{\dot{\rvac} V}{\rvac \dot{V}} =
   - \frac{1}{3} \frac{\dot{\rvac}}{\rvac} \frac{R}{\dot{R}} \; .
\eeq
Differentiating Eqs.~(\ref{Rtvac}) and (\ref{rvacDefn}) with
respect to time, we find
\beqa
\frac{\dot{R}}{R} & = & \left\{ \begin{array}{lll}
   \bigfrac{2}{3t} 
      & \qquad & (\Omato=1) \\ \\*[-3mm]
   \bigfrac{2}{3\tao} \coth \left( \bigfrac{t}{\tao} \right)
      & \qquad & (0<\Omato<1)
   \end{array} \right. \label{RdotDefn} \\ \nonumber \\
\frac{\dot{\rvac}}{\rvac} & = & \left\{ \begin{array}{lll}
   -\bigfrac{8}{3t} \, (1-\xm)
      & \qquad & (\Omato=1) \\ \\*[-3mm]
   -\bigfrac{8}{3\tao} \, (1-\xm) \coth \left( \bigfrac{t}{\tao} \right)
      & \qquad & (0<\Omato<1) \; .
\end{array} \right.
\eeqa
The ratio $\LT/\LR$ is therefore constant $[=4(1-\xm)/3]$, taking values
between $\smallfrac{4}{3}$ (in the limit $\xm\rightarrow0$ where standard
cosmology is recovered) and 1 (in the opposite limit where $\xm$ takes
its maximum theoretical value of $\smallfrac{1}{4}$).
There is thus little difference between the two scenarios in practice,
at least where this model of decaying dark energy is concerned.
We will proceed using the relativistic definition~(\ref{LRdefn})
which gives lower intensities and hence more conservative limits on the
theory.  At the end of the section it will be a small matter to calculate
the corresponding intensity for the thermodynamical case~(\ref{LTdefn})
by multiplying through by $\smallfrac{4}{3}(1-\xm)$.

We now turn to the question of the branching ratio $\beta$,
or fraction of decaying dark energy which goes into photons as
opposed to other forms of radiation such as massless neutrinos.  This
is model-dependent in general.  If the vacuum-decay radiation reaches
equilibrium with that already present, however, then we may reasonably
set this equal to the ratio of photon-to-total radiation energy
densities in the CMB:
\beq
\beta=\Ogam/\Orado \; .
\label{betaDefn}
\eeq
The density parameter $\Ogam$ of CMB photons is given in terms of
their blackbody temperature $T$ by Stefan's law.
Using the {\sc Cobe} value $\Tcmb=2.728$~K \cite{Fix96}, we get
\beq
\Ogam = \frac{4\sigSB T^{\,4}}{c^{\,3}\rcrito} =
    2.48 \times 10^{-5} \ho^{-2} \; .
\label{OgamValue}
\eeq
The total radiation density $\Orado=\Ogam+\Onu$
is harder to determine,
since there is little prospect of detecting the neutrino component directly.
What is done in standard cosmology is to calculate the size of neutrino
contributions to $\Orado$ under the assumption of
entropy conservation.
With three light neutrino species, this leads to
\beq
\Orado = \Ogam \left[ 1 + 3 \times \frac{7}{8} \left( \frac{\Tnu}{T}
   \right)^4 \right] \; ,
\label{OradDefn}
\eeq
where $\Tnu$ is the blackbody temperature of the relic
neutrinos and the
factor of $7/8$ arises from the fact that these particles obey Fermi rather
than Bose-Einstein statistics \cite{Pee93}.  During the early stages of
the radiation-dominated era, neutrinos were in thermal equilibrium with
photons so that $\Tnu=T$.  They dropped out of equilibrium, however,
when the temperature of the expanding fireball dropped below about
$kT\sim$~1~MeV (the energy scale of weak interactions).  Shortly
thereafter, when the temperature dropped to $kT\sim m_e c^{\,2}=0.5$~MeV,
electrons and positrons began to annihilate,
transferring their entropy to the remaining photons in
the plasma.  This raised the photon temperature by a factor of
$(1+2\times\smallfrac{7}{8}=\smallfrac{11}{4})^{1/3}$
relative to that of the neutrinos.  In standard cosmology, the ratio of
$\Tnu/T$ has remained at $(4/11)^{1/3}$ down to the present day,
so that (\ref{OradDefn}) gives
\beq
\Orado = 1.68 \, \Ogam = 4.17 \times 10^{-5} \ho^{-2} \; .
\label{OradValue}
\eeq
Using (\ref{betaDefn}) for $\beta$, this would imply:
\beq
\beta= 1/1.68 = 0.595 \; .
\label{betaValue}
\eeq
We will take these as our ``standard values'' of $\Orado$ and $\beta$
in what follows.  They are conservative ones, in the sense that most
alternative lines of argument would imply higher values of $\beta$.
Birkel and Sarkar \cite{Bir97}, for instance, have argued that
vacuum decay (with a constant value of $\xr$) would be easier to reconcile
with processes such as electron-positron annihilation if the vacuum coupled
to photons but not neutrinos.  This would complicate the theory, breaking
the radiation density $\rrad$ in (\ref{MainVacDE}) into a photon part 
$\rgam$ and a neutrino part $\rnu$ with different dependencies on $R$.
One need not solve this equation, however, in order to appreciate
the main impact that such a modification would have.
Decay into photons alone would pump entropy into the photon component
relative to the neutrino component in an effectively {\em ongoing version\/}
of the electron-positron annihilation argument described above.
The neutrino temperature $\Tnu$ (and density $\rnu$) would continue
to be driven down relative to $T$ (and $\rgam$) throughout the
radiation-dominated era and into the matter-dominated one.
In the limit $\Tnu/T\rightarrow 0$ one sees from (\ref{betaDefn})
and (\ref{OradDefn}) that such a scenario would lead to
\beq
\Orado = \Ogam = 2.48 \times 10^{-5} \ho^{-2} \qquad
   \beta = 1 \; .
\label{ScenarioB}
\eeq
In other words, the present energy density of radiation would be lower,
but it would effectively {\em all\/} be in the form of photons.
Insofar as the decrease in $\Orado$ is precisely offset by the increase
in $\beta$, these changes cancel each other out.  The drop in $\Orado$,
however, has an added consequence which is not cancelled: it pushes
$\teq$ farther into the past, increasing the length of time over which
decaying dark energy has been contributing to the background.  This raises the
latter's intensity, particularly at longer wavelengths.  The effect can be
significant, and we will return to this possibility at the end of the
section.  For the most part, however, we will stay with the values of
$\Orado$ and $\beta$ given by Eqs.~(\ref{OradValue}) and
(\ref{betaValue}).

Armed with a definition for vacuum luminosity, Eq.~(\ref{LRdefn}),
and a value for $\beta$, Eq.~(\ref{betaValue}), we are in a position
to calculate the luminosity of decaying dark energy.
Noting that $\dot{V}=3(R/\Ro)^3(\dot{R}/R)\Vo$ and substituting
Eqs.~(\ref{rvacDefn}) and (\ref{RdotDefn}) into (\ref{LRdefn}),
we find that
\beq
L_v(t) = \curlyLvo \Vo \times \left\{ \begin{array}{l}
   \left( \bigfrac{t}{\too} \right)^{-(5-8\xm)/3} \\
   \left[ \bigfrac{\cosh(t/\tao)}{\cosh(\too/\tao)} \right] \!
      \left[ \bigfrac{\sinh(t/\tao)}{\sinh(\too/\tao)}
      \right]^{-(5-8\xm)/3} \; .
\label{LvacDefn}
\end{array} \right.
\eeq
The first of these solutions corresponds to models with $\Omato=1$ while
the the second holds for the general case ($0<\Omato<1$).  Both results
reduce at the present time $t=\too$ to
\beq
L_{v,0} = \curlyLvo \Vo \; ,
\eeq
where $\curlyLvo$ is the {\em comoving luminosity density of decaying
dark energy\/}
\beqa
\curlyLvo & = & \frac{9 c^{\,2}\!\Ho^{\,3}\,\Orado\beta\,\xm}{8\pi G(1-\xm)}
   \nonumber \\
          & = & 4.1 \times 10^{-30} \ho \mbox{ erg s}^{-1} \mbox{ cm}^{-3} 
   \left( \frac{\xm}{1-\xm} \right) \; .
\label{curlyLvoDefn}
\eeqa
Numerically, we find for example that
\beq
\curlyLvo = \left\{ \begin{array}{ll}
   3.1 \times 10^{-31} \ho \mbox{ erg s}^{-1} \mbox{ cm}^{-3} 
      & \qquad (\xm=0.07) \\
   1.4 \times 10^{-30} \ho \mbox{ erg s}^{-1} \mbox{ cm}^{-3} 
      & \qquad (\xm=0.25) \; .
\label{curlyLvoValue}
\end{array} \right.
\eeq
In principle, dark-energy decay can produce a background
10 or even 50 times more luminous than that of galaxies, as given by
(\ref{curlyLoValue}).  Raising the value of the branching ratio $\beta$ to 1
instead of 0.595 does not affect these results, since this must be
accompanied by a proportionate drop in the value of $\Orado$ as argued above.
The numbers in (\ref{curlyLvoValue}) do go up if one replaces the relativistic
definition~(\ref{LRdefn}) of vacuum luminosity with the thermodynamical
one ~(\ref{LTdefn}) but the change is modest, raising $\curlyLvo$ by no
more than a factor of 1.2 (for $\xm=0.07$).  The primary reason for the
high luminosity of the decaying vacuum lies in the fact that it converts
nearly 60\% of its energy density into photons.  By comparison, less
than 1\% of the rest energy of ordinary luminous matter has gone into
photons so far in the history of the Universe.

\subsection{Bolometric intensity}

We showed in Sec.~\ref{ch2} that the bolometric intensity of an arbitrary
distribution of sources with comoving number density $n(t)$ and luminosity
$L(t)$ could be expressed as an integral over time by (\ref{QtDefn}).  Let us
apply this result here to regions of decaying dark energy, for which
$n_v(t)$ and $L_v(t)$ are given by~(\ref{nvacDefn}) and (\ref{LvacDefn})
respectively.  Putting these equations into (\ref{QtDefn}) along with
(\ref{Rtvac}) for the scale factor, we find that
\beq
Q = c \curlyLvo \times \left\{ \begin{array}{l}
   \bigint{\teq}{\too} \left( \bigfrac{t}{\too} \right)^{-(1-8\xm)/3} 
      dt \\
   \bigint{\teq}{\too} \left[ \bigfrac{\cosh(t/\tao)}{\cosh(\too/\tao)}
      \right] \!  \left[ \bigfrac{\sinh(t/\tao)}{\sinh(\too/\tao)}
      \right]^{-(1-8\xm)/3} dt \; .
\end{array} \right.
\eeq
The first of these integrals corresponds to models with $\Omato=1$ while
the second holds for the general case ($0<\Omato<1$).  The latter may be
simplified with a change of variables to
$y\equiv[\sinh(t/\tao)]^{8\xm/3}$.  Using the facts that
$\sinh(\too/\tao)=\sqrt{(1-\Omato)/\Omato}$ and
$\cosh(\too/\tao)=1/\sqrt{\Omato}$ along with the definition~(\ref{teqDefn})
of $\teq$, both integrals reduce to the same formula:
\beq
Q = \Qv \left[ 1 - \left( \frac{\Orado}{\Omato}
   \right)^{4\xm/(1-4\xm)}
   \right] \; .
\eeq
Here $\Qv$ is found with the help of (\ref{curlyLvoDefn}) as
\beq
\Qv \equiv \frac{c\curlyLvo}{4\Ho\,\xm} =
   \frac{9 c^{\,3} \Ho^{\,2} \Orado \beta}{32\pi G (1-\xm)} =
   \frac{0.0094 \mbox{ erg cm}^{-2} \mbox{ s}^{-1}}{(1-\xm)} \; .
\eeq
There are several points to note about this result.  First, it does
not depend on $\Vo$, as expected.  There is also no dependence on the 
uncertainty $\ho$ in Hubble's constant, since the two factors of $\ho$
in $\Ho^{\,2}$ are cancelled out by those in $\Orado$.
In the limit $\xm\rightarrow0$ one sees that $Q\rightarrow0$ as expected.
In the opposite limit where $\xm\rightarrow\smallfrac{1}{4}$,
decaying dark energy attains a maximum possible bolometric intensity of
$Q\rightarrow\Qv=0.013$~erg~cm$^{-2}$~s$^{-1}$.  This is 50~times
the bolometric intensity due to galaxies, as given by (\ref{QstarValue}).

The matter density $\Omato$ enters only
weakly into this result, and plays no role at all in the limit
$\xm\rightarrow\smallfrac{1}{4}$.  Based on our experience with the EBL
due to galaxies, we might have expected that $Q$ would rise significantly
in models with smaller values of $\Omato$ since these have longer ages,
giving more time for the Universe to fill up with light.
What is happening here, however, is that the larger values of $\too$
are offset by larger values of $\teq$ (which follow from the fact that
smaller values of $\Omato$ imply smaller ratios of $\Omato/\Orado$).
This removes contributions from the early matter-dominated era and thereby
{\em reduces\/} the value of $Q$.  In the limit 
$\xm\rightarrow\smallfrac{1}{4}$ these two effects cancel each other out.
For smaller values of $\xm$, the $\teq$-effect proves to be the stronger
of the two, and one finds an overall decrease in $Q$ for these cases.
With $\xm=0.07$, for instance, the value of $Q$ drops by 2\% when
moving from the EdS model to \LCDM, and by another 6\% when moving
from \LCDM\ to \LBDM.

\subsection{Spectral energy distribution}

To obtain limits on the parameter $\xm$, we would like to calculate 
the spectral intensity of the background due to dark-energy decay, just as
we did for galaxies in Sec.~\ref{ch3}.  For this we need to know the
spectral energy distribution (SED) of the decay photons.  As discussed in
Sec.~\ref{sec:VarLam}, theories in which the these photons are distributed
with a {\em non\/}-thermal spectrum can be strongly constrained by means
of distortions in the CMB.  We therefore restrict ourselves to the
case of a blackbody SED, as given by Eq.~(\ref{bbodySED}):
\beq
F_v(\lambda,t) = \frac{C(t)/\lambda^5}
   {\exp \left[ hc/kT(t)\lambda \right] - 1} \; ,
\label{VacSED}
\eeq
where $T(t)$ is the blackbody temperature.  The function $C(t)$ is found
as usual by normalization, Eq.~(\ref{Fnorm}).  Changing integration
variables from $\lambda$ to $\nu=c/\lambda$, we find
\beq
L_v(t)=\frac{C(t)}{c^{\,4}} \int_0^{\infty} \frac{\nu^{\,3} d\nu}
      {\exp \left[ h\nu/kT(t) \right] - 1} 
    =\frac{C(t)}{c^{\,4}} \left[ \frac{h}{kT(t)} \right]^{-4}
     \!\!\!\!\! \Gamma(4) \, \zeta(4) \; .
\eeq
Inserting our result~(\ref{LvacDefn}) for $L_v(t)$ and using the facts that
$\Gamma(4)=3\mbox{!}=6$ and $\zeta(4)=\pi^4/90$, we then obtain for $C(t)$:
\beq
C(t) = \frac{15\curlyLvo\Vo}{\pi^4} \! \left[ \frac{hc}{kT(t)} \right]^4
   \!\!\times \left\{ \begin{array}{l}
   \left( \bigfrac{t}{\too} \right)^{-(5-8\xm)/3} \\
   \left[ \bigfrac{\cosh(t/\tao)}{\cosh(\too/\tao)} \right]
      \left[ \bigfrac{\sinh(t/\tao)}{\sinh(\too/\tao)}
      \right]^{-(5-8\xm)/3} \; . \hspace{-7mm}
\end{array} \right.
\label{CtDefn}
\eeq
Here the upper expression refers as usual to the EdS case ($\Omato=1$),
while the lower applies to the general case ($0<\Omato<1$).  The 
temperature of the photons can be specified if we assume thermal
equilibrium between those created by vacuum decay and those already
present.  Stefan's law then relates $T(t)$ to the radiation energy
density $\rrad(t) c^{\,2}$ as follows:
\beq
\rrad(t) \, c^{\,2} = \frac{4 \sigSB}{c} \left[ T(t) \right]^4 \; .
\eeq
Putting Eq.~(\ref{rradDefn}) into this expression and expanding
the Stefan-Boltzmann constant, we find that
\beq
\frac{hc}{kT(t)} = \lamv \times \left\{ \begin{array}{ll}
   \left( \bigfrac{t}{\too} \right)^{2(1-\xm)/3}
      & (\Omato=1) \\
   \left[ \bigfrac{\sinh(t/\tao)}{\sinh(\too/\tao)}
      \right]^{2(1-\xm)/3}
      & (0<\Omato<1) \; ,
\label{TvacDefn}
\end{array} \right.
\eeq
where the constant $\lamv$ is given by
\beq
\lamv \equiv \left( \frac{8\pi^5 hc}{15 \rrado \, c^{\,2}}
   \right)^{1/4} \!\!\!\!
   = 0.46 \mbox{ cm} \left( \frac{\Orado \ho^2}{4.17 \times 10^{-5}}
   \right)^{-1/4} \; .
\eeq
This value of $\lamv$ tells us that the peak of the observed spectrum of
decay radiation lies in the microwave region as expected, near that of the
CMB ($\lamcmb=0.11$~cm).  Putting (\ref{TvacDefn}) back into (\ref{CtDefn}),
we obtain
\beq
C(t) = \frac{15 \lamv^4 \curlyLvo \Vo}{\pi^{\,4}} \times \left\{
      \begin{array}{l}
   \left( \bigfrac{t}{\too} \right) \\ \\*[-3mm]
   \left[ \bigfrac{\cosh(t/\tao)}{\cosh(\too/\tao)} \right] \!
      \left[ \bigfrac{\sinh(t/\tao)}{\sinh(\too/\tao)} \right] \; .
\end{array} \right.
\eeq
These two expressions refer to models with $\Omato=1$ and $0<\Omato<1$
respectively.  This specifies the SED~(\ref{VacSED}) of decaying dark
energy.

\subsection{The microwave background}

The spectral intensity of an arbitrary distribution of sources with
comoving number density $n(t)$ and an SED $F(\lambda,t)$ is expressed
as an integral over time by Eq.~(\ref{ItDefn}).  Putting
Eqs.~(\ref{Rtvac}), (\ref{nvacDefn}) and (\ref{VacSED}) into this
equation, we obtain
\beq
\Ilam(\lamo) = I_v (\lamo) \times \left\{ \begin{array}{l}
   \bigint{\teq/\too}{1} \bigfrac{\tau^{-1}\,d\tau}
      {\exp \left[ \left( \bigfrac{\lamv}{\lamo} \right)
      \tau^{-2\xm/3} \right] - 1} \\
   \bigint{\teq/\tao}{\too/\tao}
      \bigfrac{\coth\tau\,d\tau} {\exp \left[
      \bigfrac{\lamv}{\lamo} \!\left( \bigfrac{\sqrt{\Omato}\,
      \sinh\tau}{\sqrt{1-\Omato}} \right)^{-2\xm/3}
      \right] - 1} \; .
\label{IvacDefn}
\end{array} \right.
\eeq
Here we have used integration variables $\tau\equiv t/\too$
in the first case ($\Omato=1$) and $\tau\equiv t/\tao$ in the second
($0<\Omato<1$).  The dimensional content of both integrals is contained
in the prefactor $I_v(\lamo)$, which reads
\beq
I_v(\lamo) \equiv \frac{5\curlyLvo}{2\pi^{\,5}h\Ho}
   \left( \frac{\lamv}{\lamo} \right)^4 =
   15,500 \mbox{ CUs} \left( \frac{\xm}{1-\xm} \right) \!
   \left( \frac{\lamv}{\lamo} \right)^4 \; .
\eeq
We have divided this quantity through by photon energy $hc/\lamo$
so as to express the results in continuum units (CUs) as usual, where
1~CU~$\equiv 1$~photon s$^{-1}$~cm$^{-2}$~\AA$^{-1}$~ster$^{-1}$.
We use CUs throughout this review, for the sake of uniformity as
well as the fact that these units carry several advantages from the
theoretical point of view (Sec.~\ref{sec:LumDens}).  The reader who
consults the literature, however, will soon find that each
part of the electromagnetic spectrum has its own ``dialect'' of 
preferred units.  In the microwave region intensities are commonly
reported in terms of $\nuInu$, the integral of flux per unit frequency
{\em over\/} frequency, and usually expressed in units of
nW~m$^{-2}$~ster$^{-1}=10^{-6}$~erg~s$^{-1}$~cm$^{-2}$~ster$^{-1}$.
To translate a given value of $\nuInu$ (in these units) into CUs, one
need only multiply by a factor of
$10^{-6}/(hc)=50.34$~erg$^{-1}$~\AA$^{-1}$.  The Jansky (Jy) is also
often encountered, with
1~Jy~$=10^{-23}$~erg~s$^{-1}$~cm$^{-2}$~Hz$^{-1}$.
To convert a given value of $\nuInu$ from Jy~ster$^{-1}$ into CUs,
one multiplies by a factor of
$10^{-23}/h\lambda=(1509$~Hz~erg$^{-1})/\lambda$ with $\lambda$ in \AA.

\begin{figure}[t!]
\begin{center}
\includegraphics[width=\textwidth]{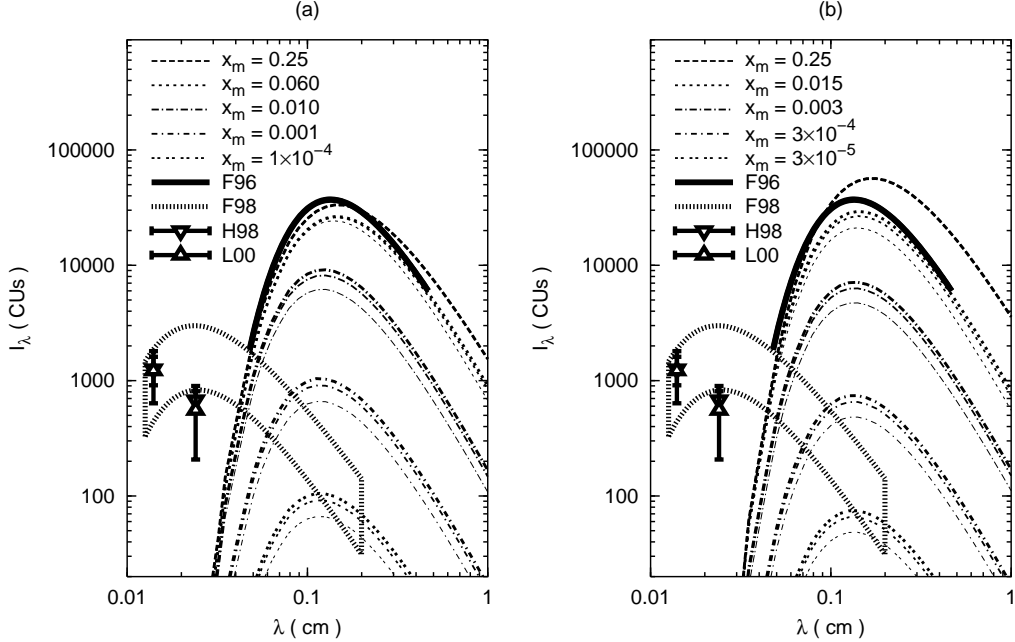}
\end{center}
\caption{The spectral intensity of background radiation due to the
   decaying vacuum for various values of $\xm$, compared with observational
   data in the microwave region (heavy solid line) and far infrared
   (heavy dotted line and squares).  For each value of $\xm$ there are
   three curves representing cosmologies with $\Omato=1$ (heaviest lines),
   $\Omato=0.3$ (medium-weight lines) and $\Omato=0.03$ (lightest lines).
   Panel~(a) assumes $L=\LR$ and $\beta=0.595$, while panel~(b) assumes
   $L=\LT$ and $\beta=1$.}
\label{fig5.2}
\end{figure}

Eq.~(\ref{IvacDefn}) gives the combined intensity of decay photons which have
been emitted at many wavelengths and redshifted by various amounts,
but reach us in a waveband centered on $\lamo$.  The arbitrary volume
$\Vo$ has dropped out of the integral as expected, and this result is
also independent of the uncertainty $\ho$ in Hubble's constant since
there is a factor of $\ho$ in both $\curlyLvo$ and $\Ho$.
Results are plotted in Fig.~\ref{fig5.2}.
over the waveband 0.01-1~cm, together with existing observational data
in this part of the spectrum.  The most celebrated of these is the
{\sc Cobe} detection of the CMB \cite{Fix96} which we have shown
as a heavy solid line (F96).  The experimental uncertainties in this
measurement are far smaller than the thickness of the line.  The other
observational limits shown in Fig.~\ref{fig5.2} have been obtained in the
far infrared (FIR) region, also from analysis of data from the {\sc Cobe}
satellite.  These are indicated with heavy dotted lines (F98 \cite{Fix98})
and open triangles (H98 \cite{Hau98} and L00 \cite{Lag00}).

Fig.~\ref{fig5.2}(a) shows the spectral intensity of background radiation
from vacuum decay under our standard assumptions, including the
relativistic definition~(\ref{LRdefn}) of vacuum luminosity and the values
of $\Orado$ and $\beta$ given by (\ref{OradValue}) and (\ref{betaValue})
respectively.  Five groups of curves are shown, corresponding to values
of $\xm$ between $3\times10^{-5}$ and the theoretical maximum of 0.25.
For each value of $\xm$ three curves are plotted: one each for the EdS,
\LCDM\ and \LBDM\ cosmologies.  As noted above in connection with the
bolometric intensity $Q$, the choice of cosmological model is less
important in determining the background due to vacuum decay than the
background due to galaxies.  In fact, the intensities here are actually
slightly {\em lower\/} in vacuum-dominated models.  The reason for this,
as before, is that these models have smaller values of $\Omato/\Orado$
and hence larger values of $\teq$, reducing the size of contributions
from the early matter-dominated era when $L_v$ was large.

In Fig.~\ref{fig5.2}(b), we have exchanged the relativistic definition
of vacuum luminosity for the thermodynamical one~(\ref{LTdefn}),
and set $\beta=1$ instead of 0.595.  As discussed in Sec.~\ref{sec:vacLum},
the increase in $\beta$ is partly offset by a drop in $\Orado$.
There is a net increase in intensity, however, because smaller
values of $\Orado$ push $\teq$ back into the past, leading to
additional contributions from the early matter-dominated era.  These
contributions particularly push up the long-wavelength part of
the spectrum in Fig.~\ref{fig5.2}(b) relative to Fig.~\ref{fig5.2}(a),
as seen most clearly in the case $\xm=0.25$.  Overall, intensities in
Fig.~\ref{fig5.2}(b) are higher than those in Fig.~\ref{fig5.2}(a)
by about a factor of four.

These figures show that {\em the decaying-vacuum hypothesis is strongly
constrained by observations of the microwave background.\/}  The parameter
$\xm$ cannot be larger than 0.06 or the intensity of the decaying vacuum
would exceed that of the CMB itself under the most conservative assumptions,
as represented by Fig.~\ref{fig5.2}(a).  This limit tightens to
$\xm\leqslant0.015$ if different assumptions are made about the
luminosity of the vacuum, as shown by Fig.~\ref{fig5.2}(b).
These numbers are comparable to the limit of $x\leqslant0.07$
obtained from entropy conservation under the assumption that
$x=\xr=\xm$ \cite{Fre87}.  And insofar as the CMB is usually
attributed entirely to relic radiation from the big bang, the real limit
on $\xm$ is probably several orders of magnitude smaller than this.

With these upper bounds on $\xm$, we can finally inquire about the
potential of the decaying vacuum as a dark-energy candidate.  Since
its density is given by (\ref{xDefn}) as a fraction $x/(1-x)$ of that
of radiation, we infer that its present density parameter ($\Ovaco$) 
satisfies:
\beq
\Ovaco = \left( \frac{\xm}{1-\xm} \right) \Orado \leqslant \left\{
     \begin{array}{ll}
  7 \times 10^{-6} & \qquad \mbox{(a)} \\
  1 \times 10^{-6} & \qquad \mbox{(b)} \; .
\label{OvacoValues}
\end{array} \right.
\eeq
Here, (a) and (b) refer to the scenarios represented by Figs.~\ref{fig5.2}(a)
and \ref{fig5.2}(b), with the corresponding values of $\Orado$ as defined
by Eqs.~(\ref{OradValue}) and (\ref{ScenarioB}) respectively.  We have
assumed that $\ho\geqslant0.6$ as usual.  It is clear from the
limits~(\ref{OvacoValues}) that a decaying vacuum of the kind we have
considered here does not contribute significantly to the density
of the dark energy.

It should be recalled, however, that there are good reasons from
quantum theory for expecting some kind of instability for the vacuum
in a universe which progressively cools.  (Equivalently, there are
good reasons for believing that the cosmological ``constant'' is not.)
Our conclusion is that if the vacuum decays, it either does so very
slowly, or in a manner that does not upset the isotropy of the
cosmic microwave background.

\section{Axions} \label{ch6}

\subsection{``Invisible'' axions} \label{sec:AxDensity}

Axions are hypothetical particles whose existence would explain what
is otherwise a puzzling feature of quantum chromodynamics (QCD),
the leading theory of strong interactions.  QCD contains a dimensionless
free parameter ($\Theta$) whose value must be ``unnaturally'' small in
order for the theory not to violate a combination of charge conservation
and mirror-symmetry known as charge parity or CP.
Upper limits on the electric dipole moment of the neutron currently
constrain the value of $\Theta$ to be less than about $10^{-9}$.  The
strong CP problem is the question: ``Why is $\Theta$ so small?''
This is reminiscent of the cosmological-constant problem
(Sec.~\ref{sec:vacenergy}), though less severe by many orders of magnitude.
Proposed solutions have similarly focused on making $\Theta$,
like $\Lambda$, a dynamical variable whose value could have been driven
toward zero in the early Universe.  In the most widely-accepted scenario,
due to Peccei and Quinn in 1977 \cite{Pec77}, this is accomplished by
the spontaneous breaking of a new global symmetry (now called PQ symmetry)
at energy scales $\fpq$.  As shown by Weinberg \cite{Wei78} and
Wilczek \cite{Wil78} in 1978, the symmetry-breaking gives rise to
a new particle which eventually acquires a rest energy
$\ma \propto \fpq^{-1}$.  This particle is the axion ($a$).

Axions, if they exist, meet all the requirements of a successful
CDM candidate (Sec.~\ref{sec:cdm}): they interact weakly with
the baryons, leptons and photons of the standard model; they are cold
(i.e. non-relativistic during the time when structure begins to form);
and they are capable of providing some or even all of the CDM density
which is thought to be required, $\Ocdm\sim0.3$.
A fourth property, and the one which is of most interest to us here,
is that {\em axions decay generically into photon pairs\/}.
The importance of this process depends on two things: the axion's rest
mass $m_a$ and its coupling strength $\gagg$.
The PQ symmetry-breaking energy scale $\fpq$ is not constrained by
the theory, and reasonable values for this parameter are such that $\ma$
might in principle lie anywhere between $10^{-12}$~eV and 1~MeV \cite{Kol90}.
This broad range of theoretical possibilities has been whittled
down by an impressive array of cosmological, astrophysical and
laboratory-based tests.  In summarizing these, it is useful to
distinguish between axions with rest energies above and below a
``critical'' rest energy $\macrit\sim3\times10^{-2}$~eV.

Axions whose rest energies lie {\em below\/} $\macrit$ arise primarily
via processes known as vacuum misalignment \cite{Pre83,Abb83,Din83} and
axionic string decay \cite{Dav86}.  These are non-thermal mechanisms,
meaning that the axions produced in this way were never in thermal
equilibrium with the primordial plasma.  Their present density would
be at least \cite{Sik00}
\beq
\Omega_a\approx \left( \frac{\ma}{4\times10^{-6}\mbox{ eV}}
   \right)^{\!\!-7/6}\!\!\!\ho^{-2} \; .
\label{OmaNontherm}
\eeq
(This number is currently under debate, and may go up by an order of
magnitude or more if string effects play an important role \cite{Bat00}.)
If we require that axions not provide {\em too much\/} CDM
($\Ocdm\leqslant0.6$) then (\ref{OmaNontherm}) implies a lower limit
on the axion rest energy:
\beq
\ma \gtrsim 7\times10^{-6} \; .
\eeq
This neatly eliminates the lower third of the theoretical axion mass window.
Upper limits on $m_a$ for non-thermal axions have come from astrophysics.
Prime among these is the fact that the weak couplings of axions to
baryons, leptons and photons allow them to stream freely from stellar
cores, carrying energy with them.  If they are massive enough, such
axions could in principle cool the core of the Sun, alter the 
helium-burning phase in red-giant stars, and shorten the duration
of the neutrino burst from supernovae such as SN1987a.  The last of
these effects is particularly sensitive and requires \cite{Jan96,Kei97}:
\beq
\ma \lesssim 6 \times 10^{-3} \mbox{ eV} \; .
\label{AxSNlimit}
\eeq
Axions with $10^{-5}\lesssim\ma\lesssim10^{-2}$ thus remain compatible
with both cosmological and astrophysical limits, and could provide much
or all of the CDM.  It may be possible to detect these particles
in the laboratory by enhancing their conversion into photons with
strong magnetic fields, as demonstrated by Sikivie in 1983 \cite{Sik83}.
Experimental search programs based on this principle are now in operation
at the Lawrence Livermore lab in the U.S.A. \cite{Hag98}, Tokyo \cite{Mor98},
the Sierra Grande mountains in Argentina ({\sc Solax} \cite{Avi98}),
the Spanish Pyrenees ({\sc Cosme} \cite{Mor02}) and CERN in Switzerland
({\sc Cast} \cite{Aal02}).  Exclusion plots from these experiments are
beginning to restrict theoretically-favoured regions of the phase space
defined by $m_a$ and $\gagg$.

\begin{figure}[t!]
\begin{center}
\includegraphics[width=45mm]{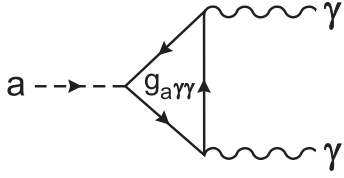}
\end{center}
\caption{The Feynman diagram corresponding to the decay of the axion ($a$)
   into two photons ($\gamma$) with coupling strength $g_{a\gamma\gamma}$.}
\label{fig6.1}
\end{figure}

Promising as they are, we will not consider non-thermal axions (sometimes
known as ``invisible axions'') further in this section.  This is because
they decay too slowly to leave any trace in the extragalactic background
light.  Axions decay into photon pairs ($a \rightarrow \gamma + \gamma$)
via a loop diagram, as illustrated in Fig.~\ref{fig6.1}.
The {\em decay lifetime\/} of this process is \cite{Kol90}
\beq
\tau_a = (6.8 \times 10^{24} \mbox{ s}) \, \mone^{-5} \, \zeta^{-2} \; .
\label{tauaValue}
\eeq
Here $\mone\equiv\ma/(1$~eV) is the axion rest energy in units of eV,
and $\zeta$ is a constant which is proportional to the coupling strength
$\gagg$ \cite{Res91}.  For our purposes, it is sufficient to treat
$\zeta$ as a free parameter which depends on the details of the axion
theory chosen.  Its value has been normalized in Eq.~(\ref{tauaValue})
so that $\zeta=1$ in the simplest grand unified theories (GUTs)
of strong and electroweak interactions.  This could drop to $\zeta=0.07$
in other theories, however \cite{Kap85}, strongly suppressing the
two-photon decay channel.  In principle $\zeta$ could even vanish
altogether, corresponding to a radiatively {\em stable\/} axion, although
this would require an unlikely cancellation of terms.  We will consider
values in the range $0.07\leqslant\zeta\leqslant1$ in what follows.
For these values of $\zeta$, and with $\mone\lesssim6\times10^{-3}$ 
as given by (\ref{AxSNlimit}), Eq.~(\ref{tauaValue}) shows that
axions decay on timescales $\taua\gtrsim9\times10^{35}$~s.  This is so
much longer than the age of the Universe that such particles would
truly be invisible.

\subsection{The multi-eV window}

We therefore shift our attention to axions with rest energies
{\em above\/} $\macrit$.  Turner showed in 1987 \cite{Tur87} that
the vast majority of these would have arisen in the early Universe via
thermal mechanisms such as Primakoff scattering and photo-production.
The Boltzmann equation can be solved to give their present comoving
number density as $n_a=(830/\gstar)$~cm$^{-3}$ \cite{Res91}, where
$\gstar\approx15$ counts the number of relativistic degrees of freedom
left in the plasma at the time when axions ``froze out'' of equilibrium.
The present density parameter $\Oma=n_a m_a/\rcrito$ of thermal axions
is thus
\beq
\Omega_a = 5.2 \times 10^{-3} \ho^{-2} \, \mone \; .
\label{OmaValue}
\eeq
Whether or not this is significant depends on the axion rest mass.
The duration of the neutrino burst from SN1987a again imposes a powerful
constraint on $\ma$.  This time, however, it is a {\em lower\/}, not an
upper bound, because axions in this range of rest energies are massive
enough to interact with nucleons in the supernova core and can no longer
stream out freely.  Instead, they are trapped in the core and radiate 
only from an ``axiosphere'' rather than the entire volume of the star.
Axions with sufficiently large $\ma$ are trapped so strongly that
they no longer interfere with the luminosity of the neutrino burst,
leading to the lower limit \cite{Tur88}
\beq
\ma \gtrsim 2.2 \mbox{ eV} \; .
\label{TurCon}
\eeq
Astrophysics also provides strong upper bounds on $\ma$ for thermal
axions.  These depend critically on whether or not axions couple only
to hadrons, or to other particles as well.  An early class of {\em hadronic\/}
axions (those coupled only to hadrons) was developed by Kim \cite{Kim79}
and Shifman, Vainshtein and Zakharov \cite{Shi80}; these particles are
often termed KSVZ axions.  Another widely-discussed model in which axions
couple to charged leptons as well as nucleons and photons has been
discussed by Zhitnitsky \cite{Zhi80} and Dine, Fischler and Srednicki
\cite{Din81}; these particles are known as DFSZ axions.  The extra lepton
coupling of these DFSZ axions allows them to carry so much energy out of
the cores of red-giant stars that helium ignition is seriously disrupted
unless $\ma\lesssim9\times10^{-3}$~eV \cite{Raf95}.  Since this upper
limit is inconsistent with the lower limit~(\ref{TurCon}), thermal
DFSZ axions are excluded.  For KSVZ or hadronic axions, red giants
impose a weaker bound \cite{Raf87}:
\beq
\ma \lesssim 0.7 \, \zeta^{-1} \mbox{ eV} \; .
\label{RafCon}
\eeq
This is consistent with the lower limit~(\ref{TurCon}) for realistic
values of the parameter $\zeta$.  For the simplest hadronic axion models
with $\zeta\geqslant0.07$, for instance, Eq.~(\ref{RafCon}) translates into
an upper limit $\ma\lesssim10$~eV.  It has been argued that axions with
$\ma\gtrsim10$~eV can be ruled out in any case because they would interact
strongly enough with baryons to produce a detectable signal in existing
\v{C}erenkov detectors \cite{Eng90}.

For thermally-produced hadronic axions, then, there remains a window of
opportunity in the multi-eV range with $2\lesssim\mone\lesssim10$.
Eq.~(\ref{OmaValue}) shows that these particles would contribute a total
density of about $0.03\lesssim\Oma\lesssim0.15$, where we take
$0.6\leqslant\ho\leqslant0.9$ as usual.  They would not be able to provide
the entire density of dark matter in the \LCDM\ model ($\Omato=0.3$),
but they could suffice in low-density models midway between \LCDM\ and \LBDM\
(Table~\ref{table3.1}).  Since such models remain compatible with most
current observational data (Sec.~\ref{ch4}), it is worth proceeding to
see how these multi-eV axions can be further constrained by their
contributions to the EBL.

\subsection{Axion halos} \label{sec:AxHalo}

Thermal axions are not as cold as their non-thermal cousins, but will
still be found primarily inside gravitational potential wells such as
those of galaxies and galaxy clusters \cite{Tur87}.  We need not be
too specific about the fraction which have settled into galaxies
as opposed to larger systems, because we will be concerned primarily
with their {\em combined\/} contributions to the diffuse background.
(Distribution could become an issue if extinction due to dust or gas
played a strong role inside the bound regions, but this is not likely
to be important for the photon energies under consideration here.)
These axion halos provide us with a convenient starting-point as
cosmological light sources, analogous to the galaxies and vacuum
source regions of previous sections.  Let us take the axions to be
cold enough that their fractional contribution ($M_h$) to the total
mass of each halo ($\Mtot$) is the same as their fractional
contribution to the cosmological matter density,
$M_h/\Mtot=\Oma/\Omato=\Oma/(\Oma+\Obar)$.
Then the mass $M_h$ of axions in each halo is
\beq
M_h = \Mtot \left( 1 + \frac{\Obar}{\Oma} \right)^{-1} \; .
\label{MaDefn1}
\eeq
(Here we have made the minimal assumption that axions constitute
{\em all\/} the nonbaryonic dark matter.)
If these halos are distributed with a mean comoving number density
$\no$, then the cosmological density of bound axions is
$\Oabound=(\no M_h)/\rcrito=(\no\Mtot/\rcrito)(1+\Obar/\Oma)^{-1}$.
Equating $\Oabound$ to $\Oma$, as given by (\ref{OmaValue}),
fixes the total mass:
\beq
\Mtot = \frac{\Oma\rcrito}{\no}
        \left( 1 + \frac{\Obar}{\Oma} \right) \; .
\label{MtotDefn}
\eeq
The comoving number density of galaxies at $z=0$ is \cite{Pee93}
\beq
\no = 0.010 \, \ho^3 \mbox{ Mpc}^{-3} \; .
\label{noValue}
\eeq
Using this together with (\ref{OmaValue}) for $\Oma$, and setting
$\Obar\approx0.016\ho^{-2}$ from Sec.~\ref{sec:baryons}, we find from
(\ref{MtotDefn}) that
\beq
\Mtot = \left\{ \begin{array}{ll}
   9 \times 10^{11} \Msun \ho^{-3} & \qquad (\mone=3) \\
   1 \times 10^{12} \Msun \ho^{-3} & \qquad (\mone=5) \\
   2 \times 10^{12} \Msun \ho^{-3} & \qquad (\mone=8)
\end{array} \right. \; .
\label{MtotValues}
\eeq
Let us compare these numbers with dynamical data on the mass of the Milky
Way using the motions of Galactic satellites.  These assume a Jaffe profile
\cite{Jaf83} for halo density:
\beq
\rtot(r) = \frac{v_c^2}{4\pi G \, r^2} \frac{r_j^2}{(r+r_j)^2} \; ,
\label{JaffeDefn}
\eeq
where $v_c$ is the circular velocity,
$r_j$ the Jaffe radius, and $r$ the radial distance from the center of
the Galaxy.  The data imply that $v_c=220\pm30$~km~s$^{-1}$ and
$r_j=180\pm60$~kpc \cite{Koc96a}.
Integrating over $r$ from zero to infinity gives
\beq
\Mtot = \frac{v_c^2 r_j}{G} = (2 \pm 1) \times 10^{12} \Msun \; .
\label{MhValue}
\eeq
This is consistent with (\ref{MtotValues}) for most values of $\mone$ and
$\ho$.  So axions of this type could in principle make up all
the dark matter which is required on Galactic scales.

Putting (\ref{MtotDefn}) into (\ref{MaDefn1}) gives the mass
of the axion halos as
\beq
M_h = \frac{\Oma\rcrito}{\no} \; .
\label{MaDefn}
\eeq
This could also have been derived as the mass of a region of space
of comoving volume $\Vo=\no^{-1}$ filled with homogeneously-distributed
axions of mean density $\rho_a=\Oma\rcrito$.  (This is the approach 
that we adopted in defining vacuum regions in Sec.~\ref{sec:vacLum}.)

To obtain the halo luminosity, we sum up the rest energies of all the
decaying axions in the halo and divide by the decay
lifetime~(\ref{tauaValue}):
\beq
L_h = \frac{M_h c^2}{\taua} \; .
\eeq
Inserting Eqs.~(\ref{tauaValue}) and (\ref{MaDefn}), we find
\beqa
L_h & = & (3.8 \times 10^{40} \mbox{ erg s}^{-1}) \, \ho^{-3} \zeta^2 \mone^6
          \nonumber \\
    & = & \left\{ \begin{array}{ll}
          7 \times 10^{9} \Lsun \, \ho^{-3} \zeta^2 & \qquad (\mone=3) \\
          2 \times 10^{11} \Lsun \, \ho^{-3} \zeta^2 & \qquad (\mone=5) \\
          3 \times 10^{12} \Lsun \, \ho^{-3} \zeta^2& \qquad (\mone=8)
\end{array} \right. \; .
\label{AxHaloLum}
\eeqa
The luminosities of the galaxies themselves are of order
$L_0=\curlyLo/\no=2\times10^{10}\ho^{-2}\Lsun$, where we have used
(\ref{curlyLoValue}) for $\curlyLo$.  Thus axion halos could in principle
outshine their host galaxies, unless axions are either very light
($\mone\lesssim3$) or weakly-coupled ($\zeta<1$).  This already
suggests that they will be strongly constrained by observations
of EBL intensity.

\subsection{Bolometric intensity}

Substituting the halo comoving number density~(\ref{noValue}) and
luminosity~(\ref{AxHaloLum}) into Eq.~(\ref{QzDefn}), we find that the
combined intensity of decaying axions at all wavelengths is given by
\beq
Q = Q_a \int_0^{\, z_f} \frac{dz}{(1+z)^2 \Htil(z)} \; .
\label{QaDefn}
\eeq
Here the dimensional content of the integral is contained in the
prefactor $Q_a$, which takes the following numerical values:
\beqa
Q_a & = & \frac{\Oma\rcrito\, c^3}{\Ho\taua} = 
          (1.2 \times 10^{-7} \mbox{ erg s}^{-1} \mbox{ cm}^{-2}) \,
          \ho^{-3} \zeta^2 \mone^6 \label{QaValue} \\
    & = & \left\{ \begin{array}{ll}
          9 \times 10^{-5} \mbox{ erg s}^{-1} \mbox{ cm}^{-2}) \,
            \ho^{-1} \zeta^2 & \qquad (\mone=3) \\
          2 \times 10^{-3} \mbox{ erg s}^{-1} \mbox{ cm}^{-2}) \,
            \ho^{-1} \zeta^2 & \qquad (\mone=5) \\
          3 \times 10^{-2} \mbox{ erg s}^{-1} \mbox{ cm}^{-2}) \,
            \ho^{-1} \zeta^2 & \qquad (\mone=8)
\end{array} \right. \; . \nonumber
\eeqa
There are three things to note about this quantity.  First, it is
comparable in magnitude to the {\em observed\/} EBL due to galaxies,
$\Qstar\approx3\times10^{-4}$~erg~s$^{-1}$~cm$^{-2}$ (Sec.~\ref{ch2}).
Second, unlike $\Qstar$ for galaxies or $Q_v$ for decaying vacuum
energy, $Q_a$ depends explicitly on the uncertainty $\ho$ in Hubble's
constant.  Physically, this reflects the fact that the axion density
$\rho_a=\Oma\rcrito$ in the numerator of (\ref{QaValue}) comes to us from
the Boltzmann equation and is independent of $\ho$, whereas the density
of luminous matter such as that in galaxies is inferred from its
luminosity density $\curlyLo$ (which is proportional to $\ho$, thus
cancelling the $\ho$-dependence in $\Ho$).  The third thing to note
about $Q_a$ is that it is independent of $\no$.  This is because the
collective contribution of decaying axions to the diffuse background
is determined by their mean density $\Oma$, and does not depend on
how they are distributed in space.

\begin{figure}[t!]
\begin{center}
\includegraphics[width=100mm]{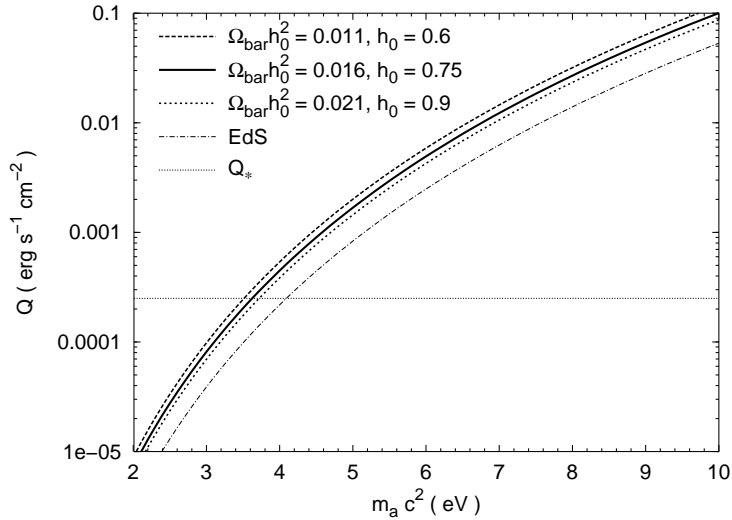}
\end{center}
\caption{The bolometric intensity $Q$ of the background radiation from
         decaying axions as a function of their rest masses $m_a$.
         The faint dash-dotted line shows the equivalent intensity in an
         EdS model (in which axions alone cannot provide the required CDM).
         The dotted horizontal line indicates the characteristic bolometric
         intensity ($\Qstar$) of the observed EBL.}
\label{fig6.2}
\end{figure}

To evaluate~(\ref{QaDefn}) we need to specify the cosmological model.
If we assume a spatially flat Universe, as increasingly suggested by
the data (Sec.~\ref{ch4}), then Hubble's parameter~(\ref{FL2}) reads
\beq
\Htil(z) = \left[ \Omato (1+z)^3 + 1- \Omato \right]^{1/2} \; ,
\label{HtilFlat}
\eeq
where we take the most economical approach and require axions to make up
all the cold dark matter so that $\Omato=\Oma+\Obar$.  Putting this into
Eq.~(\ref{QaDefn}) along with (\ref{QaValue}) for $Q_a$, we obtain the
plots of $Q(\mone)$ shown in Fig.~\ref{fig6.2} for $\zeta=1$.
The three heavy lines in this plot show the range of intensities obtained
by varying $\ho$ and $\Obar\ho^2$ within the ranges
$0.6\leqslant\ho\leqslant0.9$ and $0.011\leqslant\Obar\ho^2\leqslant0.021$
respectively.  We have set $z_f=30$, since axions were presumably decaying
long before they became bound to galaxies.  (Results are insensitive to
this choice, rising by less than 2\% as $z_f\rightarrow1000$ and dropping
by less than 1\% for $z_f=6$.)  The axion-decay background is faintest
for the largest values of $\ho$, as expected from the fact that
$Q_a\propto\ho^{-1}$.  This is partly offset by the fact
that larger values of $\ho$ also lead to a drop in $\Omato$,
extending the age of the Universe and hence the length of time over which
axions have been contributing to the background.  (Smaller values of
$\Obar$ raise the intensity slightly for the same reason.)
Fig.~\ref{fig6.2} shows that axions with $\zeta=1$ and $\ma\gtrsim3.5$~eV
produce a background brighter than that from the galaxies themselves.

\subsection{The infrared and optical backgrounds} \label{sec:SpecAx}

\begin{figure}[t!]
\begin{center}
\includegraphics[width=100mm]{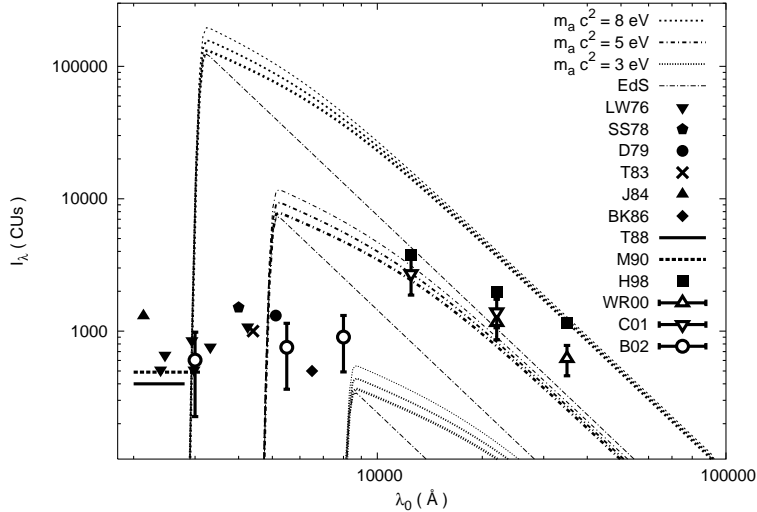}
\end{center}
\caption{The spectral intensity $\Ilam$ of the background radiation from
         decaying axions as a function of observed wavelength $\lamo$. 
         The four curves for each value of $m_a$ (labelled) correspond
         to upper, median and lower limits on $\ho$ and $\Obar$ together
         with the equivalent intensity for the EdS model (as in
         Fig.~\ref{fig6.2}).  Also shown are observational upper limits
         (solid symbols and heavy lines) and reported detections
         (empty symbols) over this waveband.}
\label{fig6.3}
\end{figure}

To go further and compare our predictions with observational data, we would
like to calculate the intensity of axionic contributions to the EBL as a
function of wavelength.  The first step, as usual, is to specify the
spectral energy distribution or SED of the decay photons in the rest frame.
Each axion decays into two photons of energy $\smallfrac{1}{2}\ma$
(Fig.~\ref{fig6.1}), so that the decay photons are emitted at or
near a peak wavelength
\beq
\lama = \frac{2hc}{\ma} = \frac{24,800\mbox{ \AA}}{\mone} \; .
\eeq
Since $2\lesssim\mone\lesssim10$, the value of this parameter tells us
that we will be most interested in the {\em infrared and optical\/}
bands (roughly 4000--40,000~\AA).
We can model the decay spectrum with a Gaussian SED as in~(\ref{gaussSED}):
\beq
F(\lambda) = \frac{L_h}{\sqrt{2\pi}\,\siglam} \exp \left[ -\frac{1}{2}
   \left( \frac{\lambda-\lama}{\siglam} \right)^2 \right] \; .
\eeq
For the standard deviation of the curve, we can use the velocity
dispersion $v_c$ of the bound axions
\cite{Kep87}.  This is $220\mbox{ km~s}^{-1}$ for the Milky Way, implying that
$\siglam\approx\mbox{40 \AA}/\mone$ where we have used $\siglam=2(v_c/c)\lama$
(Sec.~\ref{sec:gauss}).  For axions bound in galaxy clusters, $v_c$ rises to
as much as $1300\mbox{ km s}^{-1}$ \cite{Res91}, implying that
$\siglam\approx\mbox{220 \AA}/\mone$.  Let us parametrize $\siglam$
in terms of a dimensionless quantity
$\sigfifty\equiv\siglam/(50\mbox{ \AA}/\mone)$ so that
\beq
\siglam = (50\mbox{ \AA}/\mone) \, \sigfifty \; .
\eeq
With the SED $F(\lambda)$ thus specified along with Hubble's
parameter~(\ref{HtilFlat}), the
spectral intensity of the
background radiation produced by axion decays is given by (\ref{IzDefn}) as
\beq
\Ilam(\lamo) = I_a \bigint{0}{z_f} \, \bigfrac{
   \exp \left\{ -\bigfrac{1}{2} \left[ \bigfrac{\lamo/(1+z)-\lama}{\siglam}
   \right]^2 \right\} dz}{(1+z)^3 [\Omato (1+z)^3 + 1 - \Omato]^{1/2}}
   \; .
\label{IaDefn}
\eeq
The dimensional prefactor in this case reads
\beqa
I_a & = & \frac{\Oma\rcrito\, c^2}{\sqrt{32\pi^3}\,h\,\Ho\,\taua}
       \!\left( \frac{\lamo}{\siglam} \right) \nonumber \\
    & = & (95\mbox{ CUs}) \, \ho^{-1} \zeta^2 \mone^7 \, \sigfifty^{-1}
       \left( \frac{\lamo}{24,800\mbox{ \AA}} \right) \; .
\eeqa
We have divided through by the photon energy $hc/\lamo$ to put
results into continuum units or CUs as usual (Sec.~\ref{sec:LumDens}).
The number density in (\ref{IzDefn}) cancels out the factor
of $1/\no$ in luminosity~(\ref{AxHaloLum}) so that results are
independent of axion distribution, as expected.  Evaluating
Eq.~(\ref{IaDefn}) over 2000~\AA$\leqslant\lamo\leqslant$20,000~\AA\
with $\zeta=1$ and $z_f=30$, we obtain the plots of $\Ilam(\lamo)$
shown in Fig.~\ref{fig6.3}.
Three groups of curves are shown, corresponding to $\ma=3$~eV, 5~eV and
8~eV.  For each value of $m_a$ there are four curves; these assume
$(\ho,\Obar\ho^2)=(0.6,0.011),(0.75,0.016)$ and $(0.9,0.021)$ respectively,
with the fourth (faint dash-dotted) curve representing the equivalent
intensity in an EdS universe (as in Fig.~\ref{fig6.2}).  Also plotted
are many of the reported observational constraints
on EBL intensity in this waveband.  Most have been encountered already
in Sec.~\ref{ch3}.  They include data from the {\sc Oao}-2 satellite
(LW76 \cite{Lil76}), several ground-based telescope observations
(SS78 \cite{Spi78}, D79 \cite{Dub79}, BK86 \cite{Bou86}),
the Pioneer~10 spacecraft (T83 \cite{Tol83}), sounding rockets
(J84 \cite{Jak84}, T88 \cite{Ten88}), the Space Shuttle-borne
Hopkins~{\sc Uvx} experiment (M90 \cite{Mur90}),
the {\sc Dirbe} instrument aboard the {\sc Cobe} satellite
(H98 \cite{Hau98}, WR00 \cite{Wri00}, C01 \cite{Cam01}), and
combined {\sc Hst}/Las Campanas telescope observations (B02 \cite{Ber02a}).

Fig.~\ref{fig6.3} shows that 8~eV axions with $\zeta=1$ would produce a
hundred times more background light at $\sim\!3000$~\AA\ than is actually seen.
The background from 5~eV axions would similarly exceed observed levels
by a factor of ten at $\sim\!5000$~\AA, colouring the night sky green.
Only axions with $\ma\leqslant3$~eV are compatible with observation
if $\zeta=1$.  These results are brighter than ones obtained assuming
an EdS cosmology \cite{Ove93b}, especially at wavelengths
longward of the peak.  This reflects the fact that the background in
a low-$\Omato$, high-$\Olamo$ universe like that considered here
receives many more contributions from sources at high redshift.

\begin{figure}[t!]
\begin{center}
\includegraphics[width=100mm]{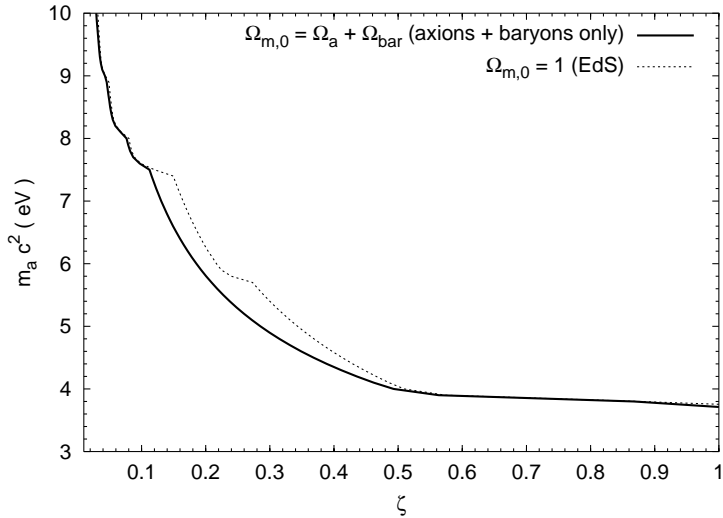}
\end{center}
\caption{The upper limits on the value of $\ma$ as a function of the
         coupling strength $\zeta$ (or vice versa).  These are derived
         by requiring that the minimum predicted axion intensity
         (as plotted in Fig.~\ref{fig6.3}) be less than or equal to the
         observational upper limits on the intensity of the EBL.}
\label{fig6.4}
\end{figure}

To obtain more detailed constraints, we can instruct a computer to
evaluate the integral~(\ref{IaDefn}) at more finely-spaced intervals in $m_a$.
Since $\Ilam\propto\zeta^{-2}$, the value of $\zeta$ required to reduce
the minimum predicted axion intensity $\Ith$ below a given observational
upper limit $\Iobs$ at any wavelength $\lamo$ in Fig.~\ref{fig6.3} is
$\zeta\leqslant\sqrt{\Iobs/\Ith}$.  The upper limit on $\zeta$
(for a given value of $m_a$) is then the smallest such value of $\zeta$;
i.e. that which brings $\Ith$ down to $\Iobs$ or below at {\em each\/}
wavelength $\lamo$.  From this procedure we obtain a function which can
be regarded as an upper limit on the axion rest mass $m_a$ as a function
of $\zeta$ (or vice versa).  Results are plotted in Fig.~\ref{fig6.4}
(heavy solid line).
This curve tells us that even in models where the axion-photon is strongly
suppressed and $\zeta=0.07$, the axion cannot be more massive than
\beq
\ma \leqslant 8.0 \mbox{ eV} \qquad (\zeta=0.07) \; .
\label{OurCon1}
\eeq
In the simplest axion models with $\zeta=1$, this limit tightens to
\beq
\ma \leqslant 3.7 \mbox{ eV} \qquad (\zeta=1) \; .
\label{OurCon2}
\eeq
As expected, these bounds are stronger than those obtained in an EdS model,
for which some other CDM candidate would have to be postulated besides the
axions (Fig.~\ref{fig6.4}, faint dotted line).  This is a small effect,
however, because the strongest constraints tend to come from the region
near the peak wavelength ($\lama$), whereas the difference between
matter- and vacuum-dominated models is most pronounced at wavelengths
longward of the peak where the majority of the radiation originates at
high redshift.  Fig.~\ref{fig6.4} shows that cosmology in this case has
the most effect over the range $0.1\lesssim\zeta\lesssim0.4$, where upper
limits on $\ma$ are weakened by about 10\% in the EdS model relative to
one in which the CDM is assumed to consist only of axions.

Combining Eqs.~(\ref{TurCon}) and (\ref{OurCon2}), we conclude that
axions in the simplest models are confined to a slender range of
viable rest masses:
\beq
2.2 \mbox{ eV} \lesssim \ma \leqslant 3.7 \mbox{ eV} \; .
\label{AxCon}
\eeq
Background radiation thus complements the red-giant bound~(\ref{RafCon})
and {\em closes off most, if not all of the multi-eV window for thermal
axions.\/}  The range of values~(\ref{AxCon}) can be further narrowed
by looking for the enhanced signal which might be expected to emanate
from concentrations of bound axions associated with galaxies and
clusters of galaxies, as first suggested by Kephart and Weiler in 1987
\cite{Kep87}.  The most thorough search along these lines was reported
in 1991 by Bershady \etal\ \cite{Ber91}, who found no
evidence of the expected signal from three selected clusters, further
tightening the upper limit on the multi-eV axion window to 3.2~eV in
the simplest models.  Constraints obtained in this way for {\em non\/}-thermal
axions would be considerably weaker, as noticed by several workers
\cite{Kep87,Lod87}, but this does not affect our results since axions
in the range of rest masses considered here are overwhelmingly thermal ones.
Similarly, ``invisible'' axions with rest masses near the upper limit given
by Eq.~(\ref{AxSNlimit}) might give rise to detectable microwave signals
from nearby mass concentrations such as the Local Group of galaxies;
this is the premise for a recent search carried out by Blout \etal\
\cite{Blo01} which yielded an independent lower limit on the
coupling parameter $\gagg$.

Let us turn finally to the question of how much dark matter can be provided
by light thermal axions of the type we have considered here.  With rest
energies given by~(\ref{AxCon}), Eq.~(\ref{OmaValue}) shows that 
\beq
0.014 \lesssim \Oma \leqslant 0.053 \; .
\eeq
Here we have taken $0.6\leqslant\ho\leqslant0.9$ as usual.  This is
comparable to the density of baryonic matter (Sec.~\ref{sec:baryons}),
but falls well short of most expectations for the density of cold 
dark matter.

Our main conclusions, then, are as follows:  thermal axions in the
multi-eV window remain (only just) viable at the lightest end of the
range of possible rest-masses given by Eq.~(\ref{AxCon}).  They
may also exist with slightly higher rest-masses, up to the limit given
by Eq.~(\ref{OurCon1}), but only in certain axion theories where
their couplings to photons are weak.  In either of these two scenarios,
however, their contributions to the density of dark matter in the Universe
are so feeble as to remove much of their motivation as CDM candidates. 
If they are to provide a significant portion of the dark matter,
then axions must have rest masses in the ``invisible'' range where
they do not contribute significantly to the light of the night sky.

\section{Neutrinos} \label{ch7}

\subsection{The decaying-neutrino hypothesis}

Experiments now indicate that neutrinos possess nonzero rest mass and
make up at least part of the dark matter (Sec.~\ref{sec:neutrinos}).
If different neutrino species have different rest masses, then heavier
ones can decay into lighter ones plus a photon.  These decay photons
might be observable, as first appreciated by Cowsik in 1977 \cite{Cow77}
and de~Rujula and Glashow in 1980 \cite{deR80}.
The strength of the expected signal depends on the way in which neutrino
masses are incorporated into the standard model of particle physics.
In minimal extensions of this model, radiative neutrino decays are
characterized by lifetimes on the order of $10^{29}$~yr or more
\cite{Pal82}.  This is so much longer than the age of the Universe
that neutrinos are effectively {\em stable\/}, and would not produce
a detectable signal.  In other theories, however, such as those involving
supersymmetry, their decay lifetime can drop to as low as $10^{15}$~yr
\cite{Bow95}.  This is within five orders of magnitude of the age of
the Universe and opens up the possibility of significant contributions
to the background light.

\begin{figure}[t!]
\begin{center}
\includegraphics[width=80mm]{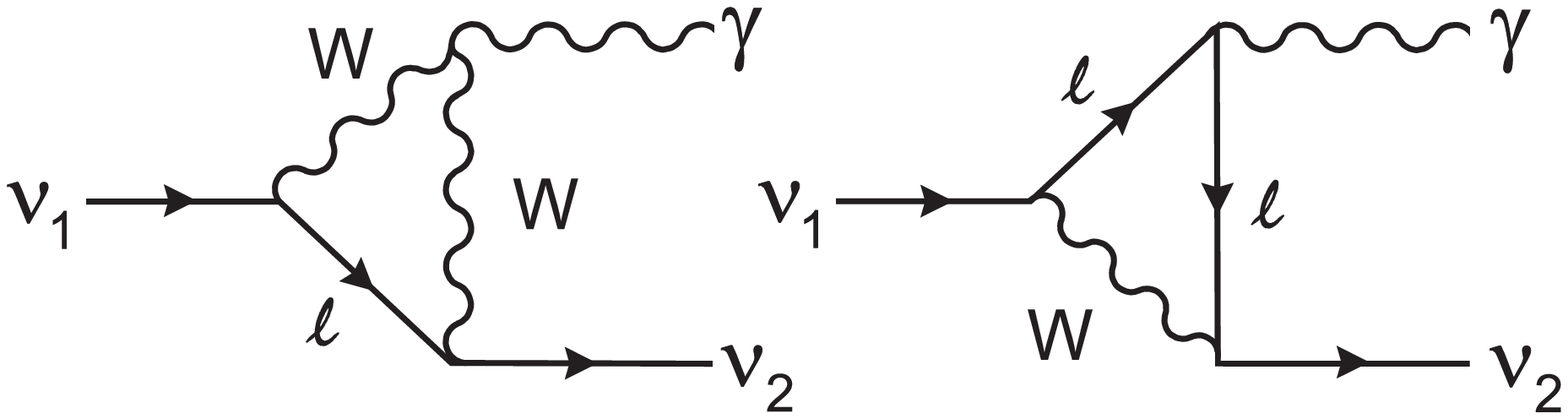}
\end{center}
\caption{Feynman diagrams corresponding to the decay of a massive
   neutrino ($\nu_1$) into a second, lighter neutrino species ($\nu_2$)
   together with a photon ($\gamma$).  The process is mediated by charged
   leptons ($\ell\,$) and the W boson (W).}
\label{fig7.1}
\end{figure}

Decay photons from neutrinos with lifetimes this short are also interesting
for another reason: their existence might resolve a number of longstanding
astrophysical puzzles involving the ionization of hydrogen and nitrogen
in the interstellar and intergalactic medium \cite{Mel81,Sci82}.
As first pointed out by Melott \etal\ in 1988 \cite{Mel88}, these
would be particularly well explained by neutrinos decaying on timescales
of order $\taunu\sim10^{24}$~s with rest energies $\mnu\sim 30$~eV.
This latter value fits awkwardly with current thinking on large-scale
structure formation in the early Universe (Sec.~\ref{sec:neutrinos}).
Neutrinos of this kind could help with so many other problems, however,
that they have continued to draw the interest of cosmologists.
Sciama and his colleagues, in particular, were led on this basis to
develop a detailed scenario known as the {\em decaying-neutrino hypothesis\/}
\cite{Sci93a,Sci90,Sci97}, in which the rest energy and decay lifetime
of the massive $\tau$-neutrino are given respectively by
\beq
\Etau = 28.9 \pm 1.1 \mbox{ eV} \qquad
   \taunu = (2\pm1) \times 10^{23} \mbox{ s} \; .
\label{taunuValue}
\eeq
The $\tau$ neutrino decays into a $\mu$ neutrino plus a photon
(Fig.~\ref{fig7.1}).  Assuming that $\mtau\gg\mmu$, conservation of
energy and momentum require this photon to have an energy
$\Egam=\smallfrac{1}{2}\Etau=14.4\pm0.5$~eV.

The concreteness of this proposal has made it eminently testable.  
Some of the strongest bounds come from searches for line emission
near 14~eV that would be expected from concentrations of decaying dark
matter in clusters of galaxies.  No such signal has been seen in the
direction of the galaxy cluster surrounding the quasar 3C~263
\cite{Fab91}, or in the direction of the rich cluster Abell~665
which was observed using the Hopkins Ultraviolet Telescope in 1991
\cite{Dav91}.  It may be, however, that absorption plays a
stronger role than expected along the lines of sight to these clusters,
or that most of their dark matter is in another form \cite{Bow95,Sci93b}.
A potentially more robust test of the decaying-neutrino hypothesis comes
from the {\em diffuse\/} background light.  This has been looked at in a
number of studies \cite{Ste80,Kim81,Sci91,Ove93c,Dod94,Ove97a,Ove99b}.
The task is a challenging one for several reasons.
Decay photons of energy near 14~eV are strongly absorbed by both dust
and neutral hydrogen, and the distribution of these quantities in
intergalactic space is uncertain.  It is also notoriously difficult,
perhaps more so in this part of the spectrum than any other,
to distinguish between those parts of the background which are truly
extragalactic and those which are due to a complex mixture of
competing foreground signals \cite{Bow91,Hen91}.  We reconsider the
problem here with the help of the formalism developed in Secs.~\ref{ch2}
and \ref{ch3}, adapting it to allow for absorption by gas and dust.

\subsection{Neutrino halos}

To begin with, we take as our sources of background radiation the
neutrinos which have become trapped in the gravitational potential wells
surrounding individual galaxies.  (Not all the neutrinos will be bound in
this way; and we will deal with the others separately.)  The comoving
number density of these galactic neutrino halos is just that of the
galaxies themselves: $\no=0.010\,\ho^3\mbox{ Mpc}^{-3}$ from
Eq.~(\ref{noValue}).

The wavelength of the neutrino-decay photons at emission (like those
from axion decay in Sec.~\ref{sec:SpecAx}) can be taken to be distributed
normally about the peak wavelength corresponding to $\Egam$:
\beq
\lamnu = \frac{hc}{\Egam} = 860 \pm 30 \mbox{ \AA} \; .
\eeq
This lies in the extreme ultraviolet (EUV) portion of the spectrum,
although the redshifted tail of the observed photon spectrum will
stretch across the far ultraviolet (FUV) and near ultraviolet (NUV) bands.
(Universal conventions regarding the boundaries between these wavebands have
yet to be established.  For definiteness, we take them as follows:
EUV=100--912~\AA, FUV=912--2000~\AA\ and NUV=2000--4000~\AA.)
The spectral energy distribution (SED) of the neutrino halos is
then given by Eq.~(\ref{gaussSED}):
\beq
F(\lambda) = \frac{L_h}{\sqrt{2\pi} \, \siglam}
             \exp \left[ -\frac{1}{2} \left( 
             \frac{\lambda-\lambda_{\gamma}}{\siglam}
             \right)^2 \right] \; ,
\label{nuSEDdefn}
\eeq
where $L_h$, the halo luminosity, has yet to be determined.
For the standard deviation $\siglam$ we can follow the same procedure
as with axions and use the velocity dispersion in the halo, giving
$\siglam=2\lamnu v_c/c$.  We parametrize this for convenience using
the range of uncertainty in the value of $\lamnu$,
so that $\sigthirty\equiv\siglam/$(30~\AA).

The halo luminosity is just the ratio of the number of decaying neutrinos
($N_{\tau}$) to their decay lifetime ($\taunu$), multiplied by the energy
of each decay photon ($\Egam$).  Because the latter is just above the
hydrogen-ionizing energy of 13.6~eV, we also need to multiply the result
by an {\em efficiency factor\/} ($\epsilon$) between zero and one,
to reflect the fact that some of the decay photons are absorbed by
neutral hydrogen in their host galaxy before they can leave the halo
and contribute to its luminosity.  Altogether, then:
\beq
L_h = \frac{\epsilon N_{\tau} E_{\gamma}}{\taunu}
    = \frac{\epsilon \, M_h c^2}{2 \taunu} \; .
\label{nuLhDefn}
\eeq
Here we have expressed $N_{\tau}$ as the number of neutrinos with
rest mass $\mtau=2\Egam/c^2$ per halo mass $M_h$.

To calculate the mass of the halo, let us follow reasoning similar
to that adopted for axion halos in Sec.~\ref{sec:AxHalo} and assume that
the ratio of baryonic to total mass in the halo is comparable to the
ratio of baryonic to total matter density in the Universe,
$(\Mtot-M_h)/\Mtot=\Mbar/\Mtot=\Obar/(\Obar+\Onu)$.
Here we have made also the economical assumption that there are no
{\em other\/} contributions to the matter density, apart from those of
baryons and massive neutrinos.  It the follows that
\beq
M_h = \Mtot \left( 1 + \frac{\Obar}{\Onu} \right)^{-1} \; .
\label{nuMhDefn}
\eeq
We take $\Mtot=(2\pm1)\times10^{12}\Msun$ following Eq.~(\ref{MhValue}).
For $\Obar$ we use
the value $(0.016\pm0.005)\ho^{-2}$ quoted in Sec.~\ref{sec:baryons}.
And to calculate $\Onu$ we put the neutrino rest mass $\mtau$
into Eq.~(\ref{OnuDefn}), giving
\beq
\Onu = (0.31 \pm 0.01) \, \ho^{-2} \; .
\label{nuOnuValue}
\eeq
Inserting these values of $\Mtot$, $\Obar$ and $\Onu$ into (\ref{nuMhDefn}),
we obtain
\beq
M_h = (0.95 \pm 0.01) \Mtot = (1.9 \pm 0.9) \times 10^{12} \Msun \; .
\label{nuMhValue}
\eeq
The uncertainty $\ho$ in Hubble's constant scales out of this result.
Eq.~(\ref{nuMhValue}) implies a baryonic mass
$\Mbar=\Mtot-M_h\approx1\times10^{11}\Msun$, in good agreement with the
observed sum of contributions from disk, bulge and halo stars plus the
matter making up the interstellar medium in our own Galaxy.

The neutrino density~(\ref{nuOnuValue}), when combined with that of baryons,
leads to a total present-day matter density of
\beq
\Omato = \Obar + \Onu = (0.32 \pm 0.01) \, \ho^{-2} \; .
\eeq
As pointed out by Sciama \cite{Sci93a}, massive neutrinos are thus consistent
with a critical-density Einstein-de~Sitter Universe ($\Omato=1$) if
\beq
\ho = 0.57 \pm 0.01 \; .
\label{hoSciama}
\eeq
This is just below the range of values which many workers now consider
observationally viable for Hubble's constant (Sec.~\ref{sec:baryons}).  But it
is a striking fact that the same neutrino rest mass which resolves
several unrelated astrophysical problems also implies a reasonable expansion
rate in the simplest cosmological model.  In the interests of testing
the decaying-neutrino hypothesis in a self-consistent way, we will follow
Sciama in adopting the narrow range of values~(\ref{hoSciama}) for
Sec.~\ref{ch7} only.  

\subsection{Halo luminosity} \label{sec:HaloAbs}

To evaluate the halo luminosity~(\ref{nuLhDefn}), it remains
to find the fraction $\epsilon$ of decay photons which escape from the
halo.  The problem is simplified by recognizing that the photo-ionization
cross-section and distribution of neutral hydrogen in the Galaxy are
such that effectively all of the decay photons striking the disk are
absorbed.  The probability of absorption for a single decay photon is
then proportional to the solid angle subtended by the Galactic disk,
as seen from the point where the photon is released.

We model the distribution of $\tau$~neutrinos (and their decay photons)
in the halo with a flattened ellipsoidal profile which has been advocated
in the context of the decaying-neutrino scenario by Salucci and Sciama
\cite{Sal90}.  This has
\beq
\rnu(r,z) = 4 \nsun \mtau {\mathcal N}_{\nu}(r,\theta) \; ,
\label{rnuDefn} 
\eeq
with
\beqa
{\mathcal N}_{\nu}(r,\theta) & \equiv & \left[ 1 +
   \sqrt{(r/\rsun)^2\sin^2\!\theta + (r/h)^2\cos^2\!\theta} \,
   \right]^{-2} \; . \nonumber
\eeqa
Here $r$ and $\theta$ are spherical coordinates,
$n_{\odot}=5 \times 10^7 \mbox{ cm}^{-3}$ is the local neutrino number
density, $\rsun=8$~kpc is the distance of the Sun from the Galactic center,
and $h=3$~kpc is the scale height of the halo.  Although this function
has essentially been constructed to account for the ionization structure
of the Milky Way, it agrees reasonably well with dark-matter halo
distributions which have derived on strictly dynamical grounds \cite{Gat95}.

Defining $x\equiv r/\rsun$, one can use~(\ref{rnuDefn}) to express
the mass $M_h$ of the halo in terms of a {\em halo radius\/} ($r_h$) as
\beq
M_h(r_h) = M_{\nu} \int_{\theta=0}^{\pi/2} \,
   \int_{x=0}^{x_{\mbox{\tiny max}}(r_h,\theta)}
   {\mathcal N}_{\nu}(x,\theta) \, x^2 \sin\theta \, dx \, d\theta \; ,
\label{nuMhDefn2}
\eeq
where
\beqa
\xmax(r_h,\theta) & = & (r_h/\rsun) / \sqrt{\sin^2\theta + (\rsun/h)^2
	\cos^2\theta} \nonumber \\
M_{\nu} & \equiv & 16\pi\,\nsun\mtau\rsun^3 = 9.8 \times 10^{11} \Msun
   \nonumber \; .
\eeqa
Outside $x>\xmax$, we assume that the halo density drops off exponentially
and can be ignored.  Using (\ref{nuMhDefn2}) it can be shown that halos
whose masses $M_h$ are given by (\ref{nuMhValue}) have scale radii
$r_h=(70\pm25)$~kpc.  This is consistent with evidence from the motion
of Galactic satellites \cite{Koc96a}.

\begin{figure}[t!]
\begin{center}
\includegraphics[width=120mm]{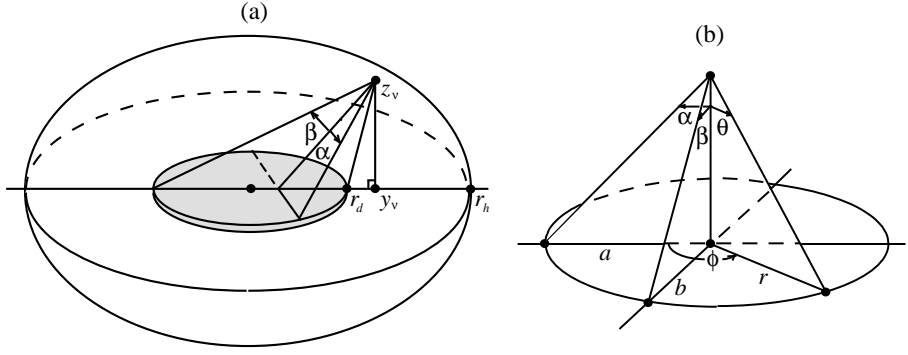}
\end{center}
\caption{Panel~(a): absorption of decay photons inside the halo.
   For neutrinos decaying at $(\ynu,\znu)$, the probability that decay
   photons will be absorbed inside the halo (radius $r_h$) is essentially
   the same as the probability that they will strike the Galactic disk
   (radius $r_d$, shaded).  Panel~(b): an ellipse of semi-major axis $a$,
   semi-minor axis $b$ and radial arm $r(\phi)$ subtends angles of
   $\alpha$, $\beta$ and $\theta(\phi)$ respectively.}
\label{fig7.2}
\end{figure}

We now put ourselves in the position of a decay photon released at
cylindrical coordinates $(\ynu,\znu)$ inside the halo, Fig.~\ref{fig7.2}(a).
It may be seen that the disk of the Galaxy presents an approximately
elliptical figure with maximum angular width $2\alpha$ and angular height
$2\beta$, where
\beqa
\alpha & = & \tan^{-1} \sqrt{ (r_d^2 - d^{\,2})/[(\ynu-d)^2+\znu^2] }
   \nonumber \\
\beta & = & \frac{1}{2} \left[ \tan^{-1} \left( \frac{\ynu+r_d}{\znu}
   \right) - \tan^{-1} \left( \frac{\ynu-r_d}{\znu} \right) \right]
   \label{AlphaBetaDefns} \; .
\eeqa
Here $r_d$ is the disk radius,
$d=[(\ynu^2+\znu^2+r_d^2)-\sqrt{(\ynu^2+\znu^2+r_d^2)^2-4\ynu^2r_d^2}
\,]/2\ynu$, $\ynu=r\sin\theta$ and $\znu=r\cos\theta$.
In spherical coordinates centered on the photon, the solid angle
subtended by an ellipse is
\beq
\Omega_e = \int_0^{2\pi} d\phi \int_0^{\theta(\phi)} \sin\theta^{\prime} \,
   d\theta^{\prime} = \int_0^{2\pi} \left[ 1 - \cos\theta(\phi) \right] \,
   d\phi \; ,
\eeq
where $\theta(\phi)$ is the angle subtended by a radial arm of the ellipse,
as shown in Fig.~\ref{fig7.2}(b).  The cosine of this angle is expressed
in terms of $\alpha$ and $\beta$ by
\beq
\cos\theta(\phi) = \left[ 1 + \frac{\tan^2\!\alpha \, \tan^2\!\beta}
    {\tan^2\!\alpha \, \sin^2\!\phi + \tan^2\!\beta \, \cos^2\!\phi}
   \right]^{-1/2} \; .
\eeq
The single-point probability that a photon released at $(x,\theta)$
will escape from the halo is then
${\mathcal P}_e = 1 - \Omega_e(\alpha,\beta)/4\pi$.
For a given halo size $r_h$, we obtain a good approximation to $\epsilon$
by averaging ${\mathcal P}_e$ over all locations $(x,\theta)$ in the halo
and weighting by the neutrino number density ${\mathcal N}_{\nu}$.
Choosing $r_d=36$~kpc as an effective disk radius \cite{OW03}, we obtain:
\beq
\epsilon = \left\{ \begin{array}{ll}
   0.63 & \qquad (r_h=45 \mbox{ kpc}) \\
   0.77 & \qquad (r_h=70 \mbox{ kpc}) \\
   0.84 & \qquad (r_h=95 \mbox{ kpc}) \; .
\end{array} \right.
\label{epsValues}
\eeq
As expected, the escape fraction of decay photons goes up as the scale
size of the halo increases relative to that of the disk.  As $r_h\gg r_d$
one gets $\epsilon\rightarrow1$, while a small halo with $r_h\lesssim r_d$
leads to $\epsilon\approx0.5$.

With the decay lifetime $\taunu$, halo mass $M_h$ and efficiency factor
$\epsilon$ all known, Eq.~(\ref{nuLhDefn}) gives for the luminosity of
the halo:
\beq
L_h = (6.5 \times 10^{42} \mbox{ erg s}^{-1}) f_h \ftau^{-1} \; .
\label{nuLhValue}
\eeq
Here we have introduced two dimensionless constants $f_h$ and $\ftau$
in order to parametrize the uncertainties in $\epsilon M_h$ and $\taunu$.
For the ranges of values given above, these take the values
$f_h=1.0\pm0.6$ and $\ftau=1.0\pm0.5$ respectively.  Setting
$f_h=\ftau=1$ gives a halo luminosity of about $2\times10^9\Lsun$,
or less than 5\% of the optical luminosity of the Milky Way
($L_0=2\times10^{10}\ho^{-2}\Lsun$), with $\ho$ as specified
by Eq.~(\ref{hoSciama}).

The combined bolometric intensity of decay photons from all the
bound neutrinos out to
a redshift $z_f$ is given by (\ref{QzDefn2}) as usual:
\beq
\Qbound = Q_h \int_0^{\, z_f} \frac{dz}{(1+z)^2 \, \Htil(z)} \; ,
\label{QboundDefn}
\eeq
where
\beqa
Q_h & \equiv & \frac{c\no L_h}{\Ho} = (2.0 \times 10^{-5}
   \mbox{ erg s}^{-1} \mbox{ cm}^{-2}) \, \ho^2 f_h \ftau^{-1} \; .
\nonumber
\eeqa
The $\ho$-dependence in this quantity comes from the fact that we have so
far considered only neutrinos in galaxy halos, whose number density $\no$
goes as $\ho^3$.  Since we follow Sciama in adopting the
Einstein-de~Sitter (EdS)
cosmology in this section, the Hubble expansion rate~(\ref{FL2}) is
\beq
\Htil(z) = (1+z)^{3/2} \; .
\label{HtilEdS}
\eeq
Putting this into (\ref{QboundDefn}), we find
\beq
\Qbound \approx \smallfrac{2}{5}Q_h = (8.2 \times 10^{-6}
   \mbox{ erg s}^{-1} \mbox{ cm}^{-2}) \, \ho^2 f_h \ftau^{-1} \; .
\eeq
(The approximation is good to better than 1\% if $z_f\geqslant8$.)
Here we have neglected absorption {\em between\/} the galaxies, an issue
we will return to below.  Despite their mass and size, dark-matter
halos in the decaying-neutrino hypothesis are not very bright.  Their
combined intensity is about 1\% of that of the EBL due to galaxies,
$\Qstar\approx3\times10^{-4} \mbox{ erg s}^{-1} \mbox{ cm}^{-2}$.
This is primarily due to the long decay lifetime of the neutrinos,
five orders of magnitude longer than the age of the galaxies.

\subsection{Free-streaming neutrinos}

The cosmological density of decaying $\tau$~neutrinos in dark-matter
halos is small: $\Onubound=\no M_h/\rho_{crit} =(0.068 \pm 0.032)\ho$.
With $\ho$ as given by (\ref{hoSciama}), this amounts to less
than 6\% of the total neutrino density, Eq.~(\ref{nuOnuValue}).
Therefore, as expected for hot dark matter particles, the bulk of the
EBL contributions in the decaying-neutrino scenario come from neutrinos
which are distributed on larger scales.  We will refer to these 
collectively as free-streaming neutrinos, though some of them may
actually be associated with more massive systems such as clusters of
galaxies.  (The distinction is not critical for our purposes, since
we are concerned with combined contributions to the diffuse background.)
Their cosmological density is found using~(\ref{nuOnuValue}) as
$\Onufree=\Onu-\Onubound=0.30 h_0^{-2} f_f$, where the dimensionless
constant $f_f=1.00 \pm 0.05$ parametrizes the uncertainties in this
quantity.

To identify sources of radiation in this section we follow the same
procedure as with vacuum energy (Sec.~\ref{sec:vacLum}) and divide
the Universe into regions of comoving volume $\Vo=\no^{-1}$.
The mass of each region is
\beq
M_f = \Onufree \, \rcrito \Vo = \Onufree\, \rcrito/\no \; .
\eeq
The luminosity of these sources has the same form as
Eq.~(\ref{nuLhDefn}) except that we put $M_h\rightarrow M_f$ and
drop the efficiency factor $\epsilon$ since the density of intergalactic
hydrogen is too low to absorb a significant fraction of the decay
photons within each region.  Thus,
\beq
L_f = \frac{\Onufree \, \rcrito \, c^2}{2\no \taunu} \; .
\label{nuLfValue}
\eeq
With the above values for $\Onufree$ and $\taunu$, and with $\rcrito$
and $\no$ given by (\ref{rcritoDefn}) and (\ref{noValue}) respectively,
Eq.~(\ref{nuLfValue}) implies a comoving luminosity density
due to free-streaming neutrinos of
\beq
\curlyLf = \no L_f = (1.2 \times 10^{-32} \mbox{ erg s}^{-1}
      \mbox{ cm}^{-3}) f_f \ftau^{-1} \; .
\eeq
This is $0.5\ho^{-1}$ times the luminosity density of the Universe,
as given by Eq.~(\ref{curlyLoValue}).  To calculate the bolometric intensity
of the background radiation due to free-streaming neutrinos, we replace
$L_h$ with $L_f$ in (\ref{QboundDefn}), giving
\beq
\Qfree = \frac{2c\no L_f}{5\Ho} = (1.2 \times 10^{-4}
   \mbox{ erg s}^{-1} \mbox{ cm}^{-2}) \, \ho^{-1} f_f \ftau^{-1} \; .
\eeq
This is of the same order of magnitude as $\Qstar$, and goes as
$\ho^{-1}$ rather than $\ho^2$.  Taking into account the uncertainties
in $\ho$, $f_h$, $f_f$ and $\ftau$, the bolometric intensity of bound
and free-streaming neutrinos together is
\beq
Q = \Qbound + \Qfree = (0.33 \pm 0.17) \, \Qstar \; .
\eeq
In principle, then, these particles are capable of shining as brightly
as the galaxies themselves, Eq.~(\ref{QQstarEdS}).  Most of this light
is due to free-streaming neutrinos, which are both more numerous than
their halo-bound counterparts and unaffected by absorption at source.

\subsection{Extinction by gas and dust}

To obtain more quantitative constraints, we would like to determine
neutrino contributions to the EBL as a function of wavelength.
This is accomplished as in previous sections by putting the source
luminosity ($L_h$ for the galaxy halos or $L_f$ for the free-streaming
neutrinos) into the SED~(\ref{nuSEDdefn}), and substituting the latter
into Eq.~(\ref{IzDefn}).  Now, however, we also wish to take into account
the fact that decay photons encounter significant amounts of absorbing
material as they travel through the {\em intergalactic medium\/}.
The wavelength of neutrino decay photons, $\lamnu=860\pm30$~\AA,
is just shortward of the Lyman-$\alpha$ line at 912~\AA, which means
that these photons are absorbed almost as strongly as they can be by
neutral hydrogen (this, of course, is one of the prime motivations of
the theory).  It is also very close to the waveband of peak extinction
by dust.  The simplest way to handle both these types of absorption is
to include an opacity term $\tau(\lamo,z)$ inside the argument of the
exponential, so that intensity reads
\beqa
\Ilam(\lamo) = \Inu \int_0^{\, z_f} (1+z)^{-9/2} \exp \bigg\{
   & - & \frac{1}{2} \left[ \frac{\lamo/(1+z)-\lamnu}{\siglam} \right]^2
      \nonumber \\
   & - & \tau(\lamo,z) \bigg\} \, dz \; .
\label{InuDefn}
\eeqa
Here we have used (\ref{HtilEdS}) for $\Htil(z)$.  The prefactor $\Inu$
is given with the help of (\ref{nuLhValue}) for bound neutrinos and
(\ref{nuLfValue}) for free-streaming ones as
\beqa
\Inu & = & \frac{c\no}{\sqrt{32\pi^3} \Ho \siglam} \times 
   \left\{ \begin{array}{l}
      L_h \\
      L_f \\
   \end{array} \right. \label{InuValues} \\
     & = &  \left\{ \begin{array}{lr}
      (940\mbox{ CUs}) \, \ho^2 \, f_h \, \ftau^{-1} \sigthirty^{-1}
         (\lamo/\lamnu) & \mbox{ (bound) } \\
      (5280\mbox{ CUs}) \, \ho^{-1} \, f_f \, \ftau^{-1} \sigthirty^{-1}
         (\lamo/\lamnu) & \mbox{ (free) }
   \end{array} \right. \; . \nonumber
\eeqa
The {\em optical depth\/} $\tau(\lambda_0,z)$ can be broken into separate
terms corresponding to hydrogen gas and dust along the line of sight:
\beq
\tau(\lamo,z) = \taugas(\lamo,z) + \taudust(\lamo,z) \; .
\label{OpDepthDefn}
\eeq
Our best information about both of these quantities comes from observations
of quasars at high redshifts.  The fact that these are visible at all already
places a limit on the degree of attenuation in the intergalactic medium.

We begin with the gas component.  Zuo and Phinney \cite{Zuo93} have
developed a formalism to describe the absorption due to randomly distributed
clouds such as quasar absorption-line systems and normalized this to the
number of Lyman-limit systems at $z=3$.  We use their model~1, which gives
the highest absorption below $\lamo\lesssim2000$~\AA, making it conservative
for our purposes.  Assuming an EdS cosmology, the optical depth at $\lamo$
due to neutral hydrogen out to a redshift $z$ is given by
\beq
\taugas(\lambda_0,z) = \left\{ \begin{array}{ll}
   \tauZP \left( \bigfrac{\lamo}{\lamL} \right)^{3/2} \!\!
      \ln (1+z) & \; \; (\lamo\leqslant\lamL) \\
   \tauZP \left( \bigfrac{\lamo}{\lamL} \right)^{3/2} \!\!
      \ln\left( \bigfrac{1+z}{\lamo/\lamL} \right)
      & \; \; [\lamL<\lamo<\lamL(1+z)] \\
   0 & \; \; [\lamo\geqslant\lamL(1+z)]
\end{array} \right.
\label{ZPopacity}
\eeq
where $\lamL=912$~\AA\ and $\tauZP=2.0$.

Dust is a more complicated and potentially more important issue,
and we pause to discuss this critically before proceeding.
The simplest possibility, and the one which should be most effective
in obscuring a diffuse signal like that considered here,
would be for dust to be spread uniformly through intergalactic space.
A quantitative estimate of opacity due to a uniform dusty intergalactic
medium has in fact been suggested \cite{Ost81}, but is regarded as an
extreme upper limit because it would lead to excessive reddening of quasar
spectra \cite{Wri81}.  Subsequent discussions have tended to treat
intergalactic dust as clumpy \cite{Ost84}, with significant debate about
the extent to which such clumps would redden and/or hide background
quasars, possibly helping to explain the observed ``turnoff'' in quasar
population at around $z \sim 3$ \cite{Wri86,Wri87,Hei88,Wri90}.
Most of these models assume an EdS cosmology.  The effects of dust
extinction could be enhanced if $\Omato<1$ and/or $\Olamo>0$ \cite{Hei88},
but we ignore this possibility here because neutrinos (not vacuum energy)
are assumed to make up the critical density in the decaying-neutrino scenario.

We will use a formalism due to Fall and Pei \cite{Fal93} in which dust
is associated with damped Ly$\alpha$~absorbers whose numbers and density
profiles are sufficient to obscure a portion of the light reaching us
from $z\sim3$, but not to account fully for the turnoff in quasar
population.  Obscuration is calculated based on the column density
of hydrogen in these systems, together with estimates of the dust-to-gas
ratio, and is normalized to the observed quasar luminosity function.
The resulting mean optical depth at $\lamo$ out to redshift $z$ is
\beq
\taudust(\lambda_0,z) = \int_0^{\, z} \,
   \frac{\tauFP (z^{\prime})
   (1+z^{\prime})}{(1+\Omato z^{\prime})^{1/2}} \;
   \xi \left( \frac{\lambda_0}{1+z^{\prime}} \right) \, dz^{\prime} .
\label{FPopacity}
\eeq
Here $\xi(\lambda)$ is the {\em extinction\/} of light by dust at
wavelength $\lambda$ relative to that in the B-band (4400~\AA).
If $\tauFP(z)=$~constant and $\xi(\lambda)\propto\lambda^{-1}$,
then $\taudust$ is proportional to $\lambda_0^{-1}[(1+z)^3-1]$ or
$\lambda_0^{-1}[(1+z)^{2.5}-1]$, depending on cosmology
\cite{Ost81,Ost84}.  In the more general treatment of Fall and Pei
\cite{Fal93}, $\tauFP(z)$ is parametrized as a function of redshift so that
\beq
\tauFP(z)=\tauFP(0) \, (1+z)^{\delta} ,
\eeq
where $\tauFP(0)$ and $\delta$ are adjustable parameters. 
Assuming an EdS cosmology ($\Omato=1$), the observational data
are consistent with lower limits of $\tau_{\ast}(0)=0.005$, $\delta =0.275$
(model~A); best-fit values of $\tau_{\ast}(0)=0.016$, $\delta =1.240$
(model~B); or upper limits of $\tau_{\ast}(0)=0.050$, $\delta =2.063$
(model~C).  We will use all three models in what follows.

The shape of the extinction curve $\xi(\lambda)$ in the 300--2000~\AA\
range can be computed using numerical Mie scattering routines in conjunction
with various dust populations.  Many people have constructed dust-grain
models that reproduce the average extinction curve for the
diffuse interstellar medium (DISM) at $\lambda>912$~\AA\ \cite{Mat90},
but there have been fewer studies at shorter wavelengths.  One such study
was carried out by Martin and Rouleau \cite{Mar91a}, who extended earlier
calculations of Draine and Lee \cite{Dra84} assuming: (1)~two populations
of homogeneous spherical dust grains composed of graphite and silicates
respectively; (2)~a power-law size distribution of the form $a^{-3.5}$
where $a$ is the grain radius; (3)~a range of grain radii from
50--2500~\AA; and (4)~solar abundances of carbon and silicon \cite{Mey79}.

\begin{figure}[t!]
\begin{center}
\includegraphics[width=100mm]{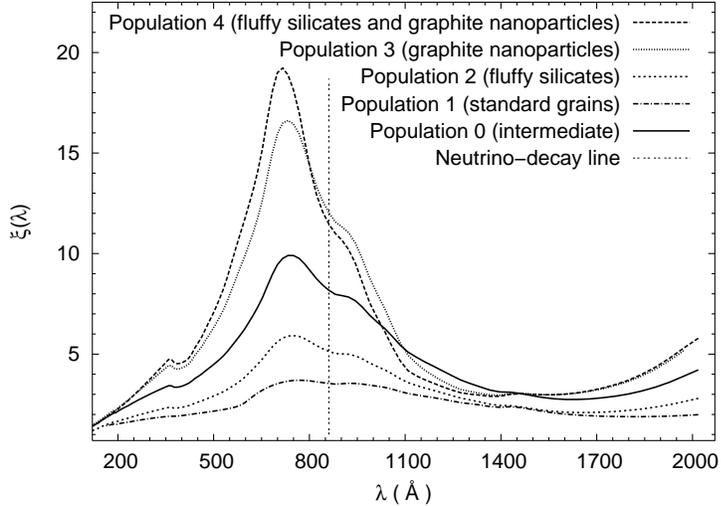}
\end{center}
\caption{The FUV extinction (relative to that in the B-band)
   produced by five different dust-grain populations.  Standard grains
   (population~1) produce the least extinction, while PAH-like carbon
   nanoparticles (population~3) produce the most extinction near the
   neutrino-decay line at 860~\AA\ (vertical line).}
\label{fig7.3}
\end{figure}

The last of these assumptions is questionable in light of recent work
suggesting that heavy elements are less abundant in the DISM than they
are in the Sun.  Snow and Witt \cite{Sno96} report interstellar
abundances of $214 \times 10^{-6}/$H and $18.6 \times 10^{-6}/$H
for carbon and silicon respectively (relative to hydrogen).  This reduces
earlier values by half and actually makes it difficult for a simple
silicate/graphite model to reproduce the observed DISM extinction curve.
We therefore use new dust-extinction curves based on the revised
abundances.  In the interests of obtaining conservative bounds on the
decaying-neutrino hypothesis, we also consider four different grain
populations, looking in particular for those that provide optimal
extinction efficiency in the FUV without drifting too far from the average
DISM curve in the optical and NUV bands.  We describe the general
characteristics of these models below and show the resulting extinction
curves in Fig.~\ref{fig7.3}; details can be found in \cite{Ove99b}.

Our population~1 grain model (Fig.~\ref{fig7.3}, dash-dotted line) assumes
the standard grain model employed by other workers, but uses the new,
lower abundance numbers together with dielectric functions due to
Draine \cite{Dra95}.  The shape of the extinction curve provides a
reasonable fit to observation at longer wavelengths (reproducing for
example the absorption bump at 2175~\AA); but its magnitude is too low,
confirming the inadequacies of the old dust model.  Extinction in the
vicinity of the neutrino-decay line at 860~\AA\ is also weak,
so that this model is able to ``hide'' very little of the light from 
decaying neutrinos.  Insofar as it almost certainly underestimates the
true extent of extinction by dust, this grain model provides a good
{\em lower limit\/} on absorption in the context of the decaying-neutrino
hypothesis.

The silicate component of our population~2 grain model (Fig.~\ref{fig7.3},
short-dashed line) is modified along the lines of the ``fluffy silicate''
model which has been suggested as a resolution of the heavy-element abundance
crisis in the DISM \cite{Mat96}.  We replace the standard silicates of
population~1 by silicate grains with a 45\% void fraction, assuming a
silicon abundance of $32.5 \times 10^{-6}$/H \cite{Ove99b}.  We also
decrease the size of the graphite grains ($a=50-250$~\AA) and reduce
the carbon depletion to 60\% to better match the DISM curve.  This
mixture provides a better match to the interstellar data at optical
wavelengths, and also shows significantly more FUV extinction than
population~1.

For population 3 (Fig.~\ref{fig7.3}, dotted line), we retain the standard
silicates of population~1 but modify the graphite component as an
approximation to the polycyclic aromatic hydrocarbon (PAH) nanostructures
which have been proposed as carriers of the 2175~\AA\ absorption bump
\cite{Dul98}.  PAH nanostructures consist of stacks of molecules such
as coronene (C$_{24}$H$_{12}$), circumcoronene (C$_{54}$H$_{18}$) and
larger species in various states of edge hydrogenation.  They have been
linked to the 3.4~$\mu$m absorption feature in the DISM \cite{Dul99} as
well as the extended red emission in nebular environments \cite{Sea99}. 
With sizes in the range $7-30$~\AA, these structures are much smaller
than the canonical graphite grains.  Their dielectric functions,
however, go over to that of graphite in the high-frequency limit
\cite{Dul98}.  So as an approximation to these particles, we use
spherical graphite grains with extremely small radii ($3-150$~\AA).
This greatly increases extinction near the neutrino-decay peak.

Our population~4 grain model (Fig.~\ref{fig7.3}, long-dashed line) 
combines both features of populations~2 and 3.  It has the same fluffy
silicate component as population~2, and the same graphite component as
population~3.  The results are not too different from those obtained with
population~3, because extinction in the FUV waveband is dominated by
small-particle contributions, so that silicates (whatever their void
fraction) are of secondary importance.  Neither the population~3 nor
the population~4 grains fit the average DISM curve as well as those of
population~2, because the Mie scattering formalism cannot accurately
reproduce the behaviour of nanoparticles near the 2175~\AA\ resonance.
However, the high levels of FUV extinction in these models --- especially
model~3 near 860~\AA\ --- suit them well for our purpose, which is to
set the most conservative possible limits on the decaying-neutrino
hypothesis.

\subsection{The ultraviolet background}

We are now ready to specify the total optical depth~(\ref{OpDepthDefn})
and hence to evaluate the intensity integral~(\ref{InuDefn}).  We will
use three combinations of the dust models just described,
with a view to establishing lower and upper bounds on the EBL intensity
predicted by the theory.  A {\em minimum-absorption\/} model is obtained
by combining Fall and Pei's model~A with the extinction curve of the
population~1 (standard) dust grains.  At the other end of the spectrum,
model~C of Fall and Pei together with the population~3 (nanoparticle)
grains provides the most conservative {\em maximum-absorption\/} model
(for $\lamo\gtrsim800$~\AA).  Finally, as an intermediate model, we
combine model~B of Fall and Pei with the extinction curve labelled
as population~0 in Fig.~\ref{fig7.3}.

\begin{figure}[t!]
\begin{center}
\includegraphics[width=100mm]{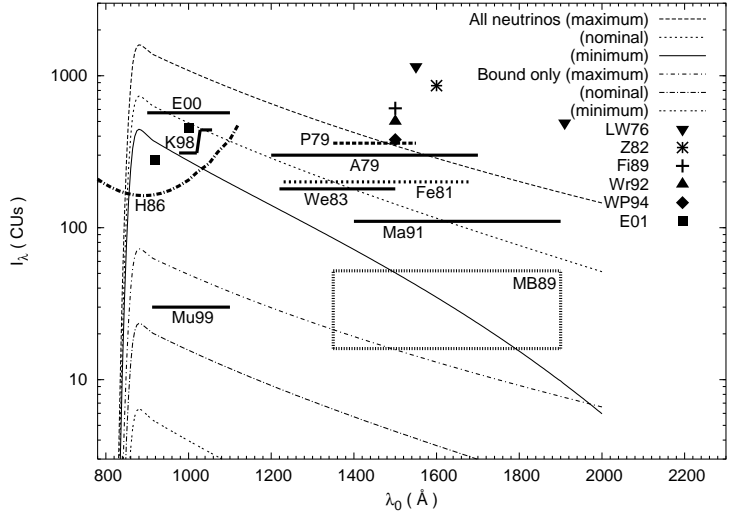}
\end{center}
\caption{The spectral intensity $\Ilam$ of background radiation from
   decaying neutrinos as a function of observed wavelength $\lamo$
   (light curves), plotted together with observational upper limits
   on EBL intensity in the far ultraviolet (points and heavy curves).
   The bottom four theoretical curves refer to bound neutrinos only,
   while the top four refer to bound and free-streaming neutrinos
   together.  The minimum predicted signals consistent with the theory
   are combined with the highest possible extinction in the intergalactic
   medium, and vice versa.}
\label{fig7.4}
\end{figure}

The resulting predictions for the spectral intensity of the FUV background
due to decaying neutrinos are plotted in Fig.~\ref{fig7.4} (light lines)
and compared with observational limits (heavy lines and points).
The curves in the bottom half of this figure refer to EBL contributions
from bound neutrinos only, while those in the top half correspond to
contributions from both bound and free-streaming neutrinos together. 

We begin our discussion with the bound neutrinos.  The key results
are the three widely-spaced curves in the lower half of the figure, with
peak intensities of about 6, 20 and 80~CUs at $\lamo\approx900$~\AA.
These are obtained by letting $\ho$ and $f_h$ take their minimum,
nominal and maximum values respectively in (\ref{InuDefn}), with the
reverse order applying to $\ftau$.  Simultaneously we have adopted the
maximum, intermediate and minimum-absorption models for intergalactic dust,
as described above.  Thus the highest-intensity model is paired with the
lowest possible dust extinction, and vice versa.  These curves should be
seen as extreme upper and lower bounds on the theoretical intensity of
EBL contributions from decaying neutrinos in galaxy halos.

They are best compared with an experimental measurement by Martin and
Bowyer in 1989 \cite{Mar89}, labelled ``MB89'' in Fig.~\ref{fig7.4}.
These authors used data from a rocket-borne imaging camera to search for
small-scale fluctuations in the FUV EBL, and deduced from this that the
combined light of external galaxies (and their associated halos) reaches the
Milky Way with an intensity of 16--52~CUs over 1350--1900~\AA.  There is
now some doubt as to whether this was really an extragalactic signal, and
indeed whether it is feasible to detect such a signal at all, given the
brightness and fluctuations of the Galactic foreground in this waveband
\cite{Sas95}.  Viable or not, however, it is of interest to see what
a detection of this order would mean for the decaying-neutrino hypothesis.
Fig.~\ref{fig7.4} shows that it would constrain the theory only weakly.
The expected signal in this waveband lies below 20~CUs in even the most
optimistic scenario where signal strength is highest and absorption is
weakest.  In the nominal ``best-fit'' scenario this drops to less than
7~CUs.  As noted already (Sec.~\ref{sec:HaloAbs}), the low intensity
of the background light from decaying neutrinos is due to their long
decay lifetime.  In order to place significant constraints on the theory,
one needs the stronger signal which comes from free-streaming, as well
as bound neutrinos.  This in turn requires limits on the intensity of
the total background rather than that associated with fluctuations.

The curves in the upper half of Fig.~\ref{fig7.4} (with peak intensities
of about 300, 700 and 2000~CUs at $\lamo\approx900$~\AA) represent
the combined EBL contributions from {\em all\/} decaying neutrinos.
We let $f_h$ and $f_f$ take their minimum, nominal and maximum values
respectively in (\ref{InuDefn}), with the reverse order applying to
$\ftau$ as well as $\ho$ (the latter change being due to the fact that
the dominant free-streaming contribution goes as $\ho^{-1}$ rather than
$\ho^2$).  Simultaneously we adopt the maximum, intermediate and
minimum-absorption models for intergalactic dust, as above.
Intensity is reduced very significantly in the maximum-absorption case
(solid line): by 11\% at 900~\AA, 53\% at 1400~\AA\ and 86\% at 1900~\AA.
The bulk of this reduction is due to dust, especially at longer
wavelengths where most of the light originates at high redshifts.
Comparable reduction factors in the intermediate-absorption case
(short-dashed line) are 9\% at 900~\AA, 28\% at 1400~\AA\ and 45\% at
1900~\AA.  In the minimum-absorption case (long-dashed line), most of the
extinction is due to gas rather than dust at shorter wavelengths, and
intensity is reduced by a total of 9\% at 900~\AA, 21\% at 1400~\AA\
and 31\% at 1900~\AA.

The most conservative constraints on the theory are obtained by comparing
the lowest predicted intensities (solid line) with observational upper
limits on total EBL intensity in the FUV band (Fig.~\ref{fig7.4}).
A word is in order about these limits, which can be usefully divided
into two groups: those above and below the Lyman~$\alpha$-line at 1216~\AA.
At the longest wavelengths we include two datapoints from Lillie and Witt's
analysis of {\sc Oao}-2 satellite data (\cite{Lil76}, labelled ``LW76''
in Fig.~\ref{fig7.4}); these were already encountered in Sec.~\ref{ch3}.
Close to them is an upper limit from the Russian Prognoz satellite by
Zvereva \etal\ (\cite{Zve82}; ``Z82'').
Considerably stronger broadband limits have come from rocket experiments
by Paresce \etal\ (\cite{Par79}; ``P79''), Anderson \etal\ (\cite{And79};
``A79'') and Feldman \etal\ (\cite{Fel81}; ``Fe81''), as well as
an analysis of data from the Solrad-11 spacecraft by Weller
(\cite{Wel83}; ``We83'').

A number of other studies have proceeded by establishing a correlation between
background intensity and the column density of neutral hydrogen inside the
Milky Way, and then extrapolating this out to zero column density to obtain
the presumed extragalactic component.  Martin \etal\ \cite{Mar91b}
applied this correlation method to data taken by the Berkeley
{\sc Uvx} experiment, setting an upper limit of 110~CUs on the
intensity of any unidentified EBL contributions over 1400--1900~\AA\
(``Ma91'').  The correlation method is subject to uncertainties
involving the true extent of scattering by dust, as well as absorption
by ionized and molecular hydrogen at high Galactic latitudes.
Henry \cite{Hen91} and Henry and Murthy \cite{Hen93} approach these
issues differently and raise the upper limit on background intensity to
400~CUs over 1216--3200~\AA.  A good indication of the complexity of the
problem is found at 1500~\AA, where Fix \etal\ \cite{Fix89} used data from
the {\sc De}-1 satellite to identify an isotropic background flux of
$530\pm80$~CUs (``Fi89''), the highest value reported so far.
The same data were subsequently reanalyzed by Wright \cite{Wri92}
who found a much lower best-fit value of 45~CUs, with a conservative upper
limit of 500~CUs (``Wr92'').  The former would rule out the decaying-neutrino
hypothesis, while the latter does not constrain it at all.  A third
treatment of the same data has led to an intermediate result of
$300\pm80$~CUs (\cite{Wit94}; ``WP94'').

Limits on the FUV background shortward of Ly$\alpha$ have been even
more controversial.  Several studies have been based on data from the
Voyager~2 ultraviolet spectrograph, beginning with that of Holberg
\cite{Hol86}, who obtained limits between 100 and 200~CUs over
500--1100~\AA\ (labelled ``H86'' in Fig.~\ref{fig7.4}).  An analysis
of the same data over 912--1100~\AA\ by Murthy \etal\ \cite{Mur91}
led to similar numbers.  In a subsequent reanalysis, however,
Murthy \etal\ \cite{Mur99} tightened this bound to 30~CUs over the
same waveband (``Mu99'').  The statistical validity of these results
has been debated vigorously \cite{Ede00,Mur01}, with a second group
asserting that the original data do not justify a limit smaller than
570~CUs (``E00'').  Of these Voyager-based limits, the strongest
(``Mu99'') is incompatible with the decaying-neutrino hypothesis,
while the weakest (``E00'') constrains it only mildly.
Two new experiments have yielded results midway between these extremes:
the {\sc Duve} orbital spectrometer \cite{Kor98} and the {\sc Eurd}
spectrograph aboard the Spanish {\sc Minisat}~01 \cite{Ede01}.
Upper limits on continuum emission from the former instrument are
310~CUs over 980--1020~\AA\ and 440~CUs over 1030--1060~\AA\ (``K98''),
while the latter has produced upper bounds of 280~CUs at 920~\AA\ and
450~CUs at 1000~\AA\ (``E01'').

What do these observational data imply for the decaying-neutrino
hypothesis?  Longward of Ly$\alpha$, Fig.~\ref{fig7.4} shows that they
span very nearly the same parameter space as the minimum and
maximum-intensity predictions of the theory (solid and long-dashed lines).
Most stringent are Weller's Solrad-11 result (``We83'') and the
correlation-method constraint of Martin \etal\ (``Ma91'').  Taken on their
own, these data constrain the decaying-neutrino hypothesis rather severely,
but do not rule it out.  Absorption (by dust in particular) plays a
critical role in reducing the strength of the signal.

Shortward of Ly$\alpha$, most of the signal originates nearby and 
intergalactic absorption is far less important.  Ambiguity here 
comes rather from the spread in reported limits, which in turn reflects
the formidable experimental challenges in this part of the spectrum.
Nevertheless it is clear that both the Voyager-based limits of
Holberg (``H86'') and Murthy \etal\ (``Mu99''), as well as the new
{\sc Eurd} measurement at 920~\AA\ (``E01'') are incompatible with
the theory.  These upper bounds are violated by even the weakest
predicted signal, which assumes the strongest possible extinction
(solid line).  The easiest way to reconcile theory with observation
is to increase the neutrino decay lifetime.  If we require that
$\Ith<\Iobs$, then the abovementioned {\sc Eurd} measurement
(``E01'') implies a lower bound of $\taunu>3\times10^{23}$~s. 
This rises to $(5\pm3)\times10^{23}$~s and
$(26\pm10)\times10^{23}$~s for the Voyager limits (``H86'' and
``Mu99'' respectively).  All these numbers lie outside the range of
lifetimes required in the decaying-neutrino scenario,
$\taunu=(2\pm1)\times10^{23}$~s.  The {\sc Duve} constraint (``K98'')
is more forgiving but still pushes the theory to the edge of its available
parameter space.  {\em Taken together, these data may safely be said
to exclude the decaying-neutrino hypothesis\/}.  This conclusion is
in accord with current thinking on the value of Hubble's constant
(Sec.~\ref{sec:baryons}) and structure formation (Sec.~\ref{sec:neutrinos}),
as well as more detailed analysis of the {\sc Eurd} data \cite{Bow99}.

These limits would be weakened (by a factor of up to nearly one-third)
if the value of Hubble's constant $\ho$ were allowed to exceed $0.57\pm0.01$,
since the dominant free-streaming contributions to $\Ilam(\lamo)$ go
as $\ho^{-1}$.  A higher expansion rate would however exacerbate
problems with structure formation and the age of the Universe,
the more so because the dark matter in this theory is hot.
It would also mean sacrificing the critical density of neutrinos.
Another possibility would be to consider lower neutrino rest masses,
a scenario that does not conflict with other observational data until
$m_{\nu} c^2 \lesssim 2$~eV \cite{Raf98}.
This would however entail a proportional reduction in decay photon energy,
which would have to drop below the Lyman or hydrogen-ionizing limit,
thus removing the whole motivation for the proposed neutrinos
in the first place.  Similar considerations apply to neutrinos with
longer decay lifetimes.

Our conclusions, then, are as follows.  Neutrinos with rest masses and
decay lifetimes as specified by the decaying-neutrino scenario produce
levels of ultraviolet background radiation very close to, and in some
cases above experimental upper limits on the intensity of the EBL.
At wavelengths longer than 1200~\AA, where intergalactic absorption
is most effective, the theory is marginally compatible with
observation --- {\em if\/} one adopts the upper limits on dust density
consistent with quasar obscuration, and {\em if\/} the dust grains are
extremely small.  At wavelengths in the range 900--1200~\AA, predicted
intensities are either comparable to or higher than those actually seen.
Thus, while there is now good experimental evidence that some of the
dark matter is provided by massive neutrinos, the light of the night sky
tells us that these particles cannot have the rest masses and decay
lifetimes attributed to them in the decaying-neutrino hypothesis.

\section{Weakly interacting massive particles} \label{ch8}

\subsection{The lightest supersymmetric particle}

Weakly interacting massive particles (WIMPs) are as-yet undiscovered
particles whose rest masses far exceed those of baryons, but whose
interaction strengths are comparable to those of neutrinos.
The most widely-discussed examples arise in the context of supersymmetry
(SUSY), which is motivated independently of the dark-matter problem
as a theoretical framework for many attempts to unify the forces of nature.
SUSY predicts that, for every known fermion in the standard model,
there exists a new bosonic ``superpartner'' and vice versa (more than
doubling the number of fundamental degrees of freedom in the simplest
models; see \cite{Jun96} for a review).  These superpartners were
recognized as potential dark-matter candidates in the early 1980s
by Cabibbo \etal\ \cite{Cab81}, Pagels and Primack \cite{Pag82},
Weinberg \cite{Wei82} and others \cite{Ell82,Gol83,Iba84,Sci84,Ell84a},
with the generic term ``WIMP'' being coined in 1985 \cite{Ste85}.

There is, as yet, no firm experimental evidence for SUSY WIMPs.
This means that their rest energies, if they exist, lie beyond the
range currently probed by accelerators (and in particular beyond the
rest energies of their standard-model counterparts).  Supersymmetry is,
therefore, not an exact symmetry of nature.  The masses of the
superpartners, like that of the axion (Sec.~\ref{ch6}), must
have been generated by a symmetry-breaking event in the early Universe.
Subsequently, as the temperature of the expanding fireball dropped
below their rest energies, heavier species would have dropped out of
equilibrium and begun to disappear by pair annihilation, leaving
progressively lighter ones behind.  Eventually, only one massive
superpartner would have remained: the {\em lightest supersymmetric
particle\/} (LSP).  It is this particle which plays the role of the
WIMP in SUSY theories.  Calculations using the Boltzmann equation show
that the collective density of relic LSPs today lies within one or two
orders of magnitude of the required CDM density across much of the
parameter space of most SUSY theories \cite{Ell01b}.  In this respect,
SUSY WIMPs are more natural DM candidates than axions (Sec.~\ref{ch6}),
whose cosmological density ranges {\em a priori\/} over many orders of
magnitude.

SUSY WIMPs contribute to the cosmic background radiation in at least
three ways.  The first is by pair annihilation to photons.  This process
occurs even in the simplest, or minimal SUSY model (MSSM), but is very
slow because it takes place via intermediate loops of charged particles
such as leptons and quarks and their antiparticles.
The underlying reason for the stability of the LSP in the MSSM is an
additional new symmetry of nature, known as R-parity, which is necessary
(among other things) to protect the proton from decaying via
intermediate SUSY states.  The other two types of background contributions
occur in {\em non\/}-minimal SUSY theories, in which R-parity is not
conserved (and in which the proton can decay).  In these theories,
LSPs can decay into photons directly via loop diagrams, and also
indirectly via tree-level decays to secondary particles which then
scatter off pre-existing background photons to produce a signal.

The first step in assessing the importance of each of these processes
is to choose an LSP.  Early workers variously identified this as the
photino ($\tilde{\gamma}$) \cite{Cab81},
the gravitino ($\tilde{g}$) \cite{Pag82},
the sneutrino ($\tilde{\nu}$) \cite{Iba84}
or the selectron ($\tilde{e}$) \cite{Sci84}.
(SUSY superpartners are denoted by a tilde and take the same names
as their standard-model counterparts, with a prefix ``s'' for
superpartners of fermions and a suffix ``ino'' for those of bosons.)
In a landmark study, Ellis \etal\ \cite{Ell84a} showed in 1984 that most
of these possibilities are disfavoured, and that the LSP is in fact most
likely to be a {\em neutralino\/} ($\tilde{\chi}$), a linear superposition
of the photino ($\tilde{\gamma}$), the zino ($\tilde{Z}$) and two neutral
higgsinos ($\tilde{h}^0_1$ and $\tilde{h}^0_2$).  (These are the SUSY
spin-$\frac{1}{2}$ counterparts of the photon, $Z^0$ and Higgs bosons
respectively.)  There are four neutralinos, each a mass eigenstate made
up of (in general) different amounts of photino, zino, etc., although
in special cases a neutralino could be ``mostly photino,'' say,
or ``pure zino.''  The LSP is by definition the lightest such
eigenstate.  Accelerator searches place a lower limit on its
rest energy which currently stands at $\mchi>46$~GeV \cite{Ell00}.

In minimal SUSY, the density of neutralinos drops only by way of the
(slow) pair-annihilation process, and it is quite possible for these
particles to ``overclose'' the Universe if their rest energy is too
high.  This does not change the geometry of the Universe, but rather
speeds up its expansion rate, which is proportional to the square root
of the total matter density from Eq.~(\ref{FL1}).  In such a situation,
the Universe would have reached its present size in too short a time.
Lower bounds on the age of the Universe thus impose an upper bound on
the neutralino rest energy which has been set at $\mchi\lesssim3200$~GeV
\cite{Gri90}.  Detailed exploration of the parameter space of minimal
SUSY theory tightens this upper limit in most cases to
$\mchi\lesssim600$~GeV \cite{Ell98}.  Much recent work is focused on
a slimmed-down version of the MSSM known as the {\em constrained minimal
SUSY model\/} (CMSSM), in which all existing experimental bounds and
cosmological requirements are comfortably met by neutralinos with
rest energies in the range 90~GeV$\lesssim\mchi\lesssim400$~GeV
\cite{Ros01}.

Even in its constrained minimal version, SUSY physics contains at least
five adjustable input parameters, making the neutralino a considerably
harder proposition to test than the axion or the massive neutrino.
Fortunately, there are several other ways (besides accelerator
searches) to look for these particles.  Because their rest energies
are above the temperature at which they decoupled from the primordial
fireball, WIMPs have non-relativistic velocities and are found
predominantly in gravitational potential wells like those of our
own Galaxy.  They will occasionally scatter against target nuclei in
terrestrial detectors as the Earth follows the Sun around the Milky Way.
Annual variations in this signal resulting from the Earth's orbital motion
through the Galactic dark-matter halo can be used to isolate a WIMP signal.
Just such a signal was reported by the {\sc Dama} team in 2000 using
detectors in Italy's Gran Sasso mountains, with an implied WIMP rest
energy of $\mchi=52^{+10}_{-8}$~GeV \cite{Ber00}.  However, subsequent
experiments under the Fr\'ejus peak in France ({\sc Edelweiss}
\cite{Ben01}) and at Stanford University \cite{Ake03}
and the Soudan mine in Minnesota \cite{Ake04} ({\sc Cdms}) have not
been able to reproduce this result.  New detectors such as
{\sc Zeplin} in England's Boulby mine \cite{Bar03} and
{\sc Igex} at Canfranc in Spain \cite{Ira03} are rapidly 
coming online to help with the search.

A second, indirect search strategy is to look for annihilation
byproducts from neutralinos which have collected inside massive bodies.
Most attention has been directed at the possibility of detecting
antiprotons from the Galactic halo \cite{Sil84} or neutrinos from the
Sun \cite{Sil85} or Earth \cite{Fre86}.  The heat generated in the
cores of gas giants like Jupiter or Uranus has also been considered as
a potential annihilation signature \cite{Kra86}.  The main challenge in
each case lies in separating the signal from the background noise.
In the case of the Earth, one can look for neutrino-induced muons
which are distinguishable from the atmospheric background by the fact
that they are travelling straight up.  The {\sc Amanda} experiment,
whose detectors are buried deep in the Antarctic ice, has recently 
reported upper limits on the density of terrestrial WIMPs based on
this principle \cite{Ahr02}.

\subsection{Pair annihilation} \label{sec:Annihilations}

Pair annihilation into {\em photons\/} provides a complementary indirect
search technique.  The photons so produced lie in the $\gamma$-ray portion
of the spectrum for the range of WIMP rest energies considered here
(50~GeV$\lesssim\mchi\lesssim1000$~GeV).  Beginning with Sciama
\cite{Sci84}, Silk and Srednicki \cite{Sil84}, many workers
have studied the possibility of $\gamma$-rays from SUSY WIMP
annihilations in the halo of the Milky Way, which gives the strongest
signal.  Prognoses for detection have ranged from very optimistic
\cite{Sil87} to very pessimistic \cite{Ste88}; converging gradually to
the conclusion that neutralino-annihilation contributions would be at or
somewhat below the level of the Galactic background, and possibly
distinguishable from it by their spectral shape \cite{Rud89,Ste89,Bou89}.
Recent studies have focused on possible enhancements of the signal in
the presence of a high-density Galactic core \cite{Ber89}, a flattened
halo \cite{Fre89}, a very extended singular halo \cite{Ber94}, a massive
central black hole \cite{Gon99}, significant substructure
\cite{Cal00,Ber03a,Kou04} and adiabatic compression due to baryons
\cite{Pra04}.  The current state of the art in this area is summarized
in Ref.~\cite{Eva04} with attention to prospects for detection by the
upcoming {\sc Glast} mission.

WIMPs at the higher end of the mass range ($\sim1$~TeV) would produce
a weaker signal, but it has been argued that this might be more than made
up for by the larger effective area of the atmospheric \v{C}erenkov
telescopes (ACTs) used to detect them \cite{Urb92}.  Not all authors are
as sanguine \cite{Cha95}, but new observations of high-energy
$\gamma$-rays from the Galactic center by the {\sc Cangaroo} \cite{Tsu04}
and {\sc Veritas} collaborations \cite{Kos04} provide tantalizing 
examples of what might be possible with this technique.  The Milagro
extensive air-shower array is another experiment that has recently set
upper limits on the density of $\sim$TeV WIMPs in the vicinity of the
Sun \cite{Atk04}.  Other teams have carried the search farther afield,
toward objects like dwarf spheroidal galaxies \cite{Lak90}, the
Large Magellanic Cloud \cite{Gon94} and the giant elliptical galaxy
M87 in Virgo \cite{Bal99}.

The possibility of neutralino-annihilation contributions to the
diffuse extragalactic background, rather than the signal from
localized concentrations of dark matter, has received less attention.
First to apply the problem to SUSY WIMPs were Cabibbo \etal\ \cite{Cab81},
who however assumed a WIMP rest energy (10--30~eV) which we now know is
far too low.  Like the decaying neutrino (Sec.~\ref{ch7}), this would
produce a background in the ultraviolet.  It is excluded, however,
by an argument due to Lee and Weinberg, which restricts WIMPs to rest
energies above 2~GeV \cite{Gol83}.  EBL contributions from SUSY WIMPs
in this range were first estimated by Silk and Srednicki \cite{Sil84}.
Their conclusion, and those of most workers who have followed them
\cite{Cli90,Gao91,Ove97b}, is that neutralino annihilations would be
responsible for no more than a small fraction of the observed
$\gamma$-ray background.  Here we review this argument, reversing our
usual procedure and attempting to set a reasonably conservative
{\em upper\/} limit on neutralino contributions to the EBL.

\begin{figure}[t!]
\begin{center}
\includegraphics[width=80mm]{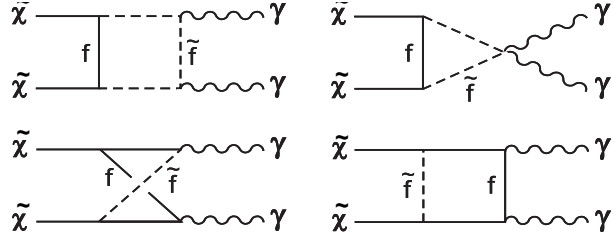}
\end{center}
\caption{Some Feynman diagrams corresponding to the annihilation of two
   neutralinos ($\tilchi$), producing a pair of photons ($\gamma$).
   The process is mediated by fermions ($f$) and their supersymmetric
   counterparts, the sfermions ($\tilde{f}$).}
\label{fig8.1}
\end{figure}

We concentrate on processes in which neutralino pairs annihilate
{\em directly\/} into photon pairs via intermediate loop diagrams
(Fig.~\ref{fig8.1}), since these provide the most distinctive signature
of new physics.  Neutralino annihilations actually produce most of their
photons indirectly, via tree-level annihilations to hadrons (mostly pions)
which then decay to photons, electrons, positrons and neutrinos.  (The 
electrons and positrons add even more to the signal by inverse Compton
scattering off low-energy CMB photons.)  However, the
energies of the photons produced in this way are broadly distributed,
resulting in a continuum $\gamma$-ray spectrum which is difficult if not
impossible to distinguish from the astrophysical background \cite{Ber98}.
By contrast, the one-loop annihilation processes in Fig.~\ref{fig8.1}
give rise to a photon spectrum that is essentially {\em monoenergetic\/},
$\Egam\approx\mchi$ (subject only to modest Doppler broadening due
to galactic rotation).  No conventional astrophysical processes
produce such a narrow peak, whose detection against the diffuse
extragalactic background would constitute compelling evidence
for dark matter.

We again take galactic dark-matter halos as our sources of background
radiation, with comoving number density $\no$.  Photon wavelengths are
distributed normally about the peak wavelength in the galaxy rest frame:
\beq
\lamann=hc/\mchi=(1.2 \times 10^{-6} \mbox{ \AA}) \, \mten^{-1} \; ,
\eeq
where $\mten\equiv\mchi/$(10~GeV) is the neutralino rest energy in units
of 10~GeV.  The standard deviation $\sigma_{\gamma}$ can be related to
the velocity dispersion of bound dark-matter particles as in previous
sections, so that $\siglam=2(v_c/c)\lamann$.  With
$v_c\!\sim\!220$~km~s$^{-1}$ and $\mten\!\sim\!1$ this is of order
$\sim\!10^{-9}$~\AA.  For convenience we specify this with the
dimensionless parameter $\signine\equiv\siglam/(10^{-9}$~\AA).
The spectral energy distribution is then given by Eq.~(\ref{gaussSED}) as
\beq
F(\lambda) = \frac{\Lann}{\sqrt{2\pi} \, \siglam}
             \exp \left[ -\frac{1}{2} \left(
             \frac{\lambda-\lamann}{\siglam}
             \right)^2 \right] \; .
\label{chiSEDdefn}
\eeq
The luminosity due to neutralino annihilations is proportional to the
rest energy of the annihilating particles times the annihilation rate,
which in turn goes as the cross-section $(\sigma v)$ times the square
of the neutralino number density, $n_{\tilde{\chi}}^2$.
The resulting expression may be written
\beq
\Lann = 2 m_{\tilde{\chi}} c^2 <\!\sigma v\!>_{\gamma\gamma} {\mathcal P} \; ,
\label{annLumDefn}
\eeq
where $<\!\sigma v\!>_{\gamma\gamma}$ is the photo-annihilation
cross-section (in cm$^3$~s$^{-1}$),
${\mathcal P} \equiv m_{\tilde{\chi}}^{-2} \int \rho_{\tilde{\chi}}^{\,2}(r)
4\pi r^2 dr$ is the radial average of $n_{\tilde{\chi}}^2$ over the halo
(in cm$^{-6}$) and $\rho_{\tilde{\chi}}(r)$ is the neutralino density
distribution (in g~cm$^{-3}$).  Berezinsky \etal\ \cite{Ber92a} have
determined $<\!\sigma v\!>_{\gamma\gamma} \approx a_{\gamma\gamma}$
for non-relativistic neutralinos as
\beq
a_{\gamma\gamma} = \frac{\hbar^2 c^3 \alpha^4 (m_{\tilde{\chi}}c^2)^2}
                        {3^6\pi (m_{\tilde{f}}c^2)^4} 
                   \left( \frac{Z_{11}}{\sin\theta_W} \right)^4
                   (45+48y+139y^2)^2 \; .
\eeq
Here $\alpha$ is the fine structure constant, $m_{\tilde{f}}$ is the
mass of an intermediate sfermion, $y \equiv (Z_{12}/Z_{11})\tan\theta_W$,
$\theta_W$ is the weak mixing angle and $Z_{ij}$ are elements of the 
real orthogonal matrix which diagonalizes the neutralino mass matrix.
In particular, the ``pure photino'' case is specified by
$Z_{11}=\sin\theta_W, y=1$ and the ``pure zino'' by
$Z_{11}=\cos\theta_W, y=-\tan^2\theta_W$.  Collecting these expressions
together and parametrizing the sfermion rest energy by
$\tilmten \equiv m_{\tilde{f}}/10$~GeV, we obtain:
\beq
<\!\sigma v\!>_{\gamma\gamma} = (8 \times 10^{-27} \mbox{ cm}^3 
                                \mbox{s}^{-1}) f_{\chi} \mten^2 
                                \, \tilmten^{-4} \; .
\label{CrossSection}
\eeq
Here $f_{\chi}$ (=1 for photinos, 0.4 for zinos) is a dimensionless
quantity whose value parametrizes the makeup of the neutralino.

Since we attempt in this section to set an upper limit on EBL
contributions from neutralino annihilations, we take $f_{\chi}\approx1$
(the photino case).  In the same spirit, we would like to use lower
limits for the sfermion mass $\tilmten$.  It is important to estimate
this quantity accurately since the cross-section goes as $\tilmten^{-4}$.
Giudice and Griest \cite{Giu89} have made a detailed study of
photino annihilations and find a lower limit on $\tilmten$
as a function of $\mten$, assuming that photinos provide at least
$0.025h_0^{-2}$ of the critical density.  Over the range
$0.1\leqslant\mten\leqslant4$, this lower limit is empirically
well fit by a function of the form $\tilmten\approx4\mten^{\,0.3}$.
If this holds over our broader range of masses, then we obtain an
upper limit on the neutralino annihilation cross-section of
$<\!\sigma v\!>_{\gamma\gamma} \; \lesssim(3\times10^{-29}\mbox{ cm}^3 
\mbox{s}^{-1})\,\mten^{0.8}$.  This expression gives results which
are about an order of magnitude higher than the cross-sections quoted by 
Gao \etal\ \cite{Gao91}.

For the WIMP density distribution $\rho_{\tilde{\chi}}(r)$ we adopt the
simple and widely-used {\em isothermal model\/} \cite{Bou89}:
\beq
\rho_{\tilde{\chi}}(r) = \rho_{\odot} \left( \frac{a^2+r_{\odot}^2}{a^2+r^2}
   \right) \; .
\label{rhochiDefn}
\eeq
Here $\rho_{\odot}=5\times10^{-25} \mbox{ g cm}^{-3}$ is the approximate
dark-matter density in the solar vicinity, assuming a spherical halo
\cite{Gat95}, $r_{\odot}=8$~kpc is the distance of the Sun from the Galactic
center \cite{Met98} and $a=(2-20)$~kpc is a core radius. 
To fix this latter parameter, we can integrate (\ref{rhochiDefn})
over volume to obtain total halo mass $M_h(r)$ inside radius $r$:
\beq
M_h(r)=4\pi\rho_{\odot} r(a^2+r_{\odot}^2) \left[ 1 - \left( \frac{a}{r} \right)
   \tan^{-1} \left( \frac{r}{a} \right) \right] \; .
\label{chiMhDefn}
\eeq
Observations of the motions of Galactic satellites imply that the total
mass inside 50~kpc is about $5\times10^{11}M_{\odot}$ \cite{Koc96a}.
This in (\ref{chiMhDefn}) implies $a=9$~kpc, which we consequently adopt.
The maximum extent of the halo is not well-constrained observationally,
but can be specified if we take $M_h=(2 \pm 1) \times 10^{12} \Msun$
as in (\ref{MhValue}).  Eq.~(\ref{chiMhDefn}) then gives a
halo radius $r_h=(170\pm80)$~kpc.  The cosmological density
of WIMPs in galactic dark-matter halos adds up to
$\Omega_h=\no M_h/\rcrito=(0.07\pm0.04)\,\ho$.

If there are no other sources of CDM, then the total matter density is 
$\Omato=\Omega_h+\Obar\approx0.1\ho$ and the observed flatness of the
Universe (Sec.~\ref{ch4}) implies a strongly vacuum-dominated cosmology.
While we use this as a lower limit on WIMP contributions to the dark
matter in subsequent sections, it is quite possible that CDM also exists
in larger-scale regions such as galaxy clusters.  To take this into account
in a general way, we define a {\em cosmological enhancement factor\/}
$f_c\equiv(\Omato-\Obar)/\Omega_h$ representing the added contributions
from WIMPs outside galactic halos (or perhaps in halos which extend
far enough to fill the space between galaxies).
This takes the value $f_c=1$ for the most conservative case just described,
but rises to $f_c=(4\pm2)\,\ho^{-1}$ in the \LCDM\ model with
$\Omato=0.3$, and $(14\pm7)\,\ho^{-1}$ in the EdS model with $\Omato=1$.

With $\rho_{\tilde{\chi}}(r)$ known, we are in a position to calculate
the quantity ${\mathcal P}$:
\beq
{\mathcal P} = \frac{2\pi\rho_{\odot}^2(a^2+r_{\odot}^2)^2}
                    {m_{\tilde{\chi}}^2 a} \left[ \tan^{-1} \left( 
               \frac{r_h}{a} \right) - \frac{(r_h/a)}{1+(r_h/a)^2} \right] \; .
\eeq
Using the values for $\rho_{\odot}$, $r_{\odot}$ and $a$ specified above and
setting $r_h=250$~kpc to get an upper limit, we find that
${\mathcal P}\leqslant(5\times10^{65}\mbox{ cm}^{-3})\,\mten^{-2}$.
Putting this result along with the cross-section~(\ref{CrossSection})
into (\ref{annLumDefn}), we obtain:
\beq
\Lann \leqslant (1 \times 10^{38} \mbox{ erg s}^{-1}) f_{\chi} 
        m_{10} \, \tilmten^{-4} \; .
\label{annLhValue}
\eeq
Inserting Giudice and Griest's \cite{Giu89} lower limit on the
sfermion mass $\tilmten$ (as empirically fit above), we find that
(\ref{annLhValue}) gives an upper limit on halo luminosity of
$\Lann\leqslant(5\times10^{35}\mbox{ erg s}^{-1}) f_{\chi} m_{10}^{-0.2}$.
Higher estimates can be found in the literature \cite{Ber92b},
but these assume a singular halo whose density drops off as only
$\rho_{\tilde{\chi}}(r)\propto r^{-1.8}$ and extends out to a very large
halo radius, $r_h=4.2h_0^{-1}$~Mpc.  For a standard isothermal
distribution of the form~(\ref{rhochiDefn}), our results confirm that
halo luminosity due to neutralino annihilations alone is very low,
amounting to less than $10^{-8}$ times the total bolometric
luminosity of the Milky Way.

The combined bolometric intensity of neutralino annihilations between
redshift $z_f$ and the present is given by substituting the comoving
number density $\no$ and luminosity $\Lann$ into Eq.~(\ref{QzDefn})
to give
\beq
Q = \Qann \int_0^{\, z_f} \frac{dz}{(1+z)^2 \,
   \left[ \Omato (1+z)^3 + (1-\Omato)\right]^{1/2}} \; ,
\label{QannDefn}
\eeq
where $\Qann=(c \no \Lann f_c)/\Ho$ and we have assumed spatial flatness.
With values for all these parameters as specified above, we find
\beq
Q = \left\{ \begin{array}{ll}
   (1\times10^{-12} \mbox{ erg s}^{-1} \mbox{ cm}^{-2}) \,
           \ho^2 \, f_{\chi} \, \mten^{-0.2} & (\mbox{if }\Omato=0.1\ho) \\
   (3\times10^{-12} \mbox{ erg s}^{-1} \mbox{ cm}^{-2}) \,
           \ho \, f_{\chi} \, \mten^{-0.2} & (\mbox{if }\Omato=0.3) \\
   (1\times10^{-11} \mbox{ erg s}^{-1} \mbox{ cm}^{-2}) \,
           \ho \, f_{\chi} \, \mten^{-0.2} & (\mbox{if }\Omato=1)
\end{array} \right. \; .
\label{QannValues}
\eeq
Here we have set $z_f=30$ (larger values do not substantially increase
the value of $Q$) and used values of $f_c=1$, $4\ho^{-1}$ and $20\ho^{-1}$
respectively.  The effects of a larger cosmological enhancement factor $f_c$
are partially offset in (\ref{QannDefn}) by the fact that a universe with
higher matter density $\Omato$ is younger, and hence contains less
background light in general.  Even the highest value of $Q$ given
in (\ref{QannValues}) is negligible in comparison to the
intensity~(\ref{QstarValue}) of the EBL due to ordinary galaxies.

\begin{figure}[t!]
\begin{center}
\includegraphics[width=100mm]{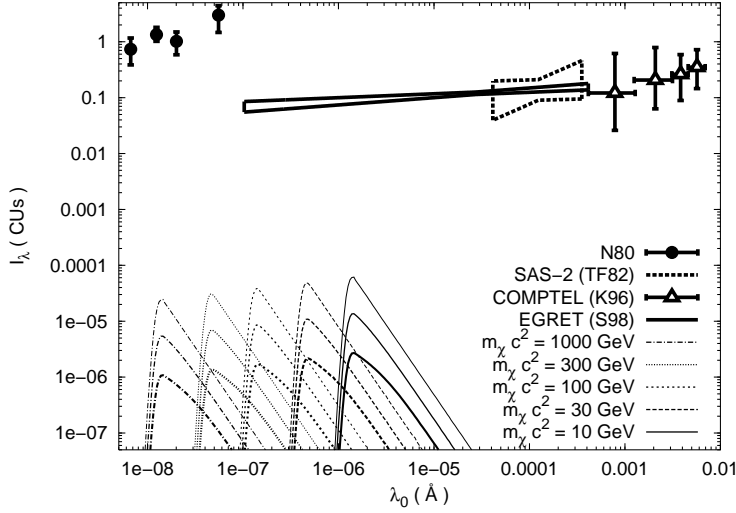}
\end{center}
\caption{The spectral intensity of the diffuse $\gamma$-ray background
   due to neutralino annihilations (lower left), compared with
   observational limits from high-altitude balloon experiments (N80),
   the {\sc Sas}-2 spacecraft and the {\sc Comptel} and {\sc Egret}
   instruments.  The three plotted curves for each value of $\mchi$
   depend on the total density of neutralinos: galaxy halos only
   ($\Omato=0.1\ho$; heavy lines), \LCDM\ model ($\Omato=0.3$;
   medium lines), or EdS model ($\Omato=1$; light lines).}
\label{fig8.2}
\end{figure}

The total spectral intensity of annihilating neutralinos is found by
substituting the SED~(\ref{chiSEDdefn}) into (\ref{IzDefn}) to give
\beq
\Ilam(\lamo) = \Iann \bigint{0}{z_f} \, \bigfrac{
   \exp \left\{ -\bigfrac{1}{2} \left[ \bigfrac{\lamo/(1+z)-\lamann}{\siglam}
   \right]^2 \right\} dz}{(1+z)^3 \, [\Omato (1+z)^3 + 1 - \Omato]^{1/2}}
   \; .
\label{IannDefn}
\eeq
For a typical neutralino with $\mten\approx10$ the annihilation spectrum
peaks near $\lamo\approx10^{-7}$~\AA.  The dimensional prefactor reads 
\beqa
\Iann & = & \frac{\no \Lann f_c}{\sqrt{32\pi^3}\,h\,\Ho}
         \!\left( \frac{\lamo}{\siglam} \right) \nonumber \\
      & = & (0.0002\mbox{ CUs}) \,
         \ho^2 \, f_{\chi} \, f_c \, \mten^{0.8} \, \signine^{-1}
         \!\left( \frac{\lamo}{10^{-7}\mbox{ \AA}} \right) \; .
\eeqa
Here we have divided through by the photon energy $hc/\lamo$ to put
results into continuum units or CUs as usual (Sec.~\ref{sec:LumDens}).
Eq.~(\ref{IannDefn}) gives the combined intensity of radiation from
neutralino annihilations, emitted at various wavelengths and redshifted
by various amounts, but observed at wavelength $\lamo$.  Results are
plotted in Fig.~\ref{fig8.2} together with observational constraints.
We defer detailed discussion of this plot (and the data) to
Sec.~\ref{sec:WIMPdiscussion}, for better comparison with 
results for the other WIMP-related processes.

\subsection{One-loop decays} \label{sec:OneLoop}

\begin{figure}[t!]
\begin{center}
\includegraphics[width=80mm]{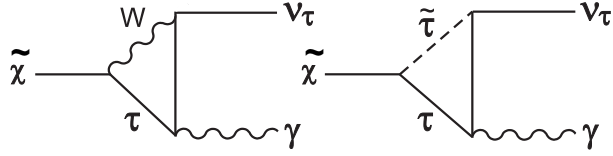}
\end{center}
\caption{Some Feynman diagrams corresponding to one-loop decays of
         the neutralino ($\tilde{\chi}$) into a neutrino (here the
         $\tau$-neutrino and a photon ($\gamma$).  The process can be 
         mediated by the W-boson and a $\tau$-lepton, or by the $\tau$
         and its supersymmetric counterpart ($\tilde{\tau}$).}
\label{fig8.3}
\end{figure}

We turn next to {\em non\/}-minimal SUSY theories in which R-parity is
not necessarily conserved and the LSP (in this case the neutralino) can
decay.  The cosmological consequences of R-parity breaking have been
reviewed by Bouquet and Salati \cite{Bou87}.  There is one {\em direct\/}
decay mode into photons, $\tilde{\chi} \rightarrow \nu + \gamma$.
Feynman diagrams for this process are shown in Fig.~\ref{fig8.3}.
Because these decays occur via loop diagrams, they are again subdominant.
We consider theories in which R-parity breaking is accomplished
spontaneously.  This means introducing a scalar sneutrino with a nonzero
vacuum expectation value $\vR\equiv\ <\!\tilde{\nu}_{\tau_R}\!>$,
as discussed by Masiero and Valle \cite{Mas90}.  Neutralino decays
into photons could be detectable if $m_{\tilde{\chi}}$ and
$\vR$ are large \cite{Ber91a}.

The photons produced in this way are again monochromatic, with
$E_{\gamma}=\smallfrac{1}{2}\mchi$.  In fact the SED here is the
same as (\ref{chiSEDdefn}) except that peak wavelength is doubled,
$\lamloop=2hc/\mchi=(2.5\times 10^{-6} \mbox{ \AA})\,\mten^{-1}$.
The only parameter that needs to be recalculated is the halo
luminosity $L_h$.  For one-loop neutralino decays of lifetime
$\tauchi$, this takes the form:
\beq
\Lloop = \frac{N_{\tilde{\chi}} b_{\gamma} E_{\gamma}}{\tauchi}
    = \frac{b_{\gamma} M_h c^2}{2\tauchi} \; .
\label{LloopDefn}
\eeq
Here $N_{\tilde{\chi}}=M_h/m_{\tilde{\chi}}$ is the number of neutralinos
in the halo and $b_{\gamma}$ is the {\em branching ratio\/}, or fraction
of neutralinos that decay into photons.  This is estimated by
Berezinsky \etal\ \cite{Ber91a} as
\beq
b_{\gamma} \approx 10^{-9} \fR^{\,2} \mten^2 \; ,
\eeq
where the new parameter $\fR\equiv\vR/(100$~GeV).  The requirement
that SUSY WIMPs not carry too much energy out of stellar cores implies
that $\fR$ is of order ten or more \cite{Mas90}.  We take $\fR>1$ as
a lower limit.

We adopt $M_h=(2\pm1)\times10^{12}\Msun$ as usual, with $r_h=(170\pm80)$~kpc
from the discussion following (\ref{chiMhDefn}).
As in the previous section, we parametrize our
lack of certainty about the distribution of neutralinos on larger scales
with the cosmological enhancement factor $f_c$.  Collecting these results
together and expressing the decay lifetime in dimensionless form as
$\ftau\equiv\tauchi/(1$~Gyr), we obtain for the luminosity of one-loop
neutralino decays in the halo:
\beq
\Lloop = (6\times10^{40} \mbox{ erg s}^{-1}) \, \mten^2 \,
      \fR^{\,2} \, \ftau^{-1} \; .
\label{loopLhValue}
\eeq
With $\mten\!\sim\!\fR\!\sim\!\ftau\!\sim\!1$, Eq.~(\ref{loopLhValue})
gives $\Lloop\sim2\times10^7\Lsun$.  This is considerably brighter than the
halo luminosity due to neutralino annihilations in minimal SUSY models,
but still amounts to less than $10^{-3}$ times the bolometric
luminosity of the Milky Way.

Combined bolometric intensity
is found as in the previous section, but with $\Lann$ in (\ref{QannDefn})
replaced by $\Lloop$ so that
\beq
Q = \left\{ \begin{array}{ll}
   (1\times10^{-7} \mbox{ erg s}^{-1} \mbox{ cm}^{-2}) \, \ho^2 \,
      \mten^2 \, \fR^{\,2} \ftau^{-1} & (\mbox{if }\Omato=0.1\ho) \\
   (4\times10^{-7} \mbox{ erg s}^{-1} \mbox{ cm}^{-2}) \, \ho \,
      \mten^2 \, \fR^{\,2} \ftau^{-1} & (\mbox{if }\Omato=0.3) \\
   (2\times10^{-6} \mbox{ erg s}^{-1} \mbox{ cm}^{-2}) \, \ho \, 
      \mten^2 \, \fR^{\,2} \ftau^{-1} & (\mbox{if }\Omato=1)
\end{array} \right. \; .
\eeq
This is again small.  However, we see that massive ($\mten\gtrsim10$)
neutralinos which provide close to the critical density ($\Omato\sim1$)
and decay on timescales of order 1~Gyr or less ($\ftau\lesssim1$)
could in principle rival the intensity of the conventional EBL.

\begin{figure}[t!]
\begin{center}
\includegraphics[width=100mm]{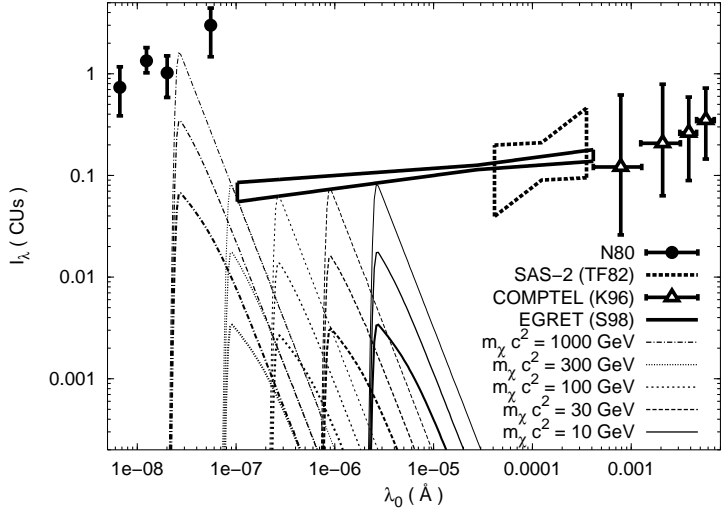}
\end{center}
\caption{The spectral intensity of the diffuse $\gamma$-ray background
   due to neutralino one-loop decays (lower left), compared with
   observational upper limits from high-altitude balloon experiments
   (filled dots), {\sc Sas}-2, {\sc Egret} and {\sc Comptel}.  The three
   plotted curves for each value of $\mchi$ correspond to models with
   $\Omato=0.1\ho$ (heavy lines), $\Omato=0.3$ (medium lines)
   and $\Omato=1$ (light lines).  For clarity we have assumed decay
   lifetimes in each case such that highest theoretical intensities
   lie just under the observational constraints.}
\label{fig8.4}
\end{figure}

To obtain more quantitative constraints, we turn to spectral intensity.
This is given by Eq.~(\ref{IannDefn}) as before, except that the dimensional
prefactor $\Iann$ must be replaced by
\beqa
\Iloop & = & \frac{\no \Lloop f_c}{\sqrt{32\pi^3}\,h\,\Ho}
         \!\left( \frac{\lamo}{\siglam} \right) \nonumber \\
      & = & (30\mbox{ CUs}) \,
         \ho^2 \, \mten^3 \, \fR^{\,2} \ftau^{-1} f_c \, \signine^{-1}
         \!\left( \frac{\lamo}{10^{-7}\mbox{ \AA}} \right) \; .
\eeqa
Results are plotted in Fig.~\ref{fig8.4}
for neutralino rest energies $1\leqslant\mten\leqslant100$.  While
their bolometric intensity is low, these particles are capable of
significant EBL contributions in narrow portions of the $\gamma$-ray
background.  To keep the diagram from becoming too cluttered, we have
assumed values of $\ftau$ such that the highest predicted intensity
in each case stays just below the {\sc Egret} limits.  Numerically,
this corresponds to lower bounds on the decay lifetime $\tauchi$ of
between 100~Gyr (for $\mchi=10$~GeV) and $10^5$~Gyr (for $\mchi=300$~GeV).
For rest energies at the upper end of this range, these limits are
probably optimistic because the decay photons are energetic enough to
undergo pair production on CMB photons.  Some would not reach us from
cosmological distances, instead being re-processed into lower energies
along the way.  As we show in the next section, however, stronger limits
arise from a different process in any case.  We defer further discussion
of Fig.~\ref{fig8.4} to Sec.~\ref{sec:WIMPdiscussion}.

\subsection{Tree-level decays} \label{sec:TreeLevel}

\begin{figure}[t!]
\begin{center}
\includegraphics[width=80mm]{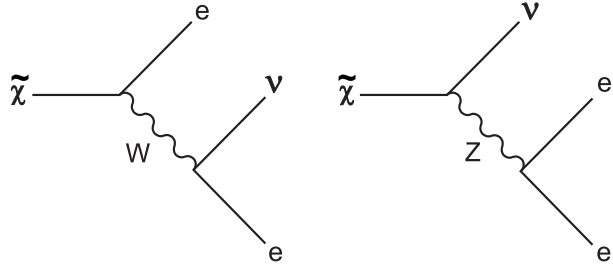}
\end{center}
\caption{The Feynman diagrams corresponding to tree-level decays of
         the neutralino ($\tilde{\chi}$) into a neutrino ($\nu$) and a
         lepton-antilepton pair (here, the electron and positron).
         The process can be mediated by the W or Z-boson.}
\label{fig8.5}
\end{figure}

The dominant decay processes for the LSP neutralino in non-minimal SUSY
(assuming spontaneously broken R-parity) are {\em tree-level\/} decays to
leptons and neutrinos, $\tilde{\chi} \rightarrow \ell^+ + \ell^- + \nu_{\ell}$.
Of particular interest is the case $\ell=e$; Feynman diagrams for this
process are shown in Fig.~\ref{fig8.5}.
Although these processes do not contribute directly to the EBL, they do
so indirectly, because the high-energy electrons undergo inverse Compton
scattering (ICS) off the CMB photons via
$e+\gamma_{cmb}\rightarrow e+\gamma$.  This gives rise to a flux of
high-energy photons which can be at least as important as those from
the direct (one-loop) neutralino decays considered in the previous
subsection \cite{Bar88}.

The spectrum of photons produced in this way depends on the rest energy
of the original neutralino.  We consider first the case $\mten\lesssim10$,
which is more or less pure ICS.  The input (``zero-generation'') electrons
are monoenergetic, but after multiple scatterings they are distributed
like $E^{-2}$ \cite{Blu70}.  From this the spectrum of outgoing photons
can be calculated as \cite{Ber92c}
\beq
\Nics(E) \propto \left\{ \begin{array}{ll}
   E^{-3/2} & \qquad (E \leqslant \Emax) \\
   0        & \qquad (E > \Emax)
      \end{array} \right. \; ,
\eeq
where
\beqa
\Emax & = & \frac{4}{3} \left( \frac{E_e}{m_e c^2} \right)^2 \Ecmb \; .
   \nonumber
\eeqa
Here $E_e=\smallfrac{1}{3}\mchi=(\mbox{3.3 GeV})\,\mten$ is the energy
of the input electrons, $m_e$ is their rest mass, and $\Ecmb=2.7 k \Tcmb$
is the mean energy of the CMB photons.  Using $m_e c^2=0.51$~MeV and
$\Tcmb=2.7$~K, and allowing for decays at arbitrary redshift $z$ (after
Berezinsky \cite{Ber92c}), we obtain the expression
$\Emax(z)=(\mbox{36 keV})\,\mten^2 (1+z)^{-1}$.

The halo SED may be determined as a function of wavelength by setting
$F(\lambda) d\lambda = E N(E) dE$ where $E=hc/\lambda$.  Normalizing
the spectrum so that $\int_0^{\infty} F(\lambda) d\lambda=\Ltree$, we find:
\beq
\Fics(\lambda) = \frac{\Ltree}{2} \times \left\{ \begin{array}{ll}
   \sqrt{\lambda_{\gamma}} \, \lambda^{-3/2} & \qquad
      (\lambda \geqslant \lambda_{\gamma}) \\
   0 & \qquad (\lambda < \lambda_{\gamma})
      \end{array} \right. \; ,
\label{ICS-SED}
\eeq
where $\lambda_{\gamma}=hc/\Emax=(0.34 \mbox{ \AA})\,\mten^{-2}(1+z)$ 
and $\Ltree$ is the halo luminosity due to tree-level decays.

In the case of more massive neutralinos with $\mten\gtrsim10$, the
situation is complicated by the fact that outgoing photons become
energetic enough to initiate pair production
via $\gamma+\gamcmb\rightarrow e^+ + e^-$.  This injects new electrons
into the ICS process, resulting in electromagnetic cascades.
For particles which decay at high redshifts ($z\gtrsim100$),
other processes such as photon-photon scattering must also be
taken into account \cite{Sve90}.  Cascades on {\em non\/}-CMB
background photons may also be important \cite{Cop97}.  A full treatment
of these effects requires detailed numerical analysis \cite{Kri97}.
Here we simplify the problem by assuming that the LSP is stable enough
to survive into the late matter-dominated (or vacuum-dominated) era.
The primary effect of cascades is to steepen the decay spectrum at
high energies, so that \cite{Ber92c}
\beq
\Ncasc(E) \propto \left\{ \begin{array}{ll}
   E^{-3/2} & \qquad (E \leqslant E_x) \\
   E^{-2}   & \qquad (E_x < E \leqslant E_c) \\
   0        & \qquad (E > E_c)
      \end{array} \right. \; ,
\eeq
where
\beqa
E_x & = & \frac{1}{3} \left( \frac{E_0}{m_e c^2} \right)^2 \Ecmb (1+z)^{-1}
   \qquad
E_c = E_0 (1+z)^{-1} \; . \nonumber
\eeqa
Here $E_0$ is a minimum absorption energy.  We adopt the numerical
expressions $E_x = (1.8 \times 10^3 \mbox{ GeV})(1+z)^{-1}$ and
$E_c = (4.5 \times 10^4 \mbox{ GeV})(1+z)^{-1}$ after Protheroe
\etal\ \cite{Pro95}.  Employing the relation
$F(\lambda) d\lambda = E N(E) dE$ and normalizing as before, we find:
\beq
\Fcasc(\lambda) = \frac{\Ltree}{[2+\ln (\lambda_x/\lambda_c)]} 
                    \times \left\{ \begin{array}{ll}
   \sqrt{\lambda_x} \, \lambda^{-3/2} & \qquad
      (\lambda \geqslant \lambda_x) \\
   \lambda^{-1}                       & \qquad
      (\lambda_x > \lambda \geqslant \lambda_c) \\
   0                                  & \qquad
      (\lambda < \lambda_c)
   \end{array} \right. \; ,
\label{CascSED}
\eeq
where the new parameters are
$\lambda_x=hc/E_x=(7 \times 10^{-9} \mbox{ \AA})(1+z)$
and $\lambda_c=hc/E_c=(3 \times 10^{-10} \mbox{ \AA})(1+z)$.

The luminosity $\Ltree$ is given by
\beq
\Ltree = \frac{N_{\tilde{\chi}} b_e E_e}{\tauchi}
    = \frac{2 \, b_e M_h c^2}{3\tauchi} \; ,
\label{LtreeDefn}
\eeq
where $b_e$ is now the branching ratio for all processes of the form
$\tilde{\chi}\rightarrow e+\mbox{ all}$ and $E_e=\smallfrac{2}{3}\mchi$
is the total energy lost to the electrons.
We assume that all of this eventually finds its way into the EBL.
Berezinsky \etal\ \cite{Ber91a} supply the following branching ratio:
\beq
b_e \approx 10^{-6} f_{\chi} \fR^{\,2} \mten^2 \; .
\eeq
Here $f_{\chi}$ parametrizes the composition of the neutralino, taking
the value 0.4 for the pure higgsino case.
With the halo mass specified by~(\ref{MhValue}) and 
$\ftau\equiv\tauchi/(1$~Gyr) as usual, we obtain:
\beq
\Ltree = (8 \times 10^{43} \mbox{ erg s}^{-1}) \, \mten^2 \,
         f_{\chi} \, \fR^{\,2} \, \ftau^{-1} \; .
\eeq
This is approximately four orders of magnitude higher than the halo
luminosity due to one-loop decays, and provides for the first time the
possibility of significant EBL contributions.  With all adjustable
parameters taking values of order unity, we find that
$\Ltree\sim\!2\times10^{10}\Lsun$, which is comparable to the
bolometric luminosity of the Milky Way.

The combined bolometric intensity of all neutralino halos is computed
as in the previous two sections.  Replacing $\Lloop$ in (\ref{QannDefn})
with $\Ltree$ leads to
\beq
Q = \left\{ \begin{array}{ll} 
   (2\times10^{-4} \mbox{ erg s}^{-1} \mbox{ cm}^{-2}) \, \ho^2 \,
      \mten^2 f_{\chi} \fR^{\,2} \ftau^{-1} & (\mbox{if }\Omato=0.1\ho) \\
   (5\times10^{-4} \mbox{ erg s}^{-1} \mbox{ cm}^{-2}) \, \ho \,
      \mten^2 f_{\chi} \fR^{\,2} \ftau^{-1} & (\mbox{if }\Omato=0.3) \\
   (2\times10^{-3} \mbox{ erg s}^{-1} \mbox{ cm}^{-2}) \, \ho \,
      \mten^2 f_{\chi} \fR^{\,2} \ftau^{-1} & (\mbox{if }\Omato=1)
\end{array} \right. \; .
\eeq
These are of the same order as (or higher than) the bolometric intensity
of the EBL from ordinary galaxies, Eq.~(\ref{QstarValue}).

\begin{figure}[t!]
\begin{center}
\includegraphics[width=100mm]{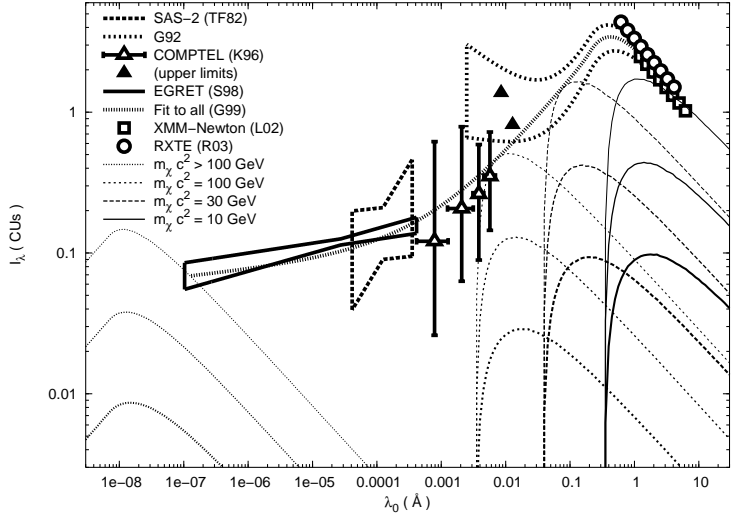}
\end{center}
\caption{The spectral intensity of the diffuse $\gamma$-ray and x-ray 
   backgrounds due to neutralino tree-level decays, compared with
   observational upper limits from {\sc Sas}-2, {\sc Egret} and
   {\sc Comptel} in the $\gamma$-ray region, from XMM-Newton and
   RXTE in the x-ray region, and from Gruber's fits to various
   experimental data (G92,G99).  The three plotted curves for each value
   of $\mchi$ correspond to models with $\Omato=0.1\ho$ (heavy lines),
   $\Omato=0.3$ (medium lines) and $\Omato=1$ (light lines).  For clarity
   we have assumed decay lifetimes in each case such that highest
   theoretical intensities lie just under the observational constraints.}
\label{fig8.6}
\end{figure}

To obtain the spectral intensity, we substitute the SEDs $\Fics(\lambda)$
and $\Fcasc(\lambda)$ into Eq.~(\ref{IzDefn}).  The results can be written
\beq
\Ilam(\lambda_0) = \Itree \int_0^{\, z_f} \frac{{\mathcal F}(z) \, dz}
   {(1+z) \left[ \Omato (1+z)^3 + (1-\Omato) \right]^{1/2}} \; ,
\label{ItreeDefn}
\eeq
where the quantities $\Itree$ and ${\mathcal F}(z)$ are defined as follows.
For neutralino rest energies $\mten\lesssim10$ (ICS):
\beqa
\Itree & = & \frac{\no \, \Ltree \, f_c}{8\pi h \, \Ho \, \mten}
             \left( \frac{0.34 \mbox{ \AA}}{\lamo} \right)^{1/2}
             \nonumber \\
       & = & (300 \mbox{ CUs}) \, \ho^2 \, \mten \, f_{\chi} \, \fR^2 \,
             \ftau^{-1} f_c \left( \frac{\lamo}{\mbox{\AA}}
             \right)^{-\smallfrac{1}{2}} \label{ItreeFdefn1} \\
{\mathcal F}(z) & = & \left\{ \begin{array}{ll}
   1 & \qquad [\lamo \geqslant \lambda_{\gamma} (1+z)] \\
   0 & \qquad [\lamo < \lambda_{\gamma} (1+z)]
      \end{array} \right. \; . \nonumber
\eeqa
Conversely, for $\mten\gtrsim10$ (cascades):
\beqa
\Itree & = & \frac{\no \, \Ltree \, f_c}{4\pi h \, \Ho \,
             [2+\ln (\lambda_x/\lambda_c)]} \left( \frac{7 \times 10^{-9}
             \mbox{ \AA}}{\lamo} \right)^{1/2} \nonumber \\
       & = & (0.02 \mbox{ CUs}) \, \ho^2 \, \mten^2 \, f_{\chi} \, \fR^2 \,
             \ftau^{-1} f_c \left( \frac{\lamo}{\mbox{\AA}}
             \right)^{-\smallfrac{1}{2}} \label{ItreeFdefn2} \\
{\mathcal F}(z) & = & \left\{ \begin{array}{cl}
   1 & \qquad [\lamo \geqslant \lambda_x (1+z)] \\
   \bigfrac{\lamo}{\lambda_x (1+z)} & \qquad
       [\lambda_x (1+z) > \lamo \geqslant \lambda_c (1+z)] \\ 
   0 & \qquad [\lamo < \lambda_c (1+z)]
      \end{array} \right. \; . \nonumber
\eeqa
Numerical integration of Eq.~(\ref{ItreeDefn}) leads to the plots
in Fig.~\ref{fig8.6}.
Cascades (like the pair annihilations we have considered already)
dominate the $\gamma$-ray part of the spectrum.  The ICS process, however,
is most important at lower energies, in the x-ray region.
We discuss the observational limits and the constraints that can
be drawn from them in more detail in Sec.~\ref{sec:WIMPdiscussion}.

\subsection{Gravitinos} \label{sec:Gravitino}

Gravitinos ($\tilde{g}$) are the SUSY spin-$\smallfrac{3}{2}$
counterparts of gravitons.  Although often mentioned along with
neutralinos, they are not favoured as dark-matter candidates
in the simplest SUSY theories.  The reason for this, known as the
{\em gravitino problem\/} \cite{Ell82}, boils down to the fact that
they interact {\em too\/} weakly, not only with other particles
but with themselves as well.  Hence they annihilate slowly and
survive long enough to ``overclose'' the Universe unless some other
way is found to reduce their numbers.  Decays are one possibility,
but not if the gravitino is a stable LSP.  Gravitino decay products must
also not be allowed to interfere with processes such as primordial
nucleosynthesis \cite{Wei82}.  Inflation, followed by a judicious
period of reheating, can thin out their numbers to almost any desired
level.  But the reheat temperature $\TR$ must satisfy
$k\TR\lesssim\!10^{12}$~GeV or gravitinos will once again become
too numerous \cite{Ell84a}.  Related arguments based on entropy
production, primordial nucleosynthesis and the CMB power spectrum
force this number down to $k\TR\lesssim(10^9-10^{10})$~GeV \cite{Ell84b}
or even $k\TR\lesssim(10^6-10^9)$~GeV \cite{Kaw95}.  These temperatures
are incompatible with the generation of baryon asymmetry in the Universe,
a process which is usually taken to require $k\TR\sim10^{14}$~GeV or higher
\cite{Kol90}.

Recent developments are however beginning to loosen the baryogenesis
requirement \cite{Giu01}, and there are alternative models in which
baryon asymmetry is generated at energies as low as $\sim\!10$~TeV
\cite{Kuz85} or even 10~MeV~--~1~GeV \cite{Dim87}.  With this in mind
we include a brief look at gravitinos here.  There are two possibilities: 
(1)~If the gravitino is {\em not\/} the LSP, then it decays early in
the history of the Universe, well before the onset of the
matter-dominated era.  In models where the gravitino decays both
radiatively and hadronically, for example, it can be ``long-lived
for its mass'' with a lifetime of 
$\taug\lesssim10^6$~s \cite{Dim88}.  Particles of this kind have important
consequences for nucleosynthesis, and might affect the shape of the
CMB if $\taug$ were to exceed $\sim\!10^7$~s.  However, they are
irrelevant as far as the EBL is concerned.  We therefore restrict our
attention to the case~(2), in which the gravitino is the LSP.
In light of the results we have already obtained for the neutralino,
we disregard annihilations and consider only models in which the LSP
can decay.

The decay mode depends on the specific mechanism of R-parity violation.
We follow Berezinsky \cite{Ber91b} and concentrate on dominant
{\em tree-level\/} processes.  In particular we consider the decay
$\tilde{g} \rightarrow e^+ + \mbox{ all }$, followed by ICS off the CMB,
as in Sec.\ref{sec:TreeLevel}.  The spectrum of photons produced by this
process is identical to that in the neutralino case, except that the
mono-energetic electrons have energy
$E_e = \smallfrac{1}{2}\mg = (\mbox{5 GeV})\,\mten$ \cite{Ber91b},
where $\mg$ is the rest energy of the gravitino and $\mten\equiv\mg/(10$~GeV)
as before.
This in turn implies that $\Emax=(81 \mbox{ keV})\,\mten^2 (1+z)^{-1}$
and $\lambda_{\gamma}=hc/E_{max}=(0.15 \mbox{ \AA})\,\mten^{-2} (1+z)$.
The values of $\lambda_x$ and $\lambda_c$ are unchanged.

The SED comprises Eqs.~(\ref{ICS-SED}) for ICS and (\ref{CascSED})
for cascades, as before.  Only the halo luminosity needs to be recalculated.
This is similar to Eq.~(\ref{LtreeDefn}) for neutralinos, except that
the factor of $\smallfrac{2}{3}$ becomes $\smallfrac{1}{2}$, and
the branching ratio can be estimated at \cite{Ber91b}
\beq
b_e \sim (\alpha/\pi)^2 = 5 \times 10^{-6} \; .
\label{Gbratio}
\eeq
Using our standard value for the halo mass $M_h$, and parametrizing
the gravitino decay lifetime by $\ftau\equiv\taug/(1$~Gyr) as before,
we obtain the following halo luminosity due to gravitino decays:
\beq
\Lgrav = (3 \times 10^{44} \mbox{ erg s}^{-1}) \ftau^{-1} \; .
\eeq
This is higher than the luminosity due to neutralino decays, and
exceeds the luminosity of the Milky Way by several times if $\ftau\sim\!1$.

The bolometric intensity of all gravitino halos is computed exactly
as before.  Replacing $\Ltree$ in (\ref{QannDefn}) with $\Lgrav$, we find:
\beq
Q = \left\{ \begin{array}{ll} 
   (7\times10^{-4} \mbox{ erg s}^{-1} \mbox{ cm}^{-2}) \, \ho^2 \,
      \ftau^{-1} & (\mbox{if } \Omato=0.1\ho) \\
   (2\times10^{-3} \mbox{ erg s}^{-1} \mbox{ cm}^{-2}) \, \ho \,
      \ftau^{-1} & (\mbox{if } \Omato=0.3) \\
   (8\times10^{-3} \mbox{ erg s}^{-1} \mbox{ cm}^{-2}) \, \ho \,
      \ftau^{-1} & (\mbox{if } \Omato=1)
\end{array} \right. \; .
\eeq
It is clear that gravitinos must decay on timescales longer than the
lifetime of the Universe ($\ftau\gtrsim16$), or they would produce a
background brighter than that of the galaxies.

\begin{figure}[t!]
\begin{center}
\includegraphics[width=100mm]{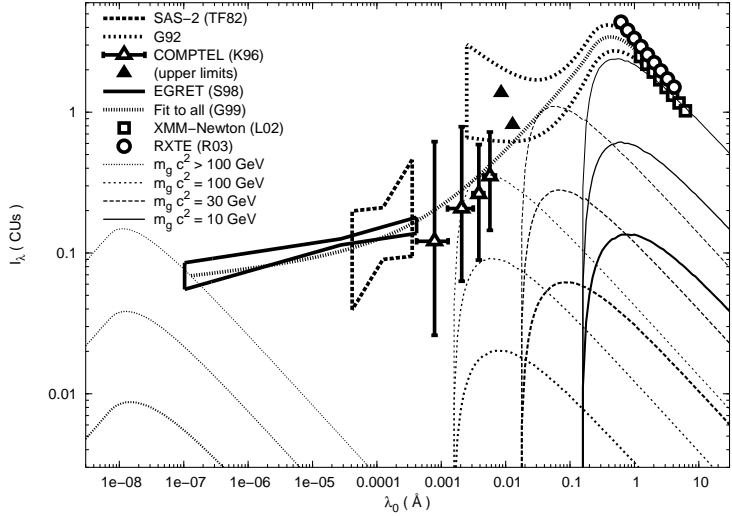}
\end{center}
\caption{The spectral intensity of the diffuse $\gamma$-ray and x-ray 
   backgrounds due to gravitino tree-level decays, compared with experimental
   data from {\sc Sas}-2, {\sc Egret}, {\sc Comptel}, XMM-Newton and RXTE,
   as well as compilations by Gruber.  The three plotted curves for each
   value of $\mchi$ correspond to models with $\Omato=0.1\ho$ (heavy lines),
   $\Omato=0.3$ (medium lines) and $\Omato=1$ (light lines).  For clarity
   we have assumed decay lifetimes in each case such that highest
   theoretical intensities lie just under the observational constraints.}
\label{fig8.7}
\end{figure}

The spectral intensity is the same as before, Eq.~(\ref{ItreeDefn}),
but with the new numbers for $\lambda_{\gamma}$ and $L_h$.  This results in
\beq
\Ilam(\lambda_0) = \Igrav \int_0^{\, z_f} \frac{{\mathcal F}(z) \, dz}
   {(1+z) \left[ \Omato (1+z)^3 + (1-\Omato) \right]^{1/2}} \; ,
\label{IgravDefn}
\eeq
where the prefactor $\Igrav$ is defined as follows.
For $\mten\lesssim10$ (ICS):
\beqa
\Igrav & = & \frac{\no \, \Lgrav \, f_c}{8\pi h \, \Ho \, \mten}
             \left( \frac{0.15 \mbox{ \AA}}{\lamo} \right)^{1/2}
             \nonumber \\
       & = & (800 \mbox{ CUs}) \, \ho^2 \, \mten^{-1} \,
             \ftau^{-1} f_c \left( \frac{\lamo}{\mbox{\AA}}
             \right)^{-\smallfrac{1}{2}} \; .
\eeqa
Conversely, for $\mten\gtrsim10$ (cascades):
\beqa
\Igrav & = & \frac{\no \, \Lgrav \, f_c}{4\pi h \, \Ho \,
             [2+\ln (\lambda_x/\lambda_c)]} \left( \frac{7 \times 10^{-9}
             \mbox{ \AA}}{\lamo} \right)^{1/2} \nonumber \\
       & = & (0.06 \mbox{ CUs}) \, \ho^2 \, 
             \ftau^{-1} f_c \left( \frac{\lamo}{\mbox{\AA}}
             \right)^{-\smallfrac{1}{2}} \; .
\eeqa
The function ${\mathcal F}(z)$ has the same form as in
Eqs.~(\ref{ItreeFdefn1}) and (\ref{ItreeFdefn2}) and does not need
to be redefined (requiring only the new value for the cutoff wavelength
$\lambda_{\gamma}$).  Because the branching ratio~$b_e$ in (\ref{Gbratio})
is independent of the gravitino rest mass, $\mten$ appears in these results
only through $\lambda_{\gamma}$.  Thus the ICS part of the spectrum goes
as $\mten^{-1}$ while the cascade part does not depend on $\mten$ at all.
As with neutralinos, cascades dominate the $\gamma$-ray part of the
spectrum, and the ICS process is most important in the x-ray region.
Numerical integration of Eq.~(\ref{IgravDefn}) leads to the results
plotted in Fig.~\ref{fig8.7}.
We proceed in the next section to discuss these, comparing them to our
previous results for neutralinos, and beginning with an overview of the
observational constraints.

\subsection{The x-ray and $\gamma$-ray backgrounds} \label{sec:WIMPdiscussion}

The experimental situation as regards EBL intensity in the x-ray and
$\gamma$-ray regions is more settled than that in the optical and ultraviolet.
Detections (as opposed to upper limits) have been made in both bands,
and are consistent with expectations based on known astrophysical sources.
The constraints that we derive here are thus conservative ones, in the
sense that the EBL flux which could plausibly be due to decaying WIMPs
is almost certainly smaller than the levels actually measured.

At the lowest or soft x-ray energies, which lie roughly between
0.1--3~keV (4--100~\AA), new measurements have been reported from the
Chandra spacecraft \cite{Mar03}.  (Universal conventions have not been
established regarding the boundaries between different wavebands; we
follow most authors and define these according to the different detection
techniques that must be used in each region.)  These data appear as a
small bowtie-shaped box near $\lambda_0\sim10$~\AA\ in Fig.~\ref{fig1.2},
where it can be seen that they interpolate beautifully between previous
detections at shorter and longer wavelengths.
The small rectangle immediately to the right of the Chandra bowtie (near
$\lambda_0\sim100$~\AA) in Fig.~\ref{fig1.2} comes from measurements by
the {\sc Euve} satellite in 1993 \cite{Lie93}.

The hard x-ray background (3--800~keV, or 0.02--4~\AA) is
crucial in constraining the decays of low-mass neutralinos and
gravitinos via the ICS process, as can be seen in Figs.~\ref{fig8.6} and
\ref{fig8.7}.  We have plotted two compilations of observational data in the
hard x-ray band, both by Gruber \cite{Gru92,Gru99}.  The first (labelled
``G92'' in Figs.~\ref{fig8.6} and \ref{fig8.7}) is an empirical fit to
various pre-1992 measurements, including those from the Kosmos and Apollo
spacecraft, {\sc Heao-1} and balloon experiments.  The range of uncertainty
in this data increases logarithmically from 2\% at 3~keV to 60\% at 3~MeV
\cite{Gru92}.  The second compilation (labelled ``G99'') is a revision of
this fit in light of new data at higher energies, and has been extended
deep into the $\gamma$-ray region.  New results from XMM-Newton \cite{Lum02}
and the Rossi X-ray Timing Explorer \cite{Rev03} confirm the accuracy of
this revised fit at low energies (``L02'' and ``R03'' respectively in
Figs.~\ref{fig8.6} and \ref{fig8.7}).  The prominent peak in the range
3--300~keV (0.04--4~\AA) is widely attributed to integrated light
from active galactic nuclei (AGN) \cite{Zdz96}.

In the low-energy $\gamma$-ray region (0.8--30~MeV, or 0.0004--0.02~\AA)
we have used results from the {\sc Comptel} instrument on the Compton
Gamma-Ray Observatory ({\sc Cgro}), which was operational from 1990-2000
\cite{Kap96}.  Four data points are plotted in Figs.~\ref{fig8.2},
\ref{fig8.4}, \ref{fig8.6} and \ref{fig8.7}, and two more (upper limits only)
appear at low energies in Figs.~\ref{fig8.6} and \ref{fig8.7}.  These
experimental results, which interpolate smoothly between other data at
both lower and higher energies, played a key role in the demise of the
``MeV bump'' (visible in Figs.~\ref{fig8.6} and \ref{fig8.7} as a
significant upturn in Gruber's fit to the pre-1992 data from about
0.002--0.02~\AA).  This apparent feature in the background had attracted
a great deal of attention from theoretical cosmologists as a possible
signature of new physics.  Figs.~\ref{fig8.6} and \ref{fig8.7}
suggest that it could also have been interpreted as the signature of
a long-lived non-minimal SUSY WIMP with a rest energy near 100~GeV.
The MeV bump is, however, no longer believed to be real, as the new fit
(``G99'') makes clear.  Most of the background in this region is now
suspected to be due to Type Ia supernovae (SNIa) \cite{The93}.

We have included two measurements in the high-energy $\gamma$-ray band
(30~MeV--30~GeV, or $4\times10^{-7}$--$4\times10^{-4}$~\AA): one from
the {\sc Sas}-2 satellite which flew in 1972-3 \cite{Tho82} and one
from the {\sc Egret} instrument which was part of the {\sc Cgro} mission
along with {\sc Comptel} \cite{Sre98}.  As may be seen in
Figs.~\ref{fig8.2}, \ref{fig8.4}, \ref{fig8.6} and \ref{fig8.7},
the new results essentially extend the old ones to 120~GeV
($\lamo=10^{-7}$~\AA), with error bars which have been reduced by
a factor of about ten.  Most of this extragalactic background is
thought to arise from unresolved blazars, highly variable AGN whose
relativistic jets point in our direction \cite{Mcn95}.  Some authors have
recently argued that Galactic contributions to the background were
underestimated in the original {\sc Egret} analysis \cite{Kes04,Str04};
if so, the true extragalactic background intensity would be lower than
that plotted here, strengthening the constraints we derive below.

Because the extragalactic component of the $\gamma$-ray background has
not been reliably detected beyond 120~GeV, we have fallen back on measurements
of {\em total\/} flux in the very high-energy (VHE) region (30~GeV--30~TeV,
or $4\times10^{-10}$--$4\times10^{-7}$~\AA).  These were obtained from a
series of balloon experiments by Nishimura \etal\ in 1980 \cite{Nis80},
and appear in Figs.~\ref{fig8.2} and \ref{fig8.4} as filled dots
(labelled ``N80'').  They constitute a very robust upper limit on EBL flux,
since much of this signal must have originated in the upper atmosphere.
At the very highest energies, in the ultra high-energy (UHE) region
($>$~30~TeV), these data join smoothly to upper limits on the diffuse
$\gamma$-ray flux from extensive air-shower arrays such as {\sc Hegra}
(20-100~TeV \cite{Aha02}) and {\sc Casa-Mia} (330~TeV-33~PeV \cite{Cha97}).
Here we reach the edge of the EBL for practical purposes, since 
$\gamma$-rays with energies of $\sim10-100$~PeV are attenuated by pair
production on CMB photons over scales $\sim30$~kpc \cite{Gou66}.

Some comments are in order here about units.  For experimental reasons,
measurements of x-ray and $\gamma$-ray backgrounds are often expressed
in terms of integral flux $\EIE$, or number of photons with energies
above $E_0$.  This presents no difficulties since the differential
spectrum in this region is well approximated with a single
power-law component, $\IE(E_0)=\Istar(E_0/\Estar)^{-\alpha}$.
The conversion to integral form is then given by
\beq
\EIE = \int_{E_0}^{\infty} \IE(E)\,dE = \frac{\Estar\Istar}{\alpha-1}
   \left( \frac{E_0}{\Estar} \right)^{1-\alpha} \; .
\eeq
The spectrum is specified in either case by its index $\alpha$
together with the values of $\Estar$ and $\Istar$ (or $E_0$ and
$E\capsub{I}{E}$ in the integral case). Thus {\sc Sas-2} results
were reported as $\alpha=2.35^{+0.4}_{-0.3}$ with
$E\capsub{I}{E}=(5.5\pm1.3)\times10^{-5}\mbox{ s}^{-1}\mbox{ cm}^{-2}\mbox{
ster}^{-1}$ for $E_0=100$~MeV \cite{Tho82}.  The {\sc Egret} spectrum
is instead fit by $\alpha=2.10\pm0.03$ with
$\Istar=(7.32\pm0.34)\times10^{-9}\mbox{ s}^{-1}\mbox{ cm}^{-2}\mbox{
ster}^{-1}\mbox{ MeV}^{-1}$ for $\Estar=451$~MeV \cite{Sre98}.
To convert a differential flux in these units to $\Ilam$ in CUs, one
multiplies by $E_0/\lamo=E_0^{\,2}/hc=80.66E_0^{\,2}$ where $E_0$
is photon energy in MeV.

We now discuss our results, beginning with the neutralino annihilation
fluxes plotted in Fig.~\ref{fig8.2}.  These are at least four orders of
magnitude fainter than the background detected by {\sc Egret} \cite{Sre98}
(and five orders of magnitude below the upper limit set by the data of
Nishimura \etal\ \cite{Nis80} at shorter wavelengths).  This agrees
with previous studies assuming a critical density of neutralinos
\cite{Sil84,Gao91}.  Fig.~\ref{fig8.2} shows that EBL contributions
would drop by another order of magnitude in the favoured scenario with
$\Omato\approx0.3$, and by another if neutralinos are confined to
galaxy halos ($\Omato\approx0.1\ho$).  Because the annihilation rate
goes as the square of the WIMP density, it has been argued that modelling
WIMP halos with steep density cusps might raise their luminosity,
possibly enhancing their EBL contributions by a factor of as much as
$\sim10^4-10^5$ \cite{Ber01}.  While such a scenario might in principle
bring WIMP annihilations back up to the brink of observability in the
diffuse background, density profiles with the required steepness
are not seen in either our own Galaxy or those nearby.  More recent
assessments have reconfirmed the general outlook discussed above in
Sec.~\ref{sec:Annihilations}; namely, that the best place to look for
WIMP annihilations is in the direction of nearby concentrations of dark
matter such as the Galactic center and dwarf spheroidals in the Local
Group \cite{Eva04,Sto03}.  The same stability that makes minimal-SUSY
WIMPs so compelling as dark-matter candidates also makes them hard to detect.

\begin{figure}[t!]
\begin{center}
\includegraphics[width=100mm]{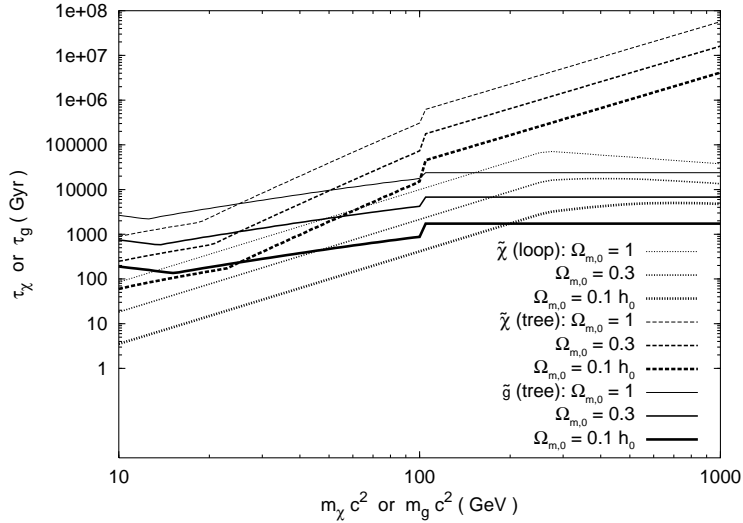}
\end{center}
\caption{The lower limits on WIMP decay lifetime derived from observations
         of the x-ray and $\gamma$-ray backgrounds.  Neutralino bounds
         are shown for both one-loop decays (dotted lines) and tree-level
         decays (dashed lines).  For gravitinos we show only the tree-level
         constraints (solid lines).  For each process there are three
         curves corresponding to models with $\Omato=1$ (light lines),
         $\Omato=0.3$ (medium lines) and $\Omato=0.1\ho$ (heavy lines).}
\label{fig8.8}
\end{figure}

Fig.~\ref{fig8.4} shows the EBL contributions from one-loop neutralino
decays in {\em non\/}-minimal SUSY.  We have put $\ho=0.75$, $z_f=30$ and
$\fR=1$.  Depending on their decay lifetime (here parametrized by $\ftau$),
these particles are capable in principle of producing a backgound
comparable to (or even in excess of) the {\sc Egret} limits.
The plots in Fig.~\ref{fig8.4} correspond to the smallest values of
$\ftau$ that are consistent with the data for $\mten=1,3,10,30$ and 100.
Following the same procedure here as we did for axions in Sec.~\ref{ch6},
we can repeat this calculation over more finely-spaced intervals in
neutralino rest mass, obtaining a lower limit on decay lifetime
$\tauchi$ as a function of $m_{\tilde{\chi}}$.  Results are shown
in Fig.~\ref{fig8.8} (dotted lines).
The lower limit obtained in this way varies from 4~Gyr for the lightest
neutralinos (assumed to be confined to galaxy halos with a total matter
density of $\Omato=0.1\ho$) to 70,000~Gyr for the heaviest (which provide
enough CDM to put $\Omato=1$).

Fig.~\ref{fig8.6} is a plot of EBL flux from {\em indirect\/} neutralino
decays via the tree-level, ICS and cascade processes described in
Sec.~\ref{sec:TreeLevel}.  These provide us with our strongest constraints
on non-minimal SUSY WIMPs.  We have set $\ho=0.75$, $z_f=30$ and
$f_{\chi}=\fR=1$, and assumed values of $\ftau$ such that the highest
predicted intensities lie just under observational limits, as before.
Neutralinos at the light end of the mass range are constrained by
x-ray data, while those at the heavy end run up against
the {\sc Egret} measurements.  Both the shape and absolute intensity of
the ICS spectra depend on the neutralino rest mass, but the cascade spectra
depend on $\mten$ through intensity alone (via the prefactor $\Itree$).
When normalized to the observational upper bound, all
curves for $\mten>10$ therefore overlap.  Normalizing across the full
range of neutralino rest masses (as for one-loop decays) gives
the lower bound on lifetime $\tauchi$ plotted in Fig.~\ref{fig8.8}
(dashed lines).  This varies from 60~Gyr to $6\times10^7$~Gyr, depending
on the values of $\mchi$ and $\Omato$.

Fig.~\ref{fig8.7}, finally, shows the EBL contributions from tree-level
{\em gravitino\/} decays.  These follow the same pattern as the neutralino
decays.  Requiring that the predicted signal not exceed the x-ray and
$\gamma$-ray observations, we obtain the lower limits on decay lifetime
plotted in Fig.~\ref{fig8.8} (solid lines).  The flatness of these curves
(relative to the constraints on neutralinos) is a consequence of the fact
that the branching ratio~(\ref{Gbratio}) is independent of $\mten$.
Our lower limits on $\taug$ range from 200~Gyr to $20,000$~Gyr.

Let us sum up our findings in this section.  We have considered
neutralinos and gravitinos, either of which could be the LSP and hence
make up the dark matter.  In the context of non-minimal SUSY theories,
these particles can decay and contribute to both the x-ray and 
$\gamma$-ray backgrounds.  We have shown that any such decay must occur
on timescales longer than $10^2$--$10^8$~Gyr (for neutralinos) or
$10^2$--$10^4$~Gyr (for gravitinos), depending on their rest masses
and various theoretical input parameters.  These results confirm that,
whether it is a neutralino or gravitino, {\em the LSP in non-minimal
SUSY theories must be very nearly stable \/}.  To the extent that an
``almost-stable'' LSP would require that R-party conservation be
violated at improbably low levels, our constraints suggest that the
SUSY WIMP either exists in the context of minimal SUSY theories,
or not at all.

\section{Black holes and solitons} \label{ch9}

\subsection{Primordial black holes} \label{sec:PBH-intro}

At first glance, black holes would appear to be among the unlikeliest
of background radiation sources.  In fact, experimental data on the
intensity of the EBL constrain black holes more strongly than any of
the dark-matter candidates we have discussed so far.  Before explaining
how this comes about, we distinguish between ``ordinary'' black holes
(which form via the gravitational collapse of massive stars at the end
of their lives) and {\em primordial black holes\/} (PBHs) which may have
arisen from the collapse of overdense regions in the early Universe.
The existence of the former is very nearly an established fact,
while the latter remain hypothetical.  However, it is PBHs which
are of interest to us as potential dark-matter candidates.

The reason for this is as follows.  Ordinary black holes come from
baryonic progenitors (i.e. stars) and are hence classified with
the baryonic dark matter of the Universe.  (They are of course not
``baryonic'' in other respects, since among other things their baryon
number is not defined.)  Ordinary black holes are therefore subject to
the nucleosynthesis bound~(\ref{ObarValue}), which limits them to
less than 5\% of the critical density.
PBHs are {\em not\/} subject to this bound because they form during
the radiation-dominated era, before nucleosynthesis begins. 
Nothing prevents them from making up most of the density in the Universe.
Moreover they constitute {\em cold\/} dark matter because their velocities
are low.  (That is, they collectively obey a dust-like equation of state,
even though they might individually be better described as
``radiation-like'' than baryonic.)  PBHs were first proposed as
dark-matter candidates by Zeldovich and Novikov in 1966 \cite{Zel66}
and Hawking in 1971 \cite{Haw71}.

Black holes contribute to the EBL via a process discovered by
Hawking in 1974 and often called {\em Hawking evaporation\/} \cite{Haw74}.
Photons cannot escape from inside the black hole, but they are produced 
at or near the horizon by quantum fluctuations in the surrounding
curved spacetime.  These give rise to a net flux of particles
which propagates outward from a black hole (of mass $M$) at a rate
proportional to $M^{-2}$ (with the black-hole mass itself dropping
at the same rate).  For ordinary, stellar-mass black holes, this process
occurs so slowly that contributions to the EBL are insignificant, and the
designation ``black'' remains perfectly appropriate over the lifetime of
the Universe.  PBHs, however, can in principle have masses far smaller
than those of a star, leading to correspondingly higher luminosities.
Those with $M\lesssim10^{15}$~g (the mass of a small asteroid) would
in fact evaporate quickly enough to shed all their mass over less than
$\sim\!10$~Gyr.  They would already have expired in a blaze of
high-energy photons and other elementary particles as $M\rightarrow0$.

We will use the PBHs themselves as sources of radiation in what follows,
taking them to be distributed homogeneously throughout space.  
The degree to which they actually cluster in the potential wells of
galaxies and galaxy clusters is not of concern here, since we are
concerned with their combined PBH contributions to the diffuse background.
Another subtlety must, however, be taken into account.  Unlike the
dark-matter halos of previous sections, PBHs cover such a wide range
of masses (and luminosities) that we can no longer treat all sources
as identical.  Instead we must define quantities like number density
and energy spectrum as functions of {\em PBH mass} as well as time,
and integrate our final results over both parameters.  

The first step is to identify the distribution of PBH masses at the
time when they formed.  There is little prospect of probing the time
before nucleosynthesis experimentally, so any theory of PBH formation
is necessarily speculative to some degree.  However, the scenario
requiring the least extrapolation from known physics would be one in
which PBHs arose via the gravitational collapse of small initial density
fluctuations on a standard Robertson-Walker background, with an equation
of state with the usual form~(\ref{EOS}), and with initial density
fluctuations distributed like
\beq
\delta = \epsilon (M_i/M_f )^{-n} \; .
\eeq
Here $M_i$ is the initial mass of the PBH, $M_f$ is the mass lying
inside the particle horizon (or causally connected Universe) at PBH
formation time, and $\epsilon$ is an unknown proportionality constant.

PBH formation under these conditions was originally investigated by Carr
\cite{Car75}, who showed that the process is favoured over an extended
range of masses only if $n=\smallfrac{2}{3}$.  Proceeding on this assumption,
he found that the initial mass distribution of PBHs formed with masses
between $M_i$ and $M_i+dM_i$ per unit comoving volume is
\beq
n(M_i)\,dM_i = \rho_f M_f^{-2} \zeta \left( \frac{M_i}{M_f}
   \right)^{-\beta} \!\!\!\! dM_i \; ,
\label{PBH-IMD}
\eeq
where $\rho_f$ is the mean density at formation time.  The parameters
$\beta$ and $\zeta$ are formally given by $2(2\gamma-1)/\gamma$ and
$\epsilon\exp[-(\gamma-1)^2/2\epsilon^2]$ respectively, where $\gamma$
is the equation-of-state parameter as usual.  However, in the interests
of lifting every possible restriction on conditions prevailing in the
early Universe, we follow Carr \cite{Car76} in treating $\beta$
and $\zeta$ as free parameters, not necessarily related to
$\gamma$ and $\epsilon$.  Insofar as the early Universe was governed
by the equation of state~(\ref{EOS}), $\beta$ takes values between
2 (dust-like or ``soft'') and 3 (stiff or ``hard''), with
$\beta=\smallfrac{5}{2}$ corresponding to the most natural situation
(i.e. $\gamma=\smallfrac{4}{3}$ as for a non-interacting relativistic
gas).  We allow $\beta$ to take values as high as 4, corresponding to
``superhard'' early conditions.  The parameter $\zeta$ represents
the fraction of the Universe which goes into PBHs of mass $M_f$ at
time $t_f$.  It is a measure of the initial inhomogeneity of the Universe.

The fact that Eq.~(\ref{PBH-IMD}) has no exponential cutoff at
high mass is important because it allows us (in principle at least)
to obtain a substantial cosmological density of PBHs.  Since
$2\leqslant\beta\leqslant 4$, however, the power-law distribution is
dominated by PBHs of {\em low\/} mass.  This is the primary reason why
PBHs turn out to be so tightly constrained by data on background radiation.
It is the low-mass PBHs whose contributions to the EBL via Hawking
evaporation are the strongest.

Much subsequent effort has gone into the identification of alternative
formation mechanisms which could give rise to a more favourable distribution
of PBH masses (i.e. one peaked at sufficiently high mass to provide the
requisite CDM density without the unwanted background radiation from the
low-mass tail).  For example, PBHs might arise from a post-inflationary
spectrum of density fluctuations which is not perfectly scale-invariant
but has a characteristic length scale of some kind \cite{Khl85}.
The parameter $\zeta$ in (\ref{PBH-IMD}) would then depend on the
inflationary potential (or analogous quantities).  This kind of
dependence has been discussed in the context of two-stage inflation
\cite{Nas85}, extended inflation \cite{Hsu90}, chaotic inflation
\cite{Car93}, ``plateau'' inflation \cite{Iva94}, hybrid inflation
\cite{Gar96} and inflation with isocurvature fluctuations \cite{Yok97}.

A narrow spectrum of masses might also be expected if PBHs formed
during a spontaneous phase transition rather than arising from
primordial fluctuations.  The quark-hadron transition \cite{Cra82},
grand unified symmetry-breaking transition \cite{Kod82} and
Weinberg-Salam phase transition \cite{Haw82} have all been
considered in this regard.  The initial mass distribution in
each case would be peaked near the horizon mass $M_f$ at transition time.
The quark-hadron transition has attracted particular attention because
PBH formation would be enhanced by a temporary softening of the equation
of state; and because $M_f$ for this case is coincidentally close to
$\Msun$, so that PBHs formed at this time might be responsible for
{\sc Macho} observations of microlensing in the halo \cite{Jed97}.
Cosmic string loops have also been explored as possible seeds for
PBHs with a peaked mass spectrum \cite{Haw89,Pol91}.
Considerable interest has recently been generated by the
discovery that PBHs could provide a physical realization of the
theoretical phenomenon known as critical collapse \cite{Nie98}.
If this is so, then initial PBH masses would no longer necessarily
be clustered near $M_f$.

While any of the above proposals can in principle concentrate the PBH
population within a narrow mass range, all of them face the same problem
of {\em fine-tuning\/} if they are to produce the desired present-day
density of PBHs.  In the case of inflationary mechanisms it is the form
of the potential which must be adjusted.  In others it is the bubble
nucleation rate, the string mass per unit length, or the fraction of
the Universe going into PBHs at formation time.  Thus, while modifications
of the initial mass distribution may weaken the ``standard'' constraints
on PBH properties (which we derive below), they do not as yet have a
compelling physical basis.  Similar comments apply to attempts to link
PBHs with specific observational phenomena.  It has been suggested,
for instance, that PBHs with the right mass could be responsible for
certain classes of $\gamma$-ray bursts \cite{Cli92,Cli97,Gre02},
or for long-term quasar variability via microlensing \cite{Haw93,Haw96}.
Other possible connections have been drawn to diffuse $\gamma$-ray emission
from the Galactic halo \cite{Wri96,Cli98} and the {\sc Macho} microlensing
events \cite{Gre99,Gre00}.  All of these suggestions, while intriguing,
would beg the question: ``Why this particular mass scale?''

\subsection{Evolution and density} \label{sec:nPBH}
 
In order to obtain the comoving number density of PBHs from their 
initial mass distribution, we use the fact that PBHs evaporate at
a rate which is inversely proportional to the square of their masses:
\beq
\frac{dM}{dt} = -\frac{\alpha}{M^2} \; . 
\label{PBHevol}
\eeq
This applies to uncharged, non-rotating black holes, which is a reasonable
approximation in the case of PBHs since these objects discharge
quickly relative to their lifetimes \cite{Car74} and also give
up angular momentum by preferentially emitting particles with
spin \cite{Pag76a}.  The parameter $\alpha$ depends in general on
PBH mass $M$ and its behaviour was worked out in detail by Page
in 1976 \cite{Pag76b}.  The most important PBHs are those with
$4.5\times10^{14}\mbox{ g}\leqslant M\leqslant 9.4\times10^{16}$~g.
Black holes in this range are light (and therefore ``hot'') enough to
emit massless particles (including photons) as well as ultra-relativistic
electrons and positrons.  The corresponding value of $\alpha$ is
\beq
\alpha = 6.9 \times 10^{25} \mbox{ g}^3 \mbox{ s}^{-1} \; .
\label{alphaValue}
\eeq
For $M>9.4\times 10^{16}$~g, the value of $\alpha$ drops to
$3.8\times10^{25}\mbox{ g}^3\mbox{ s}^{-1}$ because the larger black
hole is ``cooler'' and no longer able emit electrons and positrons.
EBL contributions from PBHs of this mass are however of lesser
importance because of the shape of the mass distribution.

As the PBH mass drops below $4.5\times10^{14}$~g, its energy
$kT$ climbs past the rest energies of progressively heavier particles,
beginning with muons and pions.  As each mass threshold is passed, the PBH
is able to emit more particles and the value of $\alpha$ increases further.
At temperatures above the quark-hadron transition ($kT\approx200$~MeV), 
MacGibbon and Webber have shown that relativistic quark and gluon jets
are likely to be emitted rather than massive elementary particles
\cite{Mac90}.  These jets subsequently fragment into stable particles,
and the photons produced in this way are actually more important 
(at these energies) than the primary photon flux.  The precise behaviour
of $\alpha$ in this regime depends to some extent on one's choice
of particle physics.  A plot of $\alpha(M)$ for the standard model is
found in the review by Halzen \etal\ \cite{Hal91}, who note that
$\alpha$ climbs to $7.8\times10^{26}\mbox{ g}^3\mbox{ s}^{-1}$ at
$kT=100$~GeV, and that its value would be at least three times higher
in supersymmetric extensions of the standard model
where there are many more particle states to be emitted.

As we will shortly see, however, EBL contributions from PBHs at these
temperatures are suppressed by the fact that the latter have already
evaporated.  If we assume for the moment that PBH evolution is
adequately described by (\ref{PBHevol}) with $\alpha=$~constant
as given by (\ref{alphaValue}), then integration gives
\beq
M(t) = ( M_i^3 - 3 \alpha t )^{\smallfrac{1}{3}} \; .
\label{PBHmass}
\eeq
The lifetime $\tpbh$ of a PBH is found by setting $M(\tpbh)=0$,
giving $\tpbh=M_i^{3}/3\alpha$.
Therefore the initial mass of a PBH which is just disappearing today
($\tpbh=\too$) is given by
\beq
\Mstar = (3\alpha\too)^{\smallfrac{1}{3}} \; .
\label{MstarDefn}
\eeq
Taking $\too=16$~Gyr and using (\ref{alphaValue}) for $\alpha$, we find
that $\Mstar=4.7\times10^{14}$~g.  A numerical analysis allowing for
changes in the value of $\alpha$ over the full range of PBH masses
with $0.06\leqslant\Omato\leqslant1$ and $0.4\leqslant\ho\leqslant1$
leads to a somewhat larger result \cite{Hal91}:
\beq
\Mstar = (5.7 \pm 1.4) \times 10^{14} \mbox{ g} \; .
\label{MstarValue}
\eeq
PBHs with $M\approx\Mstar$ are exploding at redshift $z\approx0$ and
consequently dominate the spectrum of EBL contributions.  The parameter
$\Mstar$ is therefore of central importance in what follows.

We now obtain the comoving number density of PBHs with masses
between $M$ and $M+dM$ at any time $t$.  This is the same as the
comoving number density of PBHs with {\em initial\/} masses between
$M_i$ and $M_i + dM_i$ at formation time, so
$n(M,t) \, dM = n(M_i) \, dM_i$.
Inverting Eq.~(\ref{PBHmass}) to get $M_i=(M^3+3\alpha t)^{1/3}$ and
differentiating, we find from (\ref{PBH-IMD}) that
\beq
n({\mathcal M},\tau) \, d{\mathcal M} = {\mathcal N M}^2
   \left( {\mathcal M}^3 + \tau \right)^{-(\beta+2)/3} \, d{\mathcal M} \; .
\label{nPBHdefn}
\eeq
Here we have used (\ref{MstarDefn}) to replace $\Mstar^3$ with
$3\alpha\too$ and switched to dimensionless parameters
${\mathcal M}\equiv M/\Mstar$ and $\tau\equiv t/\too$.
The quantity ${\mathcal N}$ is formally given in terms of the
parameters at PBH formation time by
${\mathcal N}=(\zeta\,\rho_f/M_f)(M_f/\Mstar)^{\beta-1}$ and has the
dimensions of a number density.  As we will see shortly, it corresponds
roughly to the comoving number density of PBHs of mass $\Mstar$.
Following Page and Hawking \cite{Pag76c}, we allow ${\mathcal N}$
to move up or down as required by observational constraints.
The theory to this point is thus specified by two free parameters: the
PBH normalization ${\mathcal N}$ and the equation-of-state
parameter $\beta$.

To convert to the present mass density of PBHs with mass ratios
between ${\mathcal M}$ and ${\mathcal M}+d{\mathcal M}$, we multiply
Eq.~(\ref{nPBHdefn}) by $M=\Mstar{\mathcal M}$ and put $\tau=1$:
\beq
\rpbh({\mathcal M},1) \, d{\mathcal M} = {\mathcal N} \Mstar {\mathcal M}^{1-\beta}
   \left( 1 + {\mathcal M}^{-3} \right)^{-(\beta+2)/3} \, d{\mathcal M} \; .
\eeq
The total mass in PBHs is then found by integrating over ${\mathcal M}$ 
from zero to infinity.  Changing variables to $x\equiv {\mathcal M}^{-3}$,
we obtain:
\beq
\rpbh = \frac{1}{3} \, {\mathcal N} \Mstar
   \int_0^{\infty} x^{a-1} (1-x)^{-(a+b)} \, dx \; ,
\eeq
where $a\equiv\smallfrac{1}{3}(\beta-2)$ and $b\equiv\smallfrac{4}{3}$.
The integral is solved to give
\beq
\rpbh = \kbeta {\mathcal N} \Mstar \qquad
   \kbeta \equiv \frac{\Gamma(a)\Gamma(b)}{3\Gamma(a+b)} \; ,
\label{rpbhDefn}
\eeq
where $\Gamma(x)$ is the gamma function.  Allowing $\beta$ to take
values from 2 through $\smallfrac{5}{2}$ (the most natural situation)
and up to 4, we find that
\beq
\kbeta = \left\{ \begin{array}{ll}
   \infty & \qquad (\mbox{if } \beta=2) \\
   1.87   & \qquad (\mbox{if } \beta=2.5) \\
   0.88   & \qquad (\mbox{if } \beta=3) \\
   0.56   & \qquad (\mbox{if } \beta=3.5) \\
   0.40   & \qquad (\mbox{if } \beta=4)
   \end{array} \right. \; .
\label{kbetaValue}
\eeq
The total mass density of PBHs in the Universe is thus
$\rpbh\approx {\mathcal N} \Mstar$.  Eq.~(\ref{rpbhDefn}) can be recast
as a relation between the characteristic number density ${\mathcal N}$
and the PBH density parameter $\Opbh=\rpbh/\rcrito$:
\beq
\Opbh = \frac{\kbeta {\mathcal N} \Mstar}{\rcrito} \; .
\label{OpbhDefn}
\eeq
The quantities ${\mathcal N}$ and $\Opbh$ are thus
interchangeable as free parameters.  If we adopt the most natural value
for $\beta$ (=2.5) together with an upper limit on ${\mathcal N}$ due to
Page and Hawking of ${\mathcal N}\lesssim10^4\mbox{ pc}^{-3}$ \cite{Pag76c},
then Eqs.~(\ref{rcritoDefn}), (\ref{MstarValue}), (\ref{kbetaValue})
and (\ref{OpbhDefn}) together imply that $\Opbh$ is at most of order
$\sim\!10^{-8}\ho^{-2}$.  If this upper limit holds (as we confirm
below, then there is little hope for PBHs to make up the dark matter.

Eq.~(\ref{kbetaValue}) shows that one way to boost their importance would be
to assume a soft equation of state at
formation time (i.e. a value of $\beta$ close to 2 as for dust-like matter,
rather than 2.5 as for radiation).  Physically this is related to the
fact that low-pressure matter offers little resistance to gravitational
collapse.  Such a softening has been shown to occur during the quark-hadron
transition \cite{Jed97}, leading to significant increases in $\Opbh$ for
PBHs which form at that time (subject to the fine-tuning problem noted in
Sec.~\ref{sec:PBH-intro}).  For PBHs which arise from primordial density
fluctuations, however, such conditions are unlikely to hold throughout
the formation epoch.  In the limit $\beta\rightarrow2$,
Eq.~(\ref{PBH-IMD}) breaks down in any case because it becomes
possible for PBHs to form on scales smaller than the horizon \cite{Car75}.

\subsection{Spectral energy distribution}

Hawking \cite{Haw75} proved that an uncharged, non-rotating black hole
emits bosons (such as photons) in any given quantum state with energies
between $E$ and $E+dE$ at the rate
\beq
d\dot{N} = \frac{\Gamma_s \, dE}{2\pi\hbar \left[ \exp (E/kT) -1
   \right] } \; .
\label{NdotDefn}
\eeq
Here $T$ is the effective black-hole temperature, and
$\Gamma_s$ is the absorption coefficient or probability that
the same particle would be absorbed by the black hole if incident upon
it in this state.  The function $d\dot{N}$ is related to the spectral
energy distribution (SED) of the black hole by
$d\dot{N}=F(\lambda,{\mathcal M}) \, d\lambda/E$, since we have defined
$F(\lambda,{\mathcal M})\,d\lambda$ as the energy emitted between
wavelengths $\lambda$ and $\lambda+d\lambda$.  We anticipate that
$F$ will depend explicitly on the PBH mass ${\mathcal M}$ 
as well as wavelength.  The PBH SED thus satisfies
\beq
F(\lambda,{\mathcal M}) \, d\lambda = \frac{\Gamma_s E \, dE}{2\pi\hbar
   \left[ \exp (E/kT) -1 \right]} \; .
\label{PBH-SED-E}
\eeq
The absorption coefficient $\Gamma_s$ is a 
function of ${\mathcal M}$ and $E$ as well as the quantum numbers $s$
(spin), $\ell$ (total angular momentum) and $m$ (axial angular momentum)
of the emitted particles.  Its form was first calculated by Page
\cite{Pag76b}.  At high energies, and in the vicinity of the peak
of the emitted spectrum, a good approximation is given by \cite{Mac91}
\beq
\Gamma_s \propto M^{\,2} E^{\,2} \; .
\label{MEapprox}
\eeq
This approximation breaks down at low energies, where it gives rise to
errors of order 50\% for $(G\!M\!E/\hbar c^3)\sim0.05$ \cite{Kri99} or
(with $E=2\pi\hbar c/\lambda$ and $M\sim\Mstar$) for $\lambda\sim10^{-3}$~\AA.
This is adequate for our purposes, as we will find that the strongest
constraints on PBHs come from those with masses $M\sim\Mstar$ at
wavelengths $\lambda\sim10^{-4}$~\AA.

Putting (\ref{MEapprox}) into (\ref{PBH-SED-E}) and making the change of
variable to wavelength $\lambda=hc/E$, we obtain the SED
\beq
F(\lambda,{\mathcal M}) \, d\lambda = \frac{C {\mathcal M}^{\,2}
   \lambda^{-5} \, d\lambda}{\exp (hc/kT\lambda)-1} \; ,
\label{PBH-SED-lam}
\eeq
where $C$ is a proportionality constant.  This has the same form as the
blackbody spectrum, Eq.~(\ref{bbodySED}).  We have made three simplifying
assumptions in arriving at this result.  First, we have neglected the
black-hole charge and spin (as justified in Sec.~\ref{sec:nPBH}).
Second, we have used an approximation for the absorption coefficient
$\Gamma_s$.  And third, we have treated all the emitted photons as if
they are in the same quantum state, 
whereas in fact the emission rate~(\ref{NdotDefn}) applies separately to
the $\ell=s\;(=1)$, $\ell=s+1$ and $\ell=s+2$ modes.  There are thus
actually {\em three distinct\/} quasi-blackbody photon spectra with
different characteristic temperatures for any single PBH.  However
Page \cite{Pag76b} has demonstrated that the $\ell=s$ mode is
overwhelmingly dominant, with the $\ell=s+1$ and $\ell=s+2$ modes
contributing less than 1\% and 0.01\% of the total photon flux 
respectively.  Eq.~(\ref{PBH-SED-lam}) is thus a reasonable approximation 
for the SED of the PBH as a whole.

To fix the value of $C$ we use the fact that the total flux of photons
(in all modes) radiated by a black hole of mass $M$ is given by \cite{Pag76b}
\beq
\dot{N} = \int d\dot{N} = \int_{\lambda=0}^{\,\infty}
   \frac{F(\lambda,{\mathcal M}) \, d\lambda}{hc/\lambda} = 5.97 
   \times 10^{34} \mbox{ s}^{-1} \left( \frac{M}{\mbox{1 g}}
   \right)^{-1} \; .
\eeq
Inserting (\ref{PBH-SED-lam}) and recalling that $M=\Mstar {\mathcal M}$,
we find that
\beq
C \int_0^{\,\infty} 
   \frac{\lambda^{-4} \, d\lambda}{\exp(hc/kT\lambda)-1} = (5.97
   \times 10^{34} \mbox{ g s}^{-1}) \frac{hc}{\Mstar {\mathcal M}^3} \; .
\label{Cintegral}
\eeq
The definite integral on the left-hand side of this equation can be
solved by switching variables to $\nu=c/\lambda$:
\beq
\int_0^{\,\infty} 
   \frac{\nu^2 \, d\nu \, / c^3}{\exp(h\nu/kT)-1} = \left(
   \frac{hc}{kT} \right)^{-3} \!\!\!\! \Gamma(3) \, \zeta(3) \; ,
\label{Mellin}
\eeq
where $\Gamma(n)$ and $\zeta(n)$ are the gamma function and Riemann
zeta function respectively.  We then apply the fact that the temperature
$T$ of an uncharged, non-rotating black hole is given by
\beq
T = \frac{\hbar\,c^3}{8\pi k G M} \; .
\label{Tdefn}
\eeq
Putting (\ref{Mellin}) and (\ref{Tdefn}) into (\ref{Cintegral}) and
rearranging terms leads to
\beq
C = (5.97 \times 10^{34} \mbox{ g s}^{-1}) \, 
   \frac{(4\pi)^6 h \, G^3 \Mstar^2}{c^5 \, \Gamma(3) \, \zeta(3)} \; .
\label{Cdefn}
\eeq
Using $\Gamma(3)=2\mbox{!}=2$ and $\zeta(3)=1.202$ along with
(\ref{MstarValue}) for $\Mstar$, we find
\beq
C = (270 \pm 120) \mbox{ erg \AA}^4 \mbox{ s}^{-1} \; .
\eeq
We can also use the definitions~(\ref{Tdefn}) to define a useful
new quantity:
\beq
\lampbh \equiv \frac{hc}{kT{\mathcal M}} = \left( \frac{4\pi}{c} \right)^2
   \!\!\! G \Mstar = (6.6 \pm 1.6) \times 10^{-4} \mbox{ \AA} \; .
\label{lampbhDefn}
\eeq
The size of this characteristic wavelength tells us that we will be
concerned primarily with the high-energy $\gamma$-ray portion of the
spectrum.  In terms of $C$ and $\lampbh$ the SED~(\ref{PBH-SED-lam})
now reads
\beq
F(\lambda,{\mathcal M}) = \frac{C {\mathcal M}^{\,2} / \lambda^5}
   {\exp ({\mathcal M}\lampbh/\lambda)-1} \; .
\label{PBH-SED}
\eeq
While this contains no explicit time-dependence, the spectrum does of
course depend on time through the PBH mass ratio ${\mathcal M}$.
To find the PBH luminosity we employ Eq.~(\ref{Fnorm}), integrating
the SED~$F(\lambda,{\mathcal M})$ over all $\lambda$ to obtain:
\beq
L({\mathcal M}) = C {\mathcal M}^{\,2} \int_{0}^{\,\infty}
   \frac{\lambda^{-5} \, d\lambda}{\exp({\mathcal M}\lampbh/\lambda)-1} \; .
\label{LpbhDefn}
\eeq
This definite integral is also solved by means of a change of variable
to frequency $\nu$, with the result that
\beq
L({\mathcal M}) = C {\mathcal M}^{\,2} \left( {\mathcal M} \lampbh \right)^{-4}
   \Gamma(4) \, \zeta(4) \; .
\eeq
Using Eqs.~(\ref{MstarValue}), (\ref{Cdefn}) and~(\ref{lampbhDefn})
along with the values $\Gamma(4)=3\mbox{!}=6$ and $\zeta(4)=\pi^4/96$,
we can put this into the form
\beq
L({\mathcal M}) = \Lpbh \, {\mathcal M}^{-2} \; ,
\label{LpbhValue}
\eeq
where 
\beqa
\Lpbh & = & \frac{(5.97 \times 10^{34} \mbox{ g s}^{-1}) \,
   \pi^2 h c^{\,3}}{512 \, \zeta(3) \, G \Mstar^{\,2}} = (1.0 \pm 0.4)
   \times 10^{16} \mbox{ erg s}^{-1} \; .
\nonumber
\eeqa
Compared to the luminosity of an ordinary star, the typical PBH
(of mass ratio ${\mathcal M}\approx 1$) is not very luminous.  A PBH of
900~kg or so might theoretically be expected to reach the Sun's
luminosity; however, in practice it would already have exploded,
having long since reached an effective temperature high enough to
emit a wide range of massive particles as well as photons.
The low luminosity of black holes in general can be emphasized by
using the relation ${\mathcal M}\equiv M/\Mstar$ to recast
Eq.~(\ref{LpbhValue}) in the form
\beq
L/\Lsun = 1.7 \times 10^{-55} (M/\Msun)^{-2} \; .
\eeq
This expression is not strictly valid for PBHs of masses near $\Msun$,
having been derived for those with $M\sim\Mstar\sim10^{15}$~g.
(Luminosity is {\em lower\/} for larger black holes, and one of
solar mass would be so much colder than the CMB that it would absorb
radiation faster than it could emit it.)  So, Hawking evaporation or not,
most black holes are indeed very black.

\subsection{Bolometric intensity} \label{sec:QPBH}

To obtain the total bolometric intensity of PBHs out to a
look-back time $t_f$, we substitute the PBH number density~(\ref{nPBHdefn})
and luminosity~(\ref{LpbhDefn}) into the integral~(\ref{QtDefn}) as usual.
Now however the number density $n(t)$ is to be replaced by
$n({\mathcal M},\tau)\,d{\mathcal M}$, $L({\mathcal M})$ takes
the place of $L(t)$, and we integrate over all PBH masses
${\mathcal M}$ as well as look-back times $\tau$:
\beq
Q = \Qpbh \, \Opbh \int_{\tauf}^1 \Rtil(\tau) \, d\tau \int_0^{\,\infty}
   \frac{d{\mathcal M}}{({\mathcal M}^3 + \tau)^{\,\varepsilon}} \; ,
\label{QpbhInt}
\eeq
where
\beq
\Qpbh = \frac{c\too\rcrito\Lpbh}{\kbeta\Mstar} \; .
\label{QpbhDefn}
\eeq
Here $\varepsilon\equiv(\beta+2)/3$ and we have used (\ref{OpbhDefn})
to replace ${\mathcal N}$ with $\Opbh$.  In principle, the integral over
${\mathcal M}$ should be cut off at a finite lower limit $\Mcut(\tau)$,
equal to the mass of the lightest PBH which has not yet evaporated at
time $\tau$.  This arises because the initial PBH mass
distribution~(\ref{PBH-IMD}) requires a nonzero minimum $\Mmin$ in order
to avoid divergences at low mass.  In practice, however, the cutoff
rapidly evolves toward zero from its its initial value of
$\Mcut(0)=\Mmin/\Mstar$.  If $\Mmin$ is of the order of the Planck mass
as usually suggested \cite{Bar91}, then $\Mcut(\tau)$ drops to zero
well before the end of the radiation-dominated era.  Since we are
concerned with times later than this, we can safely set $\Mcut(\tau)=0$.

Eq.~(\ref{QpbhInt}) can be used to put a rough upper limit on $\Opbh$
from the bolometric intensity of the background light \cite{OW92}.
Let us assume that the Universe is flat, as suggested by most
observations (Sec.~\ref{ch4}).  Then its age $\too$ can be obtained
from Eq.~(\ref{t2comp}) as
\beq
\too = 2\,\tilto/(3\,\Ho) \; .
\label{toDefn}
\eeq
Here $\tilto\equiv \tiltm(0)$ where $\tiltm(z)$ is the dimensionless function
\beq
\tiltm(z) \equiv \frac{1}{\sqrt{1-\Omato}} \, \sinh^{-1} 
   \!\!\sqrt{\frac{1-\Omato}{\Omato(1+z)^3}} \; .
\label{tiltmDefn}
\eeq
Putting (\ref{toDefn}) into (\ref{QpbhDefn}) and using Eqs.~(\ref{HoValue}),
(\ref{rcritoDefn}), (\ref{MstarValue}) and (\ref{LpbhValue}), we find:
\beqa
\Qpbh & = & \frac{(5.97 \times 10^{34} \mbox{ g s}^{-1}) \, \pi^2 h \, c^{\,4}
      \rcrito \, \tilto} {768 \, \zeta(3) G \Ho \, \kbeta\Mstar^3} \nonumber \\
   & = & (2.3 \pm 1.4) \, \ho \, \tilto \, \kbeta^{-1}
      \mbox{ erg s}^{-1} \mbox{ cm}^{-2} \; .
\label{QpbhValue}
\eeqa
We are now ready to evaluate Eq.~(\ref{QpbhInt}).  To begin with we
note that the integral over mass has an analytic solution:
\beq
\int_0^{\,\infty} \frac{d{\mathcal M}}{({\mathcal M}^3 + \tau)^{\,\varepsilon}} =
   \keps \tau^{\frac{1}{3}-\varepsilon} \qquad
\keps \equiv \frac{\Gamma(\smallfrac{1}{3}) \, \Gamma(\varepsilon-
   \smallfrac{1}{3})}{3 \, \Gamma(\varepsilon)} \; .
\label{kepsDefn}
\eeq
For the EdS case ($\Omato=1$), $\tilto=1$ and Eq.~(\ref{Rt1comp}) implies:
\beq
\Rtil(\tau)=\tau^{2/3} \; .
\label{RtilEdS}
\eeq
Putting Eqs.~(\ref{kepsDefn}) and (\ref{RtilEdS}) into (\ref{QpbhInt}),
we find that
\beq
Q = \Qpbh\,\Opbh\,\keps \int_{\tauf}^1 \tau^{1-\varepsilon} \, d\tau =
   \Qpbh\,\Opbh\,\keps \left( \frac{1 - \tauf^{\,2-\varepsilon}}{2-\varepsilon} 
   \right) \; .
\label{QEdS}
\eeq
The parameter $\tauf$ is obtained for the EdS case by inverting
(\ref{RtilEdS}) to give $\tauf=(1+z_f)^{-3/2}$.  The subscript ``$f$''
(``formation'') is here a misnomer since we do not integrate back to
PBH formation time, which occurred in the early stages of the
radiation-dominated era.  Rather we integrate out to the redshift
at which processes like pair production become significant enough to render
the Universe approximately opaque to the (primarily $\gamma$-ray) photons
from PBH evaporation.  Following Kribs \etal\ \cite{Kri99} this is
$z_f\approx700$.

Using this value of $z_f$ and substituting Eqs.~(\ref{QpbhValue}) and
(\ref{kepsDefn}) into (\ref{QEdS}), we find that the bolometric intensity
of background radiation due to evaporating PBHs in an EdS Universe is
\beq
Q = \ho \, \Opbh \times \left\{ \begin{array}{ll}
0 & \qquad (\mbox{if } \beta=2) \\
2.3 \pm 1.4 \mbox{ erg sec}^{-1} \mbox{ cm}^{-2} &
   \qquad (\mbox{if } \beta=2.5) \\
6.6 \pm 4.2 \mbox{ erg sec}^{-1} \mbox{ cm}^{-2} &
   \qquad (\mbox{if } \beta=3) \\
17 \pm 10 \mbox{ erg sec}^{-1} \mbox{ cm}^{-2} &
   \qquad (\mbox{if } \beta=3.5) \\
45 \pm 28 \mbox{ erg sec}^{-1} \mbox{ cm}^{-2} &
   \qquad (\mbox{if } \beta=4)
\end{array} \right. \; .
\label{QpbhResults}
\eeq
This vanishes for $\beta=2$ because $\kbeta\rightarrow\infty$
in this limit, as discussed in Sec.~\ref{sec:nPBH}.  The case $\beta=4$
(i.e. $\varepsilon=2$) is evaluated with the help of L'H\^opital's rule,
which gives $\lim_{\varepsilon\rightarrow2} (1-\tauf^{2-\varepsilon})/
(2-\varepsilon)=-\ln\tauf$.

The values of $Q$ in Eq.~(\ref{QpbhResults}) are far higher than the 
actual bolometric intensity of background radiation in an EdS universe,
$\smallfrac{2}{5}\Qstar=1.0\times10^{-4}\mbox{ erg s}^{-1}\mbox{ cm}^{-2}$
(Fig.~2.6).  Moreover this background is already well accounted for by
known astrophysical sources.  A firm upper bound on $\Opbh$ (for the most
natural situation with $\beta=2.5$) is therefore
\beq
\Opbh < (4.4 \pm 2.8) \times 10^{-5} \, \ho^{-1} \; .
\label{PBHbound1}
\eeq
For harder equations of state ($\beta>2.5$) the PBH density would have to
be even lower.  PBHs in the simplest formation scenario are thus eliminated
as important dark-matter candidates, even without reference to the cosmic
$\gamma$-ray background.

For models containing dark energy as well as baryons and black holes,
the integrated background intensity goes up because the Universe is older,
and down because $Q\propto\Opbh$.  The latter effect is stronger,
so that the above constraint on $\Opbh$ will be weaker in a model
such as \LCDM\ (with $\Omato=0.3,\Olamo=0.7$).  To determine the
importance of this effect, we can re-evaluate the integral~(\ref{QpbhInt})
using the general formula~(\ref{Rt2comp}) for $\Rtil(\tau)$ in place
of (\ref{RtilEdS}).  We will make the minimal assumption that PBHs
constitute the {\em only\/} CDM, so that $\Omato=\Opbh+\Obar$ with
$\Obar$ given by (\ref{ObarValue}) as usual.  Eq.~(\ref{t2comp}) shows that
the parameter $\tauf$ is given for arbitrary values of $\Omato$ by
$\tauf=\tiltm(z_f)/\tilto$ where the function $\tiltm(z)$ is defined
as before by (\ref{tiltmDefn}).

\begin{figure}[t!]
\begin{center}
\includegraphics[width=100mm]{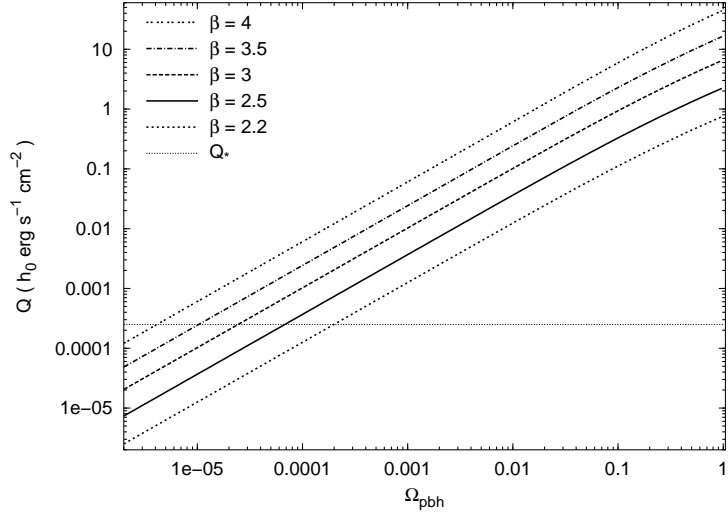}
\end{center}
\caption{The bolometric intensity due to evaporating primordial black
   holes as a function of their collective density $\Opbh$ and the
   equation-of-state parameter $\beta$.  We have assumed that 
   $\Omato=\Obar+\Opbh$ with $\Obar=0.016\ho^{-2}$, $\ho=0.75$
   and $\Olamo=1-\Omato$.  The horizontal dotted line indicates
   the approximate bolometric intensity ($\Qstar$) of the observed EBL.}
\label{fig9.1}
\end{figure}

Evaluation of Eq.~(\ref{QpbhInt}) leads to the plot of bolometric
intensity $Q$ versus $\Opbh$ in Fig.~\ref{fig9.1}.
As before, $Q$ is proportional to $\ho$ because it goes as both
$\rpbh=\Opbh\rcrito\propto\ho^2$ and $\too\propto\ho^{-1}$.
Since $Q\rightarrow0$ for $\beta\rightarrow2$ we have chosen a
minimum value of $\beta=2.2$ as representative of ``soft'' conditions.
Fig.~\ref{fig9.1} confirms that, regardless of cosmological model, PBH
contributions to the background light are too high unless $\Opbh\ll1$.
The values in Eq.~(\ref{QpbhResults}) are recovered at the right-hand
edge of the figure where $\Opbh$ approaches one, as expected.  For all
other models, if we impose a conservative upper bound $Q<\Qstar$ (as
indicated by the faint dotted line) then it follows that
$\Opbh<(6.9\pm4.2)\times 10^{-5}\ho^{-1}$ for $\beta=2.5$.
This is about 60\% higher than the limit~(\ref{PBHbound1})
for the EdS case.

\subsection{Spectral intensity}

Stronger limits on PBH density can be obtained from the
$\gamma$-ray background, where these
objects contribute most strongly to the EBL and where we have good data
(as summarized in Sec.~\ref{sec:WIMPdiscussion}).
Spectral intensity is found as usual by substituting the comoving PBH
number density~(\ref{nPBHdefn}) and SED~(\ref{PBH-SED}) into
Eq.~(\ref{ItDefn}).  As in the bolometric case, we now have to
integrate over PBH mass ${\mathcal M}$ as well as time $\tau=t/\too$, so that
\beq
\Ilam(\lamo) = \frac{c\too}{4\pi} \int_{\tauf}^{1} \Rtil^2(\tau) \, d\tau
   \int_{\Mcut(\tau)}^{\,\infty} n({\mathcal M},\tau) \,
   F(\Rtil\lamo,{\mathcal M}) \, d{\mathcal M} \; .
\label{IpbhTemp}
\eeq
Following the discussion in Sec.~\ref{sec:QPBH} we set $\Mcut(\tau)=0$.
In light of our bolometric results it is unlikely that PBHs make up
a significant part of the dark matter, so we no longer tie the value
of $\Omato$ to $\Opbh$.  Models with $\Omato\gtrsim\Obar$ must therefore
contain a second species of cold dark matter (other than PBHs) to
provide the required matter density.  Putting (\ref{nPBHdefn}) and
(\ref{PBH-SED}) into (\ref{IpbhTemp}) and using (\ref{MstarValue}),
(\ref{OpbhDefn}) and (\ref{Cdefn}), we find that
\beq
\Ilam(\lamo) = \Ipbh \, \Opbh \int_{\tauf}^1 \Rtil^{-3}(\tau) \, d\tau
   \int_0^{\,\infty} \!\!\! \frac{{\mathcal M}^4 ({\mathcal M}^3+\tau)^{-\varepsilon}
   \, d{\mathcal M}}{\exp\left[ \lampbh {\mathcal M}/\Rtil(\tau) \lamo\right] - 1}
   \; .
\label{IpbhDefn}
\eeq
Here the dimensional prefactor is a function of both $\beta$ and $\lamo$
and reads
\beqa
\Ipbh & = & \frac{(5.97 \times 10^{34} \mbox{ g s}^{-1}) (4\pi)^5 G^{\,3}
      \Mstar \rcrito \, \tilto}{3 \, \zeta(3) \, c^5 \, \kbeta \, \Ho \,
      \lamo^4} \nonumber \\
   & = & \left[ ( 2.1 \pm 0.5 ) \times 10^{-7} \mbox{ CUs} \right]
      \ho \, \kbeta^{-1} \tilto \left( \frac{\lamo}{\mbox{\AA}} \right)^{\!\!-4}
      \; .
\label{IpbhValue}
\eeqa
We have divided through by the photon energy $hc/\lamo$ to put
the results in units of CUs as usual.  The range of uncertainty in
$\Ilam(\lamo)$ is smaller than that in $Q$, Eq.~(\ref{QpbhValue}),
because $\Ilam(\lamo)$ depends only linearly on $\Mstar$ whereas $Q$
is proportional to $\Mstar^{-3}$.
(This in turn results from the fact that $\Ilam\propto C\propto\Mstar^{\,2}$
whereas $Q\propto\Lpbh\propto\Mstar^{-2}$.  One more factor of $\Mstar^{-1}$
comes from ${\mathcal N}\propto\rpbh/\Mstar$ in both cases.)
Like $Q$, $\Ilam$ depends linearly on $\ho$ since integrated intensity
in either case is proportional to both $\rpbh\propto\rcrito\propto\ho^2$
and $\too\propto\ho^{-1}$.

\begin{figure}[t!]
\begin{center}
\includegraphics[width=\textwidth]{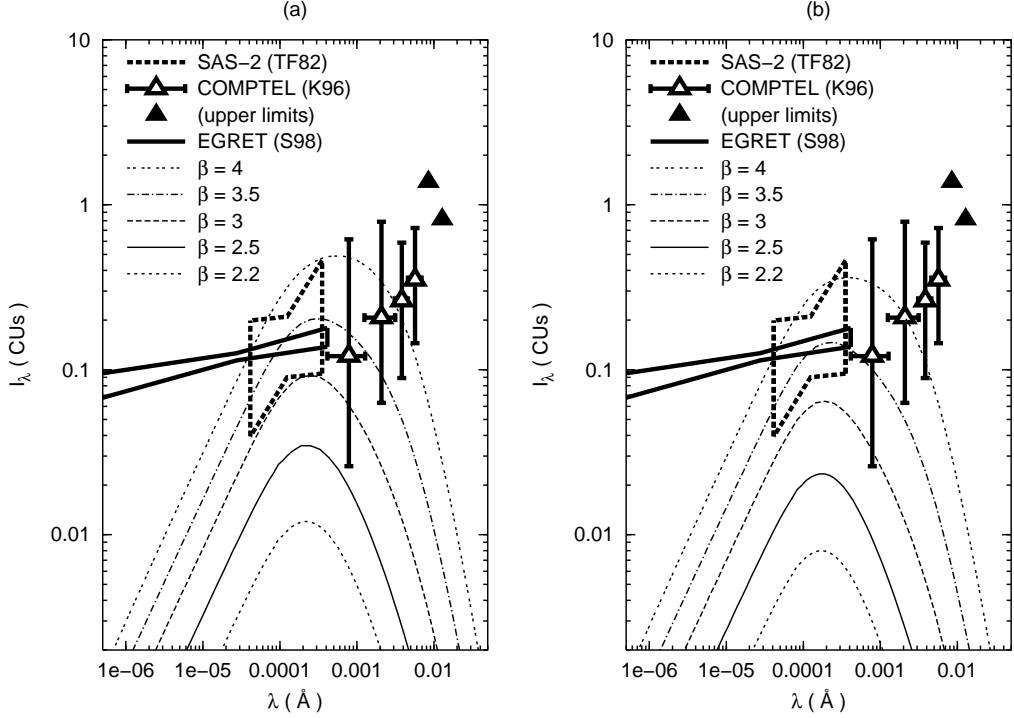}
\end{center}
\caption{The spectral intensity of the diffuse $\gamma$-ray background
   from evaporating primordial black holes in flat models, as compared
   with experimental limits from {\sc Sas}-2, {\sc Comptel} and {\sc Egret}.
   Panel~(a) assumes $\Omato=0.06$, while Panel~(b) is plotted for
   $\Omato=1$ (the EdS case).  All curves assume $\Opbh=10^{-8}$ and
   $\ho=0.75$.}
\label{fig9.2}
\end{figure}

Numerical integration of Eq.~(\ref{IpbhDefn}) leads to the
plots shown in Fig.~\ref{fig9.2}, where we have set $\Opbh=10^{-8}$.
Following Page and Hawking \cite{Pag76c} we have chosen values of
$\Omato=0.06$ in panel~(a) and $\Omato=1$ in panel~(b).  (Results are
not strictly comparable in the former case, however, since we assume
that $\Olamo=1-\Omato$ rather than $\Olamo=0$.)  Our results are in
good agreement with the earlier ones except at the longest wavelengths
(lowest energies), where PBH evaporation is no longer well described
by a simple blackbody SED, and where the spectrum begins to be affected
by pair production on nuclei.  As expected the spectra peak near
$10^{-4}$~\AA\ in the $\gamma$-ray region.  Also plotted in
Fig.~\ref{fig9.2} are the data from {\sc Sas}-2 (\cite{Tho82};
heavy dashed line), {\sc Comptel} (\cite{Kap96}; triangles) and
{\sc Egret} (\cite{Sre98}; heavy solid line).

By adjusting the value of $\Opbh$ up or down from its value of $10^{-8}$
in Fig.~\ref{fig9.2}, we can match the theoretical PBH spectra to those
measured (e.g., by {\sc Egret}), thereby obtaining the maximum value of
$\Opbh$ consistent with observation.  For $\beta=2.5$ this results in
\beq
\Opbh < \left\{ \begin{array}{ll}
   (4.2 \pm 1.1) \times 10^{-8} \, \ho^{-1}
      & \qquad (\mbox{if } \Omato=0.06) \\
   (6.2 \pm 1.6) \times 10^{-8} \, \ho^{-1}
      & \qquad (\mbox{if } \Omato=1)
\end{array} \right. \; .
\eeq
These limits are three orders of magnitude stronger than the one from
bolometric intensity, again confirming that PBHs in the simplest
formation scenario cannot be significant contributors to the dark matter.
Using (\ref{OpbhDefn}) this result can be translated into an upper limit
on ${\mathcal N}$:
\beq
{\mathcal N} < \left\{ \begin{array}{ll}
   (2.2 \pm 0.8) \times 10^4 \, \ho \mbox{ pc}^{-3}
      & \qquad (\mbox{if } \Omato=0.06) \\
   (3.2 \pm 1.1) \times 10^4 \, \ho \mbox{ pc}^{-3}
      & \qquad (\mbox{if } \Omato=1)
\end{array} \right. \; .
\label{Nbound}
\eeq
These numbers are in good agreement with the original Page-Hawking bound
of ${\mathcal N} < 1 \times 10^4 \mbox{ pc}^{-3}$ \cite{Pag76c}, which
was obtained for $\ho=0.6$.

Subsequent workers have refined the $\gamma$-ray background constraints
on $\Opbh$ and ${\mathcal N}$ in a number of ways.  MacGibbon and Webber
\cite{Mac90} pointed out that PBHs whose effective temperatures have
climbed above the rest energy of hadrons probably give off more
photons by {\em indirect\/} processes than by direct emission.
This occurs because it is not bound states (i.e. hadrons) that are
most likely to be emitted, but their elementary constituents (quarks
and gluons in the form of relativistic jets).  Accelerator experiments
and numerical simulations indicate that these jets subsequently
fragment into secondary particles whose decays (especially those of
the pions) produce a far greater flux of photons than that emitted
directly from the PBH.  The net effect is to increase the PBH
luminosity, particularly in low-energy $\gamma$-rays, strengthening
the constraint on $\Opbh$ by about an order of magnitude \cite{Mac91}.
The most recent upper limit obtained in this way using {\sc Egret} data
(assuming $\Omato=1$) is $\Opbh<(5.1\pm1.3)\times10^{-9}\ho^{-2}$
\cite{Car98b}.

Complementary upper limits on PBH contributions to the dark matter have
come from {\em direct\/} searches for those evaporating within a few kpc
of the Earth.  Such limits are subject to more uncertainty than ones
based on the EBL because they depend on assumptions about the degree
to which PBHs are clustered.  If there is no clustering then
(\ref{PBHbound1}) can be converted into a stringent upper bound
on the local PBH evaporation rate, $\dot{\mathcal N}<10^{-7}\Runits$.
This however relaxes to $\dot{\mathcal N}\lesssim10\Runits$ if PBHs
are strongly clustered \cite{Hal91}, in which case limits from direct
searches could potentially become competitive with those based on the EBL.
Data taken at energies near 50~TeV with the {\sc Cygnus} air-shower array
has led to a bound of $\dot{\mathcal N}<8.5\times10^5\Runits$ \cite{Ale93},
and a comparable limit of $\dot{\mathcal N}<(3.0\pm1.0)\times10^6\Runits$
has been obtained at 400~GeV using an imaging atmospheric \v{C}erenkov
technique developed by the Whipple collaboration \cite{Con98}.
Very strong constraints have also been claimed based on balloon observations
of cosmic-ray antiprotons \cite{Mak96}.

Other ideas have been advanced which could weaken the bounds on
PBHs as dark-matter candidates.  It might be, for instance, that these
objects leave behind stable relics rather than evaporating completely
\cite{Mac87}, a possibility that has recently been revived on the grounds
that total evaporation would be inconsistent with a generalized gravitational
version of the uncertainty principle \cite{Adl01}.  This however raises a
new problem (similar to the ``gravitino problem'' discussed in
Sec.~\ref{sec:Gravitino}) because such relics would have been overproduced
by quantum and thermal fluctuations in the early Universe.  Inflation can
be invoked to reduce their density, but must be finely tuned if the same
relics are to make up an interesting fraction of the dark matter today
\cite{Bar92}. 

A different suggestion due to Heckler \cite{Hec97a,Hec97b} 
has been that particles emitted from the black hole might interact
strongly enough above a critical temperature to form a photosphere.
This would make the PBH appear cooler as seen from a distance than its
actual surface temperature, just as the solar photosphere makes the Sun
appear cooler than its core.  (In the case of the black hole, however,
one has not only an electromagnetic photosphere but a QCD ``gluosphere.'')
The reality of this effect is still under debate \cite{Car98b},
but preliminary calculations indicate that it could reduce the intensity
of PBH contributions to the $\gamma$-ray background by 60\% at 100~MeV,
and by as much as two orders of magnitude at 1~GeV \cite{Cli99}.

Finally, as discussed already in Sec.~\ref{sec:PBH-intro}, the limits
obtained above can be weakened or evaded if PBH formation occurs in
such a way as to produce fewer low-mass objects.  The challenge
faced in such proposals is to explain how a distribution of this kind comes
about in a natural way.  A common procedure is to turn the question
around and use observational data on the present intensity of the
$\gamma$-ray background as a probe of the original PBH formation mechanism.
Such an approach has been applied, for example, to put constraints
on the spectral index of density fluctuations in the context of 
PBHs which form via critical collapse \cite{Kri99}
or inflation with a ``blue'' or tilted spectrum \cite{Kim99}.
Thus, even should they turn out not to exist, primordial black holes
provide a valuable window on conditions in the early Universe,
where information is otherwise scarce.

\subsection{Solitons}

In view of the fact that conventional black holes are disfavoured
as dark-matter candidates, it is worthwhile to consider alternatives.
One of the simplest of these is the extension of the black-hole concept
from the four-dimensional (4D) spacetime of general relativity to higher
dimensions.  Higher-dimensional relativity, also known as Kaluza-Klein
gravity, has a long history and underlies modern attempts to unify
gravity with the standard model of particle physics \cite{OW97}.
The extra dimensions have traditionally been assumed to be
compact, in order to explain their non-appearance in low-energy physics.
The past few years, however, have witnessed a surge of interest
in {\em non\/}-compactified theories of higher-dimensional gravity
\cite{Wes96,Ark98,Ran99}.  In such theories the dimensionality
of spacetime can manifest itself at experimentally accessible energies.
We focus on the prototypical five-dimensional (5D) case, although the
extension to higher dimensions is straightforward in principle.

Black holes are described in 4D general relativity by the Schwarzschild
metric, which reads (in isotropic coordinates)
\beq
ds^2 = \left( \frac{1 - GM_s/2c^{\,2}r}{1 + GM_s/2c^{\,2}r} \right)^{\!\!2} \!
   c^2 dt^2 - \left( 1 + \frac{GM_s}{2c^{\,2}r} \right)^{\!\!4} \!
   (dr^2 + r^2 \, d\Omega^2 ) \; ,
\label{Schwarz}
\eeq
where $d\Omega^2 \equiv d\theta^2 + \sin^2\theta d\phi^2$.  This is a
description of the static, spherically-symmetric spacetime around a
pointlike object (such as a collapsed star or primordial density
fluctuation) with Schwarzschild mass $M_s$.  As we have seen,
it is unlikely that such objects can make up the dark matter.

If the Universe has more than four dimensions, then the same object
must be modelled with a higher-dimensional analog of the Schwarzschild
metric.  Various possibilities have been explored over the years, with
most attention focusing on a 5D solution first discussed in detail by
Gross and Perry \cite{Gro83}, Sorkin \cite{Sor83} and Davidson and
Owen \cite{Dav85} in the early 1980s.  This is now generally known
as the {\em soliton metric\/} and reads:
\beqa
ds^2 = \left( \frac{ar-1}{ar+1} \right)^{2\xi\kappa}
       c^2 dt^2 & - & \left( \frac{a^2 r^2-1}{a^2 r^2} \right)^{2}
       \left( \frac{ar+1}{ar-1} \right)^{2\xi(\kappa-1)}
       (dr^2 + r^2 \, d\Omega^2 ) \nonumber \\
   & - & \left( \frac{ar+1}{ar-1} \right)^{2\xi}
       dy^2 \; .
\label{Soliton}
\eeqa
Here $y$ is the new coordinate and there are three metric parameters
($a,\xi,\kappa$) rather than just one ($M_s$) as in Eq.~(\ref{Schwarz}).
Only two of these are independent, however, because a consistency 
condition (which follows from the field equations) requires that
$\xi^2(\kappa^2-\kappa+1)=1$.  In the limit where $\xi\rightarrow 0$,
$\kappa\rightarrow\infty$ and $\xi\kappa\rightarrow1$, Eq.~(\ref{Soliton})
reduces to (\ref{Schwarz}) on 4D hypersurfaces $y=$~const.  In this limit
we can also identify the parameter $a$ as $a=2c^{\,2}/GM_s$ where $M_s$
is the Schwarzschild mass.

We wish to understand the physical properties of this solution in
four dimensions.  To accomplish this we do two things.  First, we
assume that Einstein's field equations in their usual form hold in the
full {\em five\/}-dimensional spacetime.  Second, we assume that the
Universe in five dimensions is {\em empty\/}, with no 5D matter fields
or cosmological constant.  The field equations then simplify to
\beq
{\mathcal R}_{AB} = 0 \; .
\label{5DEFEs}
\eeq
Here ${\mathcal R}_{AB}$ is the 5D Ricci tensor,
defined in exactly the same way as the
4D one except that spacetime indices $A,B$ run over 0--4 instead
of 0--3.  Putting a 5D metric such as (\ref{Soliton}) into the vacuum
5D field equations~(\ref{5DEFEs}), we recover the 4D field
equations~(\ref{EFEs}) with a nonzero energy-momentum tensor $\Tmat$.
{\em Matter and energy, in other words, are induced in 4D by pure
geometry in 5D.\/}  It is by studying the properties of this
induced-matter energy-momentum tensor ($\Tmat$)
that we learn what the soliton looks like in four dimensions.

The details of the mechanism just outlined \cite{Wes92} and its
application to solitons in particular \cite{Liu92,Wes94a} have been
well studied and we do not review this material here.
It is important to note, however, that the Kaluza-Klein soliton
differs from an ordinary black hole in several key respects.  It 
contains a singularity at its center, but this center is located at
$r=1/a$ rather than $r=0$.  (The point $r=0$ is, in fact, not even part
of the manifold, which ends at $r=1/a$.)  Its event horizon also
shrinks to a point at $r=1/a$.  For these reasons the soliton is better
classified as a naked singularity than a black hole.

Solitons in the induced-matter picture are further distinguished from
conventional black holes by the fact that they have an
extended matter distribution rather than having all their mass
compressed into the singularity.  It is this feature which proves to be
of most use to us in putting constraints on solitons as dark-matter
candidates \cite{Wes94b}.  The time-time component of the induced-matter
energy-momentum tensor gives us the density of the solitonic fluid as
a function of radial distance:
\beq
\rho_s(r) = \frac{c^{\,2} \xi^2 \kappa a^6 r^4}{2\pi G (ar-1)^4 (ar+1)^4}
   \left( \frac{ar-1}{ar+1} \right)^{\!\!2\xi (\kappa-1)} \; .
\label{SolDens}
\eeq
From the other elements of $\Tmat$ one finds that pressure can be
written $p_s=\smallfrac{1}{3}\rho_s c^{\,2}$, so that the soliton has
a radiation-like equation of state.  In this respect the soliton
more closely resembles a primordial black hole (which forms during the
radiation-dominated era) than one which arises as the endpoint of stellar
collapse.  The elements of $\Tmat$ can also be used to calculate the
gravitational mass of the fluid inside $r$:
\beq
M_g(r) = \frac{2 c^{\,2} \xi\kappa}{G a} \left( \frac{ar-1}{ar+1}
   \right)^{\!\!\xi} \; .
\label{SolMass}
\eeq
At large distances $r\gg 1/a$ from the center the soliton's
density~(\ref{SolDens}) and gravitational mass~(\ref{SolMass}) go over to
\beq
\rho_s(r) \rightarrow \frac{c^{\,2} \xi^2 \kappa}{2\pi G a^2 r^4} \qquad
   M_g(r) \rightarrow M_g(\infty) = \frac{2 c^{\,2} \xi\kappa}{G a} \; .
\label{RhoMg}
\eeq
The second of these expressions shows that the asymptotic value of $M_g$
is in general not the same as $M_s$ [$M_g(\infty)=\xi\kappa M_s$
for $r\gg1/a$], but reduces to it in the limit $\xi\kappa \rightarrow 1$.
Viewed in four dimensions, the soliton resembles a hole in the geometry
surrounded by a spherically-symmetric ball of ultra-relativistic matter
whose density falls off at large distances as $1/r^4$.  If the Universe
does have more than four dimensions, then objects like this should be
common, being generic to 5D Kaluza-Klein gravity in exactly the same
way that black holes are to 4D general relativity.

We therefore assess their impact on the background radiation, assuming
that the fluid making up the soliton is in fact composed of photons
(although one might also consider ultra-relativistic particles such
as neutrinos in principle).  We do not have spectral information on
these so we proceed bolometrically.  Putting the second of
Eqs.~(\ref{RhoMg}) into the first gives
\beq
\rho_s(r) \approx \frac{G M_g^{\,2}}{8\pi c^{\,2} \kappa \, r^4} \; .
\label{rhosDefn}
\eeq
Numbers can be attached to the quantities $\kappa,r$ and $M_g$ as follows.
The first ($\kappa$) is technically a free parameter.  However, a natural
choice from the physical point of view is $\kappa\sim 1$.  For this case the
consistency relation implies $\xi\sim 1$ also, guaranteeing that
the asymptotic gravitational mass of the soliton is close to its
Schwarzschild one.  To obtain a value for $r$, let us assume that
solitons are distributed homogeneously through space with average
separation $d$ and mean density
$\bar{\rho}_s = \Omega_s\rcrito = M_s/d^{\,3}$.  Since $\rho_s$ drops
as $r^{-4}$ whereas the number of solitons at a distance $r$ climbs only
as $r^3$, the local density of solitons is largely determined by the nearest
one.  We can therefore replace $r$ by $d=(M_s/\Omega_s\rcrito)^{1/3}$.
The last unknown in (\ref{rhosDefn}) is the soliton mass $M_g$ ($=M_s$
if $\kappa=1$).  The fact that $\rho_s \propto r^{-4}$ is reminiscent
of the density profile of the Galactic dark-matter halo,
Eq.~(\ref{JaffeDefn}).  Theoretical work on the classical tests
of 5D general relativity \cite{Liu00} and limits on violations of the
equivalence principle \cite{Ove00} also suggests that solitons are
likely to be associated with dark matter on galactic or larger scales.
Let us therefore express $M_s$ in units of the mass of the Galaxy,
which from (\ref{MhValue}) is $\Mgal\approx2\times10^{12}\Msun$.
Eq.~(\ref{rhosDefn}) then gives the local energy density of
solitonic fluid as
\beq
\rho_s \, c^{\,2} \approx (3 \times 10^{-17} \mbox{ erg cm}^{-3}) \, \ho^{8/3}
   \, \Omega_s^{4/3} \left( \frac{M_s}{\Mgal} \right)^{2/3} \; .
\eeq
To get a characteristic value, we take $M_s=\Mgal$ and adopt our
usual values $\ho=0.75$ and $\Omega_s=\Ocdm=0.3$.  Let us moreover 
compare our result to the average energy density of the CMB, which
dominates the spectrum of background radiation (Fig.~\ref{fig1.2}).
The latter is found from Eq.~(\ref{OgamValue}) as
$\rcmb\,c^{\,2} = \Ogam\,\rcrito\,c^{\,2} = 4 \times 10^{-13} 
\mbox{ erg cm}^{-3}$.  We therefore obtain
\beq
\rho_s/\rcmb \approx 7 \times 10^{-6} \; .
\eeq
This is of the same order of magnitude as the limit set on anomalous
contributions to the CMB by {\sc Cobe} and other experiments.
Thus the dark matter could consist of solitons, if they are
not more massive than galaxies.
Similar arguments can be made on the basis of tidal effects and
gravitational lensing \cite{Wes94b}.  To go further and put more
detailed constraints on these candidates from background radiation
or other considerations will require a deeper investigation of
their microphysical properties.

Let us summarize our results for this section.  We have noted that
standard (stellar) black holes cannot provide the dark matter insofar
as their contributions to the density of the Universe are effectively
baryonic.  Primordial black holes evade this constraint, but we have
reconfirmed the classic results of Page, Hawking and others: the
collective density of such objects must be negligible, for otherwise
their presence would have been obvious in the $\gamma$-ray background.
In fact, we have shown that their bolometric intensity alone is
sufficient to rule them out as important dark-matter candidates.
These constraints may be relaxed if primordial black holes can form
in such a way that they are distributed with larger masses, but it
is not clear that such a distribution can be shown to arise in a
natural way.  As an alternative, we have considered black hole-like
objects in higher-dimensional gravity.  Bolometric arguments do not
rule these out, but there are a number of theoretical issues to be
worked out before a more definitive assessment of their potential
can be made.

\section{Conclusions} \label{ch10}

The intensity of cosmic background radiation constitutes a rich storehouse
of information about the Universe and its contents, both seen and unseen.
At near-optical wavelengths, it guides our understanding of the way in
which galaxies formed and evolved with time.  Integrating over the known
galaxy population, with assumptions about evolution based on Hubble Deep
Field data, one obtains levels of extragalactic background light that
are within an order of magnitude of existing observational upper limits
(and tentative detections), depending on cosmological parameters.
For {\em any\/} realistic combination of parameters, the intensity of this
background light is determined to order of magnitude by the age of
the Universe, which limits the amount of light that galaxies have been
able to produce.  Expansion darkens the night sky further, but only by a
factor of two to three, depending on the details of galaxy evolution
and the ratio of dark matter to dark energy in the Universe.

The new era of ``precision cosmology'' has brought us closer to knowing
just what this ratio is.  The Universe appears to consist of roughly three
parts vacuum-like dark energy and one part pressureless cold dark matter,
with a sprinkling of hot dark matter (neutrinos) that is almost certainly
much less important than cold dark matter.  Baryons --- the stuff of which
we are made --- turn out to be mere trace elements by comparison, truly a
``second Copernican revolution'' in cosmology.
The observations do not tell us what dark energy or dark matter are made of,
nor {\em why\/} these ingredients exist in the ratios they do,
a question that is particularly nagging since their densities evolve so
differently with time.  At present it simply seems that we have stumbled
onto the cosmic stage at an extraordinarily special moment.

At wavelengths other than the optical, the spectrum of background radiation
contains an equally valuable wealth of information on the dark components
of the Universe.  The leading candidates are unstable to radiative decay,
or interact with photons in other ways that give rise to characteristic
signatures in the cosmic background radiation at various wavelengths.
Experimental data on the intensity of this background
therefore tell us what the dark matter and energy can (or cannot) be.
It cannot be dark energy decaying primarily into photons,
because this would lead to levels of microwave background radiation in
excess of those observed.  The dark matter cannot consist of axions
or neutrinos with rest energies in the eV-range,
because these would produce too much infrared, optical or ultraviolet
background light, depending on their lifetimes and coupling parameters.
It {\em could\/} consist of supersymmetric weakly interacting massive
particles (WIMPs) such as neutralinos, but data on the
x-ray and $\gamma$-ray backgrounds imply that these must be very nearly
stable.  The same data exclude a significant role for primordial black
holes, whose Hawking evaporation produces too much light at
$\gamma$-ray wavelengths.  Higher-dimensional analogs of black holes
known as solitons are more difficult to constrain,
but an analysis based on the integrated intensity of the background
radiation at all wavelengths suggests that they could be dark-matter
objects if their masses are not larger than those of galaxies.

While these are the leading candidates, the same methods can be applied
to many others as well.  We mention some of these here without going
into details.  Some of the baryonic dark matter could be bound up
in an early generation of stars with masses in the $100-10^5\Msun$ range.
These objects, sometimes termed {\em very massive objects\/} or VMOs
\cite{Car99b}, are primarily constrained by their contributions to the
infrared background \cite{San02}.  {\em Warm dark-matter\/} (WDM) particles
in the keV rest-energy range, including certain types of gravitinos and
sterile neutrinos, would help to resolve problems in structure formation
and would leave a mark in the x-ray background \cite{Aba01,DiL03}.
Decaying dark-matter particles might have partially reionized the Universe
at high redshifts, helping to explain the unexpectedly large optical depth
inferred from recent CMB observations \cite{Dor03,Han04,Kas04,Pie04}.
In a similar vein, the recent detection of 511~keV $\gamma$-rays from the
Galactic bulge by the {\sc Integral} satellite \cite{Jea03} has been
interpreted as evidence for a population of annihilating \cite{Boe04}
or decaying \cite{Hoo04} light dark-matter particles with rest energies
of $1-100$~MeV.

Numerous proposals have involved superheavy particles along the lines
of {\em ``WIMPzillas''\/} \cite{Kol99}, very massive, non-thermal WIMPs
with rest energies in the $10^{12}-10^{16}$~GeV range whose decays would
be seen at the upper reaches of the diffuse $\gamma$-ray spectrum
\cite{Zia01} and might be responsible for otherwise puzzling observations
of ultrahigh-energy cosmic rays \cite{Ber97}.
Variations on this theme include strongly-interacting WIMPzillas or
``SIMPzillas'' \cite{Alb01}, gluinos \cite{Ber02c} and axinos \cite{Kim02}
(the supersymmetric counterparts of gluons and axions),
leptonic WIMPs or ``LIMPs'' \cite{Bal03},
``superWIMPs'' \cite{Fen03} (superweakly interacting WIMPs whose existence
would only be betrayed by the decays of their parent particles,
the {\em next\/}-to-lightest SUSY particles) and
electromagnetically-coupled or ``EWIMPs'' \cite{His04}.
All these particles would affect primarily the $\gamma$-ray portion
of the EBL spectrum.

High-energy $\gamma$-rays also provide the best hunting-ground for the
dark-matter candidates that arise generically in recent theories
involving more than four spacetime dimensions \cite{OW97}.
In brane-world models \cite{Ark99}, where gravity propagates in
a higher-dimensional bulk while all other fields are restricted to the
four-dimensional brane, the graviton possesses a tower of massive
Kaluza-Klein excitations or {\em Kaluza-Klein gravitons\/} which
carry energy out of supernovae cores before eventually decaying into
photon pairs and other particles.
Radiative decays are particularly conspicuous near $\sim30$~MeV,
and the observed EBL intensity at this energy currently sets the
strongest experimental constraints on brane-world scenarios
with two and three extra dimensions \cite{Hal99,Han01}.
Massive brane fluctuations or ``branons'' are also dark-matter candidates
whose annihilations would show up in the $\gamma$-ray background \cite{Cem03}.
In ``universal-extra-dimensions'' (UED) models, where all fields can
propagate in the bulk, the {\em lightest Kaluza-Klein particle\/} or LKP
(no longer necessarily related to the graviton) becomes a natural dark-matter
candidate \cite{Che02}; such particles have a long history and were originally
known as ``pyrgons'' \cite{Kol84}.  They too turn out to be sharply
constrained by their annihilations into $\gamma$-rays
\cite{Ser03,Ber03b,Ber04}.  Higher-dimensional string and M-theories
imply the existence of other superheavy metastable states (with such
names as ``cryptons,'' ``hexons,'' ``pentons'' and ``tetrons'') which
could also be the dark matter as well as being responsible for
ultrahigh-energy cosmic rays \cite{Ben99,Ell04}.

All these possibilities are particularly interesting since it is quite
likely that the puzzles surrounding dark matter and dark energy will not
be fully understood until they are situated in the context of a fully
unified theory of all the interactions, including gravity.  In our view
such a theory will almost certainly involve extra dimensions.  A rich new
field of possibilities thus opens up for nature's most versatile
dark-matter detector: the light of the night sky.  

\section*{Acknowledgements}

For comments and discussions on dark matter and dark energy over the
years we thank many colleagues including S.~Bowyer, T.~Fukui, W.~Priester,
S.~Seahra and R.~Stabell.

\end{document}